\documentclass[twoside,12pt]{article}
\usepackage{epsfig}

\def\Journal#1#2#3#4{{#1} {#2} (#4) #3 }
\def\EPJA{{\em Eur. Phys. J.} A}

\def\NPA{{\em Nucl. Phys.} A}

\def\JPG{{\em J. Phys.} G}

\def\NPB{{\em Nucl. Phys.} B}

\def\PLB{{\em Phys. Lett.} B}

\def\PRL{\em Phys. Rev. Lett.}

\def\PREP{\em Phys. Rep.}

\def\PRD{{\em Phys. Rev.} D}
\def\PRC{{\em Phys. Rev.} C}

\def\PPNP{\em Prog. Part. Nucl. Phys.}

\def\ZPA{{\em Z. Phys.} A}
\def\ANNP{\em Ann. Phys. (N.Y.)}
\def\RMP{{\em Rev. Mod. Phys.}}

\def\INT{{\em Int. J. Mod. Phys.} E}
\def\CPC{{\em Chin. Phys.} C}
\def\JPCS{{\em J. Phys. Conf. Series}}
\def\RPP{{Rep. Prog. Phys.}}
\def\EPJWC{{\em EPJ Web Conf}}
\def\HI{{\em Hyp. Int.}}
\def\PTP{{\em Prog. Theo. Phys.}}
\def\PNL{{\em Part. Nucl. Lett.}}
\def\AIPCP{{\em AIP Conf. Proc}}
\def\JI{{\em Journal of Instrumentation}}
\def\IJMPE{{\em Int. J. of Mod. Phys.} E}
\def\PTEP{{\em Prog. Theo. Exp. Phys.}}
\def\JETPL{{\em Journal Exp. Theo. Phys. Lett.}} 
\def\APP{{\em Acta Phys. Pol.} B}
\def\APH{{\em Acta Phys. Hung.} A}
\def\ASTPHYS{{\em Astrophys. J.}}
\def\RMP{{\em Rev. Mod. Phys.}}
\def\PTPS{{\em Prog. Theo. Phys. Supp.}}
\def\r{\vec r}

\def\p{\vec p}

\newcommand{\be}{\begin{equation}}
\newcommand{\ee}{\end{equation}}
\newcommand{\bea}{\begin{eqnarray}}
\newcommand{\eea}{\end{eqnarray}}

\begin{document}

\title{ \vspace{1cm} Meson--nucleus potentials and the search for meson-nucleus bound states}
\author{V.\ Metag,$^1$ M.\ Nanova,$^1$, and  E. Ya.\ Paryev,$^{2,3}$\\
\\
$^1$ II. Physics Institute, University of Giessen, Giessen 35392, Germany\\
$^2$ Institute for Nuclear Research, Russian Academy of Sciences, Moscow 117312, Russia\\
$^3$ Institute for Theoretical and Experimental Physics, Moscow 117218, Russia}
\maketitle
\begin{abstract}
Recent experiments studying the meson-nucleus interaction to extract meson-nucleus potentials are reviewed. The real part of the potentials quantifies whether the interaction is attractive or repulsive while the imaginary part describes the meson absorption in nuclei. The review is focused on mesons which are sufficiently long-lived to potentially form meson-nucleus quasi-bound states. The presentation is confined to meson production off nuclei in photon-, pion-, proton-, and light-ion induced reactions and heavy-ion collisions at energies near the production threshold. Tools to extract the potential parameters are presented. In most cases, the real part of the potential is determined by comparing measured meson momentum distributions or excitation functions with collision model or transport model calculations. The imaginary part is extracted from transparency ratio measurements.
Results on $K^+, K^0, K^-, \eta, \eta^\prime, \omega$, and $\phi$ mesons are presented and compared with theoretical predictions. The interaction of $K^+$ and $K^0$ mesons with nuclei is found to be weakly repulsive, while the $K^-, \eta,\eta^\prime, \omega$ and $\phi$ meson-nucleus potentials are attractive, however, with widely different strengths. Because of meson absorption in the nuclear medium the imaginary parts of the meson-nucleus potentials are all negative, again with a large spread. An outlook on planned experiments in the charm sector is given. In view of the determined potential parameters, the criteria and chances for experimentally observing meson-nucleus quasi-bound states are discussed. The most promising candidates appear to be the $\eta$ and $\eta^\prime$ mesons. 
\end{abstract}
%\eject
\tableofcontents
\section{Introduction}
\subsection{\it Scope of the review}
The interaction of mesons with nuclei and the possible existence of meson--nucleus bound states have recently been in the focus of numerous theoretical and experimental studies.
This review attempts to summarise these activities and to describe the current status of the field.

Mesonic {\em atoms} where an electron is replaced by a $\pi^-$ or $K^-$ meson are well known and have been studied in high precision experiments. Information on the real and imaginary part of the meson-nucleus potential has been extracted from fitting density dependent potentials to comprehensive sets of data of strong-interaction level shifts, widths and yields across the periodic table, including more recent results on the binding energy and width of deeply bound pionic states [1-4].
%\cite{Gilg_PRC62,Itahashi_PRC62,Geissel_PRL88,Suzuki_PRL92}. 
For reviews and recent results see [5-11].
%\cite{Yamazaki_PR,Hirenzaki_PRC61,Friedman_Gal_PR,Friedman_Gal_NPA881,Friedman_Gal_NPA899,Gal_NPA914,Friedman_Gal_NPA959}. 
The dominant interaction in these cases is the Coulomb interaction while the effect of the strong interaction is only a correction. These experiments probe the meson-nucleus potentials near the nuclear surface and thus preferentially at relatively small nuclear densities.

The main focus of this review is on {\em nuclear} exotic states and thus on the interaction of nuclei with neutral mesons where only the strong interaction is relevant or where - in the case of charged mesons - the strong interaction is dominant. Experimental studies of this type have applied different projectiles for the meson production process ranging from photons, pions, protons, light ions to heavy-ions. The experiments have been performed over a broad range of energies from reactions near the production threshold up to ultra-relativistic heavy-ion collisions (see [12-14]).
%\cite{Hayano_Hatsuda,Rapp_Wambach,LMM}). 
Here, we will concentrate on energies close to or slightly above the production threshold, the regime relevant for the possible formation of meson-nucleus bound states, since only slow mesons can be captured by the nucleus provided there is sufficient attraction. Furthermore, we concentrate on mesons sufficiently longlived to form a meson--nucleus quasi-bound state. These are the pseudoscalar mesons $K^+, K^0, K^-, \eta, \eta^\prime$ and the vector mesons $\omega$ and $\phi$ for which parameters of the meson-nucleus potential have been extracted in numerous experiments. The enormous theoretical and experimental work on the $\rho$ meson has been reviewed in [12-14]
%\cite{Hayano_Hatsuda,Rapp_Wambach,LMM} 
and will not be discussed here further since the $\rho$ meson is too short-lived to form mesic states.

Extensions of  these studies to the charm sector have been considered in a number of theoretical papers but represent a real challenge experimentally. The heavier the meson, the larger is the momentum transfer in the production process and the more difficult it is to produce mesons sufficiently slow to be captured by the nucleus. The possibilities and challenges regarding nuclear bound states involving heavy flavour hadrons have been outlined in a very recent review by Krein et al.\cite{Krein_1706.02688}.

\subsection{\it Motivation}\label{sec:motivation}
Why is it interesting to study the interaction of mesons with nuclei? This interaction is an important testing ground for our understanding of Quantum Chromodynamics (QCD) as the theory of the strong interaction in the non-perturbative regime. Research in this field was motivated in particular by theoretical predictions that meson properties might change within nuclei due to a partial restoration of chiral symmetry [16-19].
%\cite{Bernard_Meissner,Meissner,Brown_Rho,Hatsuda_Lee}. 
Mesons are considered to be excitations of the QCD vacuum which has a complicated structure with non-vanishing chiral-, gluon- and higher order quark-condensates. These condensates are predicted to change within a strongly interacting medium and, as a consequence, also the excitation energy spectrum, i.e. the mass spectrum of mesons is expected to be modified. This idea fostered widespread theoretical and experimental activities which have been summarised in recent reviews [12-14].
%\cite{Hayano_Hatsuda,Rapp_Wambach,LMM}.

\begin{figure}[h]
\centering
\includegraphics[width=12cm,clip]{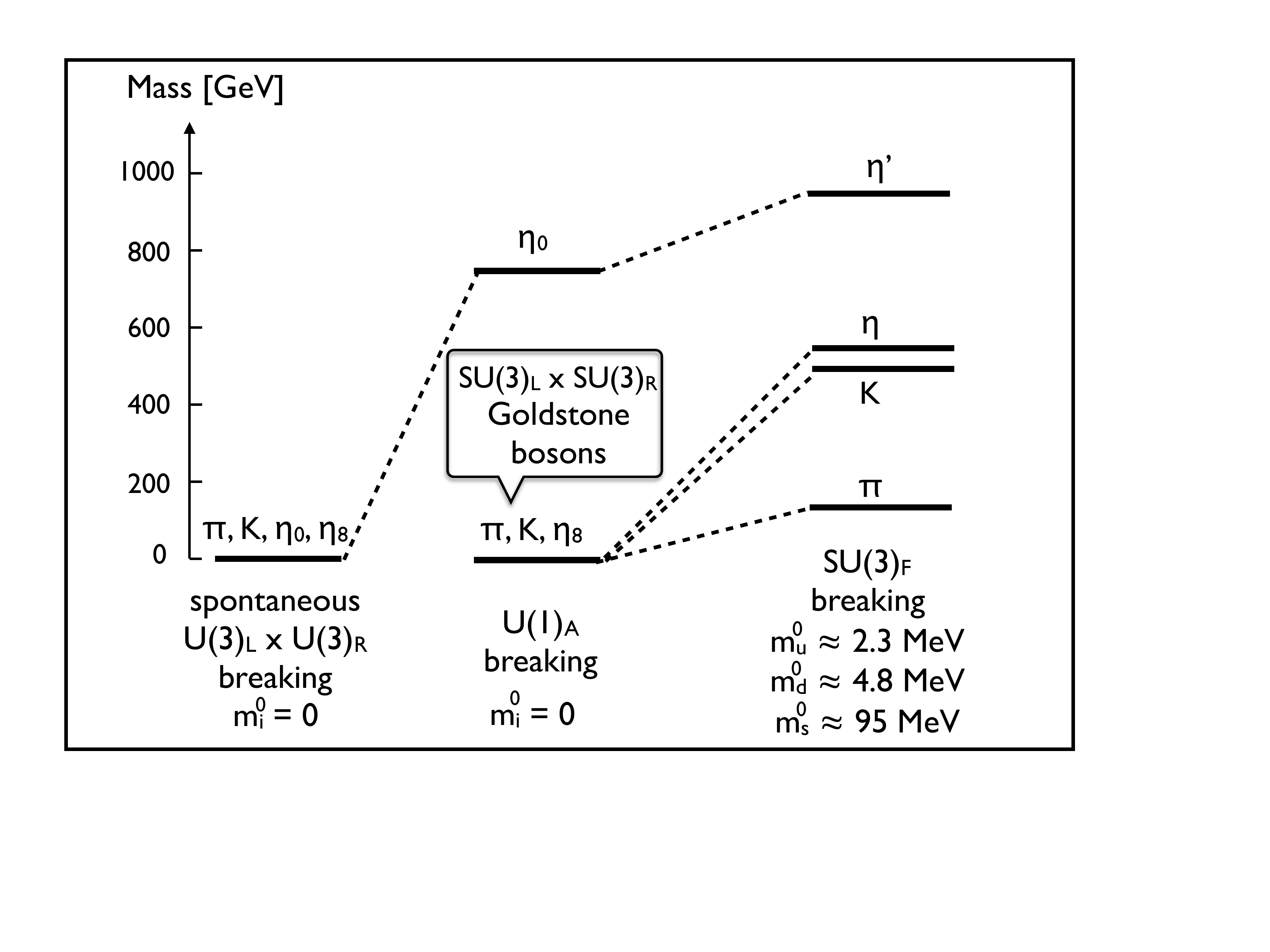}
\vspace*{-1.5cm}
\caption{Symmetry breaking pattern for pseudoscalar mesons as described in the text, adapted from~\cite{Klimt}.}
\label{fig:pseudoscalar-mesons}      
\end{figure}

Pseudoscalar mesons are particularly suited for studying in-medium modifications, as shown in Fig.~\ref{fig:pseudoscalar-mesons} \cite{Klimt}. Spontaneous chiral symmetry breaking generates a pseudoscalar nonet ($\pi, K,\bar{K},\eta,\eta^\prime$) of massless Nambu-Goldstone bosons. The explicit breaking of the U(1)$_A$ symmetry selectively shifts up the singlet $\eta_0$ mass, leaving the SU(3) flavour octet of pions, kaons and $\eta_8$ massless. Explicit chiral symmetry breaking by introducing non-zero quark masses then leads to the experimentally observed meson masses \cite{Bernard_Meissner,Meissner,Klimt}. Since symmetry breaking has such a big effect on meson masses one would expect corresponding effects in case of a partial restoration of this symmetry in a strongly interacting medium. These effects may have implications going beyond hadron- and nuclear physics. In particular, an effective reduction of the $K^-$ mass in a nuclear medium might indicate the onset of $K^-$ condensation in dense nuclear matter as e.g., in the interior of neutron stars, as proposed in \cite{Kaplan_Nelson,Brown_Bethe}. 

If spontaneously broken chiral symmetry were restored at high nuclear densities one would expect the full spectral function of vector mesons with spin parity $J^{\pi}$ = 1$^-$ to become degenerate with that of their chiral partners with the same spin but opposite parity, the axial vector mesons with $J^{\pi}$ = 1$^+$. Traces of this tendency should already show up at normal nuclear matter density and should thus become observable in photonuclear experiments and reactions induced by pions and protons.

For long-lived mesons such as pions or kaons, meson beams can be used to study the meson--nucleus interaction experimentally. This is not possible for short-lived mesons like
$\eta, \eta^\prime$ and $\omega,\phi$ mesons where such beams are not available. Here, one has to produce the mesons in a nuclear reaction and to study their interaction with nucleons or nuclei in the final-state.
%_______________________________________________________________________________________________________________
%Meson-Nucleus Potentials

\section{The meson--nucleus interaction\label{sec:optical}}

The interaction of mesons with a nuclear medium of density $\rho_N$ is usually described by a dispersion relation based on the Klein-Gordon equation (see e.g. \cite{Friedman_Gal_PR})
\begin{equation}
E^{{\prime}2} - {\bf p}^{{\prime}2} - m^2 -\Pi(E^{\prime},{\bf p}^{\prime},\rho_N(r)) = 0, \label{eq:dispersion}
\end{equation}
where $\Pi(E^{\prime},{\bf p}^{\prime},\rho_N(r))$ is the meson self-energy which summarizes all the meson interactions in the medium. Here $m, {\bf p}^{\prime}, E^{\prime} $ and $r$ are the free meson rest mass, its 3-momentum, its energy in the medium and its distance to the center of the nucleus. To leading order in density $\rho_N(r)$, 
\begin{equation}
\Pi(E^{\prime},{\bf p}^{\prime},\rho_N(r)) = -4 \pi f(E^{\prime},{\bf p}^{\prime})
\end{equation}
where  $ f(E^{\prime},{\bf p}^{\prime})$ is the (in medium generally off-shell) meson-nucleon forward scattering amplitude which may involve scalar and/or vector t-channel terms.

A meson-nucleus potential may be defined through the self-energy as
\begin{equation}
U(E^{\prime},{\bf p}^{\prime},\rho_N(r)) = \frac{\Pi(E^{\prime},{\bf p}^{\prime},\rho_N(r))}{2 \sqrt{{\bf p}^{{\prime}2}+m^2}}.\label{eq:U-E}
\end{equation}
This potential is complex:
\begin{equation}
U(r) = V(r) + iW(r) \label{eq:potential_effective}
\end{equation}
The real part $V(r)$ encodes whether the interaction of the meson with the nucleus is attractive or repulsive, while the imaginary part $W(r)$ of the potential accounts for the absorption of the meson in the medium through inelastic reactions.

Combining Eqs.~(\ref{eq:U-E}),(\ref{eq:potential_effective})  one obtains
\begin{equation}
U(r) = \frac{Re\Pi(E^{\prime},{\bf p}^{\prime},\rho_N(r))}{2\sqrt{{\bf p}^{{\prime}2}+m^2}} + i\frac{Im\Pi(E^{\prime},{\bf p}^{\prime},\rho_N(r))}{2\sqrt{{\bf p}^{{\prime}2}+m^2}}.\label{eq:U-Pi}
\end{equation}

In this review the focus is on slow mesons with momenta small compared to their mass: $|{\bf p}^{\prime}| \le m $, i.e. mesons that are so slow that they may even be captured by the nucleus in case of an attractive interaction; thus in the region of main interest Eq.~\ref{eq:U-Pi} simplifies to 
\begin{equation}
V(r) = \frac{Re\Pi(E^{\prime},{\bf p}^{\prime},\rho_N(r))}{2m},  \quad W(r) = \frac{Im\Pi(E^{\prime},{\bf p}^{\prime},\rho_N(r))}{2m}.
\end{equation}

In this approximation and assuming that the in-medium meson width is small compared to its in-medium mass \cite{Waas} (which turns out to be justified for all mesons considered in this review), the real part of Eq.~\ref{eq:dispersion} takes the form
\begin{equation}
E^{{\prime}2} - {\bf p}^{{\prime}2} - m^2 - 2mV(r)= 0 \label{eq:dispersion_real}
\end{equation}

Introducing an effective in-medium mass 
\begin{equation}
m^* = (m + \Delta m(\rho_N))\label{eq:m*},
\end{equation}
where $\Delta m$ is the density dependent mass modification of the meson due to the interaction with the nuclear medium, the real potential V(r) can be related to $\Delta m$. From Eq. \ref{eq:dispersion} one obtains
\begin{equation}
E^{{\prime}^2} - {\bf p}^{{\prime}2} - m^{*2} = E^{{\prime}2} - {\bf p}^{{\prime}2}- (m+\Delta m(\rho_N))^2 = E^{{\prime}2} - {\bf p}^{{\prime}2}- (m^2 + 2m \Delta m(\rho_N)).\label{eq:Deltam-V}
\end{equation}
Here it has been assumed that the mass modification $\Delta m$ is small compared to the meson mass, so that terms quadratic in $\Delta m$ can be neglected \cite{Nagahiro_Hirenzaki}.

Comparing Eqs.~(\ref{eq:dispersion_real}),(\ref{eq:Deltam-V}), the real part $V(r)$ of the meson-nucleus potential is found to be equal to the in-medium meson mass shift $\Delta m (\rho_N(r))$. Assuming a linear density dependence one gets
\begin{equation}
V(r) = \Delta m(\rho_N(r)) = \Delta m (\rho_N=\rho_0) \cdot \rho_N(r)/\rho_0 = V_0 \cdot \rho_N(r)/\rho_0 \label{eq:V=Dm}.
\end{equation}

Correspondingly, the imaginary part W(r) of the potential and the in-medium meson width in the nuclear rest frame can be derived from the imaginary part of the self-energy
\cite{Waas,Hernandez_Oset,Cabrera_NPA733}
\begin{equation}
W(r) = - \frac{1}{2} \Gamma(E^{\prime},{\bf p}^{\prime},\rho_N(r)) =  \frac{Im\Pi(E^{\prime},{\bf p}^{\prime},\rho_N(r))}{2m}.\label{eq:energy_self2}
\end{equation}

Assuming again a linear density dependence $\Gamma(\rho_N) = \Gamma(\rho_N=\rho_0) \cdot \rho_N(r)/\rho_0 = \Gamma_0 \cdot \rho_N(r)/\rho_0 $ one obtains
\begin{equation}
W(r) =  -\frac{1}{2}\Gamma_{0}\cdot \frac{\rho_N(r)}{\rho_{0}}.\label{eq:W}
\end{equation}

The quantities $V_0$ and $\Gamma_0$ are determined from the data as averages over a momentum range, typically $0 \le |{\bf p}^{\prime}| \le m$.

In this review we discuss the determination of the real and imaginary parts of the meson--nucleus potential for pseudo-scalar mesons $K^+, K^0, K^-, \eta, \eta^\prime$ and the vector mesons $\omega$ and $\phi$.

%_________________________________________________________________________________________
\section{Methods for determining the meson--nucleus potential}

Most of the theoretical predictions have been calculated under the idealised assumption of complete equilibrium e.g. for a meson at rest embedded in infinitely extended nuclear matter with constant density and temperature, a scenario far from reality. In meson production experiments in the 1-3 GeV energy range, the kinematics of the reaction leads to meson recoil momenta on average comparable to their mass; on the way out of the nucleus mesons see the nuclear density profile with a fall-off at the surface. Any density dependent mass shift or broadening is thereby smeared out due to the variation in density. A link between the theoretical predictions and the experimental observables is provided by either transport calculations like, e.g. Giessen-Boltzmann-Uehling-Uhlenbeck (GiBUU) \cite{GiBUU}, Coupled-Channel-Boltzmann-Uehling-Uhlenbeck (CBUU) \cite{CBUU}, Relativistic Quantum Molecular Dynamics (RQMD) \cite{RQMD_Stoecker,RQMD_Lehmann,RQMD_Fuchs}, Relativistic-Boltzmann-Uehling-Uhlenbeck (RBUU) \cite{Maruyama_NPA573}, Isospin Quantum Molecular Dynamics (IQMD) \cite{IQMD}, Hadron String Dynamics (HSD) \cite{HSD} or calculations within a collision model based on nuclear spectral functions \cite{Paryev_JPG2013}, and Monte Carlo simulations \cite{Salcedo_NPA484,Carrasco_NPA570}. For a meaningful comparison of theoretical predictions with experimental data to extract in-medium properties of hadrons, it is indispensable to investigate with such calculations how non-equilibrium effects or reaction dynamics modify the theoretically predicted initial signals. These calculations are briefly described in the following subsections.

\subsection{\it Transport simulations\label{sec:transport}}
Transport calculations simulate photon-, electron-, neutrino-, hadron- and heavy-ion-induced reactions on nuclei. The relevant degrees of freedom are mesons and baryons which propagate in mean fields and scatter according to cross sections which are provided as external input. The simulations are based on a set of semi-classical kinetic equations which explicitly describe the dynamics of a hadronic system in space and time, i.e. the hadronic degrees of freedom in nuclear reactions are taken into account, including the propagation, elastic and inelastic collisions and decays of particles. In addition to nucleons, all established nucleon resonance as well as light pseudoscalar and vector mesons are taken into account in the propagation through the nuclear medium. Here we restrict ourselves to a brief  presentation of the basic ideas and confine ourselves for simplicity to non-relativistic transport, however, including relativistic kinematics. Details can be found in the original literature cited above. As an example the concept of GiBUU model calculations \cite{Effenberger} will be discussed.

The transport equations describe the classical time evolution of a system with N particle species under the influence of a self-consistent mean-field potential and a collision term. For each particle species $i$ involved one obtains a differential equation of the type:
\begin{equation}
(\frac {\partial}{\partial t} + \frac{\partial  H_i}{\partial \vec{p}}\frac{\partial}{\partial \vec{r}} - \frac{\partial H_i}{\partial \vec{r}}\frac{\partial}{\partial \vec{p}})F_i(\vec{r},\vec{p},W,t) = G_i S_i(1\pm f_i) -L_iF_i\label{eq:BUU}\,\,\,\, (i=1,.....,N),
\end{equation}
where $\vec{r}$ and $\vec{p}$ are the spatial and momentum coordinates of the particle $i$. $F_i(\vec{r},\vec{p},W,t)$ denotes the so-called spectral phase-space density of particle species $i$ which is the product of the ordinary quasiparticle phase-space density $f_i$ and the spectral function $S_i$ \cite{GiBUU}:
\begin{equation}
F_i(\vec{r},\vec{p},W,t) = f_i(\vec{r},\vec{p},W,t)S_i(W,\vec{r},\vec{p})
\end{equation}
The spectral function $S_i$ gives directly the probability to find a particle of species $i$ and pole mass $m_i$ with invariant mass W and momentum $\vec{p}$ at position $\vec{r}$:
\begin{equation}
S_i(W,\vec{r},\vec{p}) = \frac{2}{\pi}\frac{W^2 \Gamma_{\rm tot}(W,\vec{r},\vec{p})}{(W^2-m_i^2)^2 + W^2\Gamma_{\rm tot}^2(W,\vec{r},\vec{p})}.
\end{equation}
Here, $\Gamma_{\rm tot}$ is the total width of a particle of species $i$ in its rest frame.
The Hamilton function H$_i(\vec{r},\vec{p},W,f_1,f_2,..f_i,..f_N$) is given by the relativistic expression for the single particle energy:
\begin{equation}
H_i(\vec{r},\vec{p},W,f_1,...f_i,...f_N) = \sqrt{(W + U_i(\vec{r},\vec{p},W,f_1,...f_i,...f_N))^2 + \vec{p}^2},
\end{equation}
where $U_i$ is the coordinate- and momentum-dependent effective scalar potential seen by particle $i$. Since all particle species mutually interact the integro-differential equations are coupled by the collision integrals which describe the gain $G_i$ and loss $L_i$ of each particle species in inelastic collisions. These equations are solved numerically by the so-called test-particle method.

\subsubsection{\it Off-shell transport}
Of particular importance for the study of in-medium properties of hadrons is the implementation of an off-shell transport, describing the propagation of quasi-particles with broad mass spectra, taking the density dependence of the spectral function into account. As the particle propagates through regions of varying density the spectral function changes accordingly, implying changes in the mass distribution. In particular, it has to be ensured that particles leaving the nucleus return to their vacuum spectral function, i.e. stable particles need to come back to their vacuum mass listed in \cite{PDG} even though they had a broad (and shifted) mass distribution in the medium. A practical way to simulate off-shell effects is to define a scalar potential as
\begin{equation}
\tilde{U}_i(\vec{r_i}) = (W_i^* -W_i^0)\frac{\rho_N(\vec{r_i})}{\rho_N(\vec{r_i}^{\rm cr})},
\end{equation}
where $\rho_N(\vec{r_i})$ is the local nucleon density and the asterisk refers to the mass determined at the instant of the test-particle production at the point with spatial coordinate $\vec{r_i}^{\rm cr}$. $W_i^0$ is a mass chosen randomly according to the vacuum spectral function. This potential enters the single-particle Hamiltonian as a conventional scalar potential, thus guaranteeing energy conservation. The effective in-medium mass of the test-particle during its propagation through the nucleus is then given as
\begin{equation}
W_i = W_i^0 + \tilde{U}_i(\vec{r}),
\end{equation}
which yields the correct asymptotic value of the test-particle mass since the potential vanishes outside of the nucleus. Thus, the mass spectrum of a physical particle, represented by a bunch of test-particles, takes the form of the vacuum spectral function when all test-particles have escaped from the nuclear environment. For a more rigoruos formulation of relativistic off-shell transport the reader is referred to \cite{Cassing_Juchem,Juchem_Cassing_Greiner,Leupold_NPA672}.

\subsubsection{\it The collision term}
The so-called collision term in the right hand side of Eq.~\ref{eq:BUU} influences the time evolution of the particle phase-space density via the creation and destruction of particles in scattering and decay processes. For simplicity we restrict us here to the quasiparticle limit, i.e. $S_i  \rightarrow 1$ and $F_i \rightarrow f_i$. The right hand side of  Eq.~\ref{eq:BUU} then reads
\begin{equation}
G_i(1\pm f_i) - L_i f_i,
\end{equation}
where $(1 \pm f_i)$ is a Pauli blocking or Bose enhancement factor for particle $i$ being a fermion or a boson, respectively. In this simplified case - the complete formulae for the general case are given in \cite{GiBUU} - the loss term $L_i$ can be written as
\begin{equation}
L_i(\r,\p,t) = \frac{1}{(2\pi)^3}\sum_{j}\sum_{ab}\int d^3p_j \int d\Omega_{cm}\frac{d\sigma_{ij\rightarrow ab}}{d\Omega_{cm}}v_{ij}f_j(\r,\p_j,t) P_a P_b + \sum_{cd} \int d\Omega_{cm}
\frac{d\Gamma_{i \rightarrow cd}}{d\Omega_{cm}} P_c P_d,
\end{equation}
where $d\sigma_{ij \rightarrow ab}/d\Omega_{cm}$ and $d\Gamma_{i \rightarrow cd}/d\Omega_{cm}$ are the angle differential cross sections and decay widths of the processes
$ij \rightarrow ab$  and $i \rightarrow cd$, respectively. The terms  $P_x = 1-f_x(\r,\p_x,t)$ with $X=a,b,c,d$ introduce Pauli blocking of final-states in the case where $a,b,c,d$ are fermions or $P_x =1$ otherwise; $v_{ij}$ is the relative velocity of the collision partners $i$ and $j$. The final  states $ab$ or $cd$ and the collision partners $j$ are summed over in order to obtain the total loss rate $L_i$. The expression for the particle gain $G_i$ has a similar form.

These ingredients are characteristic for any transport simulation although there are differences in the implementation, depending on the optimisation for certain classes of reactions.

Codes focusing on high-energy reaction include sub-nucleonic degrees of freedom to account for the phenomena like color transparency, hadron formation time or the transition to a quark-gluon plasma in ultra-relativistic heavy-ion reactions. For a detailed description of the different transport codes the reader is referred to the original literature [27-31,33,34,43-46].
%\cite{GiBUU,CBUU,RQMD_Stoecker,RQMD_Lehmann,RQMD_Fuchs,IQMD,HSD,Linnyk_Bratkovskaya_Cassing,Nekipelov,Iljinov,Scheinast}.

\subsubsection{\it Comparison to experiment}\label{sec:compexp}
For a comparison to experimental data it is important that transport calculations take nuclear many-body effects into account which evolve dynamically in the course of the reaction. In 
particular they treat
\vspace{0.2cm}
\\
- initial state effects: the absorption of incoming  beam particles\\
- non equilibrium effects: varying nucleon density and temperature\\
- absorption and regeneration of mesons\\
- fraction of decays outside of the nucleus\\
- final-state interactions: distortion of momenta of decay products.
\vspace{0.2cm}
\\As a result, a full space-time evolution of the many-particle system is obtained starting from the initial state of the reaction to the final-state particles as they are observed in the experiment.

As an example for the importance of transport calculations in the comparison of theoretical predictions with experimental observables we discuss the line shape of the $\omega$ meson reconstructed from the $\omega \rightarrow \pi^0 \gamma$ decay in photoproduction off nuclei. As presented in Section \ref{sec:omega}, a reduction of the $\omega$ mass at normal nuclear matter density by 120 MeV has been predicted in QCD sum rule calculations \cite{Hatsuda_Lee}. Within the Nambu-Jona-Lasinio model a similar mass shift of -(100-150) MeV has been claimed together with a broadening by $\approx$ 40 MeV \cite{Klingl}. The latter calculation has, however, been criticized in \cite{Eichstaedt_PTPS168}. In contrast, no mass shift but a broadening by 60 MeV has been obtained in a resonance coupling model \cite{Muehlich_NPA780}.

\begin{figure}[tb]
\centering
\includegraphics[width=14cm,clip]{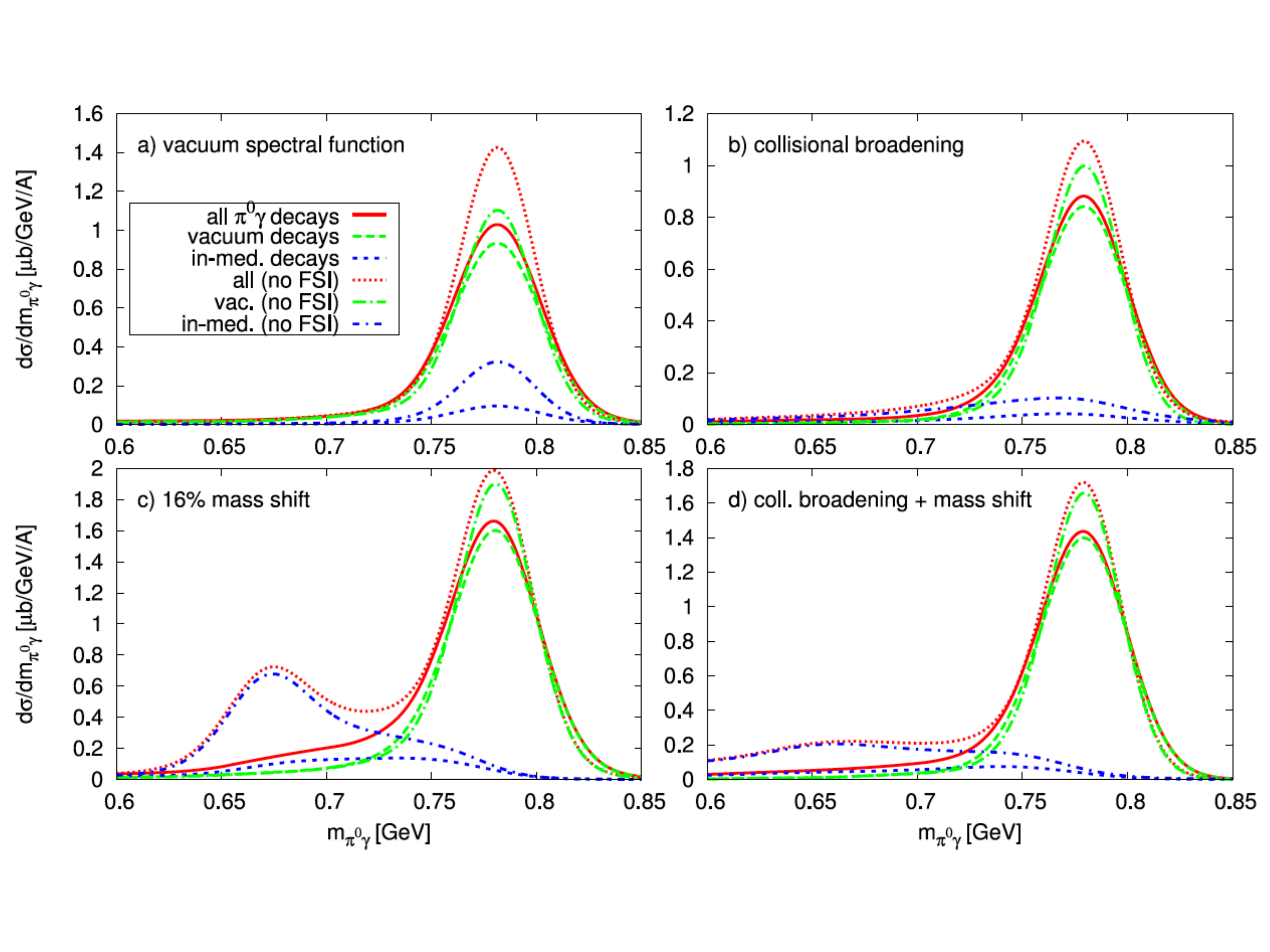}
\caption{$\pi^0 \gamma$ invariant mass spectra from $\omega$ decays for $\gamma$ + Nb reactions at incident photon energies of 0.9--1.3 GeV calculated with the GiBUU code \cite{GiBUU} for four different in-medium modification scenarios: a.) no medium modification (vacuum spectral function), b.) collisional broadening by 150 MeV \cite{Kotulla}, c.) mass shift by -16$\%$, d.) collisional broadening (150 MeV) and mass shift (-16$\%$). The total spectrum is shown as well as contributions from in-medium ($\rho_N > 0.1 \rho_0$) and vacuum ($\rho_N < 0.1 \rho_0$) decays into the $\pi^0 \gamma$ channel, with and without pion final-state interaction (FSI). The calculated curves are folded with the experimental resolution of the CBELSA/TAPS detector of $\sigma$ = 19 MeV. The figure is taken from \cite{Weil}. }
\label{fig:omega_pi^0g}       
\end{figure}

Applying the GiBUU transport code, Weil et al. \cite{Weil} have studied to what extent these predicted in-medium modifications can be observed under realistic experimental conditions in a photon-induced nuclear reaction. Fig.~\ref{fig:omega_pi^0g} a.) shows that in case of no in-medium modifications only about 20-30$\%$ of all $\omega \rightarrow \pi^0 \gamma $ decays occur in the nuclear medium
for a photoproduction experiment on a Nb target near the production threshold. Due to the reaction kinematics, the $\omega$ mesons recoil from the nucleus with average momenta comparable to their mass and the decay length $ l_{\rm dec} = \beta \gamma c \tau = \frac{p}{m} c \tau \approx c \tau $ =22 fm thus becomes larger than nuclear dimensions. Assuming a collisional broadening by 150 MeV \cite{Kotulla} the $\pi^0 \gamma $ mass distribution for in-medium decays is smeared out, leading to the $\omega$ line shape given in 
Fig.~\ref{fig:omega_pi^0g} b.). Fig.~\ref{fig:omega_pi^0g} c.) shows the $\omega$ signal to be expected without collisional broadening but for a mass shift by -16$\%$ at normal nuclear matter density and Fig.~\ref{fig:omega_pi^0g} d.) displays the expected line shape for collisional broadening and mass shift.

It should be noted that only a fraction of the in-medium decays occur in the center of the nucleus and that the invariant mass reconstructed from the 4-momenta of the decay products is sensitive to the nuclear density at the decay point. Every nucleus has a diffuse surface with a slowly dropping density distribution. Assuming a density-dependent mass shift, $\omega$ meson decays at the nuclear surface will further smear out the in-medium peak. While these distortions of the originally predicted theoretical signal apply for all $\omega$ decay modes, also for e.g. the $\omega\rightarrow e^+e^- $ decays, the $\omega \rightarrow \pi^0 \gamma $ channel has an additional drawback. The 4-momentum vectors of the $\pi^0$ and thus also the $\pi^0 \gamma$ invariant mass will be distorted by elastic scattering or the $\pi^0$ may even be absorbed within the nucleus due to the strong interaction. Information on decays near the center of the nucleus will thus not reach the detector. Only decays near the nuclear surface  will be registered where the mass shifts are smaller. The sensitivity of the meson line shape to in-medium modifications is thus reduced as evident from Fig.~\ref{fig:omega_pi^0g} d.). This example shows how crucial it is to check whether theoretically predicted signals can really be observed under realistic experimental conditions.

In parallel to these rather involved and ambitious transport simulations simplified descriptions of meson production reactions have been developed which describe the main features of these reactions but do not take into account density dependent changes of the meson spectral function; i.e. the off-shell transport, discussed above, is replaced by a simpler treatment, where instead of the local effective meson mass their average in-medium mass is used. These are the so-called collision model calculations which are described in the next session.

%-------------------------------------------------------------------------------------------
%Collision Model

\subsection{\it Collision model based on the nuclear spectral function}\label{sec:collmod}

Extensive calculations of the production of $K^+$, $K^-$, $\eta$, $\omega$ mesons and
antiprotons in proton--, pion--, and photon--nucleus reactions at incident energies near or below the free
nucleon--nucleon thresholds have been performed in the framework of folding models [52-64].
% \cite{Shor,Cassing_PLB1990,Sib_Buscher,Cassing_ZPA1994,Sib_PLB1995,Sib_ZPA1995,Akindinov_JPG,Golubeva,
%Efremov_1995,Efremov_1996,Krippa,Kiselev_nuclth,Akindinov_JETP}.
These investigations are based on both the direct
and two-step particle production mechanisms, using different paramterisations for internal
nucleon momentum distributions and for the free elementary cross sections. In these folding models only the
internal nucleon momentum distribution has been used and the off-shell propagation of the struck target
nucleon has been neglected or only crudely taken into account, but it could be significant in
processes that are limited by phase-space such as near-threshold heavy meson production. As is well known [65-70]
%\cite{Ciofi_PLB1989,Ciofi_PRC1996,Ciofi_PRC1990,Ciofi_PRC1991,Gil_NPA627,Benhar_RMP80},
the off-shell behavior of a bound nucleon is described by the nucleon (nuclear) spectral function
$P_A({\bf p}_t,E)$, which represents the probability to find in the nucleus a nucleon with momentum
${\bf p}_t$ and removal (binding) energy $E$ and thus contains all the information on the structure of a target
nucleus. The knowledge of the spectral function $P_A({\bf p}_t,E)$ is needed for calculations of cross sections of various kinds of nuclear reactions. In particular, it has been widely adopted for the analysis
of inclusive and exclusive quasielastic electron scattering by nuclei [71-73].
%\cite{Benhar_NPA1994,Weinstein_PRC1994,Sick_PLB1994}. 
It was found that even
at very high momentum and energy transfer in the scattering process, the target nucleus cannot be simply described
as a collection of $A$ on-shell nucleons subject only to Fermi motion, but the full nucleon momentum and
binding energy distribution has to be considered. The one- and two-step folding model, based on the nucleon
spectral function, has been developed in [35,74-87] 
%\cite{Paryev_JPG2013,Efremov_1998,Paryev_EPJA1999,Paryev_EPJA72000,
%Paryev_EPJA92000,Paryev_EPJA2003,Paryev_EPJA2005,Paryev_JPG36,Paryev_JPG2010,Paryev_PAN2012,Paryev_JPG2016,Kiselev_IJMPE,Paryev_YF2017,Paryev_K-,Paryev_nuclth2016}
to analyze the kaon, $\phi$, ${\eta^\prime}$, $J/\psi$
meson as well as $\Lambda(1115)$ and $\Lambda(1520)$ hyperon production in proton-- and photon--nucleus
collisions in the near-threshold energy regime with the aim of obtaining information about
in-medium hadron properties. The main ingredients of this model are illustrated by discussing as an example the inclusive
production of ${\eta^\prime}$ mesons off nuclei in photon-induced reactions. A detailed comparison of the calculated ${\eta^\prime}$ meson production cross sections with the corresponding experimental data will be given in Sec.\ref{sec:etaprime} of this review.

%%%%%%%%%%%%%%%%%%%%%%%%%%%%%%%%%%%%%%
\subsubsection{\it Near-threshold  ${\eta^\prime}$ meson photoproduction off nuclei}
\subsubsection{\it One-step  ${\eta^\prime}$ production mechanism}

In the incident photon energy range up to 2.6 GeV, the following direct
elementary processes, which have the lowest free production threshold ($\approx$ 1.446 GeV), contribute
to ${\eta^\prime}$ photoproduction on nuclei:

\begin{equation}
\gamma+p \to \eta^\prime+p, \label{eq:primary_1}
\end{equation}

\begin{equation}
\gamma+n \to \eta^\prime+n. \label{eq:primary_2}
\end{equation}
The medium modification of the ${\eta^\prime}$ mesons, participating in the production processes (\ref{eq:primary_1}),(\ref{eq:primary_2}), is accounted for by using  for simplicity  their average in-medium mass $<m^*_{{\eta^\prime}}>$ defined as:

\begin{equation}
<m^*_{{\eta^\prime}}>=\int d^3r{\rho_N({\bf r})}m^*_{{\eta^\prime}}({\bf r})/A,\label{eq:m_etap_av}
\end{equation}
where ${\rho_N({\bf r})}$ and $m^*_{{\eta^\prime}}({\bf r})$ are the local nucleon density and
${\eta^\prime}$ effective mass inside the nucleus, respectively. Assuming in line with
\cite{Paryev_JPG2013} that
\begin{equation}
m^*_{{\eta^\prime}}({\bf r})=m_{{\eta^\prime}}+V_0\frac{{\rho_N({\bf r})}}{{\rho_0}},
\end{equation}
Eq.~\ref{eq:m_etap_av} can be readily rewritten in the form
\begin{equation}
<m^*_{{\eta^\prime}}>=m_{{\eta^\prime}}+V_0\frac{<{\rho_N}>}{{\rho_0}}.\label{eq:m*etapr}
\end{equation}
Here, $m_{{\eta^\prime}}$ is the ${\eta^\prime}$ vacuum mass, $<{\rho_N}>$ and $\rho_0$ are the
average and saturation nucleon densities, respectively. $V_0$ is the ${\eta^\prime}$ scalar potential
depth (or the ${\eta^\prime}$ in-medium mass shift) at density $\rho_0$ to be extracted from the data for $\eta^\prime$ photoproduction off nuclei (see Section \ref{sec:etaprime}).

The total energy $E^\prime_{{\eta^\prime}}$ of the ${\eta^\prime}$ meson inside the nuclear medium is
expressed via its average effective mass $<m^*_{{\eta^\prime}}>$ and its in-medium momentum
${\bf p}^{\prime}_{{\eta^\prime}}$ by:
\begin{equation}
E^\prime_{{\eta^\prime}}=\sqrt{({\bf p}^{\prime}_{{\eta^\prime}})^2+(<m^*_{{\eta^\prime}}>)^2}.
\end{equation}
Since in the considered case of $\gamma A$ $\eta^\prime$ photoproduction the created $\eta^\prime$ meson is expected to propagate out of the nucleus in the field of conservative nuclear forces, this energy is assumed to be equal to the total vacuum energy $E_{{\eta^\prime}}$ of the $\eta^\prime$ meson,
having the vacuum momentum ${\bf p}_{{\eta^\prime}}$ \cite{Paryev_JPG2013}:

\begin{equation}
E^\prime_{{\eta^\prime}}=E_{{\eta^\prime}}=
\sqrt{{\bf p}^2_{{\eta^\prime}}+m^2_{{\eta^\prime}}}.
\end{equation}
In the collision model \cite{Paryev_JPG2013,Paryev_JPG2016}
the medium modification of the final nucleons has also been taken into account
by using, by analogy with equation Eq.~\ref{eq:m*etapr}, their average in-medium momentum-dependent mass
$<m^*_{N}({\bf p}_N^{'2})>$:
\begin{equation}
<m^*_{N}({\bf p}_N^{'2})>=m_{N}+V_{NA}^{\rm SC}({\bf p}_N^{'2})\frac{<{\rho_N}>}{{\rho_0}}.
\end{equation}
Here, $m_{N}$ is the nucleon vacuum mass and $V_{NA}^{\rm SC}({\bf p}_N^{'2})$ is the scalar momentum-dependent nuclear nucleon-potential at saturation density. It depends on the in-medium nucleon momentum ${\bf p}_N^{'}$ and can be determined from the relation \cite{Paryev_JPG2016}
\begin{equation}
V_{NA}^{\rm SC}({\bf p}_N^{'2})=\frac{\sqrt{m_N^2+{{\bf p}_N^{'2}}}}{m_N}V_{NA}^{\rm SEP}({\bf p}_N^{'2}),
\end{equation}
where $V_{NA}^{\rm SEP}({\bf p}_N^{'2})$ is the Schroedinger-equivalent potential for nucleons
at normal nuclear matter density. This potential is momentum-dependent and can be parametrized
as a function of the momentum relative to the nuclear matter at rest by
\cite{Paryev_JPG2016}:
\begin{equation}
V_{NA}^{\rm SEP}({\bf p}_N^{'2})=\left(V_1-V_2{\rm e}^{-2.3{\bf p}_N^{'2}}\right);\\\\ V_1=50~{\rm MeV},
\\\\V_2=120~{\rm MeV},
\end{equation}
where the momentum $|{\bf p}_N^{'}|$ is measured in GeV/c.
In the reactions (\ref{eq:primary_1}),(\ref{eq:primary_2}), the total energy $E_N^{'}$ of the outgoing nucleon in the nuclear interior can be expressed in terms of its
effective mass $<m^*_{N}({\bf p}_N^{'2})>$ defined above and its in-medium momentum ${\bf p}_N^{'}$ as
in the free particle case, i.e.:
\begin{equation}
E_{N}^{'}=\sqrt{{\bf p}_{N}^{'2}+[<m^*_{N}({\bf p}_{N}^{'2})>]^2}.
\end{equation}
When the nucleon escapes from the nucleus with momentum ${\bf p}_{N}$ its total energy $E_N$ becomes
equal to that corresponding to its bare mass $m_N$: $E_{N}=\sqrt{{\bf p}_{N}^{2}+m_{N}^2}$, i.e. it is assumed that $E_N^{'}=E_N$. In the production of slow mesons the recoiling protons are highly energetic with momenta much larger than the momenta of the surrounding nucleons. In this case it is natural to assume that the energy required to bring the nucleon back on shell is taken from the energy of this nucleon itself and not from the nucleus. 

Neglecting the distortion of the incident photon in nuclear matter
and describing the ${\eta^\prime}$ meson final-state absorption by the in-medium cross section
$\sigma_{\rm {\eta^\prime}N}$, the inclusive differential
cross section for the production of ${\eta^\prime}$ mesons with vacuum momentum
${\bf p}_{{\eta^\prime}}$ off nuclei in the primary photon--induced reactions (\ref{eq:primary_1},\ref{eq:primary_2})
can be represented  as a product of the effective number of participating target nucleons $I_{V}[A]$, defined by Eq. (11) in \cite{Paryev_JPG2013}, and the off-shell differential cross sections of these reactions, averaged over the internal momentum and the binding energy of the struck nucleon \cite{Paryev_JPG2013}:
\begin{equation}
\frac{d\sigma_{{\gamma}A\to {\eta^\prime}X}^{({\rm prim})}
(E_{\gamma})}
{d{\bf p}_{\eta^\prime}}=I_{V}[A]
\times
\left[\frac{Z}{A}\left<\frac{d\sigma_{{\gamma}p\to {\eta^\prime}p}({\bf p}_{\gamma},
{\bf p}^{\prime}_{\eta^\prime})}{d{\bf p}^{\prime}_{\eta^\prime}}\right>_A+
\frac{N}{A}\left<\frac{d\sigma_{{\gamma}n\to {\eta^\prime}n}({\bf p}_{\gamma},{\bf p}^{\prime}_{\eta^\prime})}{d{\bf p}^{\prime}_{\eta^\prime}}\right>_A\right]\frac{d{\bf p}^{\prime}_{{\eta^\prime}}}{d{\bf p}_{{\eta^\prime}}}. \label{primary}
\end{equation}

In the collision model \cite{Paryev_JPG2013} it is assumed that these cross sections
are equivalent to the respective on-shell cross sections calculated for the off-shell kinematics of the elementary processes (\ref{eq:primary_1} ,\ref{eq:primary_2}), using the ${\eta^\prime}$ and nucleon in-medium masses $<m^*_{{\eta^\prime}}>$ and $<m^*_{N}({\bf p}_{N}^{'2})>$, respectively. For the on-shell differential cross sections
for $\eta^\prime$ production off the "free" proton and neutron the respective paramterisations of the
measured cross sections \cite{Crede_PRC2009,Jaegle_EPJA2011} are used in \cite{Paryev_JPG2013}.
Before considering the two-step $\eta^\prime$ production mechanism, the nucleon spectral function will be discussed, which is a crucial point in the collision model [35,74-87].
%\cite{Paryev_JPG2013,Efremov_1998,Paryev_EPJA1999,Paryev_EPJA72000,
%Paryev_EPJA92000,Paryev_EPJA2003,Paryev_EPJA2005,Paryev_JPG36,Paryev_JPG2010,Paryev_PAN2012,Paryev_JPG2016,Kiselev_IJMPE,Paryev_YF2017,Paryev_K-,Paryev_nuclth2016}.

%--------------------------------------------------------------------------
%Nucleon Spectral Function

\subsubsection{\it Nucleon spectral function}

Considering the ground-state nucleon-nucleon ($NN$) correlations, which are generated by the short-range
and tensor parts of the realistic $NN$ interaction, the spectral function $P_A({\bf p}_t,E)$ can be represented
in the following form \cite{Ciofi_PLB1989,Ciofi_PRC1996,Benhar_RMP80}:
\begin{equation}
  P_A({\bf p}_t,E)=P_0({\bf p}_t,E)+P_1({\bf p}_t,E),
\end{equation}
where $P_0$ includes the ground and one-hole states of the residual $(A-1)$ nucleon system and $P_1$
the more complex configurations (mainly 1$p$--2$h$ states) that arise from the 2$p$--2$h$ excited states
generated in the ground state of the target nucleus by $NN$ correlations.
For the single-particle (uncorrelated) part $P_0({\bf p}_t,E)$ of the nucleon spectral function the harmonic oscillator spectral function 
and the Fermi-gas model spectral function have been employed in case of $^{12}$C and for other medium or heavy target nuclei, respectively.

\begin{figure*}
\centering
\includegraphics[width=14cm,clip]{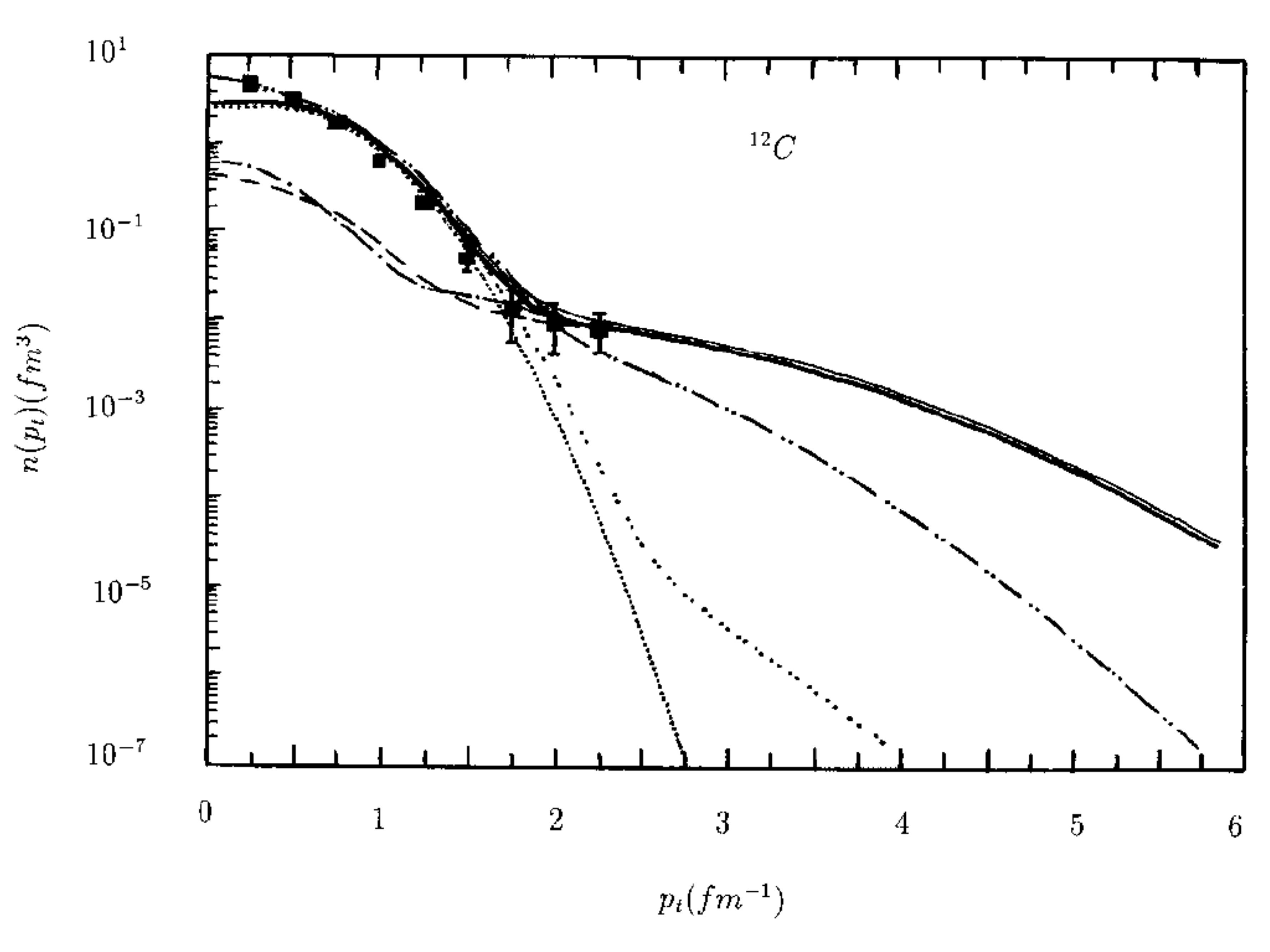}
\caption{Internal nucleon momentum distribution for $^{12}$C. The dashed and dotted lines are the many-body correlated and uncorrelated momentum distributions, presented
in \cite{Ciofi_PLB1989,Ciofi_PRC1990} and defined as $\int P_1({\bf p}_t,E) dE$ and $\int P_0({\bf p}_t,E) dE$, respectively. 
Light solid curve: total nucleon momentum distribution given by the sum of the shell-model momentum distribution \cite{Efremov_1998} and the correlated momentum distribution parametrized in \cite{Efremov_1998} by two exponents and shown by the dot-dashed curve. Heavy solid line: total many-body momentum distribution, presented in \cite{Ciofi_PLB1989,Benhar_PLB1986}. Squares represent the existing experimental data \cite{Ciofi_PRC431991}. For further details see \cite{Efremov_1998}.
The momentum distributions $n(p_t)$ are normalised according to $\int \limits_{0}^{\infty}n(p_t)p_t^2dp_t=1$. For heavier target nuclei the total nucleon momentum distribution in the considered collision model is given by the sum of the Fermi-gas model momentum distribution and that given by the dot-dashed line shown above. The figure is taken from \cite{Efremov_1998}. With kind permission of The European Physical Journal (EPJ).}
\label{fig:Nmom_dist}      
\end{figure*}

For the correlated part $P_1({\bf p}_t,E)$ of the nucleon spectral function the simple analytical expression, given in \cite{Efremov_1998,Paryev_EPJA72000}, is adopted.
The internal nucleon momentum distributions used in calculations of specific particle production off carbon
are shown in Fig.~\ref{fig:Nmom_dist}. Recent work on nucleon-nucleon short-range correlations and the nucleon spectral function has been summarised in \cite{Hen_arXiv_1611.09748}.

%---------------------------------------------------------------------
%Two Step Processes

\subsubsection{\it Two-step ${\eta^\prime}$ production mechanism}

In the incident photon energy range of interest, the following two-step
production process may contribute to the ${\eta^\prime}$ production in
${\gamma}A$ interactions \cite{Paryev_JPG2013}.
In the first inelastic collision with an intranuclear nucleon an initial photon can produce pions through the elementary reaction

\begin{equation}
\gamma+N_1 \to 2\pi+N. \label{eq:secondary_1}
\end{equation}
Then, these pions, which are assumed to be on-shell, produce the ${\eta^\prime}$ meson
on another nucleon of the target nucleus via the elementary subprocess with the lowest free
production threshold momentum (1.43 GeV/c)

\begin{equation}
\pi+N_2 \to {\eta^\prime}+N. \label{eq:secondary_2}
\end{equation}
For kinematic reasons  the elementary processes
${\gamma}N \to {\pi}N$, ${\gamma}N \to 3{\pi}N$, ${\pi}N \to {\eta^\prime}N{\pi}$ are expected to play
a minor role in ${\eta^\prime}$ production in ${\gamma}A$ reactions at photon energies $E_{\gamma} \approx 2$ GeV \cite{Paryev_JPG2013}, however,
the elementary reactions ${\gamma}p \to {\pi^+}{\pi^-}p$, ${\gamma}p \to {\pi^+}{\pi^0}n$,
${\gamma}n \to {\pi^+}{\pi^-}n$ and ${\gamma}n \to {\pi^-}{\pi^0}p$ have been included
in the calculations \cite{Paryev_JPG2013} of the ${\eta^\prime}$ production on nuclei.

     Accounting for the medium effects on the ${\eta^\prime}$ mass on the same footing as 
employed in calculating the ${\eta^\prime}$ production cross section off the target nucleus in the primary
processes (\ref{eq:primary_1},\ref{eq:primary_2}), and ignoring, for the sake of numerical simplicity, the medium modifications of the
outgoing  nucleon in the subprocess (\ref{eq:secondary_2}), the ${\eta^\prime}$ production total cross section for ${\gamma}A$ reactions from this subprocess can be expressed by:

\begin{equation}
\sigma_{{\gamma}A\to {\eta^\prime}X}^{({\rm sec})}
(E_{\gamma})=I_{V}^{({\rm sec})}[A]
\sum_{\pi'=\pi^+,\pi^0,\pi^-}\int d{\bf p}_{\pi} \label{eq:secondary}
\times
\end{equation}
$$
\times
\left[\frac{Z}{A}\left<\frac{d\sigma_{{\gamma}p\to {\pi'}X}({\bf p}_{\gamma},
{\bf p}_{\pi})}{d{\bf p}_{\pi}}\right>_A+
\frac{N}{A}\left<\frac{d\sigma_{{\gamma}n\to {\pi'}X}({\bf p}_{\gamma},
{\bf p}_{\pi})}{d{\bf p}_{\pi}}\right>_A\right]
$$
$$
\times
\left[\frac{Z}{A}\left<\sigma_{{\pi'}p\to {\eta^\prime}N}({\bf p}_{\pi})\right>_A+
\frac{N}{A}\left<\sigma_{{\pi'}n\to {\eta^\prime}N}({\bf p}_{\pi})\right>_A\right].
$$
The cross section is given by the product of the effective number of $N_1N_2$ pairs $I_{V}^{({\rm sec})}[A]$ (defined by Eq.(44) in \cite{Paryev_JPG2013}) involved in the two-step ${\eta^\prime}$ production processes under consideration and the integral over the intermediate pion three-momentum
\begin{figure}[h!]
\begin{center}
\includegraphics[width=13.0cm]{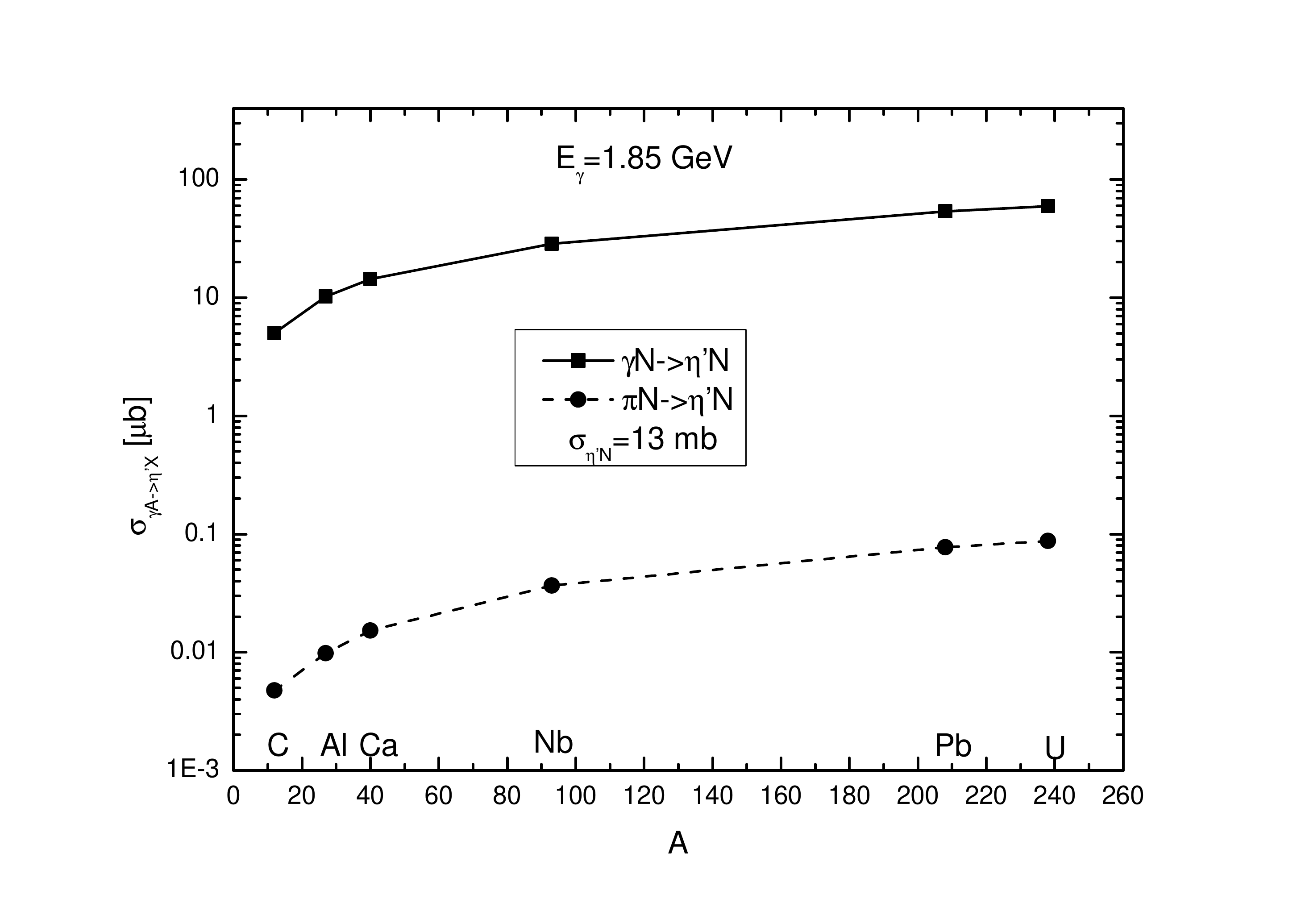}
\vspace*{-2mm} \caption{A--dependences of the total cross sections of $\eta^\prime$ production
by 1.85 GeV photons from primary ${\gamma}N \to {\eta^\prime}N$ channels and from secondary
${\pi}N \to {\eta^\prime}N$ processes in the full phase-space in the scenario without $\eta^\prime$
and nucleon mass shifts. The absorption of $\eta^\prime$ mesons in nuclear matter is calculated assuming an inelastic cross
section $\sigma_{{\eta^\prime}N}=13.0$ mb. The lines are shown to guide the eye. The figure is adapted from \cite{Paryev_JPG2013}.}
\label{fig:etaprime_A}
\end{center}
\end{figure}
of the inclusive differential cross sections for pion production from the primary photon-induced reaction channel (\ref{eq:secondary_1}) and the in-medium total cross sections for
the production of $\eta^\prime$ mesons with the reduced mass $<m^*_{{\eta^\prime}}>$ in ${\pi}N_2$ collisions  \cite{Paryev_JPG2013}.
The inclusive differential cross sections for pion production in process (\ref{eq:secondary_1}) are calculated following the approach \cite{Paryev_nuclth2011}.
The elementary $\eta^\prime$ production reactions
${\pi}^+n \to {\eta^\prime}p$,
${\pi}^0p \to {\eta^\prime}p$, ${\pi}^0n \to {\eta^\prime}n$ and
${\pi}^-p \to {\eta^\prime}n$ have been included in the calculations of the $\eta^\prime$ production on nuclei. Their total cross sections are given in \cite{Paryev_JPG2013}.

 In Fig.~\ref{fig:etaprime_A}, adapted from \cite{Paryev_JPG2013}, the resulting A--dependencies of the total $\eta^\prime$ production
cross sections for the one-step and two-step $\eta^\prime$ production mechanisms
in ${\gamma}A$ ($A=$$^{12}$C, $^{27}$Al, $^{40}$Ca, $^{93}$Nb, $^{208}$Pb, and $^{238}$U) collisions are shown, calculated for $E_{\gamma}=1.85$ GeV.
The role of the secondary pion--induced reaction channel ${\pi}N \to {\eta^\prime}N$ turns out to be negligible compared to that of the primary ${\gamma}N \to {\eta^\prime}N$
processes for all considered target nuclei.
This gives confidence that the channel ${\pi}N \to {\eta^\prime}N$ can be ignored in the analysis
of data on $\eta^\prime$ photoproduction off nuclei obtained by the CBELSA/TAPS Collaboration discussed in Section \ref{sec:etaprime}.

%------------------------------------------------------------------------------------------------
%-------------------------------------------------------------------------
%Imaginary Potential
\subsection{\it Determination of the imaginary part of the meson--nucleus potential from measurements of the transparency ratio}
\label{sec:imag}
The imaginary part of the meson--nucleus potential is a measure for the absorption of the meson in the medium. It is related to the in-medium width of the meson $m$ through Eq.~\ref{eq:W} and can be extracted from the so-called transparency ratio, defined as \cite{Hernandez_Oset,Cabrera_NPA733},
\begin{equation}
T_A = \frac{\sigma_{\gamma A \rightarrow m X}}{A \cdot \sigma_{\gamma N \rightarrow m X}}\label{eq:transp}.
\end{equation}
It compares the production cross section per nucleon of meson $m$ off a nucleus with mass number A with the production cross section on a free nucleon $N$. The role of the nucleus is twofold: it serves as a production target as well as an absorber. $T_A$ quantifies the loss of the meson flux in a nuclear
target which, in turn, is governed by the imaginary part of the meson in-medium self-energy or width.
In case of no absorption, $T_A$ = 1, if secondary production processes can be neglected. In the low density approximation the in-medium width $\Gamma$ in the nuclear rest frame at nuclear density $\rho_N$ is related to the meson--nucleon inelastic cross section by
\begin{equation}
\Gamma(\rho_N) = \Gamma_0 \cdot \frac{\rho_N}{\rho_0} = \hbar c \cdot \beta \cdot \sigma_{inel} \cdot \rho_N.\label{eq:lindens}
\end{equation}
Throughout this review the in-medium width of mesons will always refer to the nuclear rest frame. The width in the meson eigen-system is then given by 
\begin{equation} 
\Gamma_{eigen} = \gamma \cdot \Gamma,
\end{equation}
where $\gamma$ is the relativistic factor $\gamma = E/m_0$, the ratio of the particle energy $E$ to its restmass $m_0$.

To avoid systematic uncertainties in the comparison with experimental data, e.g. due to differences in the initial state interaction, partially unknown meson production cross sections on the neutron or secondary meson production, the transparency ratio for a heavy target with mass number A is frequently normalised to the transparency ratio for a light nucleus like Carbon:
\begin{equation}
T_A^{\rm C} = \frac{12 \cdot \sigma_{\gamma A \rightarrow m X}}{A \cdot \sigma_{\gamma {\rm C} \rightarrow m X}}\label{eq:transp_C}.
\end{equation}
Thereby nuclear effects not related to the absorption of mesons largely cancel.
Transport calculations \cite{Muehlich_Mosel}, Monte Carlo simulations \cite{Kaskulov} and collisional model calculations \cite{Paryev_JPG2013} have been performed to study the sensitivity of
the transparency ratio for $\omega$ and $\eta^{\prime}$ mesons to the in-medium width and inelastic cross sections.
\begin{figure}
\begin{center}
\includegraphics[width=12.0cm]{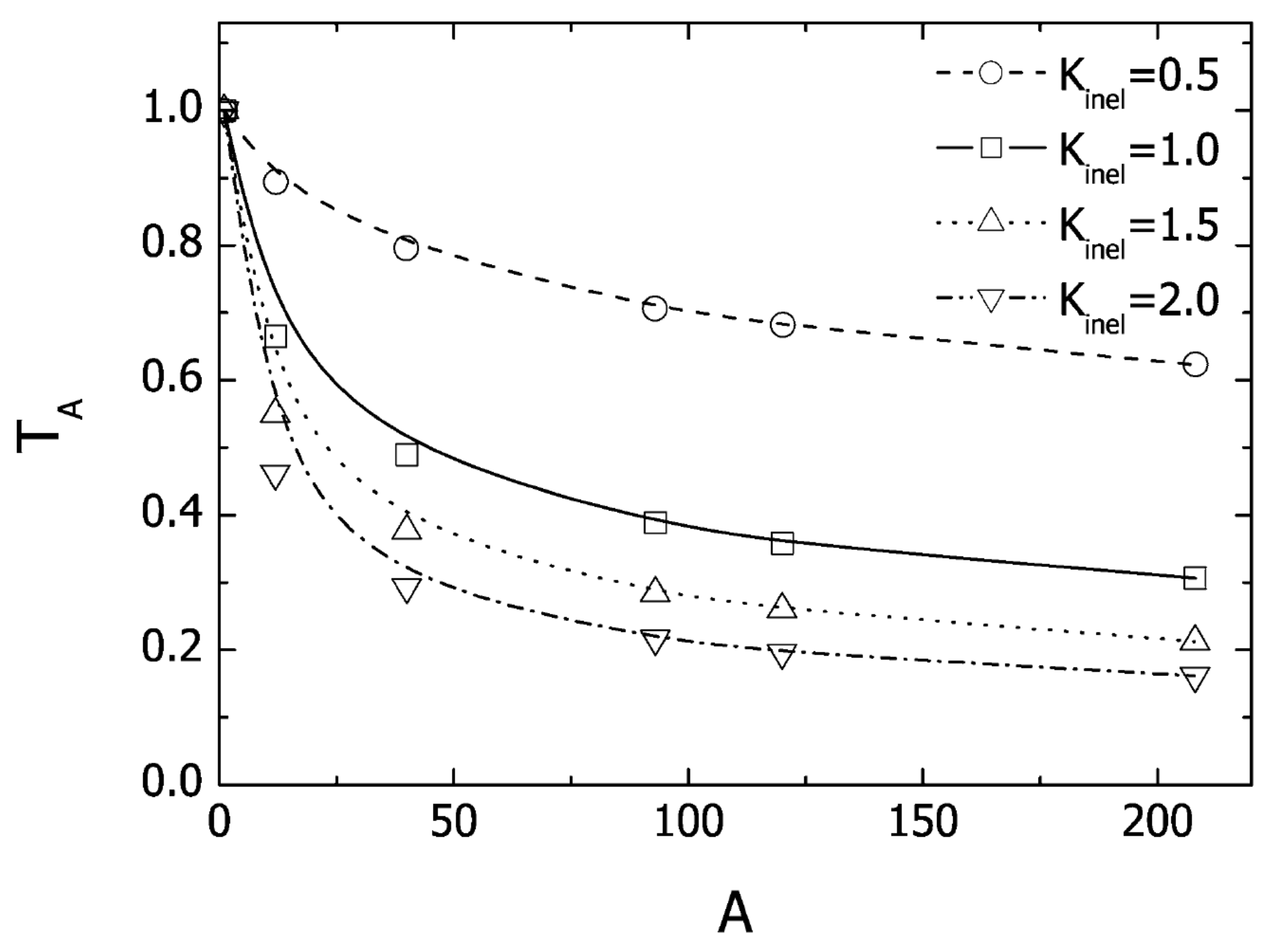}
\vspace*{-2mm} \caption{Nuclear transparency ratio as a function of the nuclear mass number A  for $\omega$ photoproduction for different multiplicative factors K$_{inel}$ to vary the in-medium width, starting from $\Gamma_0$ = 37 MeV for K$_{inel}$ = 1.0. The calculation has been performed for an incident photon energy of 1.5 GeV. The figure is taken from \cite{Muehlich_Mosel}.  }
\label{fig:TA_Muehlich}
\end{center}
\end{figure}
As an example Fig.~\ref{fig:TA_Muehlich} shows the transparency ratio (\ref{eq:transp}) obtained from GiBUU simulations for $\omega$ photoproduction as a function of the target mass number for different $\omega$ in-medium widths. After a rather steep fall off for $A \le $ 50, the transparency ratio almost levels off for higher masses A. In the latter mass region,
a variation in the in-medium width by a factor 4 roughly leads to a change in the transparency ratio also by a factor 4.
\begin{figure}
\begin{center}
\includegraphics[width=9.0cm]{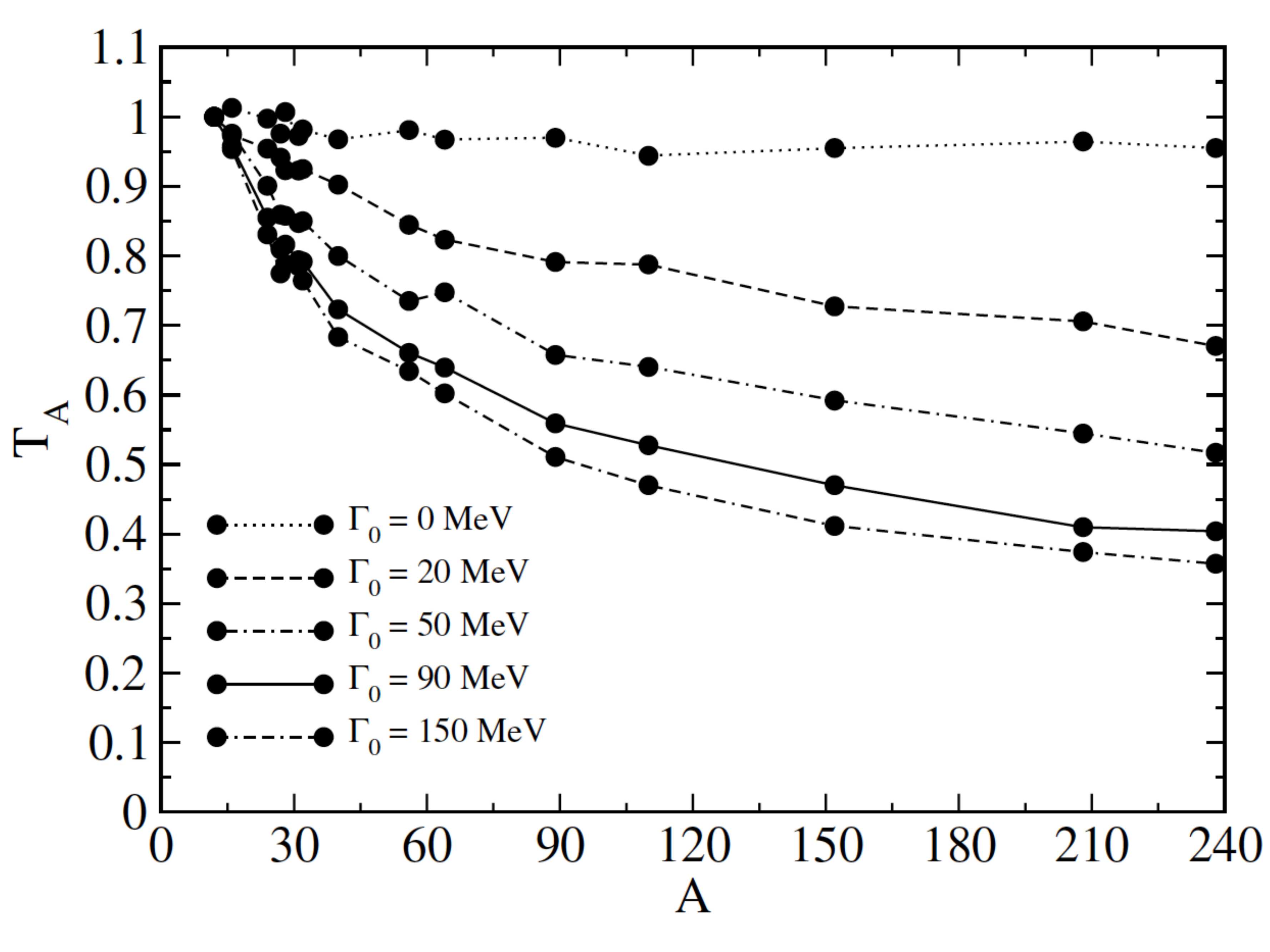}\includegraphics[width=9.0cm]{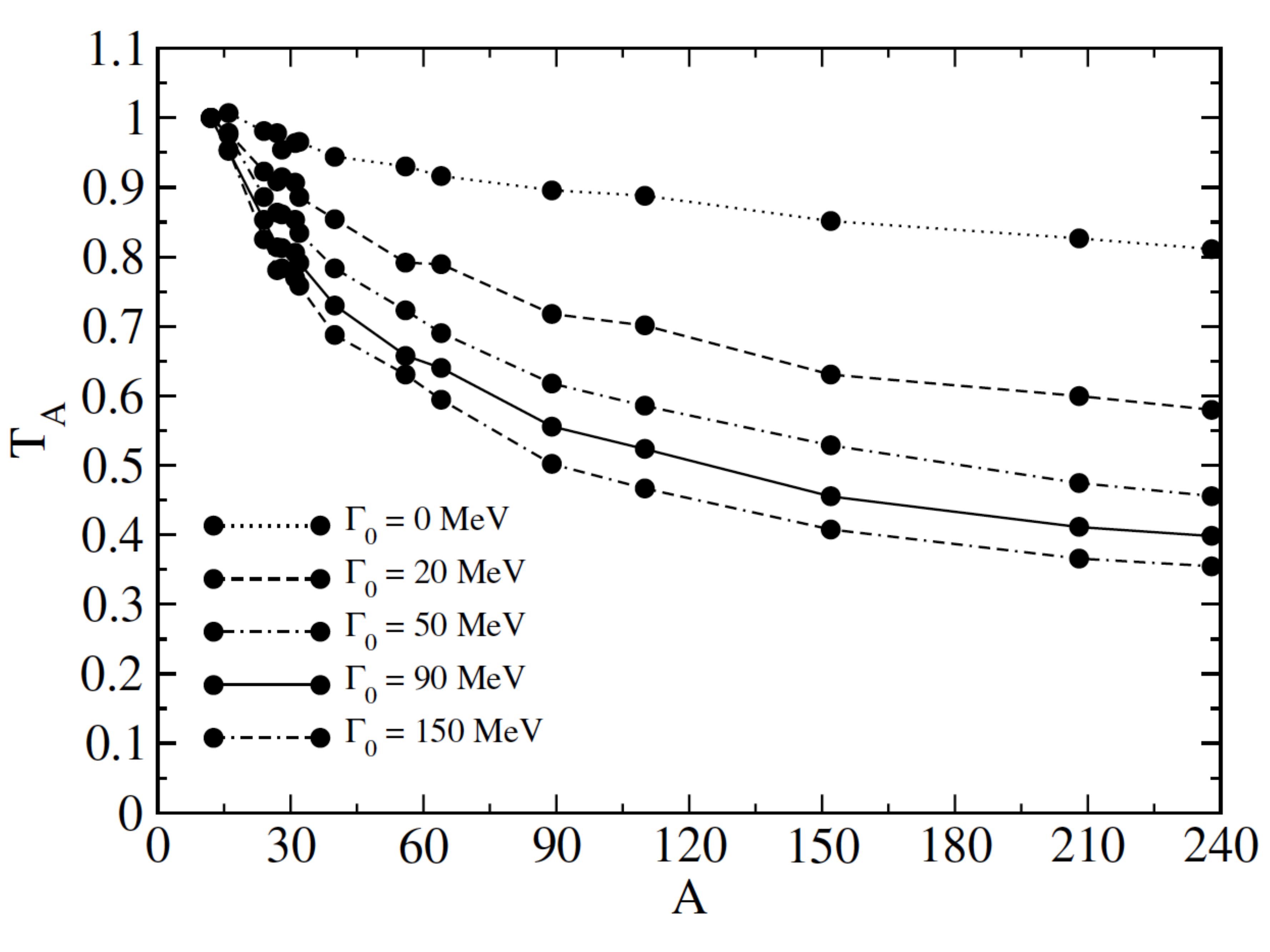}
\vspace*{-2mm} \caption{Nuclear mass dependence of the transparency ratio $T_A^{\rm C}$, normalised to Carbon, for $\omega$ photoproduction confining the photon energy range to 1.45--1.55 GeV  without (left panel) and with (right panel) taking the final-state interaction of $\pi^0$ mesons into account. A cut of $T_{\pi^0} \ge $ 150 MeV on the kinetic energy of the outgoing pion has been applied to suppress contributions from elastic $\pi^0$ scattering. The different curves correspond to different in-medium widths of the $\omega$ meson. The figure is taken from \cite{Kaskulov}. With kind permission of The European Physical Journal (EPJ).}
\label{fig:TA_Kaskulov}
\end{center}
\end{figure}
\begin{figure}
\begin{center}
\includegraphics[width=13.0cm]{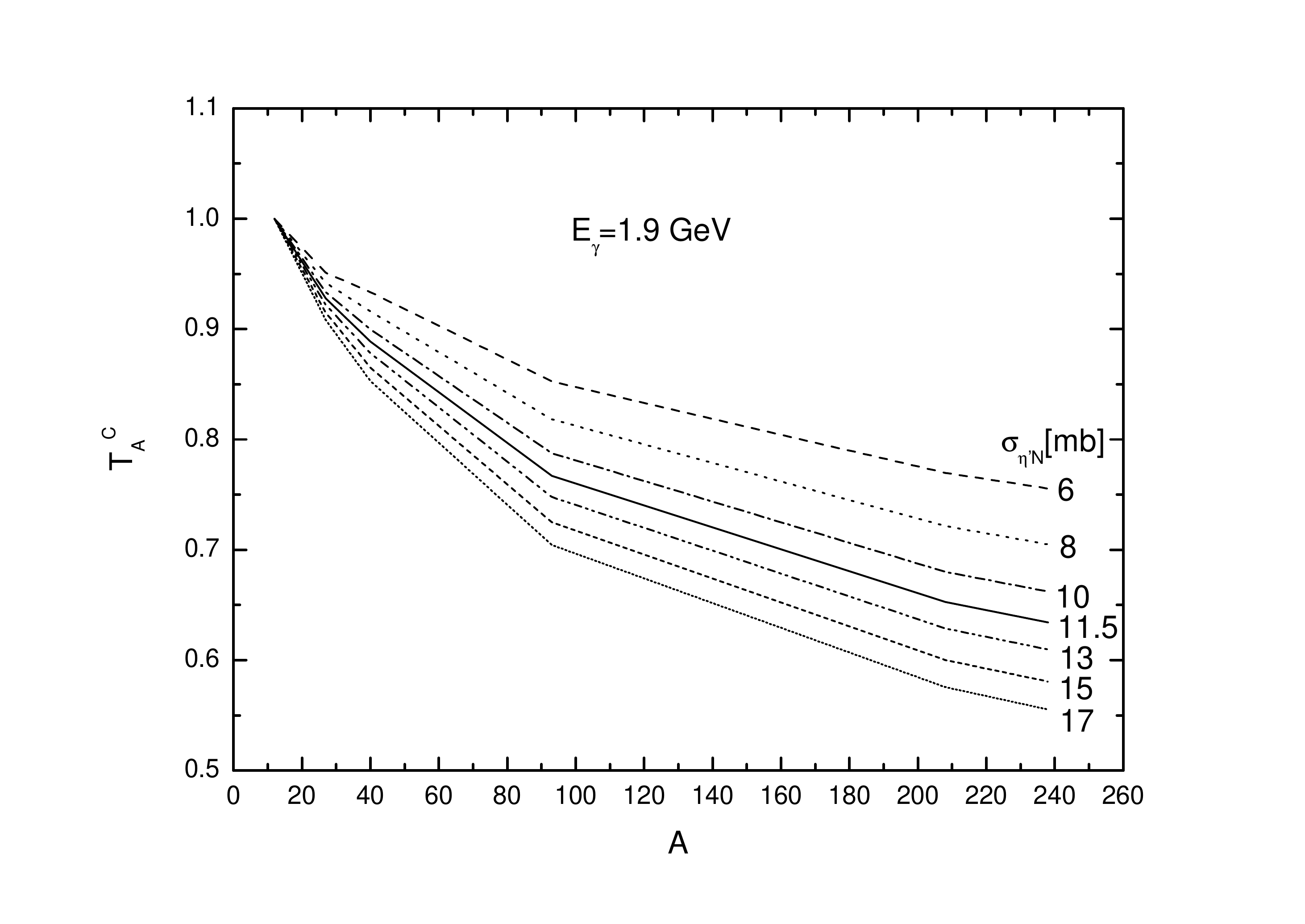}
\vspace*{-2mm} \caption{Transparency ratio $T_A^{\rm C}$, normalised to Carbon, for $\eta^\prime$ photoproduction as a function of the nuclear mass number A for different in-medium absorption cross sections at an incident photon energy of 1.9 GeV. The figure is adapted from \cite{Paryev_JPG2013}.}
\label{fig:TA_Paryev}
\end{center}
\end{figure}
In Monte Carlo simulations \cite{Kaskulov} Kaskulov et al. studied the impact on the transparency ratio by in-medium decays involving strongly interacting decay products. Specifically they investigated the $\omega \rightarrow \pi^0 \gamma$ decay.  In the experiment one tries to determine the absorption of $\omega$ mesons; however, even if the $\omega$ is not absorbed, the detector may not see the $\omega$ meson because the final-state interaction of the $\pi^0$ prevents the reconstruction of the $\omega$ invariant mass from the 4-momentum vectors of the decay products: the $\pi^0$ may either undergo elastic scattering, leading to an energy loss of the pion in the laboratory and thus to a strongly distorted $\pi^0\gamma$ invariant mass, or the $\pi^0$ meson itself may be absorbed. In both cases the $\omega$ meson is lost in the analysis although it was not absorbed in the nucleus.
It has been shown \cite{Kaskulov,Messchendorp} that the first class of events can be suppressed by a lower cut T$_{\pi^0} \ge$150 MeV on the kinetic energy of the outgoing pions.
Fig.~\ref{fig:TA_Kaskulov} demonstrates, however, that the impact of the $\pi^0$ final-state interaction on the $\omega$ transparency ratio is only relevant for small $\omega$ in-medium widths of $ \le 20 $ MeV. Note, however, that all experimentally deduced $\omega$ in-medium widths are much larger (see Section \ref{sec:omega}).

Collision model calculations have been performed \cite{Paryev_JPG2013} to study the dependence of the $\eta^\prime$ transparency ratio on the size of the inelastic cross section. Varying $\sigma_{inel}$ from 6 to 17 mb, i.e. by a factor $\approx$ 3, the transparency ratio changes by about a factor 1.4, as follows from Fig.~\ref{fig:TA_Paryev}.

The examples discussed above show that transport calculations and collision model calculations show qualitatively very similar results which give confidence that a comparison with experimental data will yield reliable information on the imaginary part of the meson--nucleus potential. In the following sections we will present and discuss the potential parameters extracted from photon-, proton- and heavy-ion induced reactions for individual pseudoscalar and light vector mesons, using the approaches discussed above.

%---------------------------------------------------------------------------------------------
%Real Potential

\subsection{\it Determination of the real part of the meson--nucleus potential}
\label{sec:real}
As discussed below, the real part of the meson--nucleus potential has been determined for many mesons by comparing experimental observables like the production cross sections, momentum or transverse momentum distributions of the produced mesons or their sideward flow with corresponding predictions by transport calculations. These calculations are performed assuming different values for the real and imaginary part of the potential, thereby allowing the sensitivity of these observables to meson in-medium modifications to be studied. The potential parameters are then extracted by optimising the agreement between the experimental distributions and the transport model results.

\subsubsection{\it Real part of the meson--nucleus potential deduced from excitation function measurements}
\label{sec:real_exci}
Using transport calculations, it has been shown in Section \ref{sec:transport} and in \cite{Weil} that it may be difficult to directly determine the in-medium mass of a meson - and thus the real part of the meson-nucleus potential - by reconstructing the invariant mass from the 4-momentum vectors of the meson decay products: the small fraction of meson decays within the nuclear volume, the radial dependence of the nuclear density, the increase in width by inelastic reactions and final-state interactions may distort the true in-medium mass distribution of the meson. In \cite{Weil} alternative approaches have therefore been worked out to access the in-medium meson mass such as the measurement of the excitation function, i.e. the cross section for meson production as a function of the incident beam energy or the measurement of the meson momentum distribution. In contrast to the line shape analysis, which is sensitive to the nuclear density at the meson decay point, these methods are sensitive to the nuclear density at the production point and hence applicable for all mesons, irrespective of their lifetime.
\begin{figure*}
\begin{center}
\includegraphics[width=9.0cm]{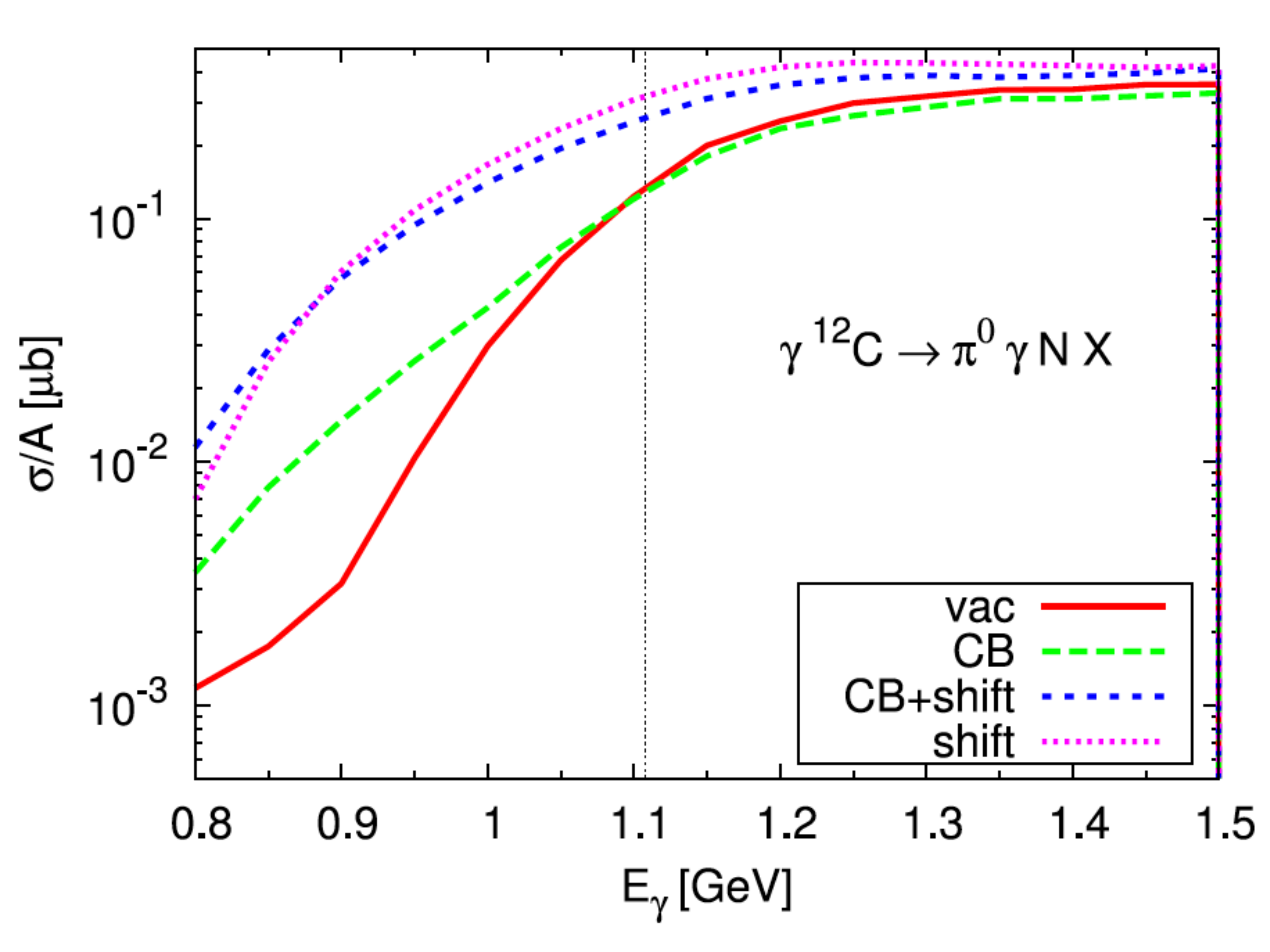}\includegraphics[width=9.0cm]{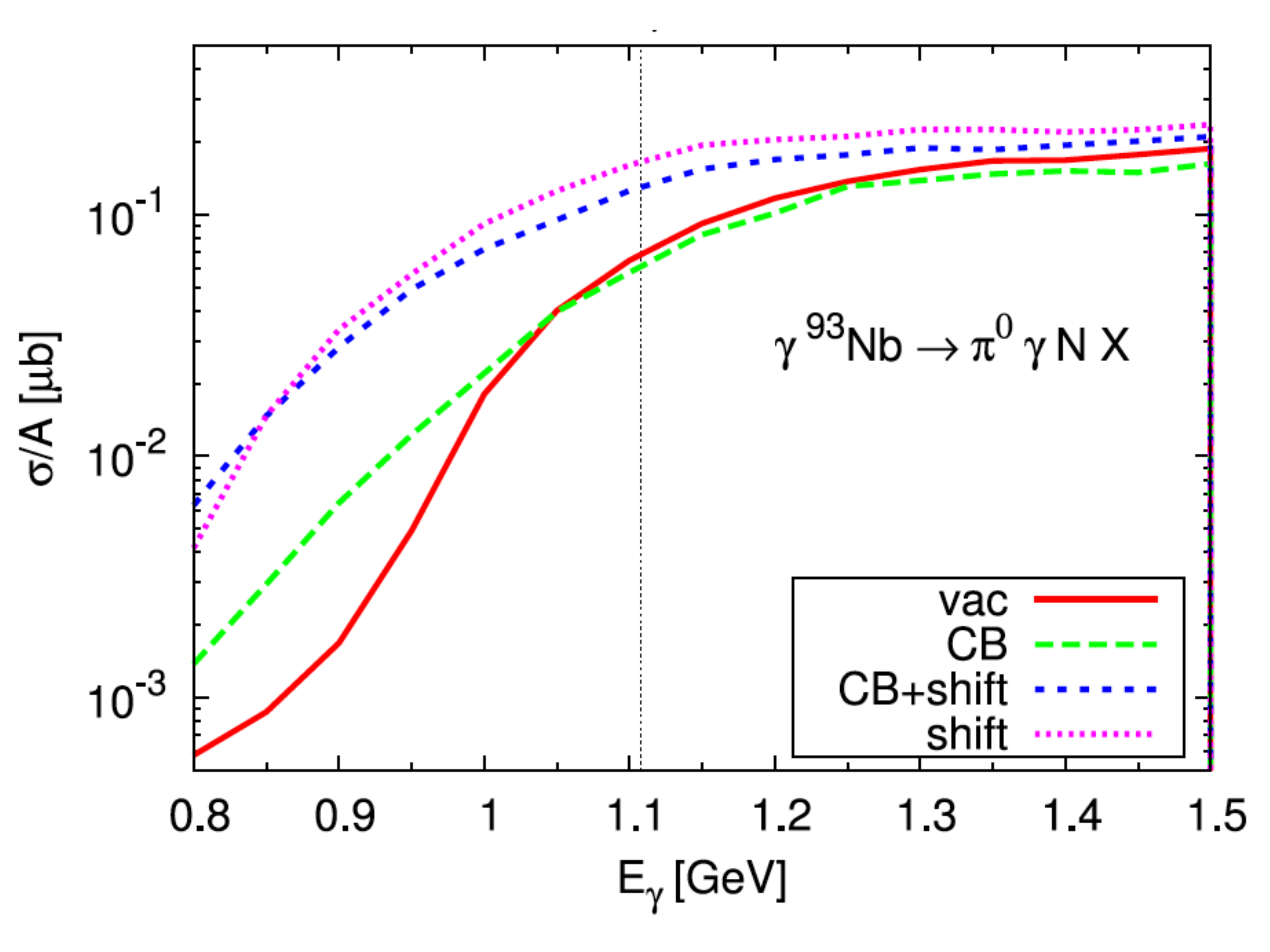}
%\vspace*{-2mm}
 \caption{Excitation function for photoproduction of $\omega$ mesons in the $\omega \rightarrow \pi^0 \gamma$ decay channel off $^{12}$C and $^{93}$Nb calculated with the GiBUU model \cite{GiBUU}. The curves show the results for four in-medium scenarios: (i) red curve: no in-medium modification, (ii) green dashed curve: collisional broadening by 150 MeV, (iii) dotted magenta curve: mass shift by -16$\%$, (iv) blue dashed curve: collisonal broadening and mass shift. The vertical dashed line indicates the $\omega$ photoproduction threshold on a free proton of E$_{\gamma} $ = 1.108 GeV. The figure is taken from \cite{Weil}.}
\label{fig:Exci_Weil}
\end{center}
\end{figure*}
\begin{figure*}
\begin{center}
\includegraphics[width=12.0cm]{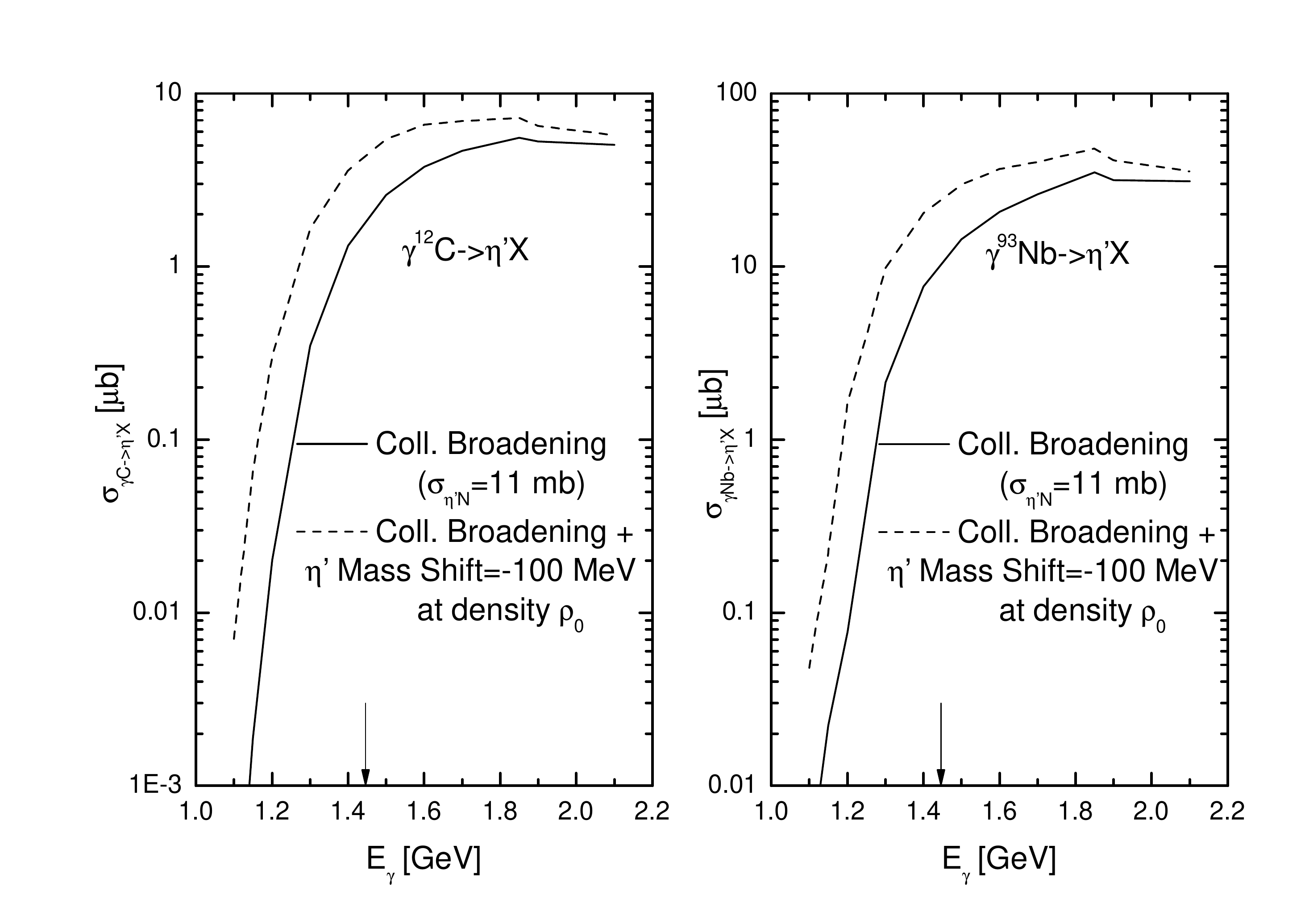}
\caption{Excitation function for photoproduction of $\eta^\prime$ mesons off $^{12}$C (left panel) and $^{93}$Nb (right panel). The solid and dashed curves are collision model calculations assuming a collisional broadening of the $\eta^\prime$ meson corresponding to an inelastic cross section of 11 mb without and with an $\eta^\prime$ mass shift of -100 MeV at normal nuclear matter density, respectively. The arrows indicate the energy of the photoproduction threshold on a free proton. The figure is adapted from \cite{Paryev_JPG2013}.}
\label{fig:Exci_Paryev}
\end{center}
\end{figure*}
As pointed out in \cite{Muehlich_Mosel}, the excitation function is sensitive to the in-medium modification of the meson since a downward mass shift would lower the meson production  threshold and increase the production cross section at a given incident beam energy due to the enlarged phase-space. Fig.~\ref{fig:Exci_Weil} shows as an example GiBUU calculations \cite{Weil} of the cross section for photoproduction of $\omega$ mesons off C and Nb for different in-medium modification scenarios. Even in case of no in-medium modification the calculations give non-vanishing cross sections below the threshold for $\omega$ photoproduction off a free nucleon ($E_{\rm thr} = 1106 $ MeV), due to the Fermi motion of nucleons in nuclei. If a photon with $E_{\gamma} \le 1106 $ MeV collides with an oncoming nucleon the available energy $\sqrt{s}$ in the centre-of-mass system may exceed the energy needed for $\omega$ production, giving rise to the subthreshold yield. The subthreshold cross section is only slightly enhanced by allowing for an in-medium broadening of the $\omega$ meson.
The scenarios including a mass shift, however, exhibit a much stronger cross section enhancement which even persists near and above the free nucleon production threshold, but diminishes for higher incident photon energies.

A corresponding result has been obtained in collision model calculations for photoproduction of $\eta^\prime$ mesons \cite{Paryev_JPG2013}. Again a sizeable cross section enhancement is predicted in the subthreshold region and somewhat above for an in-medium lowering of the $\eta^\prime$ mass (see Fig.~\ref{fig:Exci_Paryev}).

\subsubsection{\it Real part of the meson--nucleus potential deduced from measurements of the meson momentum distribution}\label{real_mom}
Transport model calculations have demonstrated that also the momentum distribution of mesons produced off nuclei is sensitive to the in-medium properties of the meson \cite{Weil}.
\begin{figure}
\begin{center}
\includegraphics[width=10.0cm]{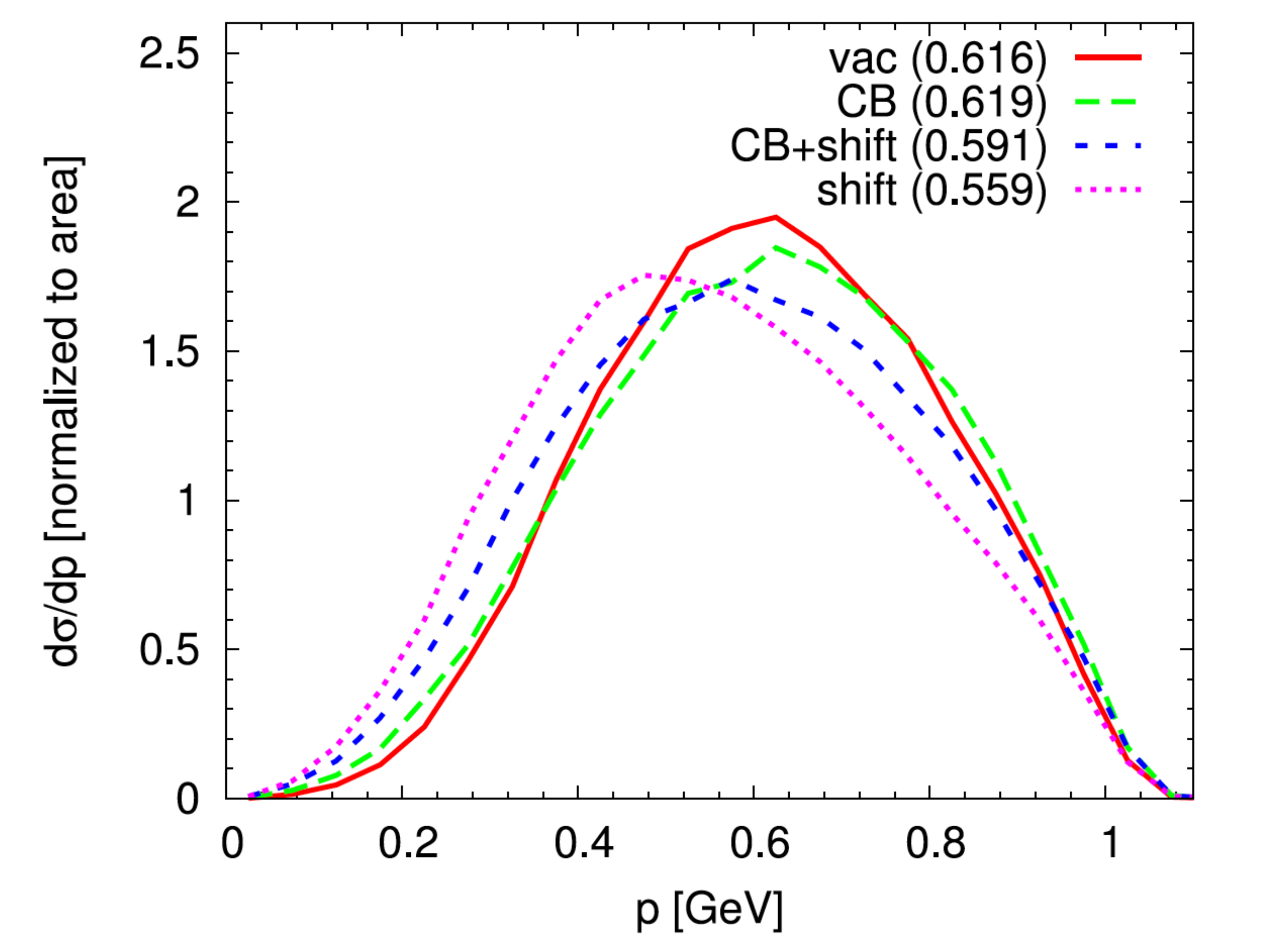}
\vspace*{-2mm} \caption{$\pi^0 \gamma$ momentum distributions from $\omega$ decays calculated with the GiBUU model \cite{GiBUU} for $\gamma + {\rm Nb}$ at $E_{\gamma} = 0.9$--1.3 GeV. The curves are normalised to have the same integral. The average momenta for the different in-medium scenarios (same as in Fig.~\ref{fig:Exci_Weil}) are given in brackets (in GeV). The figure is taken from \cite{Weil}.}
\label{fig:mom_Weil}
\end{center}
\end{figure}
\begin{figure}
\begin{center}
\includegraphics[width=12.0cm]{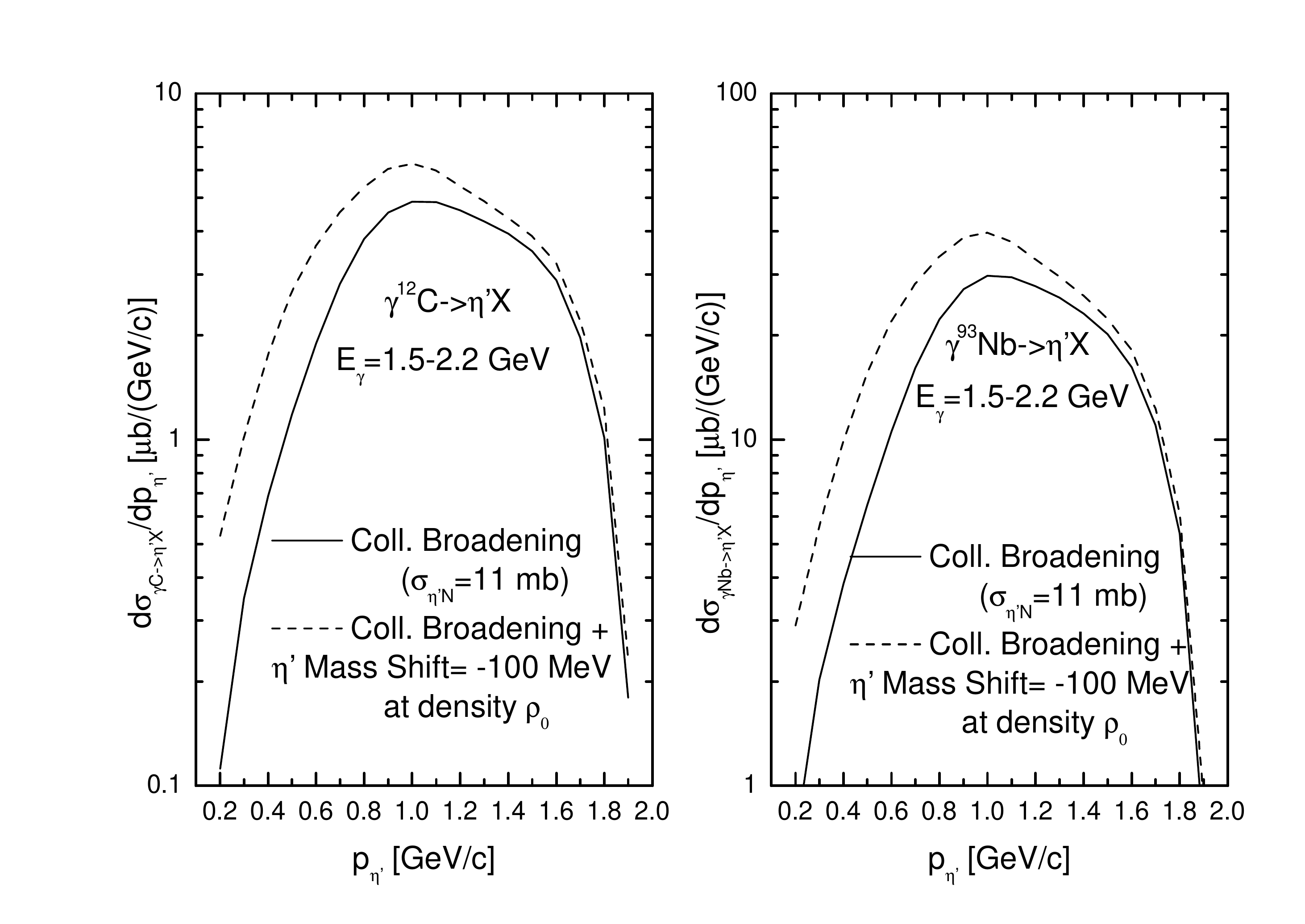}
\vspace*{-2mm} \caption{Momentum differential cross sections for $\eta^\prime$ meson photoproduction from the primary $\gamma N \rightarrow \eta^\prime N$ channel for photon energies of 1.5--2.2 GeV off $^{12}$C (left panel) and off $^{93}$Nb (right panel) calculated with the collision model \cite{Paryev_JPG2013} for the same in-medium scenarios as in 
Fig.~\ref{fig:Exci_Paryev}. The figure is adapted from \cite{Paryev_JPG2013}.}
\label{fig:mom_Paryev}
\end{center}
\end{figure}
The simulations show that, due to the kinematics of the in-medium meson production process, mesons with a lower mass are - on average - produced with a lower total energy. Furthermore, when leaving the nucleus the meson has to get back to its free vacuum mass. The missing mass has to be generated at the expense of the kinetic energy. Consequently, this energy $\rightarrow $ mass conversion lowers the meson momentum and shifts the momentum distribution to lower average values. Furthermore, the enhanced meson production cross section at and below the free production threshold in case of a mass drop will also enhance the meson yield at low momenta. It should be noted, however, that the meson momentum distribution is also sensitive to the meson angular distribution. In any analysis determining the in-medium properties of the meson from the momentum distribution one has to make sure that the experimental meson angular distribution is reproduced by the calculation. As shown in Fig.~\ref{fig:mom_Weil}, the maximum of the momentum distribution and also the average momentum are shifted by about 25-50 MeV for the scenarios involving a mass shift by -16$\%$ while the collisional broadening scenario is hardly distinguishable from the vacuum case.

The corresponding effect is also found in collision model calculations as shown in Fig.~\ref{fig:mom_Paryev}, displaying the momentum differential cross section for $\eta^\prime$ photoproduction off Carbon and Nb for scenarios with collisional broadening and with or without an additional in-medium $\eta^\prime$ mass shift  by -10$\%$ at normal nuclear matter density.

%-----------------------------------------------------------------------_pred_
% Experimental results on Meson-Nucleus Potentials

\section{Experimental results on meson--nucleus potentials}
%---------------------------------------------------------------------------------------------------------------------------------------------------------------------------------------------------------------------------------------------
%K^+
\subsection{\it K$^+$--nucleus potential}\label{sec:K+}
\subsubsection{\it $K^+$--nucleon scattering length and $K^+$--nucleus potential}\label{sec:K+theo}
\begin{figure*}]
\begin{center}
\includegraphics[width=12.0cm,clip]{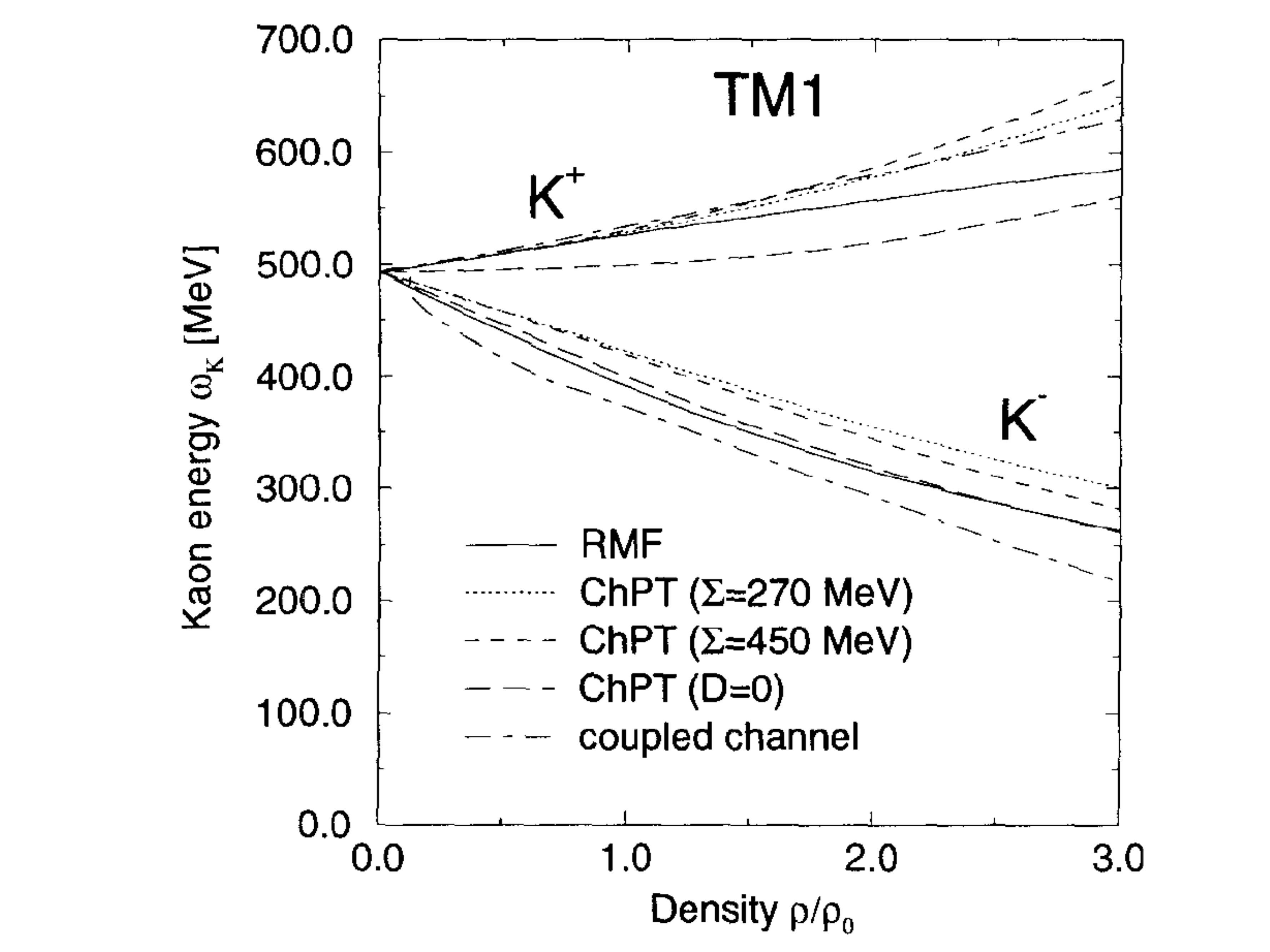}
\vspace*{-2mm} \caption{The energy of kaons and antikaons in nuclear matter
(at ${\bf p}_{K^+}^{{\prime}}=0$) as a function of density for the soft equation of state (EOS)
within different approaches. The figure is taken from \cite{Schaffner_NPA1997}.}
\label{fig:K+K-_pred-Schaffner}
\end{center}
\end{figure*}
Kaon and antikaon properties in a strongly interacting environment are a subject of considerable current interest in the hadron and nuclear physics community (see, e.g., \cite{Friedman_Gal_PR,IQMD} for recent reviews), especially in
connection with the questions of the partial restoration of chiral symmetry in hot/dense nuclear matter
and of the possible existence of a kaon condensate in neutron stars. Since the ${K^+}N$ interaction is relatively
weak and smooth at low energies ($\sigma_{K^+N}^{\rm tot}\approx$ 12 mb for $p_{lab} \le 800$ MeV/c
and there are no resonances with strangeness $S=1$), the real part of the kaon potential in a nuclear
medium at these energies can be well estimated from the isospin-averaged kaon--nucleon scattering length in free space ${\bar a}_{KN}=-0.255$ fm using the impulse ($t\rho$) approximation
\cite{Schaffner_NPA1997,Li_PLB1994}, i.e.
\begin{equation}
V_{K^+}(\rho_N)=-\frac{2\pi}{m_K}\left(1+\frac{m_K}{m_N}\right){\bar a}_{KN}\rho_N
\approx+31\mbox{\rm MeV}\frac{\rho_N}{\rho_0},
\end{equation}
where $m_K$ and $m_N$ are the vacuum kaon and nucleon masses, respectively, and the saturation density is taken to be $\rho_0=0.16$ fm$^{-3}$. With this potential, the total kaon energy $E_{K^+}^{\prime}$ in the
nuclear interior of ordinary nuclei can be expressed in terms of its in-medium mass $m_{K^+}^{*}$, defined as:
\begin{equation}
m_{K^+}^{*}(\rho_N)=m_K+V_{K^+}(\rho_N)\label{eq:mK+*}
\end{equation}
and its in-medium three-momentum ${\bf p}_{K^+}^{\prime}$ as in the free particle case [98-101]:
%\cite{Li_PLB1994,Li_NPA1995,Li_PRL1995,Li_PRC1996}:
\begin{equation}
E_{K^+}^{\prime}=\sqrt{{\bf p}_{K^+}^{{\prime}2}+m_{K^+}^{*2}}.\label{eq:EK++}
\end{equation}

\subsubsection{\it $K^+$--nucleus potential. Other theoretical approaches}\label{sec:K++}
The kaon energy in a medium can also be obtained from the mean-field approximation to the chiral
Lagrangian [21,98-100,102-104], i.e.
%\cite{Kaplan_Nelson,Li_PLB1994,Li_NPA1995,Li_PRL1995,Li_NPA1997,Li_PRL1997,Nelson}
\begin{equation}
E_{K^+}^{\prime}({\bf p}_{K^+}^{\prime},\rho_N)=\left[{\bf p}_{K^+}^{{\prime}2}+m_{K}^{2}-
\frac{\Sigma_{KN}}{f^2}\rho_S+\left(\frac{3}{8}\frac{\rho_N}{f^2}\right)^2\right]^{1/2}+
\frac{3}{8}\frac{\rho_N}{f^2},\label{eq:EK+}
\end{equation}
where $f=93$ MeV is the pion decay constant, $\Sigma_{KN}$ is the $KN$ sigma term, which depends on
the strangeness content of a nucleon, and $\rho_S$ is the scalar nucleon density.\begin{figure}[h!]
\begin{center}
\includegraphics[width=10.0cm,clip]{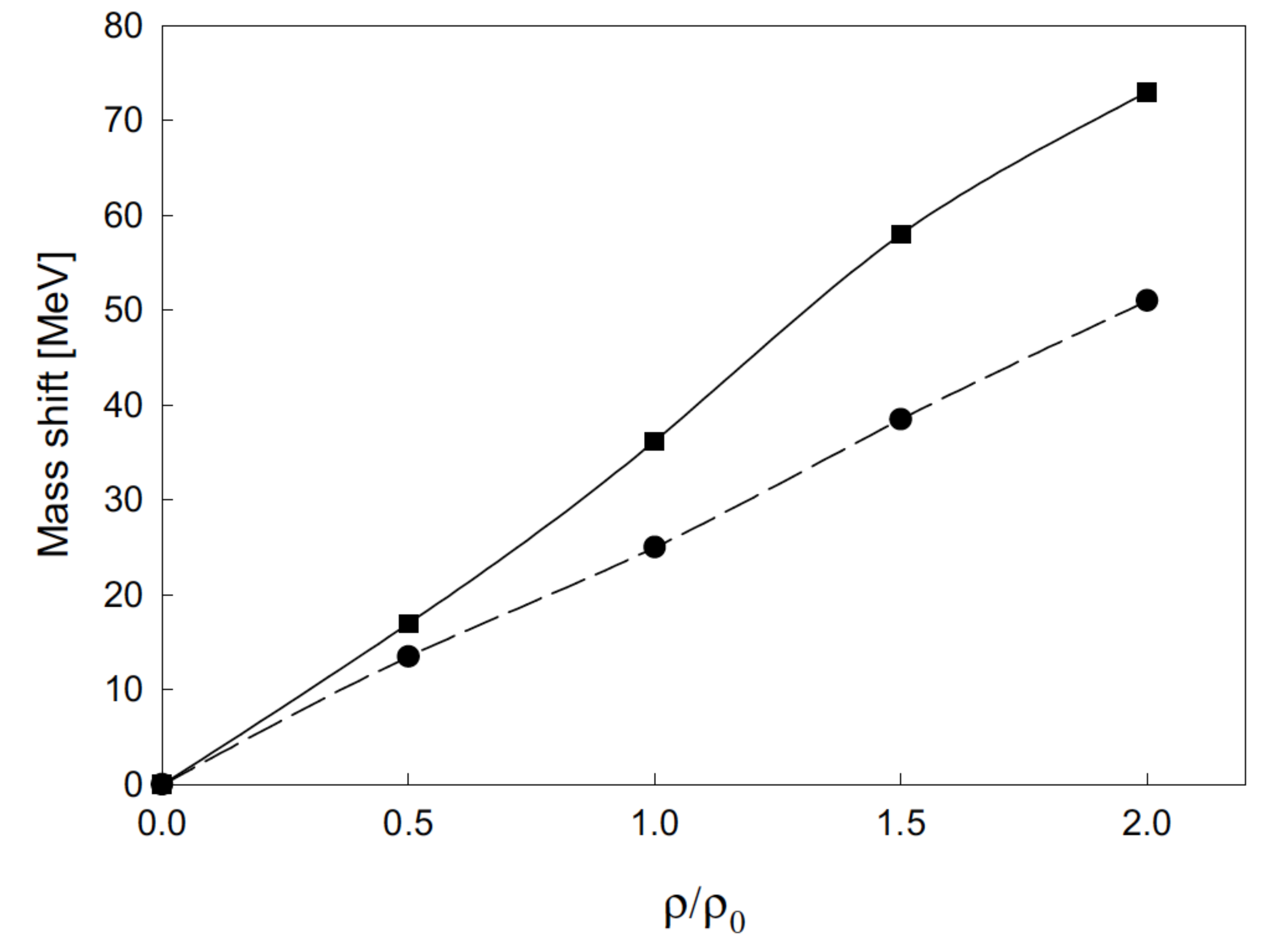}
\vspace*{-2mm} \caption{The kaon mass shift as a function of the nucleon density. The circles show the results based on the vacuum kaon-nucleon scattering amplitude, while the squares correspond to the self-consistent result. The curves smoothly connect the points. The figure is taken from \cite{KorpaLutz}.}
\label{fig:KorpaLutz4}
\end{center}
\end{figure}
 Here, it should be noted that this expression includes terms arising from a scalar and a vector interaction of kaons with nucleons. Without the vector interaction the effective in-medium $K^+$ mass would decrease due to scalar attraction. The vector interaction ultimately leads to repulsion of the $K^+$ meson, while - due to a sign change because of G-parity - it provides the main contribution to the $K^-$ attraction in the nuclear medium \cite{Schaffner_PLB334}, as will be discussed in section \ref{sec:K-}. At low densities,
the kaon in-medium mass, defined as its energy for zero three-momentum, is according to Eqs. (\ref{eq:m*}),(\ref{eq:V=Dm}) then given by
\begin{equation}
m_{K^+}^{*}(\rho_N)=m_K\left[1-\frac{\Sigma_{KN}}{f^2m_K^2}\rho_S+\left(\frac{3}{8}
\frac{\rho_N}{f^2m_K}\right)^2\right]^{1/2}+\frac{3}{8}\frac{\rho_N}{f^2}\approx{m_K+V_{K^+}(\rho_N)},\label{eq:mK+*_long}
\end{equation}
where \cite{Paryev_EPJA92000}
\begin{equation}
V_{K^+}(\rho_N)=-\frac{\Sigma_{KN}}{2f^2m_K}\rho_S+\frac{3}{8}\frac{\rho_N}{f^2}.\label{eq:VK+}
\end{equation}
\begin{figure}[h!]
\begin{center}
\includegraphics[width=10.0cm,clip]{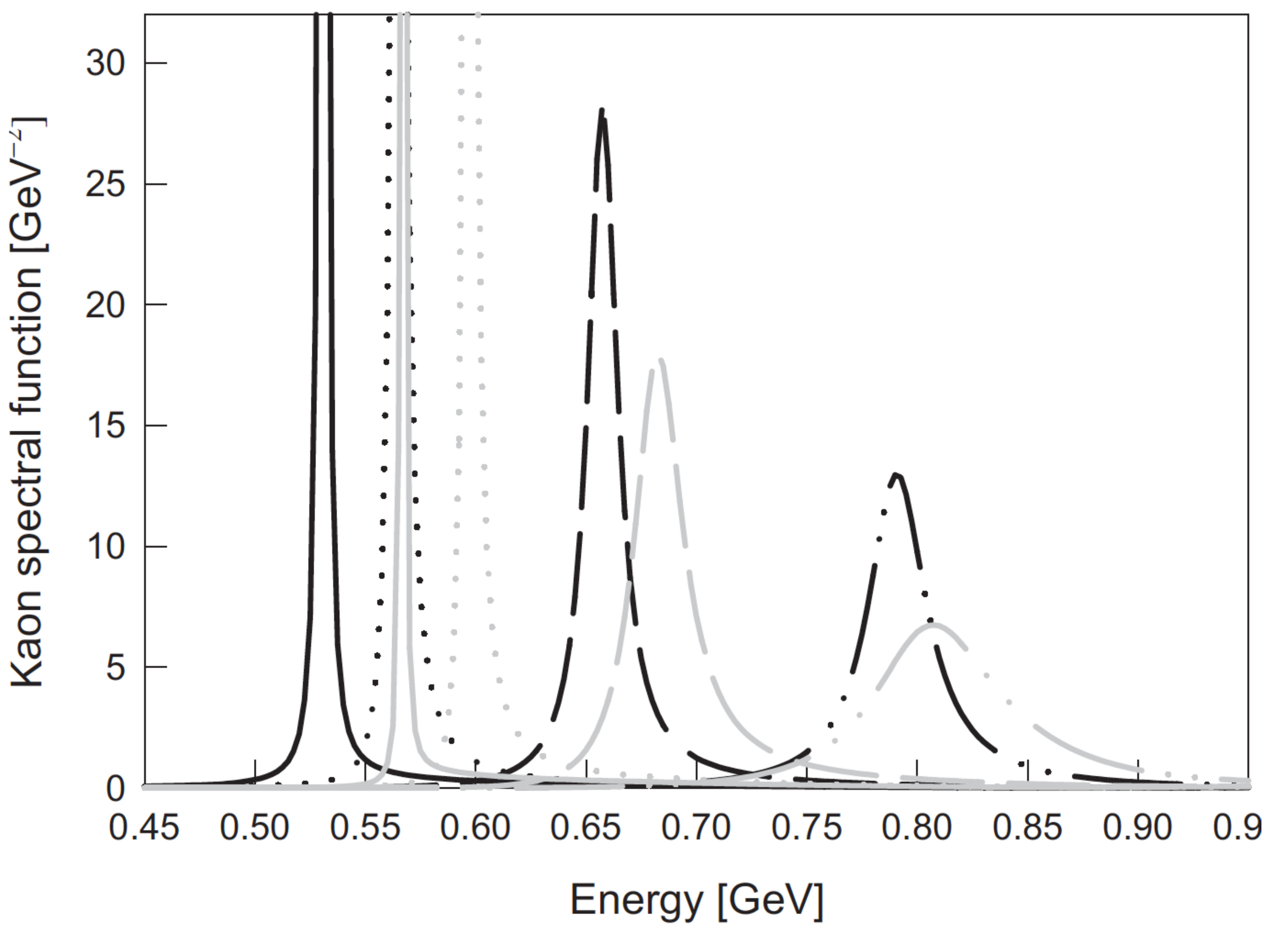}
\vspace*{-2mm} \caption{The kaon spectral function for different momenta: 0 (solid line), 200 MeV/c
(dotted line), 400 MeV/c (dashed line), and 600 MeV/c (dash-dot-dot line). Black lines show the results
at saturation density $\rho_0$, gray lines at twice saturation density.
The figure is taken from \cite{KorpaLutz}.}
\label{fig:KorpaLutz3}
\end{center}
\end{figure}
Since the exact value of $\Sigma_{KN}$ and the size of the higher-order corrections - leading to different
scalar attractions for kaon and antikaon - are not very well known (thus, $\Sigma_{KN}$ may vary from 270 to
450 MeV), the quantity $\Sigma_{KN}$ has been treated in \cite{Li_NPA1997,Li_PRL1997} as a free parameter
which is adjusted separately for $K^+$ and $K^-$ to achieve
good fits to the experimental $K^+$ and $K^-$ spectra in heavy-ion collisions  \cite{Barth_PRL1997,Senger_APP1996}. Using the value of the "empirical kaon sigma term", obtained in \cite{Li_NPA1997,Li_PRL1997}, and accounting for that $\rho_S\approx{}0.9\rho_N$ at $\rho_N \le \rho_0$ \cite{Li_NPA1995}, Eq.~\ref{eq:VK+}  can be rewritten in the form \cite{Paryev_EPJA92000}:
\begin{equation}
V_{K^+}(\rho_N)=+22\mbox{\rm MeV}\frac{\rho_N}{\rho_0}.\label{eq:VK+rhoN}
\end{equation}
Other sophisticated studies within the relativistic mean-field (RMF) approach, the chiral perturbation
theory (ChPT) incorporating different values for $\Sigma_{KN}$ \cite{Schaffner_NPA1997} and in the
framework of the coupled-channel model of Waas et al. \cite{Waas} showed that the $K^+$ mesons indeed feel a weak repulsive potential of about 30--40 MeV at saturation density $\rho_0$ for low momenta (see table 2 and Fig.~\ref{fig:K+K-_pred-Schaffner}). This value is in accordance with those given above.

\begin{figure}[h!]
\begin{center}
\includegraphics[angle = 90,width=18.0cm,clip]{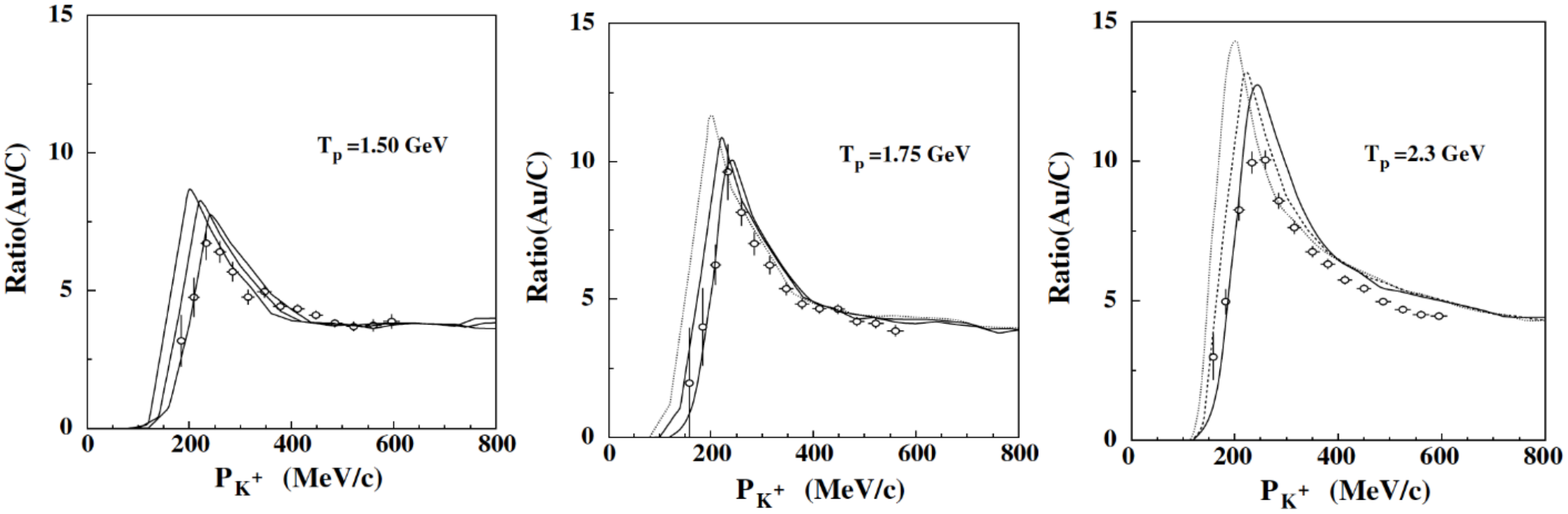}
\vspace*{-45mm} \caption{The ratio of the calculated differential $K^+$ spectra from $p+$Au to $p+$C reactions at 1.5, 1.75 and 2.3 GeV in comparison to the ANKE data \cite{Buscher_EPJA22}.
The lines, starting from the left, correspond to calculations employing repulsive kaon potentials
of 0, 10 and 20 MeV at $\rho_0$. The figure is taken from \cite{CBUU1}. With kind permission of The European Physical Journal (EPJ).}
\label{fig:Rudy4}
\end{center}
\end{figure}

Korpa and Lutz \cite{KorpaLutz} have also studied the kaon properties in cold isospin-symmetric nuclear
matter by solving self-consistently the in-medium Bethe-Salpeter equation. Their results for the $K^+$
mass shift as a function of nucleon density are shown in Fig.~\ref{fig:KorpaLutz4}. The circles here show the mass shift
obtained with vacuum kaon--nucleon scattering amplitude, while the squares correspond to the full
self-consistent solution, based on the in-medium scattering amplitude. One can see that the repulsive
self-consistent mass shift exceeds that based on the vacuum kaon--nucleon scattering amplitude and
reaches the value of 36 MeV at density $\rho_0$. There is no structure in the kaon spectral function,
which has a form of a modestly broadened quasiparticle peak (see Fig.~\ref{fig:KorpaLutz3}). Its position is a few MeV below
\begin{figure}[h!]
\begin{center}
\includegraphics[width=14.0cm,clip]{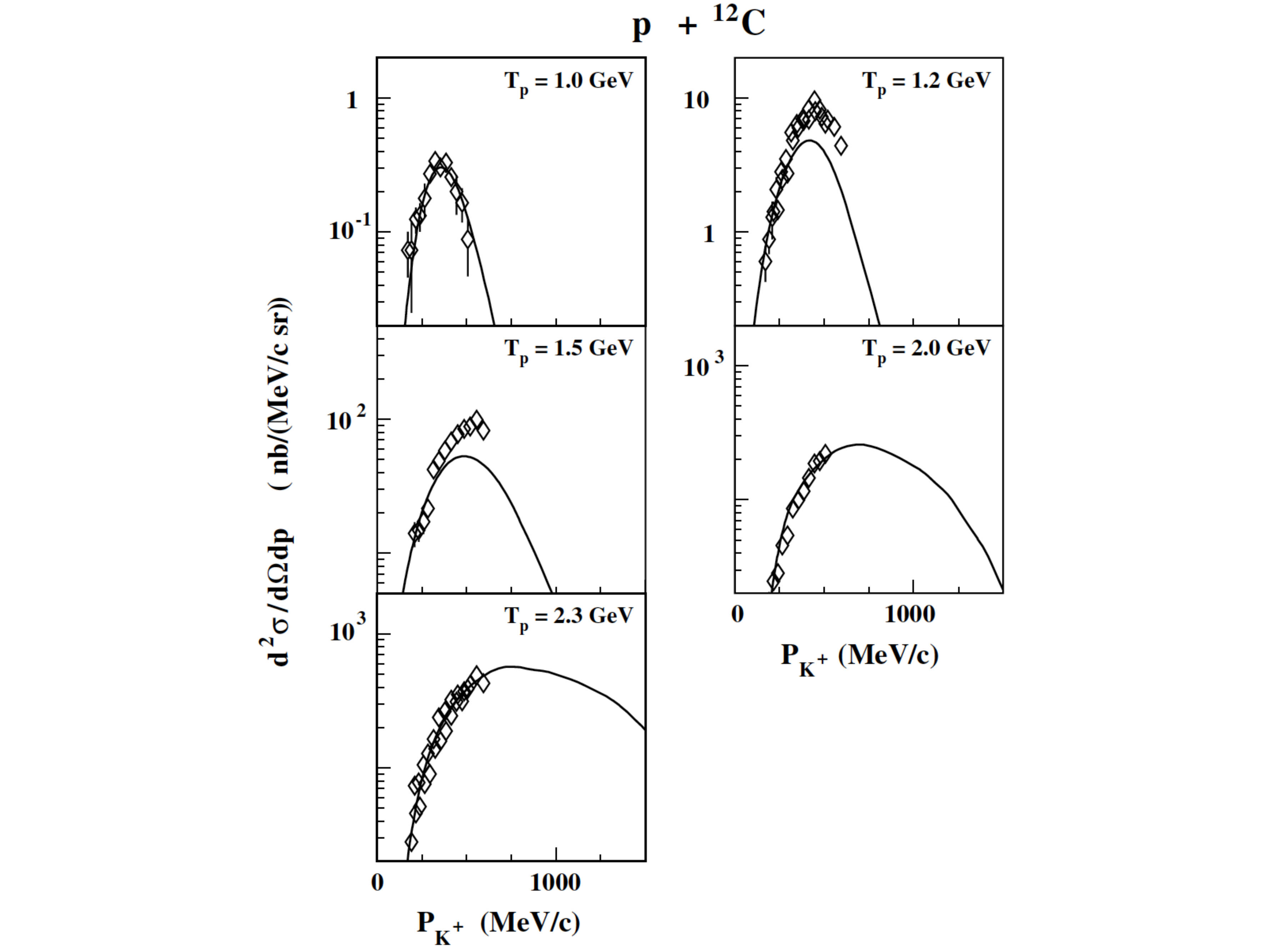}
\vspace*{-2mm} \caption{The $K^+$ spectra from CBUU calculations (solid lines) in comparison
to the data from the ANKE Collaboration \cite{Buscher_EPJA22,Koptev_PRL2001} for
$p+$C reactions at 1.0, 1.2, 1.5, 2.0 and 2.3 GeV . The calculations include both the
momentum-dependent nucleon potential and a repulsive $K^+$ potential of +20 MeV at density $\rho_0$.
The figure is taken from \cite{CBUU1}. With kind permission of The European Physical Journal (EPJ).}
\label{fig:Rudy5}
\end{center}
\end{figure}
the value given by the $\sqrt{{\bf p}_{K^+}^{{\prime}2}+m_{K^+}^{*2}}$ for
momenta $|{\bf p}_{K^+}^{{\prime}}| > 200$ MeV/c, where $m_{K^+}^{*}$ is the peak's position
for ${\bf p}_{K^+}^{{\prime}}=0$ (cf. Eq.(~\ref{eq:EK++})).
Thus, $K^+$ mesons ($K^+={\bar s}u$) at low momenta feel a slightly repulsive nuclear potential of about 20--40 MeV in nuclear matter at normal nuclear matter density $\rho_0$. The dispersion analysis of \cite{SibCas_NPA1998}, which uses as input the vacuum $K^+N$ scattering amplitude, finds that this potential is also repulsive ($\approx{20}$ MeV at density $\rho_0$) and shows only a moderate momentum dependence. Therefore, using in practical calculations
\cite{CBUU,Paryev_EPJA92000,SibCas_NPA1998,Muehlich_PRC2003,CBUU1}  the quasiparticle
dispersion relation (Eq.~\ref{eq:EK++}) with momentum-independent kaon scalar potential
$V_{K^+}(\rho_N)=V_{K^+}^0\rho_N/\rho_0$ with $V_{K^+}^0\approx$ 20--40 MeV, entering in the
 $K^+$ in-medium mass, appears well justified (cf. also the results of Ref. \cite{KorpaLutz}).
%%%%%%%%%%%%%%%%%%%%%%%%%%%%%%%%%%%%%%%%%%%%%%%%%%%%%%%%%%%%%%%%%%%%%%%%%%%%%%%%%%%%%%%%
%%%%%%%%%%%%%%%%%%%%%%%%%%%%%%%%%%%%%%%%%%%%%%%%%%%%%%%%%%%%%%%%%%%%%%%%%%%%%%%%%%%%%%%%%
%%%%%%%%%%%%%%%%%%%%%%%%%%%%%%%%%%%%%%%%%%%%%%%%%%%%%%%%%%%%%%%%%%%%%%%%%%%%%%%%%%%%%%%%%
%%%%%%%%%%%%%%%%%%%%%%%%%%%%%%%%%%%%%%%%%%%%%%%%%%%%%%%%%%%%%%%%%%%%%%%%%%%%%%%%%%%%%%%%%
\begin{figure*}
\centering
\includegraphics[width=13.0cm,clip]{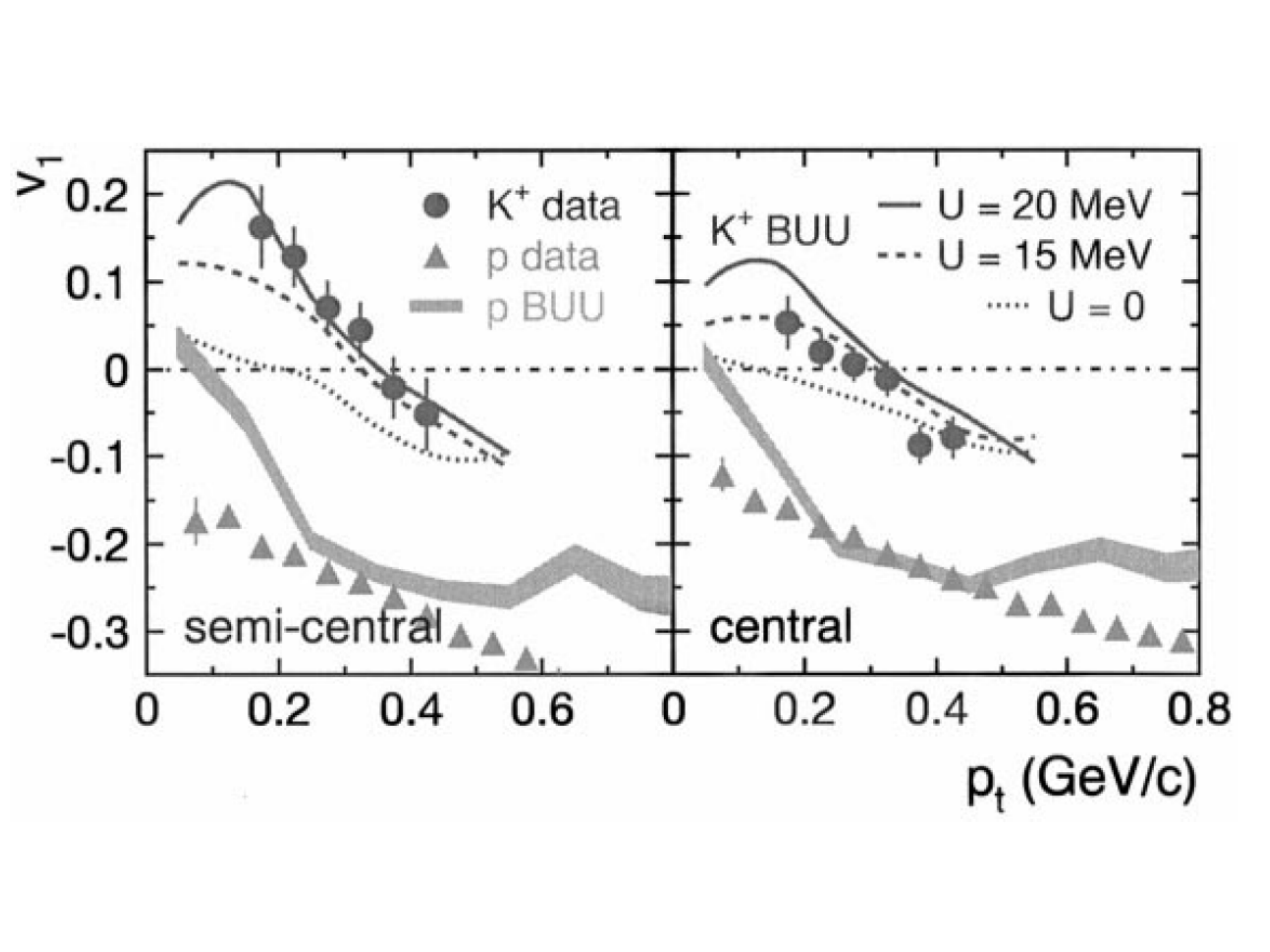}
\vspace{-15mm}\caption{Sideward flow $v_1$ versus transverse momentum $p_t$ for protons (triangles) and K$^+$ mesons (dots) in the rapidity range -1.2 $\le y^{(0)} \le$ -0.65 for semi-central (left) and central (right)
Ru+Ru collisions at 1.69 A GeV \cite{Crochet_FOPI}. The curves and shaded area show the predictions of the RBUU model \cite{Maruyama_NPA573} for K$^+$ and proton, respectively. $K^+$ potentials of 0, 15, and 20 MeV are applied in the model calculations. The figure is taken from \cite{Crochet_FOPI}.}
\label{fig:Crochet_FOPI}
\end{figure*}
%%%%%%%%%%%%%%%%%%%%%%%%%%%%%%%%%%%%%%%%%%%%%%%%%%%%%%%%%%%%%%%%%%%%%%%%%%%%%%%%%%%%%%%%%%%%%%%%%%
\begin{figure*}
\centering
\includegraphics[width=11.0cm,clip]{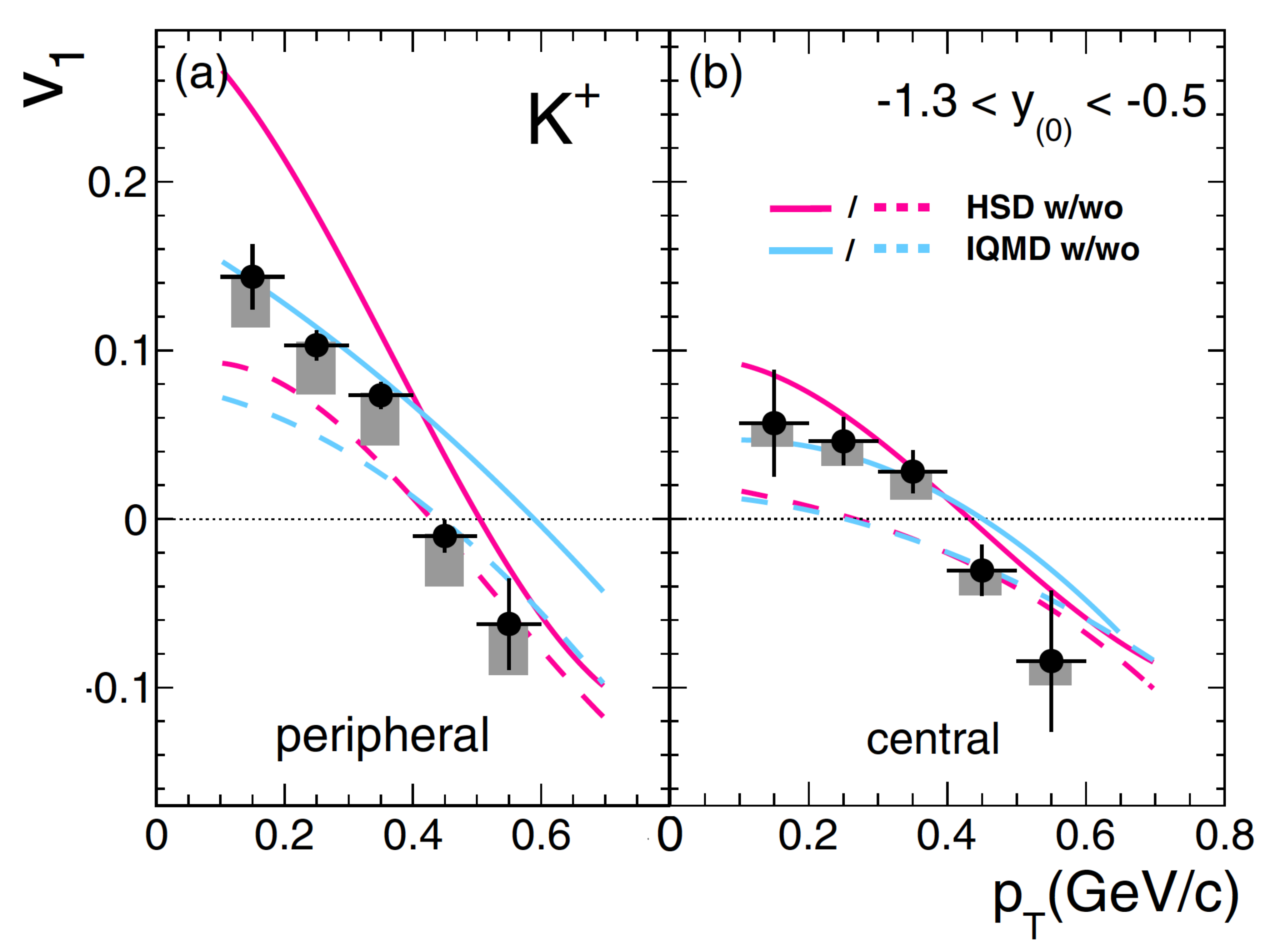}
\vspace{-2mm}\caption{Sideward flow $v_1$ versus transverse momentum $p_T$ for K$^+$ mesons (dots)
in peripheral (a) and central (b) Ni+Ni collisions at 1.91 A GeV in the rapidity range
-1.3 $< y_{(0)} < $ -0.5. The data are compared to HSD \cite{HSD}
and IQMD \cite{Hartnack_EPJA1998}
transport model predictions with (solid lines) and without (dashed lines)
an in-medium kaon potential of 20$\pm$5 MeV for kaons at rest and normal matter density. The potential is assumed to vary linearly with the baryon density. The figure is taken from \cite{Zinyuk_PRC90}.}
\label{fig:Zinyuk_FOPI}
\end{figure*}
%%%%%%%%%%%%%%%%%%%%%%%%%%%%%%%%%%%%%%%%%%%%%%%%%%%%%%%%%%%%%%%%%%%%%%%%%%%%%%%%%%%%%%%%%%%%%%%%%%
\subsubsection{\it Determination of the $K^+$--nucleus real potential}
The predictions of the theoretical approaches, outlined above, on kaon properties in dense matter can
be tested by analyzing proton--nucleus and heavy--ion collision data on kaon spectra, excitation functions
(see Section \ref{sec:real}) and collective flow [34,99-101,112-115].
%\cite{Li_NPA1995,Li_PRL1995,Li_PRC1996,HSD,Brat_NPA1997,Song_NPA1999,Wang_EPJA5,Crochet_NPA1999}.
The latter observable was also recognised as a promising probe of the kaon potential in dense matter,
since it should push the $K^+$ mesons away from the nucleons and attract $K^-$ mesons towards the nucleons.
The analysis of KaoS [106,107,116-119],
%\cite{Barth_PRL1997,Senger_APP1996,Sturm_PRL2001,Shin_PRL1998,Menzel_PLB2000,Forster_PRL2003},
FOPI [120-122] 
%\cite{Herrmann_PPNP1999,Ritman_ZPA1995,Crochet_FOPI}
and ANKE \cite{Buscher_EPJA22}
on $K^+$ production in heavy--ion and proton--nucleus reactions established in the framework of transport approaches 
\cite{HSD,Li_PRL1997,Li_NPA654,Larionov_PRC2005,Srisawad} 
and \cite{CBUU,Nekipelov,CBUU1} that the $K^+$ meson indeed feels a moderately repulsive nuclear potential of about 20--30 MeV at normal nuclear matter density $\rho_0$, in agreement with the above theoretical estimates. As an example of such analysis, 
Fig.~\ref{fig:Rudy4} shows the ratio of differential $K^+$ spectra from $p+$Au to $p+$C collisions at 1.5, 1.75 and 2.3 GeV calculated within the CBUU transport model in comparison to the data from \cite{Buscher_EPJA22}.
One can see that at low momenta ($p_{\rm lab} <$ 230 MeV/c)
the production of $K^+$ mesons on the gold target is suppressed with respect to their creation on the
carbon target. This observation can be explained by the combined repulsive effect of both the Coulomb and
nuclear fields, which accelerate kaons before they escape from the nucleus. 
The CBUU calculation, which includes a repulsive kaon potential of 20 MeV, provides a good
description of the experimental data over the full momentum range at all beam energies. A $\chi^2$-fit of the data with the calculated ratio of the differential $K^+$ spectra for the adopted scenarios for kaon potential of 0, 10 and 20 MeV gives $V_{K^+}(\rho_0)=20\pm5$ MeV \cite{CBUU1}. It should be pointed out that the systematic analysis within the CBUU transport model \cite{CBUU1} of the experimental $K^+$ spectra from C, Cu, Ag and Au targets taken at COSY-Juelich \cite{Buscher_EPJA22} also gives the same kaon potential at normal nuclear matter density (see Fig.~\ref{fig:Rudy5}).

 Another example of such an analysis is given in Fig.~\ref{fig:Crochet_FOPI} which shows the differential experimental data
on the transverse momentum and centrality dependence of the $K^+$ meson and proton sideward collective
flows, measured with the FOPI detector at SIS/GSI in the reactions Ru+Ru at 1.69 A GeV \cite{Crochet_FOPI}. The comparison to the predictions of the RBUU transport model \cite{Maruyama_NPA573} also clearly favours the existence of an in-medium weak repulsive $K^+$ mean-field potential of about 20 MeV at density $\rho_0$.
In line with this result, recent findings on the azimuthal emission pattern of $K^+$ mesons in Ni+Ni collisions at a beam kinetic energy of 1.91 A GeV, obtained by the FOPI Collaboration \cite{Zinyuk_PRC90}, also support the existence of a repulsive kaon--nucleus potential of 20$\pm$5 MeV for particles at rest and at saturation density (see Fig.~\ref{fig:Zinyuk_FOPI}). Employing this kaon potential, both the HSD and IQMD transport approaches describe correctly the data in the central event sample. The IQMD model is also in agreement with the data for the peripheral event sample, while within
the HSD approach the transverse momentum dependence of  $v_1$ is overpredicted.

In summary, the currently available theoretical predictions and experimental data on $K^+$ production
off nuclear targets indicate consistently a slightly repulsive in-medium $K^+$ potential of about 20--30 MeV at density $\rho_0$ and low momenta.
\\

%----------------------------------------------------------------------------------------------------------------------------------------------------------------------------------------------------------------------------------------------
%K^0

\subsection{\it K$^0$--nucleus potential}
\label{sec:K0}
\subsubsection{\it In-medium neutral kaon potential. Theoretical expectations}

\begin{figure}[h!]
\begin{center}
\includegraphics[width=12.0cm,clip]{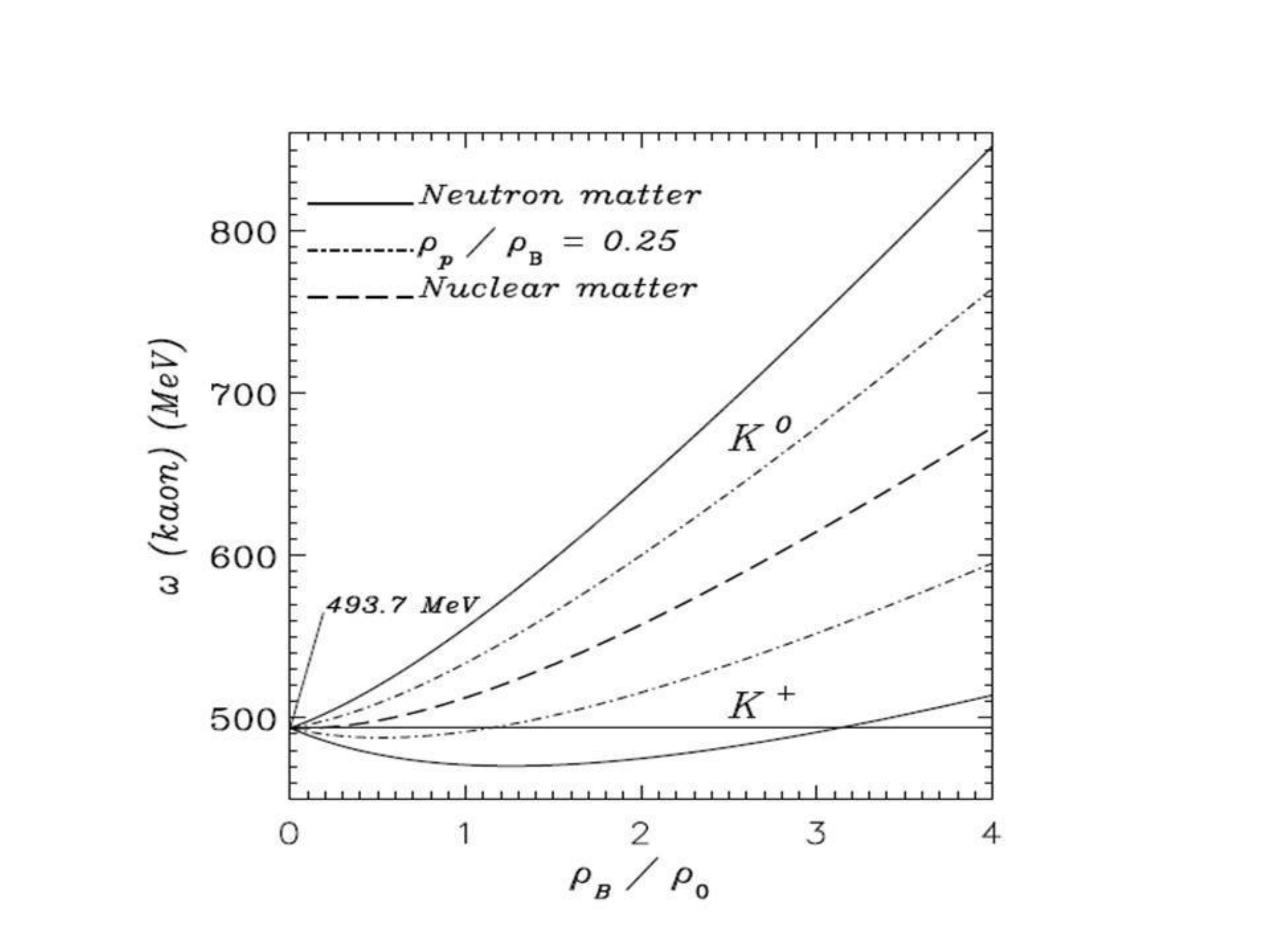}
\vspace*{-2mm} \caption{$K^+$ and $K^0$ energies at zero momentum versus baryon density for symmetric nuclear matter (dashed curves), for $\rho_p/ \rho_B$ = 0.25 (dashed dotted curves) and for neutron matter (solid curves). The figure is taken from \cite{Tsushima_PLB1998}.}
\label{fig:QMC3}
\end{center}
\end{figure}

Similar to the $K^+$ meson, the real part of the $K^0$ in-medium nuclear potential $V_{K^0}$ is expected to be repulsive as well
due to its quark content ($K^0={\bar s}d$). Both these mesons have one light quark
($u$ in $K^+$ and $d$ in $ K^0$) which repulsively interacts with the quarks ($u$ and $d$ ) in the nuclear medium due to
Pauli blocking. In this intuitive physical picture in-medium modifications of $K^0$ mesons
should be close to those for $K^+$ mesons in symmetric nuclear matter with equal densities of protons
and neutrons, $\rho_p=\rho_n=\rho_N/2$. This is also true from the hadronic point of view that the $K^0$ self-energy $\Pi_{K^0}$ in nuclear matter can be obtained from the $K^+$ self-energy $\Pi_{K^+}$ by an isospin transformation as $\Pi_{K^0}$=$\Pi_{K^+}$ \cite{Kaiser_Weise}.
Indeed, according to the predictions of the Quark-Meson Coupling (QMC)
model by Tsushima et al. \cite{Tsushima_PLB1998}, in which the scalar ($\sigma$) and the vector ($\rho$, $\omega$)
mesons are assumed to couple directly to the nonstrange quarks and antiquarks in the $K$ and ${\bar K}$ mesons,
one expects the neutral kaon in-medium mass $m_{K^0}^{*}$ to be equal to $m_{K^+}^{*}$ of $K^+$, i.e.
$m_{K^0}^{*}=m_{K^+}^{*}$, as illustrated in Fig.~\ref{fig:QMC3} (see also \cite{Klingl_PLB1998,Oset_Ramos_NPA2001}). On the other hand, as we observe here, the $K^+$ and $K^0$ mesons
are not mass degenerate in an asymmetric nuclear medium. In the QMC model, this is a consequence of the $\rho$ meson,
which induces different mean-field potentials for each member of the isodoublets, $K$ and ${\bar K}$, when they
are embedded in asymmetric nuclear matter.

\subsubsection{\it Determination of the $K^0$--nucleus real potential}
\begin{figure*}
\centering
\includegraphics[width=11.0cm,clip]{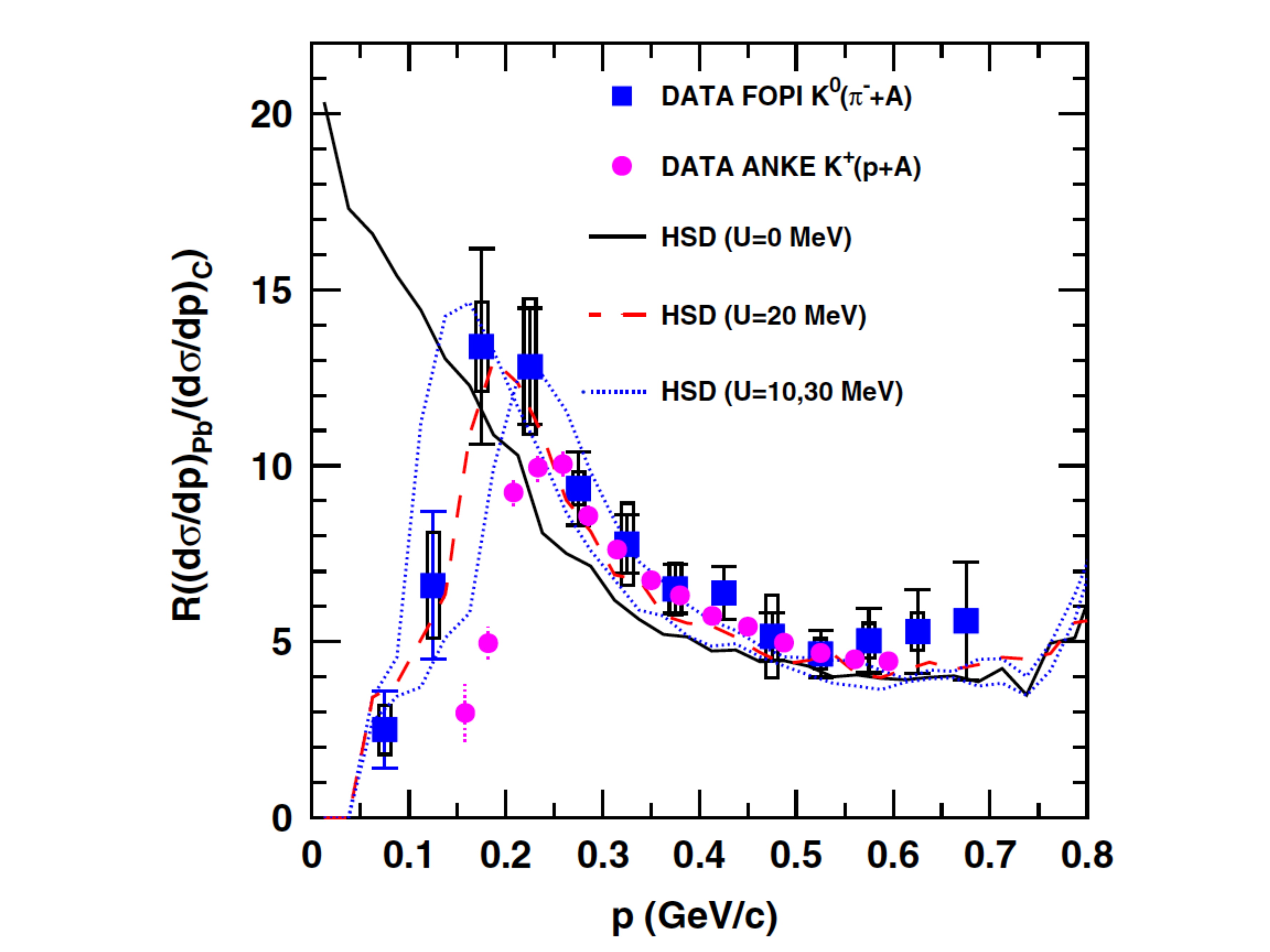}
\caption{The ratio of K$^0$ (K$^+$) yields produced by pions (protons) on heavy and light targets as a function
of the laboratory momentum. The full squares depict the ratio of K$^0_s$ produced on Pb and C targets \cite{FOPI_K0}. The ratio of K$^+$ yields measured in proton-induced reactions on Au and C targets at 2.3 GeV
is represented by full circles \cite{Buscher_EPJA22}. The results of the HSD model \cite{HSD,Cassing_HSD} for different strengths of the $K^0$ potential are depicted by solid (black), dashed (red), and dotted (blue) lines. The figure is taken from \cite{FOPI_K0}.}
\label{fig:FOPI_K0}
\end{figure*}

\begin{figure}[tb]
\begin{minipage}[t]{10 cm}
\includegraphics[width=14.0cm,clip,angle=90]{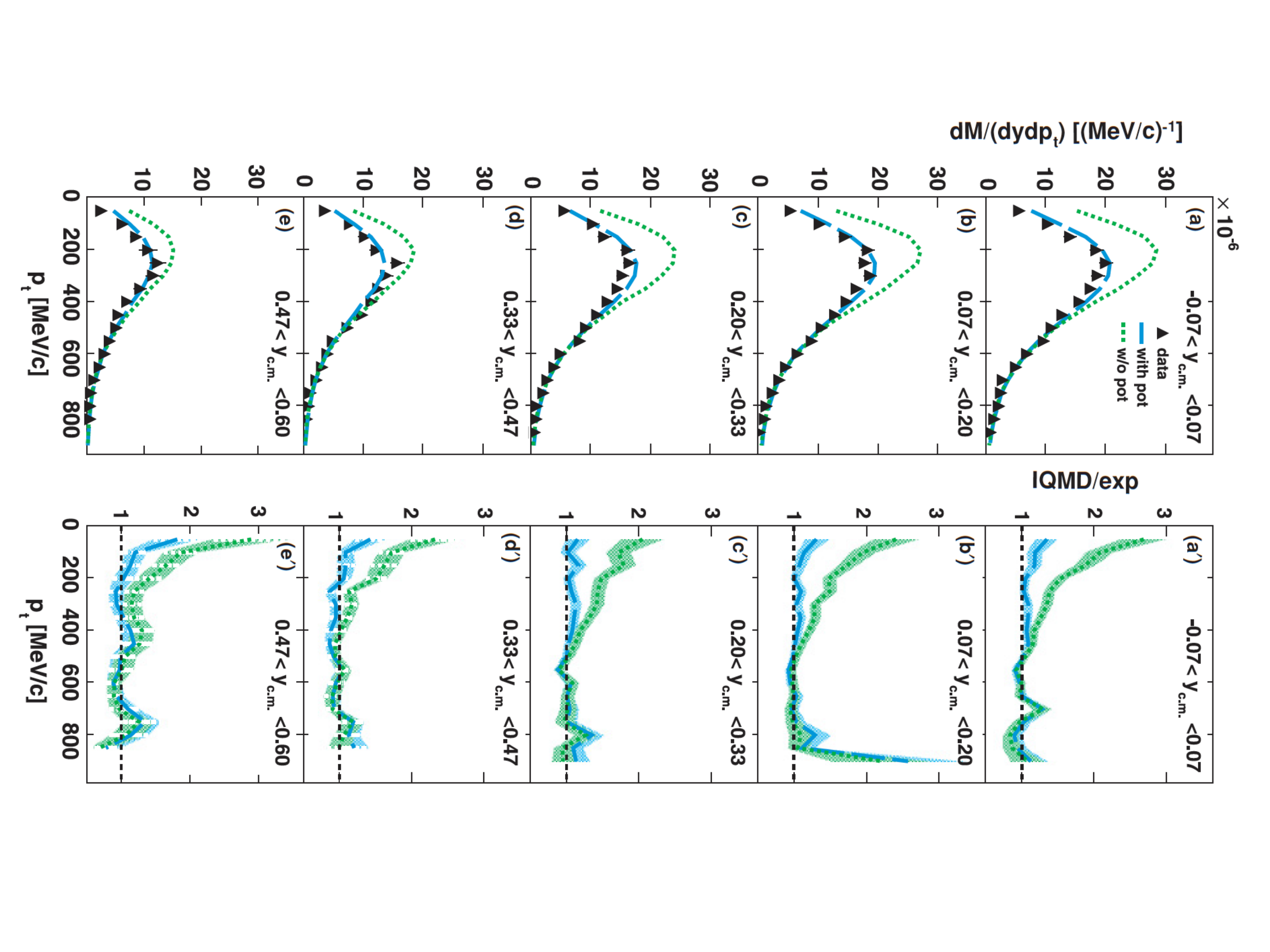}
\end{minipage}
\begin{minipage}[t]{8.5 cm}
\vspace{-10cm} 
\caption{Left: transverse momentum p$_t$ distributions of $K_s^0$ 
mesons (full triangles) for Ar + KCl collisions at 1.756 AGeV \cite{HADES_K0} in comparison with IQMD model calculations \cite{Aichelin}
for a repulsive K$^0$--nucleus potential of 46.1 MeV (dashed curves) and without potential (dotted curves) for different rapidity bins. Right: ratio of IQMD model calculations and experimental data for both scenarios as a function of p$_t$ for the same rapidity bins. The figure is taken from \cite{HADES_K0}.\label{fig:HADES_K0}}
\end{minipage}
\end{figure}

\begin{figure*}
\centering
  \includegraphics[width=10.0cm,clip]{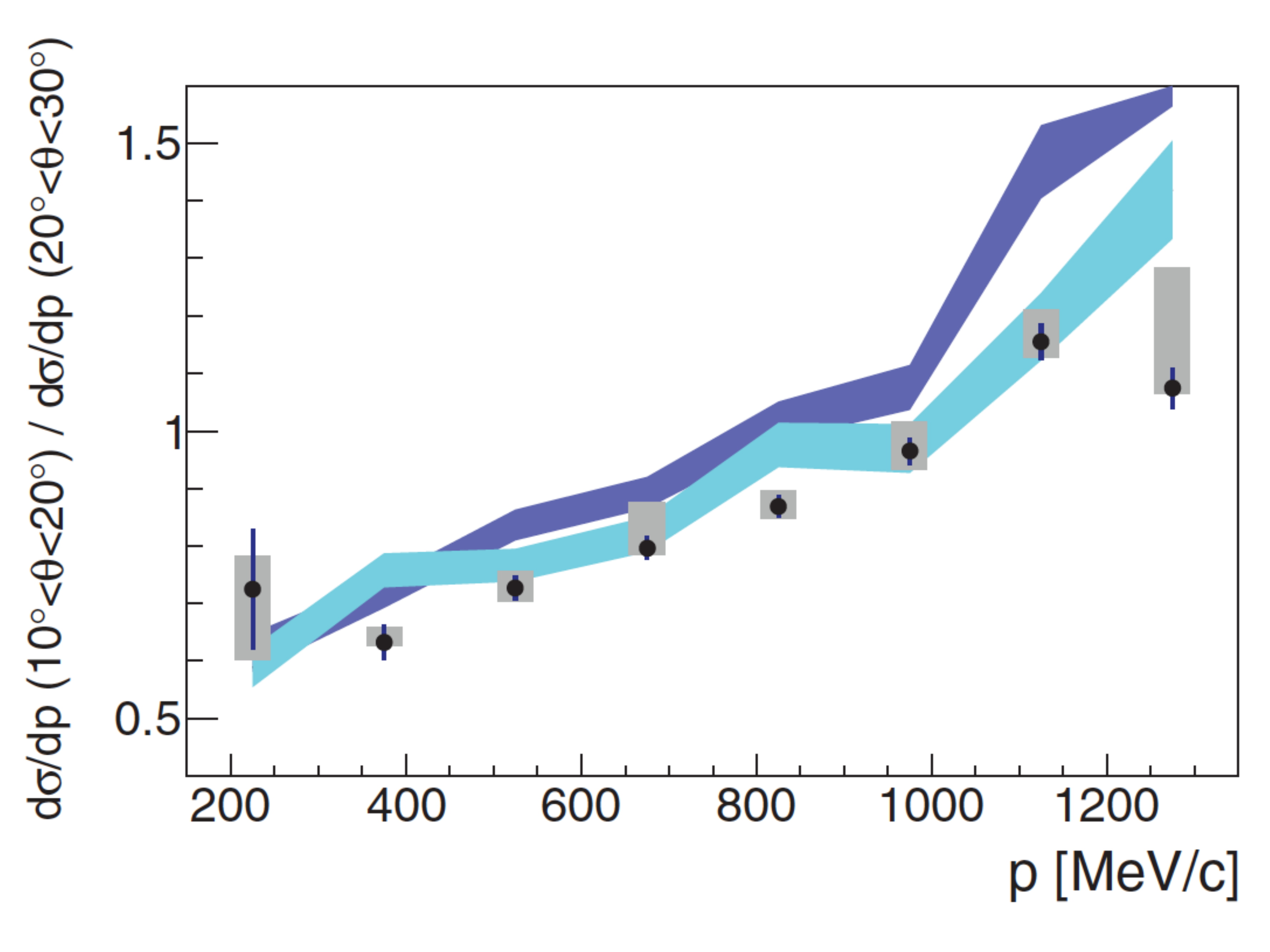}
\caption{Ratio of $K^0$ momentum spectra measured in p+Nb collisions at 3.5 GeV \cite{HADES_K0_p}
in two neighbouring bins of the polar angle (black circles) and GiBUU transport model simulations \cite{GiBUU}
with (cyan) and without (blue) an in-medium ChPT repulsive momentum-dependent
kaon potential defined as the difference of the in-medium (cf. Eq.~\ref{eq:EK+}) and vacuum kaon energies and described
in detail in \cite{HADES_K0_p} (it amounts to $\approx$ +35 MeV for the kaon at rest and for density $\rho_0$).
Both the momentum and the polar angle are in the laboratory reference frame. The figure is taken from \cite{HADES_K0_p}.}
\label{fig:HADES_K0_p}
\end{figure*}
Neutral kaons can help to get a deeper insight into the kaon potential in the nuclear medium. As compared to $K^+$
mesons, they have the essential advantage that possible medium effects are not shaded by the additional repulsive
Coulomb interaction, which for heavy nuclei like Au amounts approximately to 17 MeV, i.e. it is of the same order of
magnitude as the repulsive $K^+$ nuclear mean-field potential. These medium effects have been studied in the $K^0$
meson production on several nuclear targets in $\pi^-$-induced reactions at 1.15 GeV/c momentum by the FOPI
Collaboration at SIS/GSI \cite{FOPI_K0}. Compariing the ratio of the measured 
$K_s^0$ momentum distributions from $\pi^-$+Pb and $\pi^-$+C reactions \cite{FOPI_K0} with HSD transport model calculations
\cite{HSD,Cassing_HSD}, a repulsive $K^0$--nucleus potential of about 20 MeV at a
normal nuclear matter density is suggested, as is illustrated by Fig.~\ref{fig:FOPI_K0}.

It is clearly seen that the calculation 
including a +20 MeV $K^0$ nuclear potential at $\rho_0$ (with a linear dependence of the potential on nuclear
density), depicted by the dashed line in Fig.~\ref{fig:FOPI_K0}, reproduces qualitatively the observed dependence of the $K_s^0$ differential yield ratio both at low ($p_{\rm lab} < $ 170 MeV/c) and at high ($p_{\rm lab}\approx650$ MeV/c) kaon momenta. Whereas the HSD calculation without this potential, shown by the solid curve in Fig.~\ref{fig:FOPI_K0}, misses completely the data at low momenta. A $\chi^2$-fit of the data with the calculated ratio of the differential $K_s^0$ spectra for the employed scenarios for $K^0$ potential of 0, 10, 20 and 30 MeV (cf. also dotted lines in Fig.~\ref{fig:FOPI_K0}) gives $V_{K^0}(\rho_0)=20\pm5$ MeV \cite{FOPI_K0}. It is worth noting that the same strength of the repulsive $K^+A$ interaction has been extracted as well from $K^+$ production in $pA$ reactions as studied by the ANKE Collaboration (see Section \ref{sec:K+}).

On the other hand, the data on $K_s^0$ transverse momentum spectra and rapidity distributions in Ar+KCl
reactions at a bombarding energy of 1.756 A GeV, collected recently by the HADES Collaboration, point to
the existence of a stronger repulsive in-medium $K^0$ potential of about 40 MeV at saturation density
$\rho_0$ for kaons at rest \cite{HADES_K0} as illustrated by Fig.~\ref{fig:HADES_K0}. The IQMD simulations
reproduce rather well the measured $K_s^0$ $p_t$ distributions for all rapidity bins and over the full
momentum range assuming a repulsive $K^0$ potential of 46.1 MeV (dashed curves), whereas the calculations without
the potential overestimate the experimental data, especially at low transverse momenta ($<$ 400 MeV/c), as is
nicely seen in the ratio plots given in Fig.~\ref{fig:HADES_K0} (Right). Note, however, that the IQMD calculations involve less baryon resonances than HSD.

The HADES Collaboration has also recently reported a measurement of inclusive $K^0$ production in p+Nb
collisions at a beam kinetic energy of 3.5 GeV \cite{HADES_K0_p}. The obtained $K^0$ data
(phase-space distributions and the ratio of momentum spectra) were compared to theoretical calculations
with the GiBUU transport model \cite{GiBUU} leading to an estimate of the
strength of the repulsive $K^0$ potential of 40$\pm$5 MeV for particles at rest
and at saturation density $\rho_0$ \cite{HADES_K0_p}. This value is consistent with that inferred also by the
HADES Collaboration from the analysis of $K^0$ meson production in heavy--ion reactions \cite{HADES_K0} (see above).
As an example for such a comparison, Fig.~\ref{fig:HADES_K0_p} shows the ratios of measured and calculated $K^0$ momentum spectra (obtained within the GiBUU transport model) for p+Nb collisions at 3.5 GeV beam energy in two adjacent bins of the lab polar angle. The GiBUU calculation, which includes the ChPT repulsive kaon potential, gives indeed a better description of the considered experimental data.

Thus, the  presently available experimental data on $K^0$ meson production in pion--nucleus, proton--nucleus and
heavy--ion collisions strongly support the existence of a repulsive in-medium $K^0$ potential of about 20--40 MeV at density $\rho_0$ for neutral kaons at rest, consistent with that obtained for the $K^+$ nuclear potential (see Section \ref{sec:K+}).

%----------------------------------------------------------------------------------------------------------------------------------------------------------------------------------------------------------------------------------------------
%K^-
\subsection{\it $K^-$--nucleus potential}
\label{sec:K-}
%%%from E.P

\subsubsection{\it In-medium antikaon potential. Theoretical predictions}
The study of the ${\bar K}N$ two-body interaction at energies close to or below the kaon--nucleon threshold
both in free space and in nuclear matter as well as antikaon interaction with nuclei has received considerable
interest in recent years [7,8,10,23,108,109,131,139-149].
%\cite{Friedman_Gal_PR, Friedman_Gal_NPA881,Gal_NPA914,Waas,KorpaLutz,SibCas_NPA1998,Tsushima_PLB1998,Koch_PLB1994,Waas1,Ohnishi_PRC56,
%Lutz_PLB1998,Cieply_NPA2001,Ramos_Oset,Weise_nuclth,Ikeda_Hyodo_Weise_PLB706,Ikeda_Hyodo_Weise_NPA881,Cieply_PRC2011,Cassing_NPA727}.
\begin{figure}[h!]
\begin{center}
\includegraphics[width=18.0cm,clip]{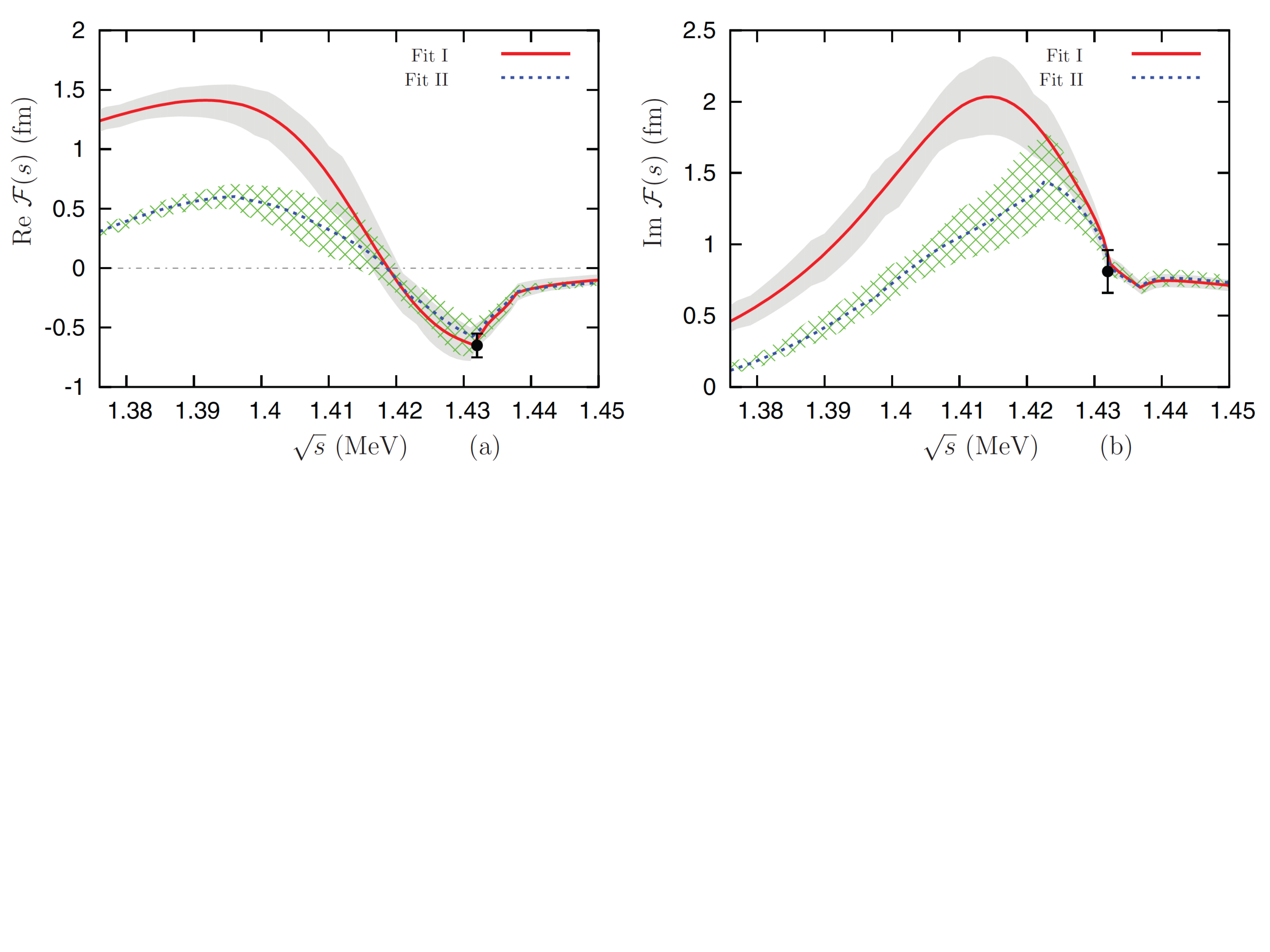}
\vspace*{-6.5cm} \caption{Extrapolation of the real (left panel) and imaginary (right panel) parts
of the amplitude $K^-$$p$$\to$$K^-$$p$ scattering to the subthreshold energy region generated in
two NLO chiral-model fits \cite{Guo_Oller}. The $K^-$$p$ threshold values marked by solid dots with error bars 
were obtained from SIDDHARTA $K^-$ hydrogen data \cite{Bazzi_PLB704,Bazzi_NPA881}. The figure is taken from \cite{Guo_Oller}.}
\label{fig:K-scat_ampl}
\end{center}
\end{figure}
This interest has been motivated by the hope of getting valuable information on a possible
$K^-$ condensation in the interior of neutron stars \cite{Kaplan_Nelson,Brown_Bethe} as well as of exotic nuclear objects such as
antikaon-nuclear bound states, i.e. $K^-$--few--nucleon systems (antikaonic clusters) or bound states of a $K^-$
in heavier nuclei (see Section \ref{sec:K-cluster}). But in spite of large efforts, contrary to the K$^+$ and K$^0$ mesons,
the $K^-$ meson ($K^-$=$s{\bar u}$) properties in nuclear matter are much less known and still controversially
debated. 

The free $K^-$$p$ interaction at threshold and above threshold energies
is repulsive and well constrained \cite{Guo_Oller} by the SIDDHARTA $K^-$ hydrogen data \cite{Bazzi_PLB704,Bazzi_NPA881} as well as by the low-energy
scattering and reaction data \cite{Evans} (and references therein). This is reflected in the negative real part of the elastic $K^-$$p$ forward
scattering amplitude (which is an average of the isospin $I=0$ and $I=1$ components) at and above threshold, as is demonstrated by the left panel of Fig.~\ref{fig:K-scat_ampl}. 
From the strong interaction induced energy shift and width of the 1s state in kaonic hydrogen, Ikeda et al. \cite{Ikeda_Hyodo_Weise_PLB706,Ikeda_Hyodo_Weise_NPA881} extract a free $K^-$$p$ scattering length of $a_{K^- p} = (-0.65+i 0.81)$ fm.

However, at subthreshold energies the free $K^-$$p$ interaction is attractive (real part of the $K^-$$p$ forward scattering amplitude is positive) and is extremely model dependent.
As demonstrated in Fig.~\ref{fig:K-scat_ampl}, two different NLO chiral-model fits to the available experimental data on
low-energy $K^-$$p$ reactions and SIDDHARTA $K^-$ hydrogen data generate essentially different $K^-$$p$ scattering amplitudes in the subthreshold region.
Here, the $K^-$$p$ interaction is dominated by the presence of the $\Lambda(1405)$ resonance
in the isospin $I=0$ channel just 27 MeV below the  $K^-$$p$ threshold. The interaction
in the isospin $I=1$ channel, corresponding to the two-body $K^-$$n$ free space interaction, is not affected
by the $\Lambda(1405)$ resonance and is moderately attractive both above and below threshold \cite{Cieply_PRC2011}. As long as the 1s energy shift and width in kaonic deuterium is not measured - corresponding experiments are in preparation \cite{Zmeskal_APP46,Curceanu_NPA914} - the $K^-$$n$ scattering length has to be calculated. Using the Kyoto $\bar{K}N$ potential, $a_{K^- n} = (0.58+i 0.78)$ fm has been obtained in \cite{Hoshino_1705.06857}. 
The existence of the subthreshold $\Lambda(1405)$ resonance leads to the pronounced peak in imaginary part of the free $K^-$$p$ forward scattering amplitude,
shown in the right panel of Fig.~\ref{fig:K-scat_ampl}, and to the change of sign of its real part at energies closely below the $K^-$$p$ threshold. 
The $\Lambda(1405)$ resonance is dynamically generated and corresponds to
two-poles in the complex energy plane of the $I=0$ $K^-$$p$ and ${\pi}$${\Sigma}$ scattering $S$-wave amplitudes
as is established in [150,157-159].
%\cite{Guo_Oller,Oller_Meissner,Jido_NPA725,Meissner_Hyodo}. 
The lower (broader) mass pole (at about 1330 MeV) mainly couples to the ${\pi}$${\Sigma}$ channel, while the higher pole located at about 1430 MeV is more coupled to the ${\bar K}N$ channel. For this reason the line shape of the $\Lambda(1405)$ resonance depends on the initial state of the reaction and on the observed decay channel. The position of its lower pole in the complex energy plane
is less well known than the position of the higher one \cite{Guo_Oller}, although the parameters of the two poles have recently been further constrained \cite{Meissner_Hyodo,Mai_Meissner_EPJA51} by including recent photoproduction data of the $\gamma p \rightarrow K^+ \Sigma \pi$ reaction \cite{Moriya_PRL112} in the analysis. The remaining uncertainty in the position of the lower pole has an important influence on the subthreshold extrapolation of the elastic $K^-$$p$ forward scattering amplitude
(cf. Fig.~\ref{fig:K-scat_ampl}) -- a crucial input for our understanding of antikaon properties in nuclear matter. 

While antikaonic atoms probe $K^-$N c.m. energies down to about 30--50 MeV below 
the $K^-N$ threshold, $K^-$ - nuclear 1s quasi-bound states would reach substantially lower energies
 \cite{Friedman_Gal_NPA881,Gal_NPA914,Friedman_Gal_NPA959,Cieply_PRC2011,Gal_EPJWeb2014}, where the role of the $\Lambda(1405)$ subthreshold resonance is
expected to be dominant in the construction of medium-modified subthreshold $K^-$N scattering amplitudes and relevant nuclear antikaon potentials.
The $\Lambda(1405)$ resonance experiences very strong modifications in the nuclear medium at densities around
(0.1--0.2) $\rho_0$ (it dissolves) mainly due to Pauli blocking of the intermediate proton states \cite{Waas,Waas1}.
Pauli effects reduce the attractive force which binds $K^-$ and $p$ in the $\Lambda(1405)$ resonance and shift it to higher energies \cite{Waas,Koch_PLB1994,Waas1}.
This strongly affects the in-medium $K^-$$p$ scattering amplitude close to threshold \cite{Waas,Koch_PLB1994,Waas1,Cassing_NPA727}, which together with the in-medium (smooth) $K^-$$n$ (I = 1) amplitude determines the antikaon self-energy in nuclear matter. Since the $\Lambda(1405)$ has a $K^-$--proton bound state component,
the $K^-$ and nucleon self-energies should be accounted for in the in-medium dynamics of the $\Lambda(1405)$ 
or in the subthreshold behavior of the in-medium ${\bar K}N$ scattering amplitudes.  This means that the $K^-$
in-medium properties and the $\Lambda(1405)$ states in matter should be treated self-consistently [8,108,142-144,148,149].
%\cite{Friedman_Gal_NPA881,KorpaLutz,Lutz_PLB1998,Cieply_NPA2001,Ramos_Oset,Cieply_PRC2011,Cassing_NPA727}.
It was shown first in \cite{Lutz_PLB1998} that such self-consistent treatment leaves the position of the
$\Lambda(1405)$ resonance practically unchanged, due to a compensation of its repulsive mass shift resulting
from Pauli effects with the attraction felt by the $K^-$ meson, however, with an increased decay width.
In this case the real part of the in-medium $K^-$$p$ scattering amplitude remains positive (attractive)
in the energy region around the threshold in agreement with phenomenological analyses of antikaonic atoms
and experiences a sharp increase when going to subthreshold energies \cite{Cieply_PRC2011}. This means that the $K^-$$p$
interaction becomes much stronger at energies $\sim$ 30 MeV below the $K^-$$p$ threshold with respect
to its strength at threshold.

The real part of the $K^-$--nuclear potential, relevant for antikaonic atoms,
can be estimated from the real part of the effective in-medium density-dependent isospin-averaged 
$K^- N$ scattering length ${\bar a}_{eff}(\rho_N)$ using the single--nucleon approximation \cite{Friedman_Gal_NPA881,Li_PLB1994,Waas1}

\begin{figure*}
\centering
\includegraphics[width=8.5cm,clip]{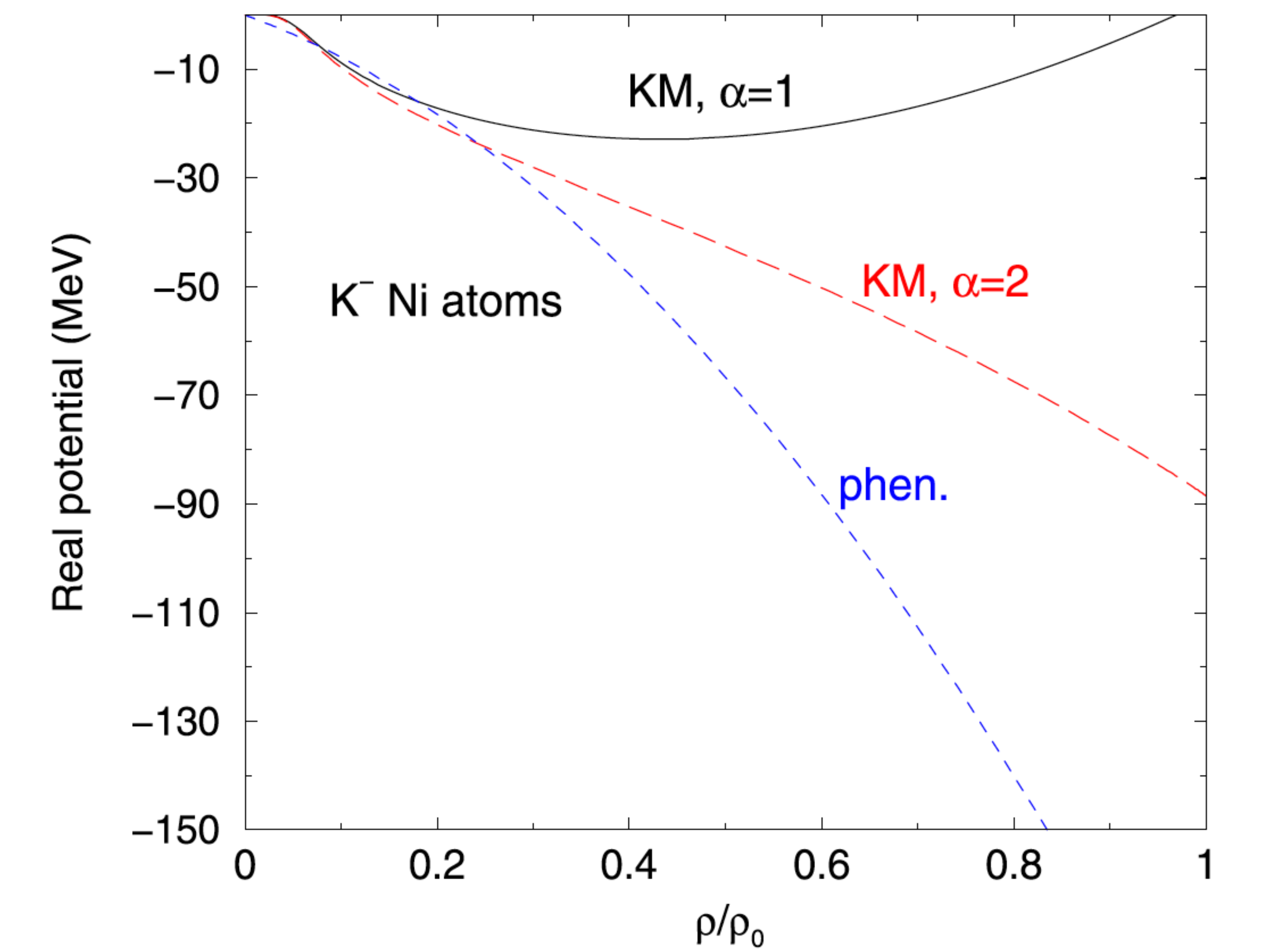}\includegraphics[width=8.5cm,clip]{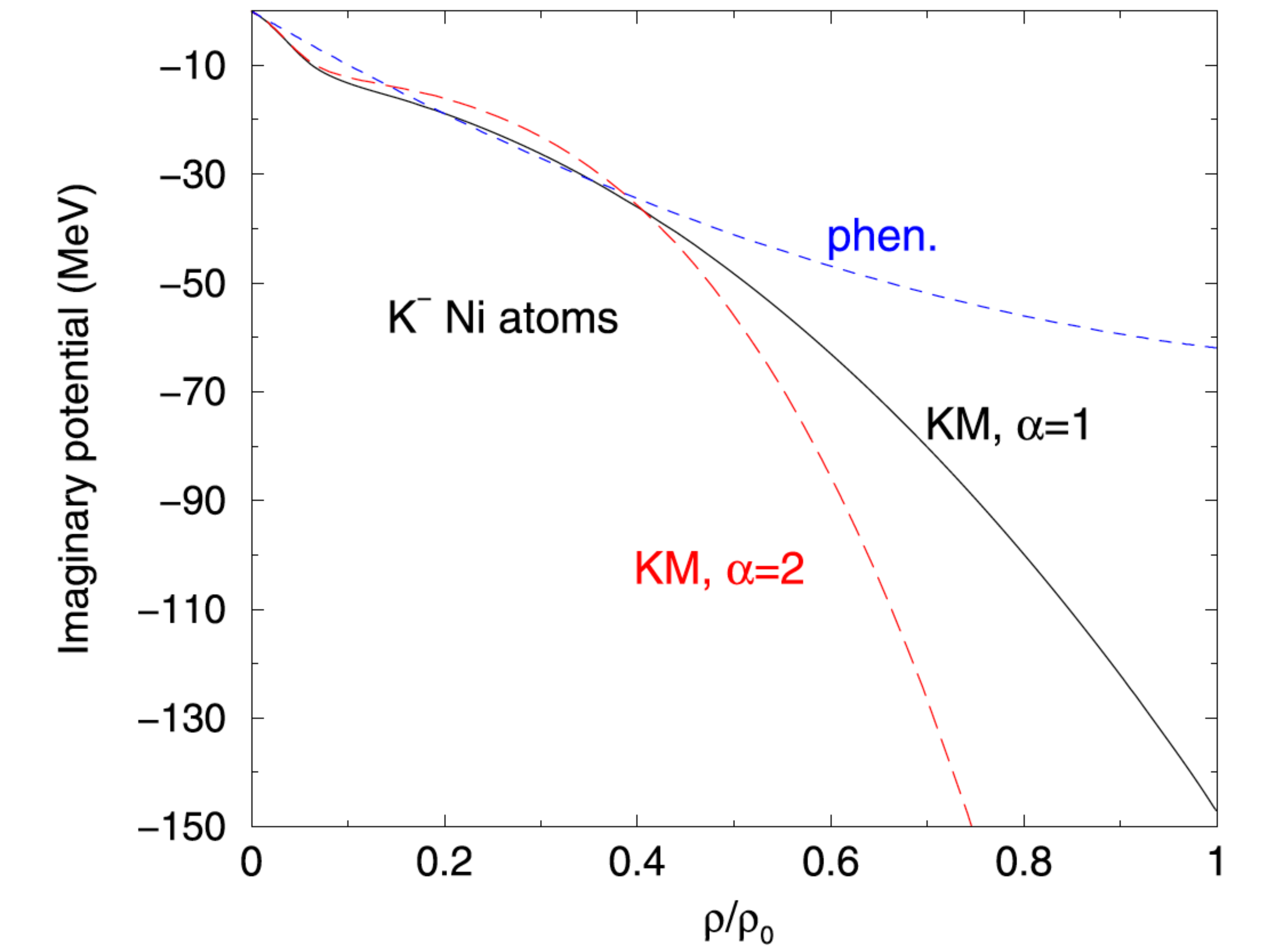}
\caption{ (Left) real part and (Right) imaginary part of best-fit $K^-$ potentials for kaonic Ni atoms as a function of nuclear density. Solid curves and long dashed curves are based on the single-nucleon amplitudes \cite{Ikeda_Hyodo_Weise_PLB706,Ikeda_Hyodo_Weise_NPA881} plus a phenomenological term proportional to $(\rho/\rho_0)^{\alpha}$, dashed curves are for a purely phenomenological density dependent best-fit potential \cite{Friedman_Gal_PR}. All three potentials lead to equally good fits to 65 kaonic atom data points. The figure is taken from \cite{Friedman_Gal_NPA959}.}
\label{fig:dens_dep}
\end{figure*}

\begin{figure*}
\begin{center}
\includegraphics[width=12.0cm,clip]{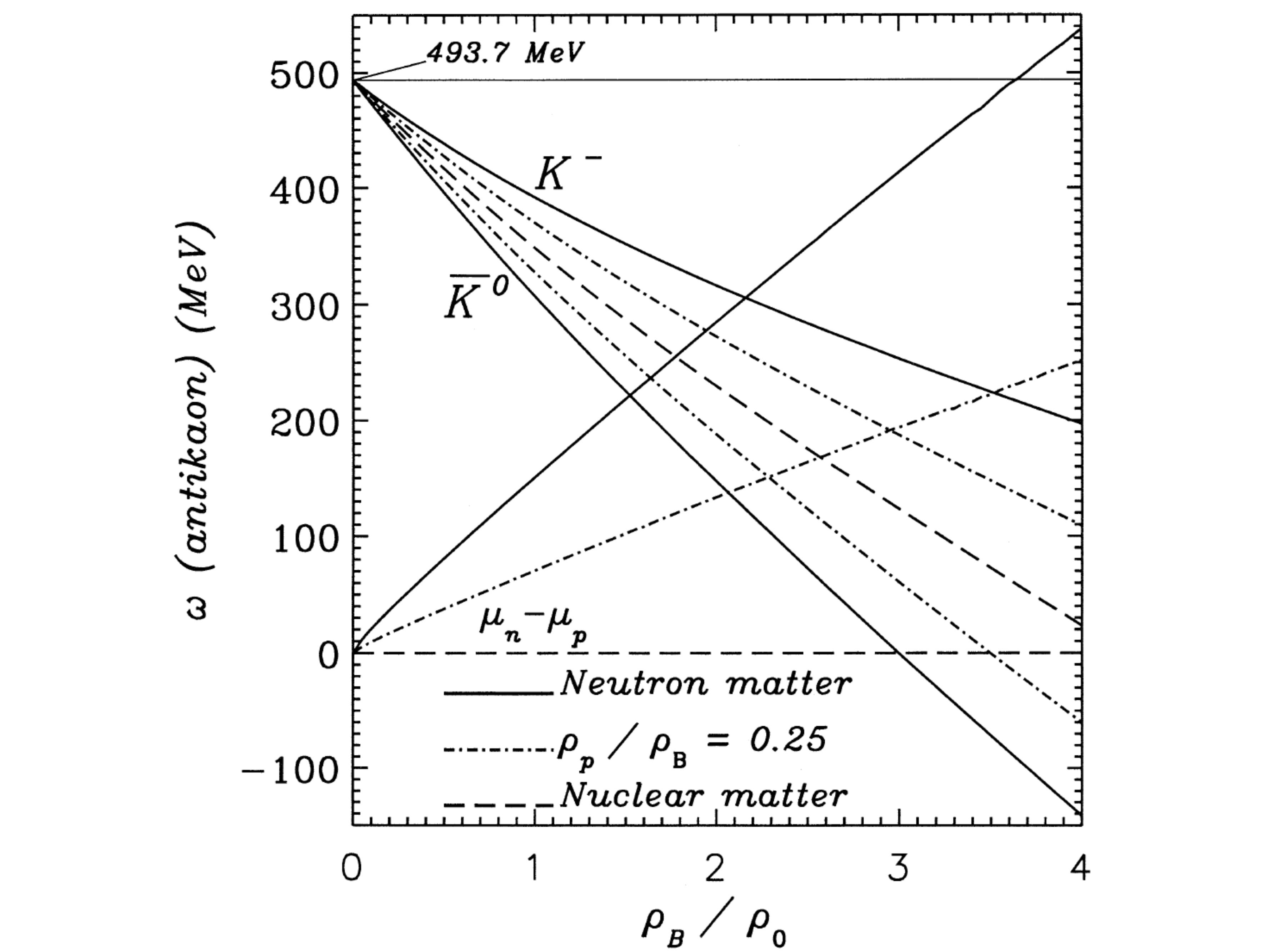}
\vspace*{-2mm} \caption{Antikaon energies at zero momentum and the difference of the calculated chemical potentials for the neutron and proton, $\mu_n-\mu_p$ versus baryon density for symmetric nuclear matter (dashed curves), for $\rho_p/ \rho_B$ = 0.25 (dashed dotted curves) and for neutron matter (solid curves).
The figure is taken from \cite{Tsushima_PLB1998}.}
\label{fig:QMC4}
\end{center}
\end{figure*}

\begin{equation}
V_{K^-}(\rho_N)=\frac{\Pi_{K^-}(\rho_N)}{2m_K}=
-\frac{2\pi}{m_K}\left(1+\frac{m_K}{m_N}\right){\bar a_{eff}}(\rho_N)\rho_N.\label{eq:pot_thr}
\end{equation}
The effective scattering length ${\bar a}_{eff}$ was extracted in \cite{FGB_PLB1993} from the analysis of antikaonic atom data, viz.
\begin{equation}
{\bar a}_{eff}(\rho_N)=\left[-0.15+1.7\left(\frac{\rho_N}{\rho_0}\right)^{0.25}\right]~\mbox{{\rm fm}}.
\end{equation}
With this effective scattering length and for saturation density $\rho_0=0.16$ fm$^{-3}$,
the $K^-$ potential is given by
%%formula (3)
\begin{equation}
V_{K^-}(\rho_N)=-122\left[-0.15+1.7\left(\frac{\rho_N}{\rho_0}\right)^{0.....25}\right]\frac{\rho_N}{\rho_0}~\mbox{{\rm MeV}}. \label{eq:pot_atoms}
\end{equation}
This potential~(\ref{eq:pot_atoms}) is strongly attractive and amounts to -190 MeV in the nuclear interior \cite{FGB_PLB1993,FGB_NPA579}, in contrast to a weakly repulsive $K^-$--nuclear potential  of $V_{K^-}(\rho_0)\approx+5$ MeV which would be obtained from Eq.~(\ref{eq:pot_thr}) for the free $K^- N$ scattering length ${\bar a}_{{\bar K}N}=\frac{1}{2}(a_{K^-p}+a_{K^-n})\approx (-0.04+i 0.80)$~fm. As explained above, this contradiction between a repulsive potential deduced from the vacuum isospin averaged antikaon-nucleon scattering length and an empirical attractive potential from heavy atom X-ray data, can be explained by the medium modification of the $K^-$$p$ scattering amplitude in the isospin I = 0 channel through the in-medium modification of the $\Lambda(1405)$ resonance.

It should be noted, however, that the antikaonic atom data probe the surface of the nucleus and thus do not provide
strong constraints on the $K^-$--nucleus potential at normal nuclear matter density (see Fig.~\ref{fig:dens_dep}).  In fact, more recently Friedman and Gal \cite{Friedman_Gal_NPA959} have investigated the model-dependence of extrapolating $K^- $ - nucleus potentials to nuclear saturation density $\rho_0$. These potentials, based on different $K^- N$ amplitudes, fit kaonic atom data equally well and reproduce the experimentally determined fraction of $K^-$ absorption on single nucleons. As shown in Fig.~\ref{fig:dens_dep} the $K^-$ -nucleus potential can reliably be determined only up to nucleon densities of about 30$\%$ of $\rho_0$ for the real part and up to 60$\%$ for the imaginary part, respectively. For higher densities the parameterisations diverge, indicating that antikaonic atom data are insensitive to the nuclear interior. The limited sensitivity of kaonic atom X-ray data to the $K^- $ - nucleus potential in the interior of the nucleus is also demonstrated by the fact that Baca et al. \cite{Baca_NPA673} - using a shallow chiral potential - reproduce the experimental energy shifts and widths over the full periodic table - although with a worse $\chi^2$ than in \cite{Friedman_Gal_PR}. For Z $\le$ 30,  Hirenzaki et al. \cite{Hirenzaki_PRC61} find a reasonable description of kaonic x-ray data with a shallow potential of $V_{K^-}(\rho_N=\rho_0) \approx$ -(30-40) MeV. 

Since the $\Lambda(1405)$ does not contribute in the isospin $I=1$ channel, the attractive
$K^-$$n$ scattering length $a_{K^-n}$ is practically density-independent \cite{Friedman_Gal_PR,Li_PLB1994,Cieply_PRC2011}
and the real part of the $K^-$ potential in pure neutron matter at low energies can be estimated from
the free $K^-$$n$ scattering length $a_{K^-n} = (0.58+i 0.78)$~fm \cite{Ikeda_Hyodo_Weise_NPA881,Hoshino_1705.06857}
using Eq.~(\ref{eq:pot_thr}), 
%formula (5)
\begin{equation}
V_{K^-}(\rho_n)= -70~\mbox{\rm  MeV}~\frac{\rho_n}{\rho_0}.\label{eq:neutron}
\end{equation}
With the $K^-$ potential~(\ref{eq:pot_thr}), the in-medium total antikaon energy $E_{K^-}^{\prime}$
at low in-medium $K^-$ momentum ${\bf p}_{K^-}^{\prime}$ can be determined as in [98-101]
%\cite{Li_PLB1994,Li_NPA1995,Li_PRL1995,Li_PRC1996}
\begin{equation}
E_{K^-}^{\prime}=\sqrt{{\bf p}_{K^-}^{{\prime}2}+m_K^2+2m_KV_{K^-}(\rho_N)}.\label{eq:energy}
\end{equation}
In the case when $2|V_{K^-}(\rho_N)|/m_K << 1$, Eq.~(\ref{eq:energy}) can be rewritten in the form
%formula (7)
\begin{equation}
E_{K^-}^{\prime}=\sqrt{{\bf p}_{K^-}^{{\prime}2}+m_{K^-}^{*2}}\label{eq:energy_free}
\end{equation}
corresponding to the free one but incorporating the effective antikaon mass $m_{K^-}^{*}$ defined as
(compare to Eq.~(\ref{eq:mK+*}) used for $K^+$ mesons)
%%formula (8)
\begin{equation}
m_{K^-}^{*}(\rho_N)=m_K+V_{K^-}(\rho_N).\label{eq:mass}
\end{equation}
It should be noted that the use of equation~(\ref{eq:energy_free}) instead of 
formula~(\ref{eq:energy}) is well proven for antikaon potentials satisfying the condition
$|V_{K^-}| \le 100$ MeV, as our estimates showed. In view of Eq.~(\ref{eq:mass}), the result~(\ref{eq:neutron})
for $K^-$ potential in neutron matter scales the $K^-$ mass as
%%formula (9)
\begin{equation}
m_{K^-}^{*}(\rho_n)=m_K\left(1-{\alpha_{\bar K}}\frac{\rho_n}{\rho_0}\right) \label{eq:mass_dep}
\end{equation}
with ${\alpha_{\bar K}}=0.14$, which is in good agreement with the respective value of 0.13 obtained in the
coupled-channel approach \cite{Waas}.

Similar to kaons, the antikaon energy in a medium can be also determined
from the mean-field approximation to the chiral Lagrangian [21,98-100,102-104].
%\cite{Li_PLB1994,Li_NPA1995,Li_PRL1995,Li_NPA1997,Li_PRL1997,Kaplan_Nelson,Nelson}.
It is expressed by the formula (\ref{eq:EK+}) in which one has to make the following substitutions:
$E_{K^+}^{\prime} \to E_{K^-}^{\prime}$, ${\bf p}_{K^+}^{\prime} \to {\bf p}_{K^-}^{\prime}$ and
$\frac{3}{8}\frac{\rho_N}{f^2} \to -\frac{3}{8}\frac{\rho_N}{f^2}$.
Then at low densities, the $K^-$ meson in-medium mass is defined by formulae (\ref{eq:mK+*_long}) and (\ref{eq:VK+}) in which
the term $\frac{3}{8}\frac{\rho_N}{f^2}$ and kaon potential $V_{K^+}$ are replaced, correspondingly,
by -$\frac{3}{8}\frac{\rho_N}{f^2}$ and the antikaon potential $V_{K^-}$.                             
Proceeding in the same way as for the $K^+$ potential (Eq.~(\ref{eq:VK+rhoN}) in Section \ref{sec:K++}) and using the empirical value for $\Sigma_{\overline{K}N}$ from \cite{Li_NPA1997,Li_PRL1997} one obtains the $K^-$--nuclear potential \cite{Paryev_EPJA92000}:
%%formula (10)
\begin{equation}
V_{K^-}(\rho_N)=-126~\mbox{\rm MeV}~\frac{\rho_N}{\rho_0}.\label{eq:pot_mean1}
\end{equation}
It should be pointed out that at densities $\rho_N \le \rho_0$ this potential is in line with
those obtained within the microscopic studies \cite{Schaffner_NPA1997,Waas,Waas1}, and shown before in Fig.~\ref{fig:K+K-_pred-Schaffner}.
It is also roughly in line with the calculations by Tsushima et al. \cite{Tsushima_PLB1998} within the
Quark-Meson Coupling model that predicts a decrease of the effective $K^-$ mass in symmetric nuclear matter
at zero momentum by $\approx$ 144 MeV, as is illustrated by Fig.~\ref{fig:QMC4}.
It is seen from this figure, as in the case of $K^+$ and $K^0$ mesons, considered above (cf. Fig.~\ref{fig:QMC3}),
that the energies of the $K^-$ and ${\bar {\rm K}^0}$ mesons are not degenerate in
asymmetric nuclear matter. The antikaon potential~(\ref{eq:pot_mean1}) scales the $K^-$ mass in nuclear
matter in line with Eq.~(\ref{eq:mass_dep})
with ${\alpha_{\bar K}}\approx 0.26$ and $\rho_n \rightarrow \rho_N$, which agrees reasonably well with that of
${\alpha_{\bar K}}\approx$ 0.2--0.25 used in transport model calculations \cite{HSD,Cassing_HSD}.
Since the actual magnitude of the antikaon scalar potential, entering into the dispersion relation~(\ref{eq:energy_free})
through the Eq.~(\ref{eq:mass}), is not reliably known also at finite momenta (see Section~\ref{sec:K-_real}), it has been adopted
in the momentum-independent form~(\ref{eq:pot_mean1}) or is treated as a free parameter to be extracted in the
analysis of inclusive \cite{Akindinov_JETP,Akindinov_ITEP} and exclusive \cite{Kiselev_K-} data on $K^-$
production in $pA$ collisions, taken at the ITEP/Moscow and COSY/J$\ddot{\rm u}$lich accelerators,
in the framework of the nuclear spectral function approach \cite{Paryev_EPJA92000,Paryev_EPJA2003,Paryev_K-}. We note in passing that the "free particle" dispersion relation~(\ref{eq:energy_free}) for the $K^-$
energy in matter with the effective momentum-independent antikaon mass (and momentum-independent scalar potential
$V_{K^-}$)
%%formula (11)
\begin{equation}
m_{K^-}^{*}(\rho_N)=m_K\left(1-0.22\frac{\rho_N}{\rho_0}\right)=m_K-110~\mbox{\rm MeV}~\frac{\rho_N}{\rho_0},
\end{equation}
has also been employed in \cite{Muehlich_PRC2003} in the study of $\phi$ photoproduction from nuclei.
However, the dispersion analysis of \cite{SibCas_NPA1998} finds that the $K^-$ scalar potential shows a strong
momentum dependence. It turns out to be $\approx$ -140 MeV at density $\rho_0$ and for zero momentum, while it decreases
rapidly in magnitude for higher momenta and saturates at $\approx$ -50 MeV for high momenta. Therefore, to achieve
a better understanding of the $K^-$ properties  in the nuclear medium and to test the predictions of the microscopic
models involved, one needs to determine the momentum dependence of the in-medium antikaon
potential experimentally. In principle, this can be done by measuring the $K^-$ meson spectra from $pA$ reactions
at bombarding energies close to the threshold in free $pN$ collisions
\cite{Paryev_EPJA92000,Paryev_EPJA2003,Paryev_K-,SibCas_NPA1998} (see also below).

A lot of self-consistent coupled-channel calculations of the $K^-$ self-energy in nuclear matter have been performed 
based on chiral Lagrangians [143,144,168-171]
%\cite{Cieply_NPA2001,Ramos_Oset,Schaffner_NPA2000,Ramos_NPA2001,Tolos_PRC74,Tolos_PRC78}
or on meson-exchange potentials \cite{Tolos_NPA690,Tolos_PRC65}.
They predict a relatively shallow low-energy $K^-$--nucleus potentials with a central
depths of the order of -50 to -80 MeV, showing different momentum dependences at finite momenta
ranging up to 500 MeV/c [168-173].
%\cite{Schaffner_NPA2000,Ramos_NPA2001,Tolos_PRC74,Tolos_PRC78,Tolos_NPA690,Tolos_PRC65}.
Recently, a chirally motivated meson-baryon coupled-channel model \cite{Cieply_PRC2011,Gazda_Mares},
which accounts for the subthreshold in-medium $K^-N$ $s$-wave scattering amplitudes,
produced $K^-$ potential depths in the range of  -(80--90) MeV in antikaonic atoms at nuclear matter density.
However, taking $K^-$ absorption on two nucleons such as $K^-NN \to YN$ into account to improve the agreement with antikaonic X-ray data by adding a ${\rho_N^2}$-dependent phenomenological term, the antikaon potential becomes twice as deep \cite{Friedman_Gal_NPA881,Gal_NPA914,Cieply_PRC2011,Gal_EPJWeb2014}. 

The situation with the imaginary part $W_{K^-}$ of the $K^-$--nuclear potential at low energies is also
controversial. Models based on a chiral Lagrangian [143,144,169-171]
%\cite{Cieply_NPA2001,Ramos_Oset,Ramos_NPA2001,Tolos_PRC74,Tolos_PRC78}
or on meson-exchange potentials \cite{Tolos_NPA690,Tolos_PRC65}, predict 
central depths of the order of $W_{K^-}(\rho_N=\rho_0)\approx$ -25 to -60 MeV at threshold.
On the other hand, adding again a ${\rho_N^2}$-dependent phenomenological term, simulating two--nucleon $K^-NN \to YN$  and multi--nucleon
absorptive processes, to the single--nucleon potential from chirally motivated
models \cite{Friedman_Gal_NPA881,Gal_NPA914,Cieply_PRC2011,Gal_EPJWeb2014} yields an imaginary $K^-$ potential depth in the
range -(70--80) MeV. The momentum dependence of this potential has been also investigated in [169-173]
%\cite{Ramos_NPA2001,Tolos_PRC74,Tolos_PRC78,Tolos_NPA690,Tolos_PRC65} 
in the momentum range from 0 to 500 MeV/c. It shows both flat \cite{Ramos_NPA2001,Tolos_PRC74,Tolos_PRC65} and non-flat
\cite{Tolos_PRC78,Tolos_NPA690} behavior at these momenta and at densities $\rho_N \le \rho_0$.

Thus, one may come to the conclusion that the situation with the $K^-$--nucleus potential both at threshold
and at finite momenta is still very unclear presently and more theoretical and experimental work is needed
to clarify it.

\subsubsection{\it Determination of the $K^-$--nucleus real potential}
\label{sec:K-_real}

\begin{figure}[tb]
\begin{minipage}[t]{10 cm}
\includegraphics[width=12cm,clip]{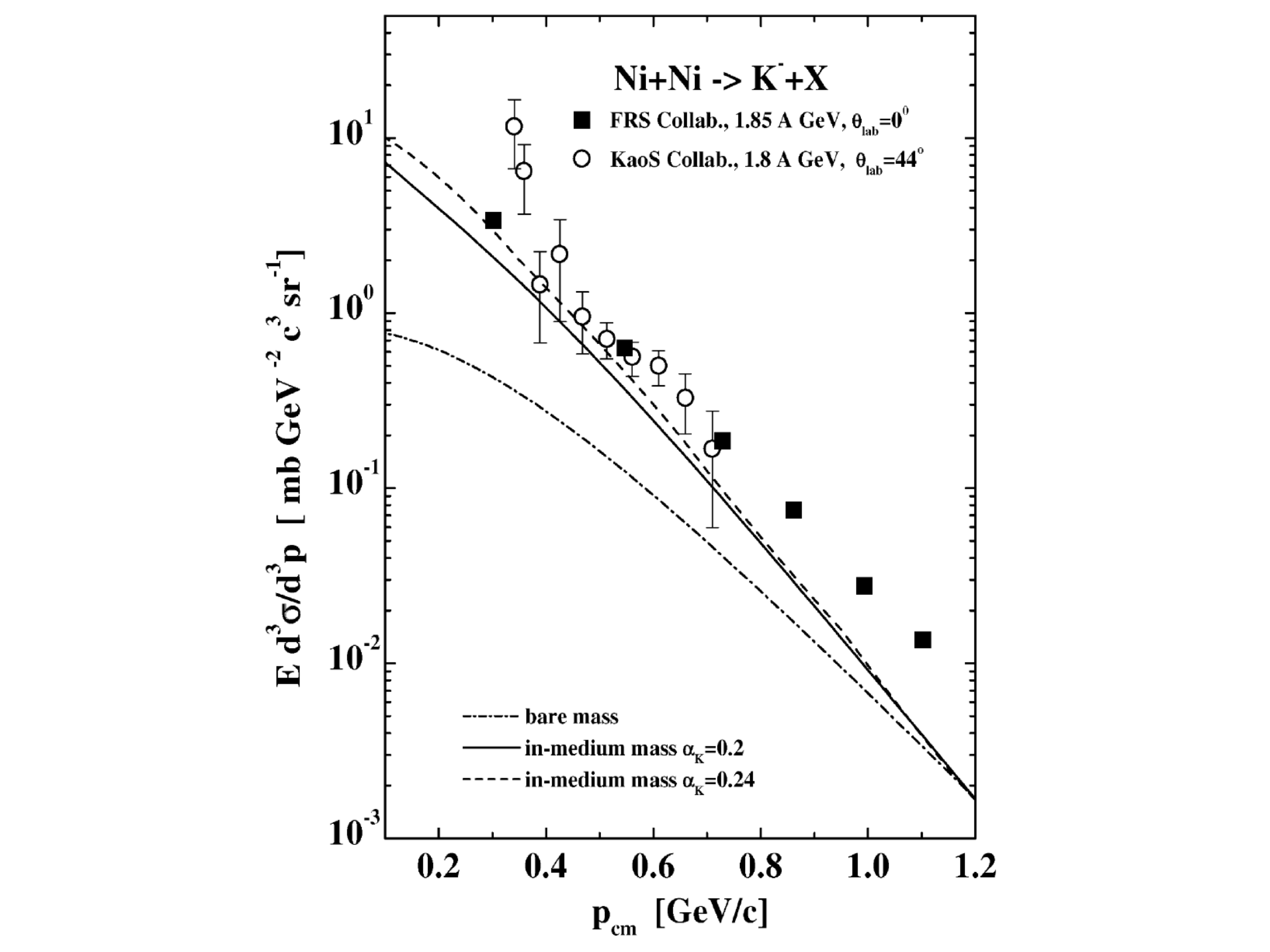}
\end{minipage}
\begin{minipage}[t]{8.5 cm}
\vspace{-9cm} 
\caption{ Inclusive
invariant $K^-$ production cross section as a function of the antikaon momentum in the nucleus--nucleus
cms for Ni+Ni collisions at 1.8 A GeV \cite{Barth_PRL1997} and 1.85 A GeV \cite{Schroter_ZPA350,Kienle} in comparison with CBUU transport model calculations \cite{HSD}. The dot-dashed line corresponds to a calculation with bare kaon masses, whereas the solid and dashed lines show the results with an antikaon
self-energy according to $m_{K^-}^*(\rho_N)=m_K\left[1-{\alpha_{\bar K}}(\rho_N/\rho_0)\right]$ with
${\alpha_{\bar K}}=0.2$ and 0.24, respectively. The adopted options for ${\alpha_{\bar K}}$
correspond to an attractive $K^-$ potentials of $V_{K^-}(\rho_0)\approx -100$ MeV and $V_{K^-}(\rho_0)\approx -120$ MeV
for antikaons at rest, respectively. The figure is taken from \cite{HSD}.}
\label{fig:CasBrat_K-}
\end{minipage}
\end{figure}

\begin{figure}[h!]
\centering
\includegraphics[width=14.0cm,clip]{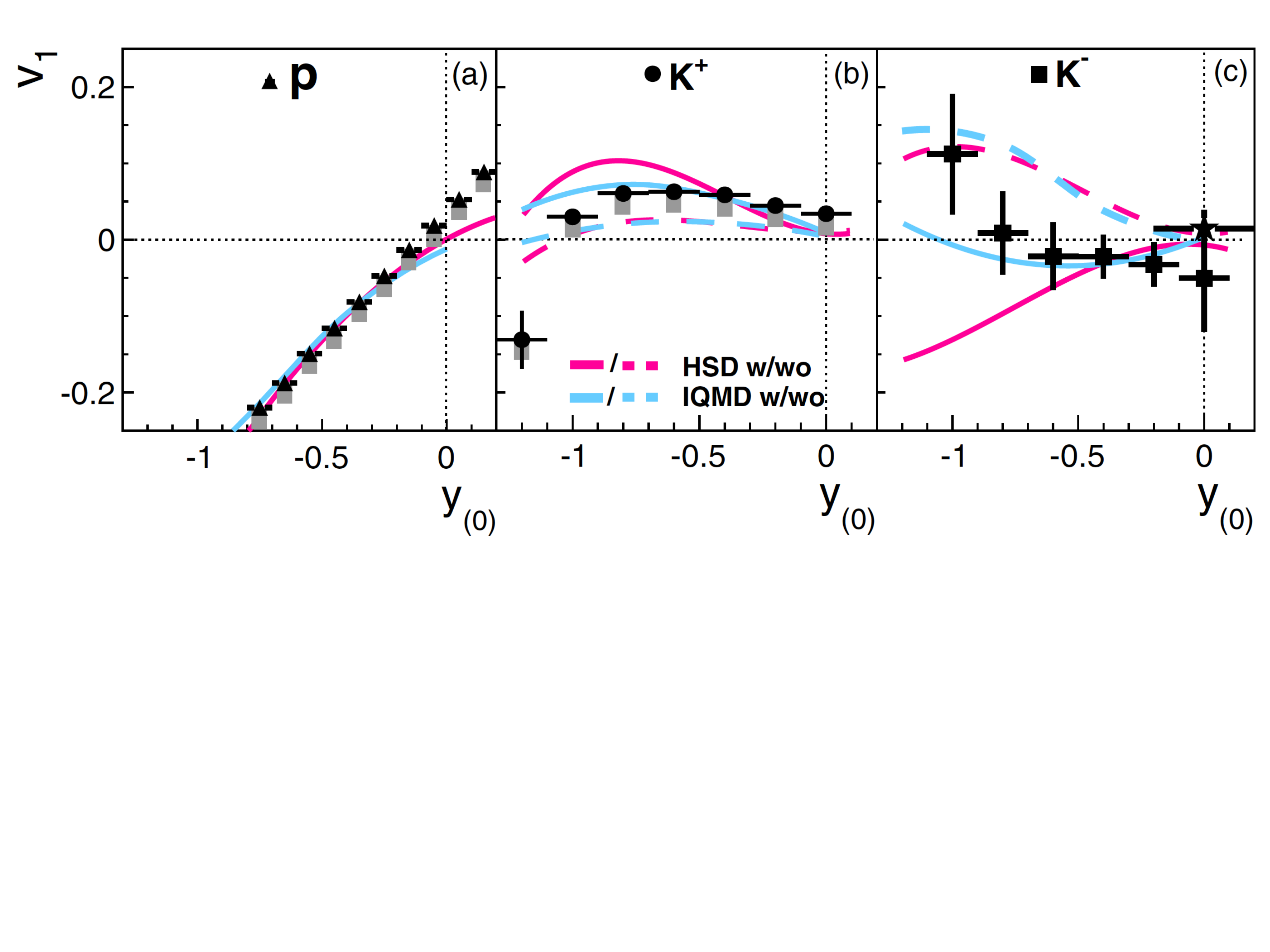}
\vspace{-45mm}\caption{Sideward flow $v_1$ versus rapidity for protons (a, triangles), $K^+$ (b, dots) and $K^-$ mesons
(c, squares) in Ni+Ni collisions at 1.91 A GeV \cite{Zinyuk_PRC90}
in comparison with HSD \cite{HSD} and IQMD \cite{Hartnack_EPJA1998}
transport model calculations with (solid lines) and without (dashed lines) in-medium particle potentials.
A repulsive $K^+$ potential of 20$\pm$5 MeV for kaons at rest at density $\rho_0$ and a linear dependence
on baryon density has been employed in both HSD and IQMD calculations. For $K^-$ mesons an attractive potential with
$V_{K^-}(\rho_0,{\bf p}_{K^-}^{\prime}=0) = -45$ MeV has been used in the IQMD model, while in the HSD approach $K^-$
mesons are treated as off-shell particles using the G-matrix formalism corresponding to
$V_{K^-}(\rho_0,{\bf p}_{K^-}^{\prime}=0) = -50$ MeV. The star symbol for $K^-$ mesons at midrapidity in (c)
is from the high statistics data in the momentum range $<$ 1.0 GeV/c with signal/background $>$ 5.
The figure is taken from \cite{Zinyuk_PRC90}.}
\label{fig:Zinyuk2_FOPI}
\end{figure}
%%%%%%%%%%%%%%%%%%%%%%%%%%%%%%%%%%%%%%%%%%%%%%%%%%%%%%%%%%%
\begin{figure}[h!]
\begin{center}
\includegraphics[width=12.0cm,clip]{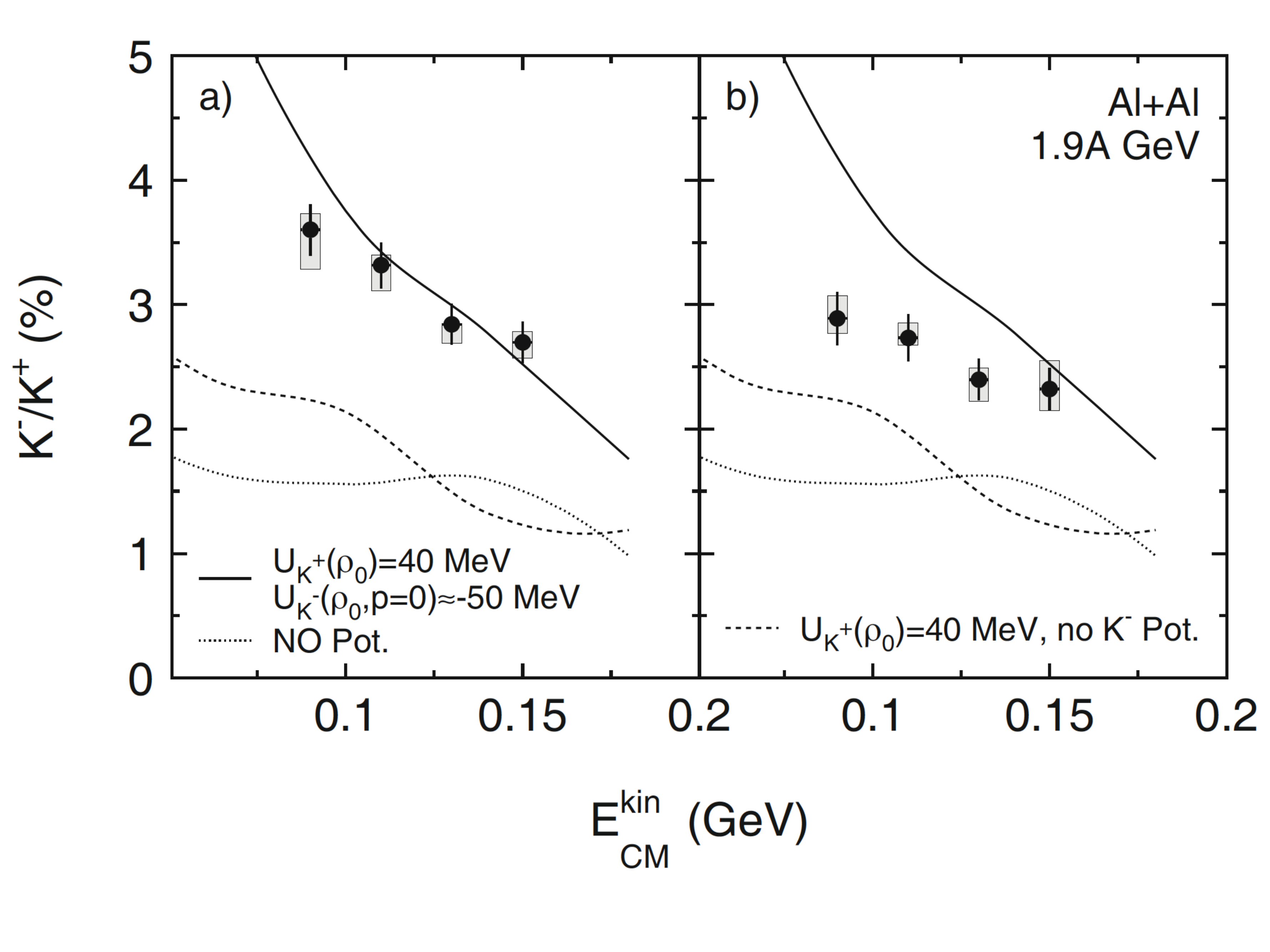}
\vspace*{-2mm} \caption{The measured $K^-$/$K^+$ ratio as a function of the kinetic energy in the center-of-mass system for Al+Al collsions at 1.9 A GeV 
 \cite{Gasik_EPJA52}: (a) not-corrected and (b) corrected for the 17\% contribution of $K^-$ mesons
from $\phi$ decays. Error bars represent the statistical uncertainties. Shaded rectangles
depict the estimation of systematic errors. Curves are the results of HSD transport model \cite{HSD}
predictions without any in-medium effects (dotted curves), with only a $K^+$ potential with a linear dependence on
density with $V_{K^+}(\rho_0) = 40$ MeV (dashed curves) as well as with the in-medium effects for both kaons and
antikaons ($V_{K^+}(\rho_0) = 40$ MeV and $V_{K^-}(\rho_0,{\bf p}_{K^-}^{\prime}=0) \approx -50$ MeV) (solid curves).
The figure is taken from \cite{Gasik_EPJA52}. With kind permission of The European Physical Journal (EPJ).}
\label{fig:Gasik9}
\end{center}
\end{figure}
%%%%%%%%%%%%%%%%%%%%%%%%%%%%%%%%%%%%%%%%%%%%%%%%%%%%%%%%%%%
\begin{figure}[tb]
\begin{minipage}[t]{10 cm}
\includegraphics[width=12.0cm,clip,angle=90]{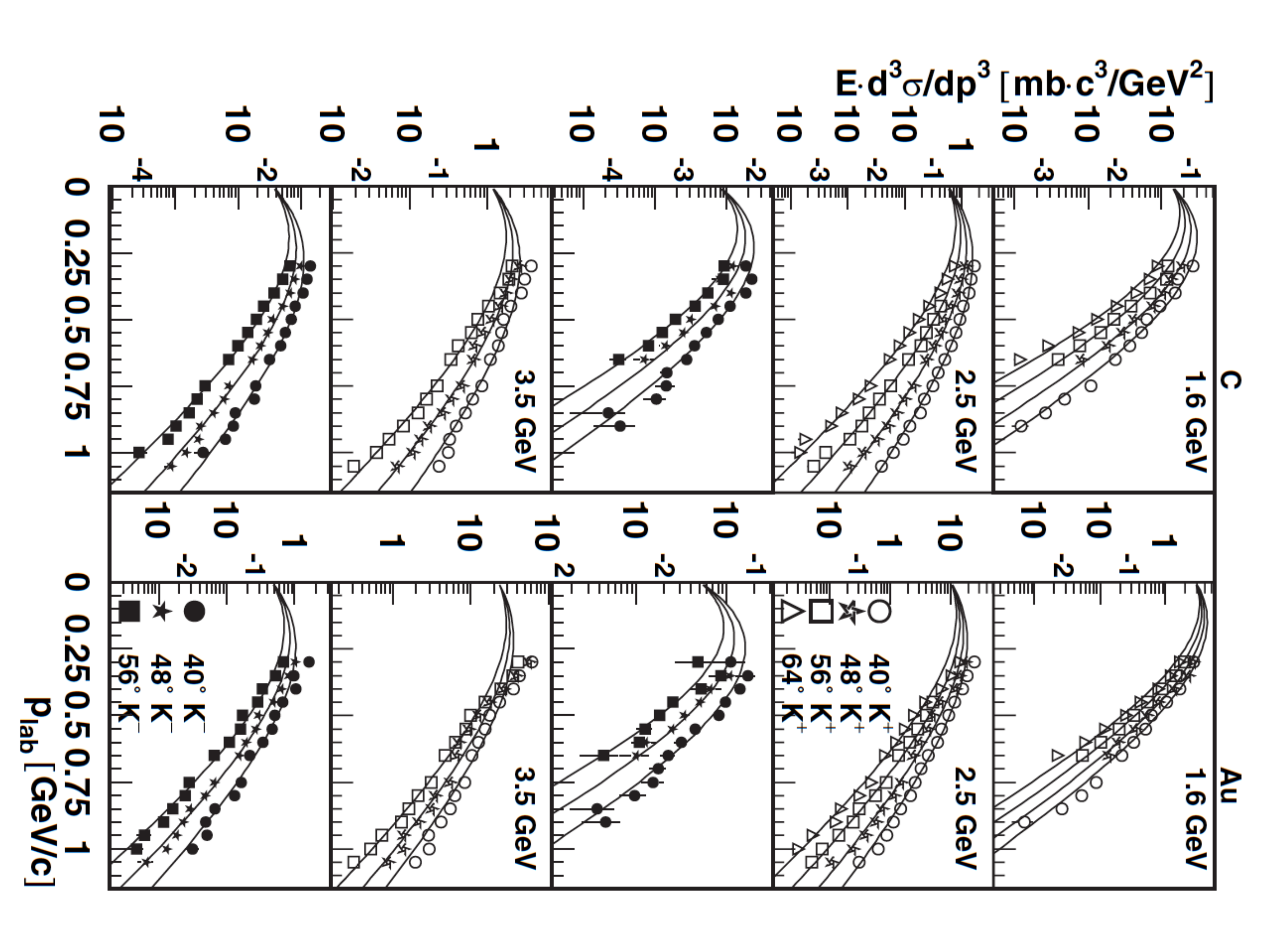}
\end{minipage}
\begin{minipage}[t]{8.5 cm}
\vspace{-9cm} 
\caption{Invariant production cross section for $K^+$ (open symbols) and $K^-$ mesons (full symbols)
in inclusive $p$--C (left) and $p$--Au (right) collisions at 1.6, 2.5, and 3.5 GeV (from top to bottom)
as a function of laboratory momentum. The curves are fits to the data \cite{Scheinast}.
The figure is taken from \cite{Scheinast}.}
\label{fig:Kaos_K-spectra}
\end{minipage}
\end{figure}

\begin{figure}[tb]
\begin{minipage}[t]{11 cm}
\includegraphics[width=10.0cm,clip]{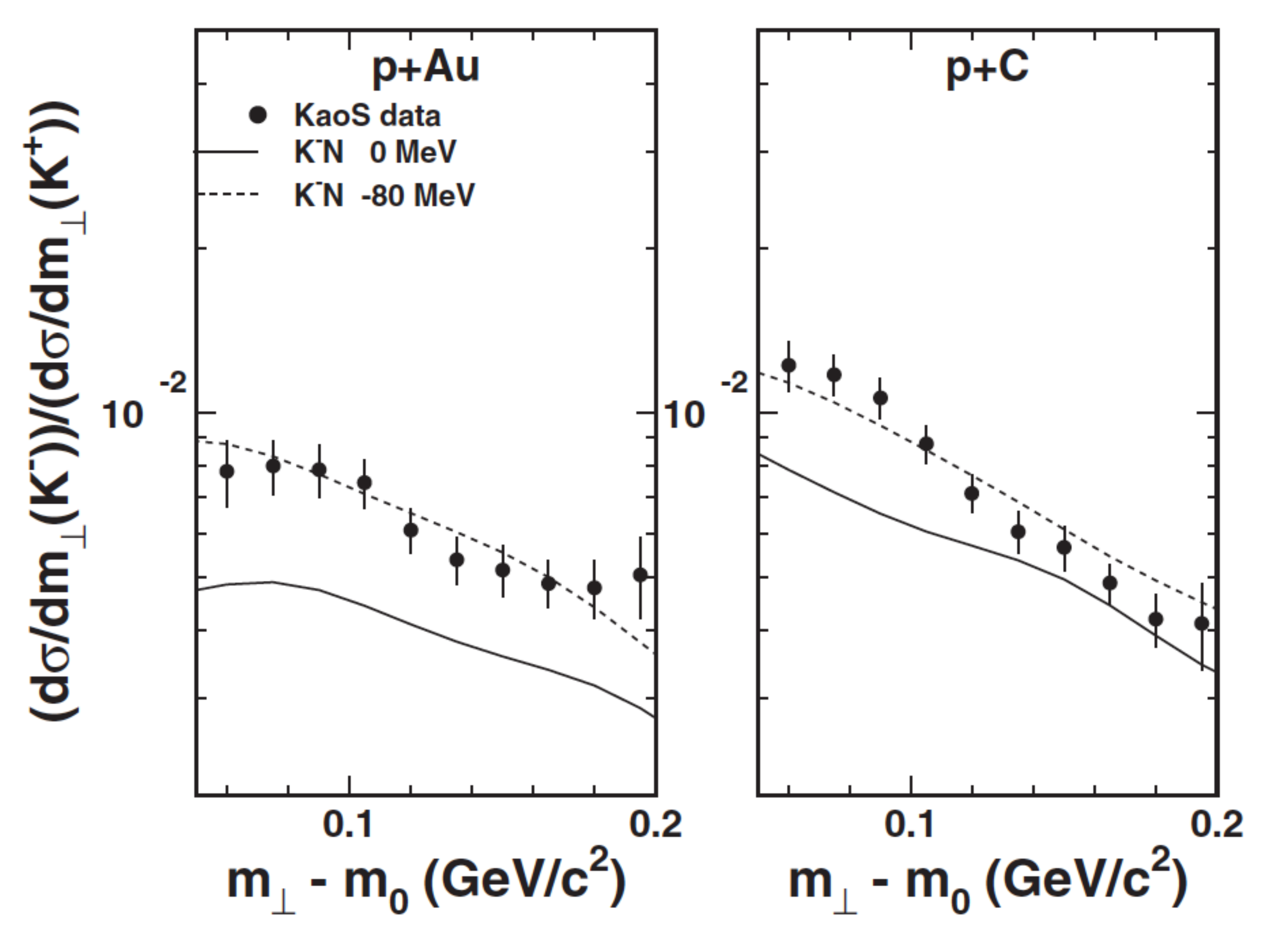}
\end{minipage}
\begin{minipage}[t]{7.0 cm}
\vspace{-8cm} 
\caption{Ratio of invariant production cross sections of $K^-$ over K$^+$ mesons for inclusive p-Au (left) and p-C (right) collisions as a function of transverse mass. The data (solid circles) were taken at 2.5 GeV and were integrated over laboratory angles between $\theta_{lab} = 36^{\circ}$ and 60$^{\circ}$. The solid and dashed curves depict results of BUU transport model calculations \cite{Barz} including strangeness exchange as well as a repulsive K$^+$ nucleus potential of +25 MeV. The $K^-$ nucleus potentials used in the calculations are indicated in the figures. The figure is taken from \cite{Scheinast}.}
\label{fig:Kaos_K-pot}
\end{minipage}
\end{figure}

\begin{figure*}
\centering
  \includegraphics[width=11.0cm,clip]{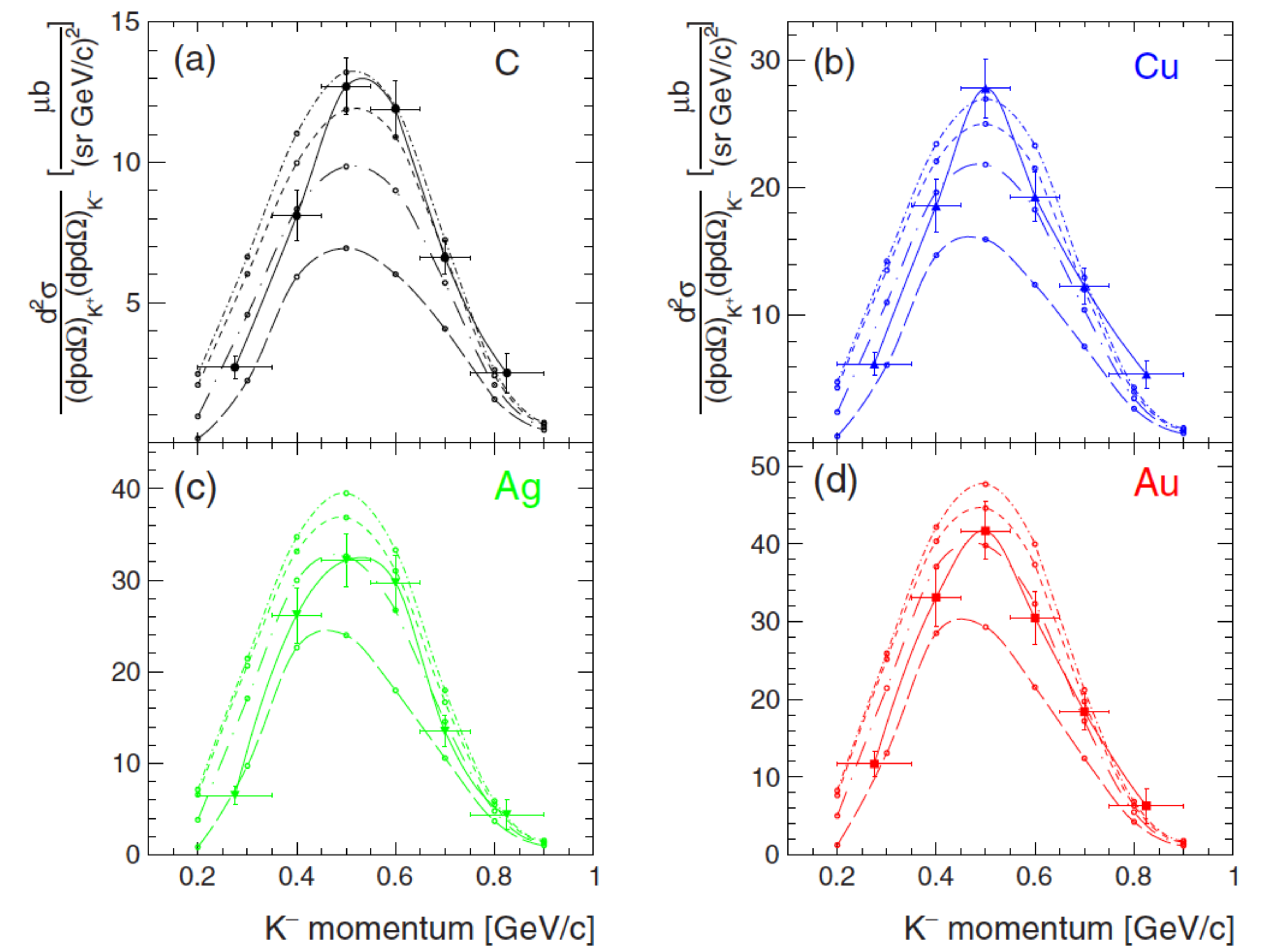}
\caption{Double-differential  cross sections for the production of non-resonant $K^+K^-$ meson pairs in the ANKE acceptance in the collisions of 2.83 GeV protons with (a) C, (b) Cu, (c) Ag, (d) Au targets as a function of the $K^-$ laboratory momentum. The data are averaged over $K^{\pm}$ angles $\theta_{K^{\pm}} \le 12^{\circ}$ and over $K^+$ momenta in the range $200 \le p_{K^+} \le 600 $MeV/$c$. The curves represent model calculations \cite{Paryev_K-} for $K^-$ potential depths $U = V_{K^-} (\rho_0) = 0$ MeV (long dashed), -60 MeV (dot long dashed), -126 MeV (short dashed), and -180 MeV (dot short dashed), respectively. The solid lines are spline functions through the experimental data points. The figure is taken from
\cite{Kiselev_K-}.}
\label{fig:Kiselev_K-spectra}
\end{figure*}
%%%%%%%%%%%%%%%%%%%%%%%%%%%%%%%%%%%%%%%%%%%%%%%%%%%%%%%%%%%%%%
\begin{figure*}
\centering
  \includegraphics[width=10.0cm,clip]{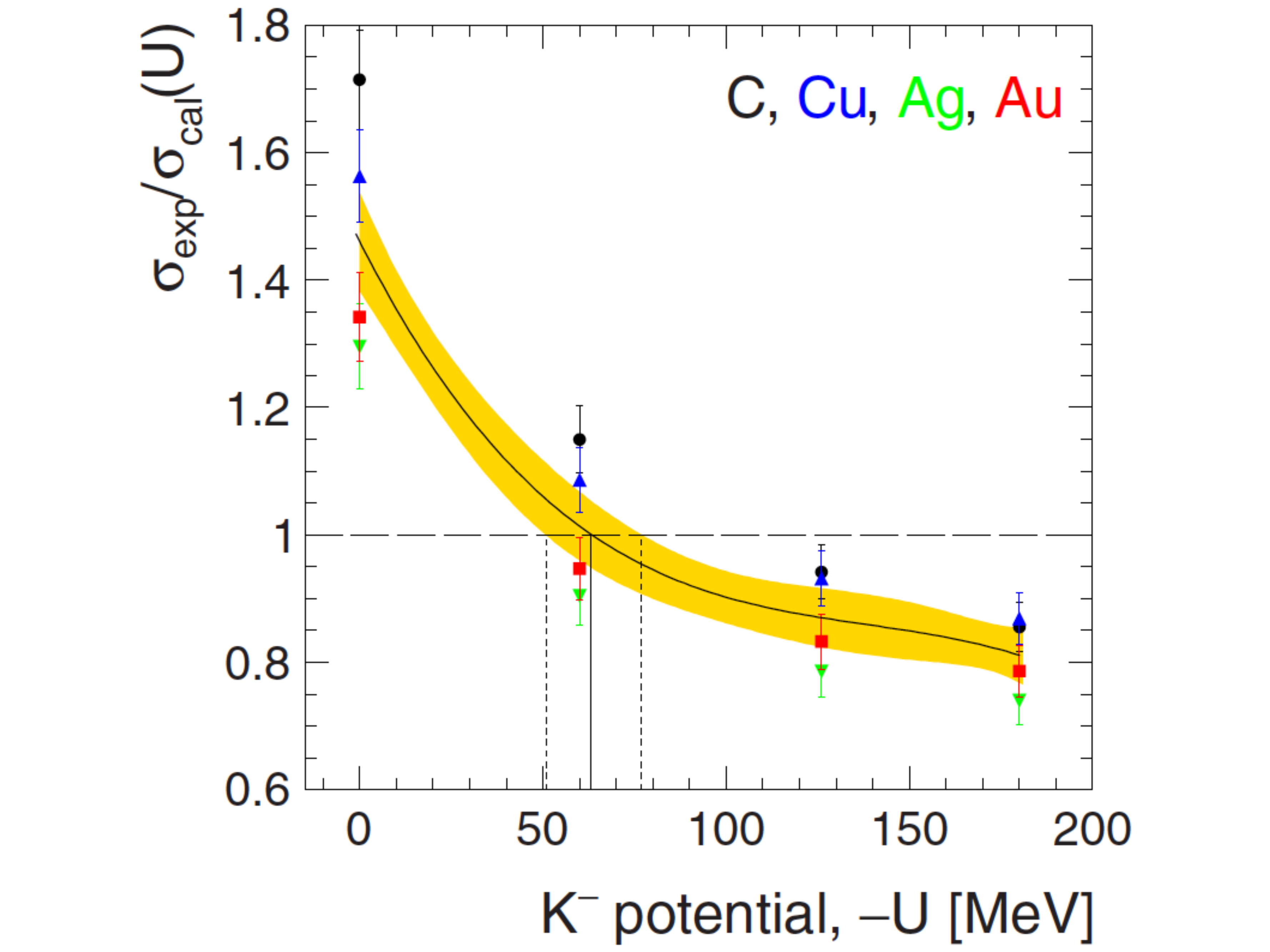}
\caption{Ratio of the measured integrated cross section for non-resonant $K^+K^-$ pair production on C, Cu, Ag, Au to the corresponding calculated cross section assuming four values of the $K^-$ potential depth at nuclear matter density:  $U = V_{K^-} (\rho_0) = $ 0, -60 ,-126, and -180 MeV. The curve represents a third-order polynomial fit of all ratios. The shaded band indicates the 1$\sigma$ confidence interval. The pair of vertical dotted lines corresponds to the regions where the ratio is unity within the errors given by the fit. The color code is identical to that in Fig. \ref{fig:Kiselev_K-spectra}. The figure is taken from \cite{Kiselev_K-}.}
\label{fig:Kiselev_K-ratio}
\end{figure*}
%%%%%%%%%%%%%%%%%%%%%%%%%%%%%%%%%%%%%%%%%%%%%%%%%%%%%%%%%%%%%%
Experimental information about in-medium properties of antikaons can be deduced from the study of their production both in
heavy--ion and proton--nucleus collisions at incident energies near or below the free nucleon--nucleon
threshold (2.5 GeV). The dropping $K^-$ mass scenario will lead to an enhancement of the $K^-$ yield
in these collisions due to in-medium shifts of the elementary production thresholds to lower energies.
Indeed, an antikaon enhancement in C+C, Ni+Ni interactions at beam energies per nucleon below the $NN$ threshold
has been observed by the KaoS and FRS Collaborations at SIS/GSI over two decades ago
\cite{Barth_PRL1997,Senger_APP1996,Laue_PRL82,Schroter_ZPA350}. This phenomenon has been attributed to the
in-medium $K^-$ mass reduction
\cite{HSD,Li_PLB1994,Li_NPA1997,Li_PRL1997,Li_NPA654,Cassing_HSD}.
An analysis of inclusive $K^-$ momentum spectra taken at KaoS \cite{Barth_PRL1997,Senger_APP1996,Laue_PRL82} and FRS \cite{Schroter_ZPA350} (see Fig.~\ref{fig:CasBrat_K-}) in the framework of the CBUU transport model \cite{HSD,Cassing_HSD} suggests a momentum-independent downward in-medium mass shift (or an attractive effective scalar $K^-$ potential $V_{K^-}(\rho_0)$) of about -(100--120) MeV at density $\rho_0$, assuming a linear dependence of the mass shift on the nuclear density. A similar $K^-$ potential of $V_{K^-}(\rho_0)\approx-110$ MeV for kaons at rest has been extracted by Li et al.
\cite{Li_NPA1997,Li_PRL1997,Li_NPA654} from a comparison of the KaoS data \cite{Barth_PRL1997,Senger_APP1996}
with relativistic transport model calculations. The extracted antikaon potential $V_{K^-}(\rho_0)$
is comparable to potentials predicted by adopting the relativistic mean-field or effective chiral
Lagrangian approach \cite{Schaffner_NPA1997} as well as the coupled-channel model \cite{Waas}
(see Fig.~ \ref{fig:K+K-_pred-Schaffner}, lower curves, and table 1).

On the other hand, recent findings on the azimuthal emission pattern of $K^-$ mesons in Ni+Ni collisions
at a beam kinetic energy of 1.91 A GeV, obtained by the FOPI Collaboration \cite{Zinyuk_PRC90}, imply an only weakly attractive $K^-$--nucleus potential (see Fig.~\ref{fig:Zinyuk2_FOPI}) as predicted by both (HSD and IQMD) transport model calculations. Furthermore,
the IQMD transport approach, which describes correctly also the data on the sideward flows of protons and
$K^+$ mesons (cf. Fig.18 as well), suggests a $K^-$--nucleus potential of $V_{K^-}(\rho_0)=-40\pm10$ MeV
for particles at rest \cite{Zinyuk_PRC90}. 

However, there is a general problem with extracting information on the $K^-$-nucleus potential from heavy-ion data. The shape of the $K^-$ momentum spectra and the $K^-$ yield are strongly distorted by feeding from $\phi \rightarrow K^+ K^-$ decays. Most of the transport model calculations underestimate the $\phi$ yield and thereby do not properly account for these distortions (the contribution of the $\phi$ decay channel $\phi \to K^+K^-$ to $K^+$ meson yield is negligible because of the much larger $K^+ $ production cross section). Recently the FOPI Collaboration has measured the $\phi/K^-$ ratio in Al + Al and Ni + Ni collisions at 1.9 AGeV to be 0.34$\pm$0.06(stat)$^{+0.04}_{-0.14}$(syst) \cite{Gasik_EPJA52} and 0.36$\pm$0.05, respectively \cite{Piasecki_PRC94}, implying that $\approx$17\% of the observed $K^-$ mesons stem from feeding through the $\phi \to K^+K^-$ decay. This observation is confirmed by a recent HADES measurement of Au-Au collisions at 1.23 AGeV who report a $\phi/K^-$ ratio of 0.52$\pm$ 0.16 \cite{HADES_AuAu}. The effect of these surprisingly large $\phi/K^-$ ratios on the determination of the $K^-$ -nucleus potential is demonstrated in Fig. \ref{fig:Gasik9} which shows the $K^-/K^+$ ratio as a function of the kaon kinetic energy. Ignoring the feeding from $\phi$ decays, a $K^-$- nucleus potential of -50 MeV would be obtained in comparison to HSD model transport calculations, assuming $V_{K^+}(\rho_0) = 40$ MeV \cite{HSD}, while a considerably weaker $K^-$ potential is needed to reproduce the data once the feeding is corrected for. The production of $K^-$ mesons via $\phi$ meson decays is sizable and should be taken into account in attempts to extract the $K^-$ potential from comparisons between $K^-$ data in nucleus - nucleus collisions and transport model simulations. This example shows how important it is to check whether transport calculations reproduce also other observables before conclusions regarding meson-nucleus potentials can be drawn.

The in-medium modifications of $K^-$ meson properties have also been studied experimentally in
proton--nucleus reactions at beam energies close to or below the production threshold in $NN$ collisions
over the last years (see, for example,
\cite{Scheinast,Akindinov_JPG,Akindinov_JETP,Akindinov_ITEP,Kiselev_K-}).
The advantage of such reactions compared to heavy--ion collisions is that 
the processes of hadron production and propagation proceed in cold static nuclear matter of 
well-defined density at zero temperature. The in-medium effects are somewhat smaller but still comparable
to those in heavy--ion collisions. 
Inclusive differential production cross sections of $K^{\pm}$ mesons and $K^-/K^+$ ratios have been measured
in $p$+C and $p$+Au reactions at 1.6, 2.5, and 3.5 GeV proton beam energy by the KaoS Collaboration
\cite{Scheinast}, see Figs.~\ref{fig:Kaos_K-spectra},~\ref{fig:Kaos_K-pot}.
An analysis within the BUU transport model \cite{Barz} of the ratios of $K^-$ and $K^+$ inclusive differential yields
presented in Fig.~\ref{fig:Kaos_K-pot}  has shown
(cf. Fig.~\ref{fig:Kaos_K-pot}) that these data are consistent with an in-medium momentum-independent $K^-$ nuclear potential of about -80 MeV at normal nuclear density \cite{Scheinast}. Measurements of inclusive antikaon
momentum distributions from 0.6 to 1.3 GeV/c at a laboratory angle of 10.5$^{\circ}$ in $p$Be and $p$Cu interactions
at 2.25 and 2.4 GeV beam energy have been performed at the ITEP/Moscow accelerator \cite{Akindinov_JPG}. A reasonable
description of these data has been achieved in the framework of a folding model, assuming vacuum $K^+$, $K^-$ masses \cite{Akindinov_JPG}. These calculations, based on the target nucleon momentum distribution and on free elementary cross sections, consider incoherent primary proton--nucleon,
secondary pion--nucleon antikaon production processes and processes associated with the creation of antikaons via
the decay $\phi \to K^+K^-$ of intermediate $\phi$ mesons. It has also been shown for the first time \cite{Akindinov_JPG} that the $K^-$ production for momenta $<$ 0.8 GeV/c
is dominated by the $pN \to pN{\phi}$, $\phi \to K^+K^-$ channel on a light nucleus like Be, while on a Cu nucleus
the main contribution to the cross sections comes from this channel and the ${\pi}N \to N K K^-$ process.
This indicates the importance of accounting for the effect of the $\phi$ feed-down in attempts
to extract the antikaon potential from $K^-$ spectra (cf. the same conclusion as drawn above). The $K^-$ excitation functions
in $p$Be and $p$Cu interactions have also been determined for a $K^-$ momentum of 1.28 GeV/c for bombarding energies
$<$ 3 GeV at the ITEP/Moscow accelerator \cite{Akindinov_JETP,Akindinov_ITEP}.
They are reasonably reproduced as well by the above folding model, assuming vacuum $K^+$, $K^-$ masses
\cite{Akindinov_JETP}. On the other hand, these data do not contradict the results of calculations within the
spectral function approach \cite{Paryev_EPJA2003} for incoherent primary proton--nucleon and
secondary pion--nucleon antikaon production processes (but without accounting for the $\phi$ feed-down effect on $K^-$ yield which is insignificant
at this high momentum \cite{Akindinov_JETP,Akindinov_JPG}), 
carried out in the scenario with zero kaon potential and antikaon potential of about -126 MeV at 
density $\rho_0$. This can be explained \cite{Paryev_EPJA2003} by the fact that at a high $K^-$ momentum
of 1.28 GeV/c the calculations are not so sensitive to the strength of the antikaon potential. 
So, one must admit that
to make progress in understanding the strength of the $K^-$ interaction in the nuclear medium, it is
necessary to carry out detailed measurements with tagged low-momentum $K^-$ mesons. In line with the above mentioned,
the $K^-$ mesons must not stem from $\phi \to K^+K^-$ decays so that they bring "genuine" information about the
$K^-$ yield. 

Such measurements have recently been performed by the ANKE Collaboration at the COSY/J$\ddot{\rm u}$lich
accelerator, where the production of $K^+K^-$ pairs with invariant masses corresponding to both the
non-$\phi$ \cite{Kiselev_K-} and $\phi$ \cite{Polyanskiy,Hartmann_PRC85} regions (see also Section 4.7 below)
has been studied in proton collisions with C, Cu, Ag, and Au targets at an initial beam energy of 2.83 GeV.
The $K^-$ momentum dependence of the coincident differential cross section has been measured for laboratory
polar angles $\theta_{K^{\pm}} \le 12^{\circ}$ over the momentum range of 0.2--0.9 GeV/c for these four target nuclei.
The data presented in Fig.~\ref{fig:Kiselev_K-spectra} are compared with detailed model calculations,
based on the nuclear spectral function for incoherent primary proton--nucleon and secondary pion--nucleon
$K^+K^-$ creation processes for different scenarios of the $K^-$ nuclear potential \cite{Paryev_K-}.
In general, the cross sections calculated for $K^-$ potential depths of -60,
-126, and -180 MeV at density $\rho_0$ follow the data for all target nuclei for antikaon momenta above about
0.4 GeV/c; the data exclude small in-medium $K^-$ mass shifts. On the other hand, the data
at lower antikaon momenta are described reasonably well with almost no $K^-$ potential and are overestimated
by all the calculations with a nonzero antikaon potential. This suggests \cite{Kiselev_K-} that the model
misses some aspects of the absorption of low-momentum $K^-$ mesons and/or their production in nuclear
matter. Therefore, to determine the antikaon nuclear potential, the ratio of the measured, momentum integrated $K^-$ cross section for the nonresonant $K^+K^-$ pair production on a given nucleus A to the corresponding cross section calculated within the model for different potential strengths,
has been considered in \cite{Kiselev_K-}, rather than the differential cross sections themselves, which are shown in Fig.~\ref{fig:Kiselev_K-spectra}. This approach has the advantage of decreasing both the statistical and systematic uncertainties. The cross section ratios are shown in Fig.~\ref{fig:Kiselev_K-ratio}.
The condition $\sigma_{exp}/\sigma_{cal}=1$ is achieved for $V_{K^-}(\rho_0) = -\left(63^{+15}_{-12}\right)$ MeV. 
Accounting for the overall systematic uncertainties in the data leads to the extended error band of $V_{K^-}(\rho_0) = -\left(63^{+50}_{-31}\right)$ MeV
\cite{Kiselev_K-}. The antikaon potential depth extracted in this way corresponds to an average $K^-$
momentum of about 0.5 GeV/c, where the main strength of the measured differential
distributions is concentrated \cite{Kiselev_K-}. Within the uncertainties quoted, this value
obtained for the antikaon potential depth is consistent with those extracted in
\cite{HSD,Li_NPA1997,Li_PRL1997,Li_NPA654,Cassing_HSD} from the KaoS 
\cite{Barth_PRL1997,Senger_APP1996,Laue_PRL82} and FRS data \cite{Schroter_ZPA350}, from the FOPI A--A collision data
\cite{Zinyuk_PRC90,Gasik_EPJA52} as well as from the $p$A KaoS data \cite{Scheinast} (see above).
It also agrees with a potential of the order of -100 MeV at saturation density (but for zero $K^-$ momentum),
predicted  in \cite{Schaffner_NPA1997,Waas} and shown in Fig.12 (lower curves). Furthermore, it is consistent  with the moderate $K^-$--nucleus potential of the order of -50 to -80 MeV predicted by calculations based on a chiral Lagrangian [143,144,168-171]
%\cite{Cieply_NPA2001,Ramos_Oset,Schaffner_NPA2000,Ramos_NPA2001,Tolos_PRC74,Tolos_PRC78}
or on meson-exchange potentials \cite{Tolos_NPA690,Tolos_PRC65}, as well as with a smaller potential of about -28 MeV
at density $\rho_0$, extracted at an antikaon momentum of 0.8 GeV/c \cite{SibCas_nuclth99}. However, it is
difficult to reconcile these values with the very deep potentials of order
-200 MeV claimed in experiments studying in-flight ($K^-$,$N$) reactions on $^{12}$C and $^{16}$O at 1 GeV/c \cite{Kishimoto_PTP118,Kishimoto}. The analysis of the latter data has, however, been questioned in \cite{Magas_PRC81}.

Summarising the above considerations, one may conclude that the $K^-$--nucleus interaction is strong and attractive but the presently available results do not permit to draw unambiguous conclusions on the exact strength of the $K^-$--nucleus potential at low energies.
%----------------------------------------------------------------------------------------------------------------------------------------------------------------------------------------------------------------------------------------------
%ETA

\subsection{\it $\eta$--nucleus potential}\label{sec:eta}
While the interaction of kaons with nucleons and nuclei can be studied using kaon beams, $\eta$ and $\eta^\prime$ mesons with lifetimes $\le 5 \cdot 10^{-19}$ s are too short-lived to generate a particle beam. Information on the $\eta$--nucleon and $\eta$ -- nucleus interaction can thus only be extracted from final-state interactions in $\eta$ production off nucleons and nuclei. 

Numerous investigations of $\eta$ production in pion, proton and photon- induced reactions have been performed along with many theoretical studies. Detailed experimental information on these reactions exists and has been summarised in several recent reviews, see e.g. \cite{Machner,Kelkar,Krusche_Wilkin}. In this review only little can be added to this wealth of information. For completeness the main results will be discussed.

\subsubsection{\it The $\eta$ -nucleon scattering length} 
\begin{figure*}
\centering
  \includegraphics[width=15cm,clip]{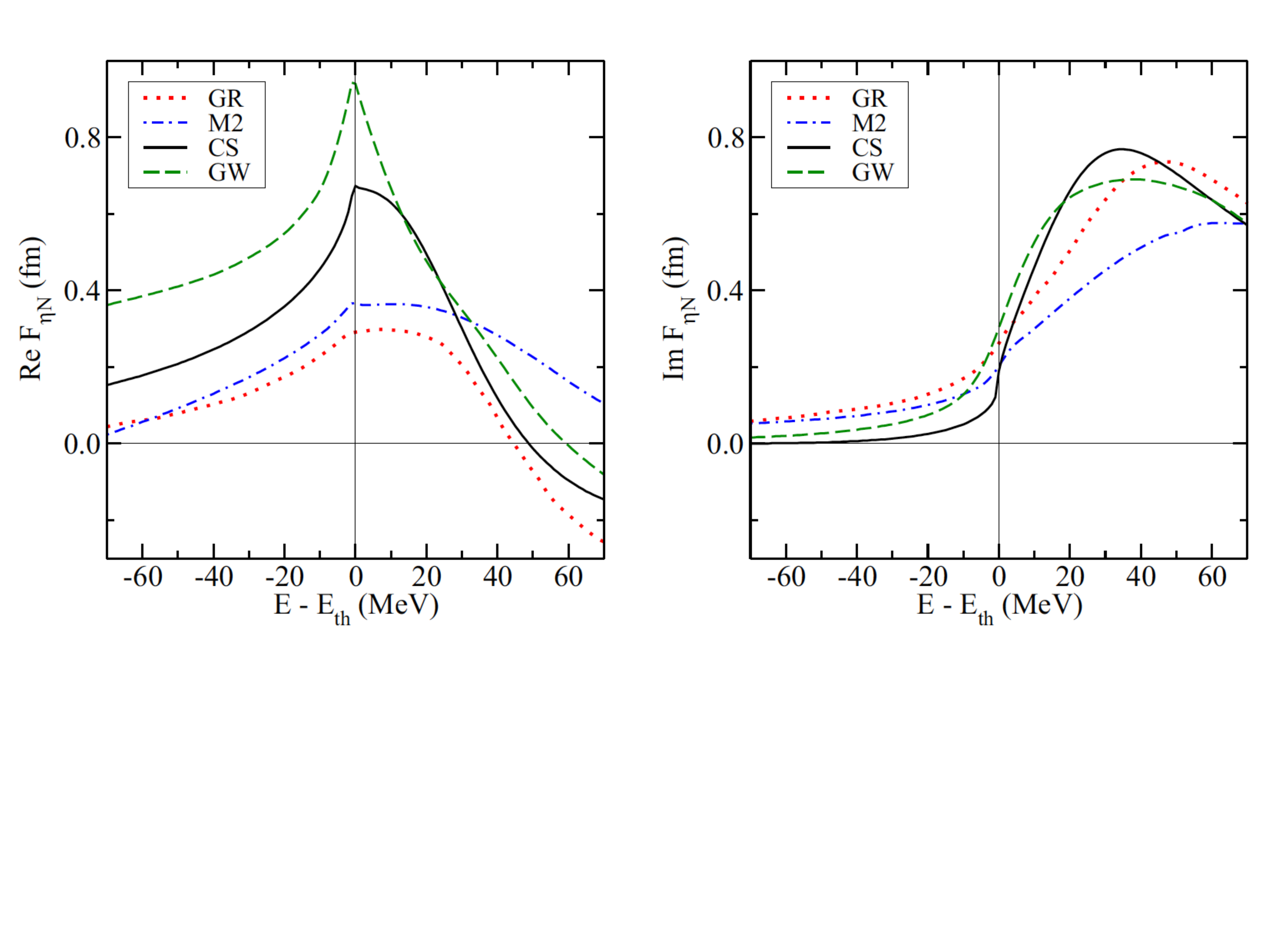} 
\vspace{-35mm}\caption{Energy dependence of the real (Left) and imaginary (Right) parts of the free $\eta N$ scattering amplitude in different interaction models GW \cite{Green_Wycech} (dashed), CS \cite{Cieply_Smejkal} (solid), M2 \cite{Mai} (dot-dashed) and GR \cite{Inoue,Garcia_Recio_PLB550} (dotted). The vertical line denotes the $\eta N$ threshold. The figure is taken from \cite{Mares_eta}.}
\label{fig:eta_scat}
\end{figure*}

The interaction of an $\eta$ meson with a nucleon near threshold is dominated by the S$_{11}(1535) 1/2^-$ resonance which is located just 49 MeV above the production threshold and has a width of 150 MeV. Since the $\eta N \rightarrow \pi N, \pi \pi N$ and $\gamma N$  channels are always open, the $\eta N$ scattering length is complex; the imaginary part reflects the absorptive part of the cross section. All these channels have to be taken into account and thus coupled channel calculations are required to extract the $\eta N$ scattering length.  Analysing data on $\eta$ production in photon-, pion-, and proton- induced reactions with different theoretical models, a broad spread of values has been deduced for the $\eta$  -- nucleon scattering length $a_{\eta N} $ in the range of 0.18 fm $\le Re(a_{\eta N} )\le$ 1.03 fm and 0.16 fm $\le Im(a_{\eta N} )\le$ 0.49 fm \cite{Kelkar}. Recent determinations of the free $\eta N$ scattering amplitude using different interaction models are compiled in Fig.~\ref{fig:eta_scat}.
An experiment is planned at the ELPH facility to determine the real part of the $\eta-N$ scattering length to a precision of $\pm$ 0.1 fm by measuring the $\gamma d \rightarrow \eta p n $ and $\gamma p \rightarrow p n $ cross sections under specific kinematic conditions \cite{Nakamura_1704.07029}. The results of these measuremnets are eagerly awaited.

The $\eta N$ interaction is found to be strong and attractive in the s-wave. The rather large values of the scattering length led Haider and Liu \cite{Haider_Liu} to predict possible exotic states of $\eta$ mesons and nuclei bound by the strong interaction. The current status of searches for $\eta$ mesic states will be discussed in Section \ref{sec:eta_mesic}.

\subsubsection{\it The imaginary part of the $\eta$ - nucleus potential}\label{sec:eta_imag}
\begin{figure*}
\centering
  \includegraphics[width=9cm,clip]{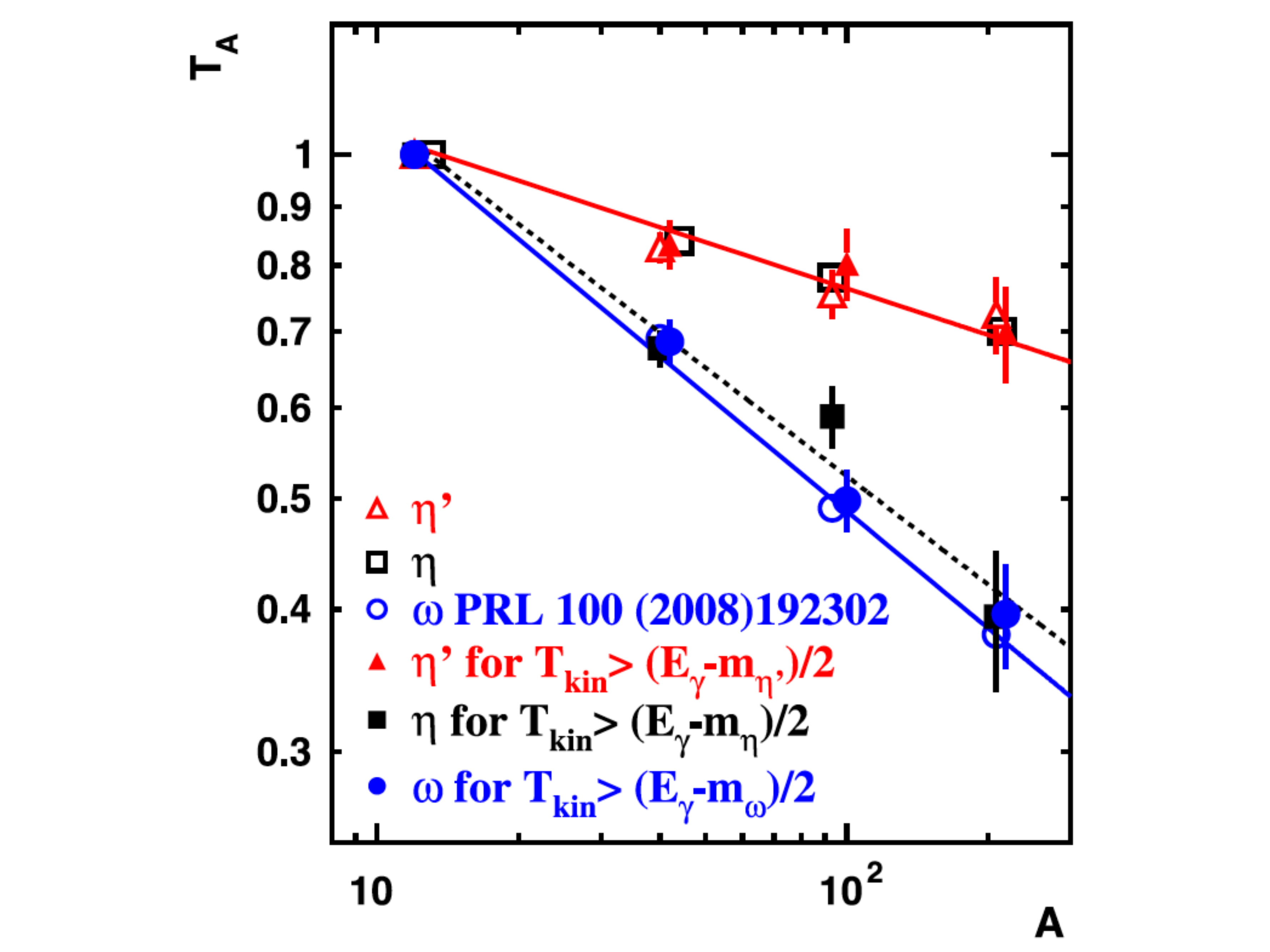} \includegraphics[width=9cm,clip]{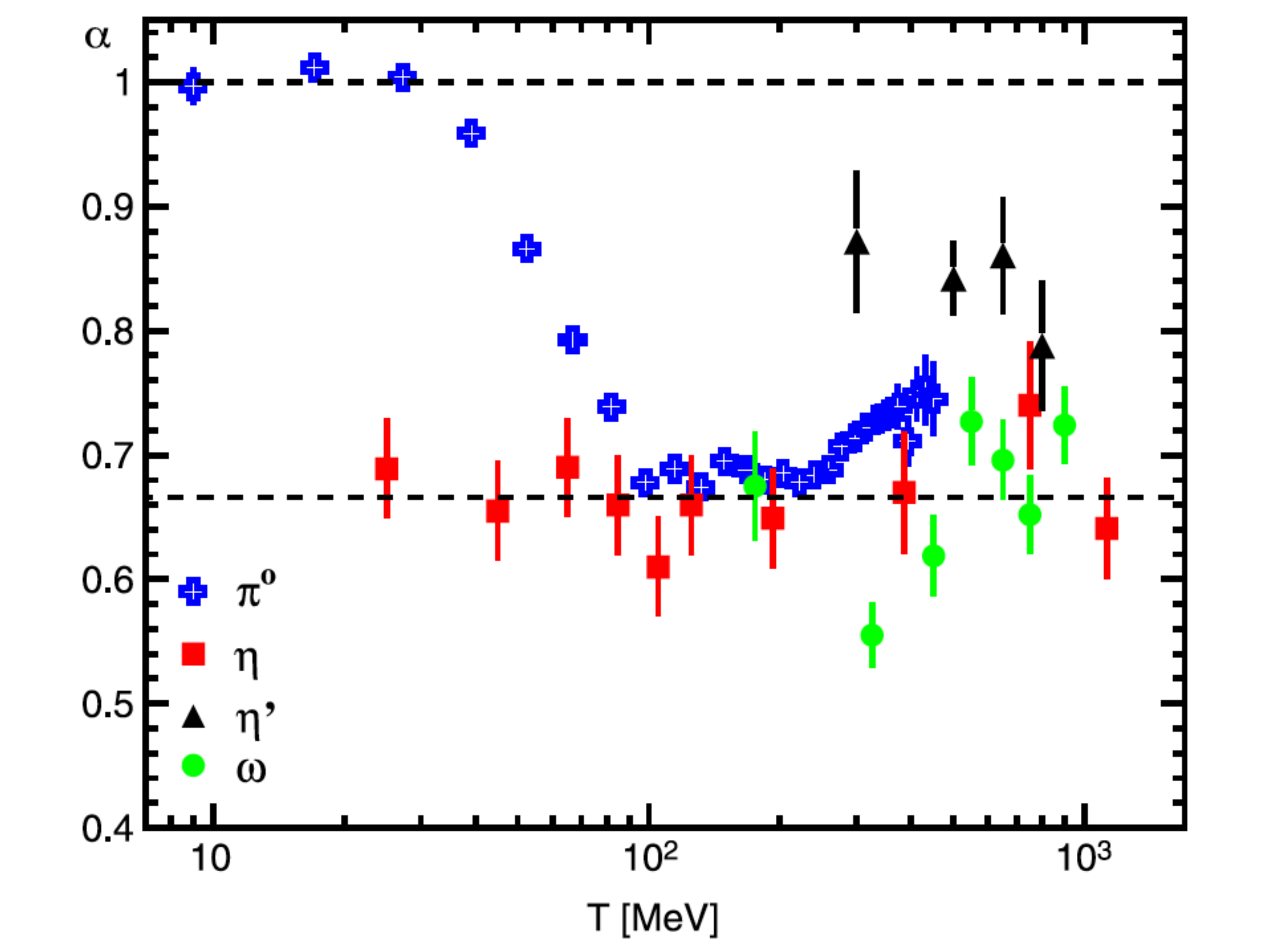}
\caption{(Left) Transparency ratio for $\eta, \omega$ and $\eta^\prime$ mesons as a function of the nuclear mass number A. The transparency ratio with a cut on the kinetic energy of the respective mesons (Eq.~(\ref{eq:Mertens_cut})) to suppress secondary production processes is shown with full symbols. The incident photon energy is in the range from 1500 to 2200 MeV. The lines are fits to the data. (Right) Dependence of the parameter $\alpha$ (Eq.~(\ref{eq:alpha})) on the kinetic energy $T$ of the mesons for $\pi^0, \eta, \omega$ and $\eta^\prime$ mesons. The figures are taken from \cite{Nanova_PLB_TA}. }
\label{fig:sigma_abs}
\end{figure*}

Inclusive $\eta$ photoproduction off nuclei $(^{12}C,^{40}Ca,^{63}Cu,^{93}Nb,^{nat}Pb)$ has been studied at MAMI, ELSA and LNS (Tohoku) [199-202]
%\cite{Roebig-Landau,Mertens,Yorita,Kinoshita} 
for incident photon energies of 0.7 to 2.2 GeV. From the measured cross sections the transparency ratios in Fig.~\ref{fig:sigma_abs} (Left) have been deduced (normalised to carbon) shown together with corresponding data for $\omega$ \cite{Kotulla} and $\eta^\prime$ \cite{Nanova_PLB_TA} mesons which will be discussed in detail in the next subsections. The data points for the $\eta$ show only a weak drop with increasing nuclear mass reaching a transparency ratio of 70$\%$ for a heavy nucleus like Pb. In interpreting this result one has to consider the possibility that the transparency ratio may be distorted by secondary production processes where in a first step the incident photon produces a pion which then produces an $\eta$ meson on another nucleon of the same nucleus with the much larger strong interaction cross section. Because of the energy balance, $\eta$ mesons produced in secondary reactions will have a smaller kinetic energy than those produced directly by the incident photon. As discussed in \cite{Mertens} secondary production processes can thus be suppressed by a cut 
\begin{equation}
T_{\eta} \ge (E_{\gamma} - m_{\eta})/2 \label{eq:Mertens_cut}
\end{equation}
on the kinetic energy $T_{\eta}$ for each incident photon energy $E_{\gamma}$. Applying this cut the transparency ratio for $\eta$ mesons drops to 40$\%$ for heavy nuclei while the transparency ratios for $\omega$ and $\eta^\prime$ mesons are not affected by this cut within errors. This shows that for the $\eta$ meson the transparency ratio has to be corrected for contributions from secondary production processes while two-step processes do not seem to play an important role in the production of the heavier $\omega$ and $\eta^\prime$ mesons. Analysing the corrected $\eta$ transparency ratio as described in Section \ref{sec:imag}, the $\eta$--nucleon inelastic cross section is found to be consistent with earlier Glauber - type analyses of near threshold data, giving $\sigma_{inel}^{\eta} \approx 30$ mb \cite{Roebig-Landau}, consistent with predictions in \cite{Bennhold_Tanabe,Cassing_eta}.This cross section corresponds to a mean free path of $\approx$ 2 fm at normal nuclear matter density. For the $\eta$ momentum range of 150 - 380 MeV/$c$ covered in \cite{Roebig-Landau} this inelastic cross section leads to an 
$\eta$ in-medium width of 25 - 55 MeV within the linear density approximation (Eq.~(\ref{eq:lindens})). The imaginary part of the $\eta$--nucleus potential is thus in the range of -10 to -30 MeV.

An equivalent approach to deduce information on the absorption of mesons in nuclei is to investigate the scaling of the meson production cross section with nuclear mass number. In quasi-free production the incident photon interacts with one individual nucleon off which the $\eta$ meson is produced while the rest of the nucleons in the nucleus only acts as spectators. As long as initial state interactions, i.e. photon shadowing \cite{Bianchi} can be neglected, which holds for $E_{\gamma} \le 1.5 $ GeV, one would expect a scaling of the $\eta$ production cross section with the number of nucleons A, i.e. with the nuclear volume. Any deviation from such a linear dependence expressed by a coefficient $\alpha$ different from 1.0 in 
\begin{equation}
\sigma = A^{\alpha (T)} \label{eq:alpha}
\end{equation}
reflects the absorption of the produced meson in the nucleus and thus provides information on the imaginary part of the $\eta$ --nucleus interaction. Since the meson absorption may be energy dependent, the coefficient $\alpha$ may vary as a function of the kinetic energy $T$ of the meson. If only nucleons on the surface of the nucleus would contribute to meson production because mesons produced in the interior of the nucleus are absorbed due to a large absorption cross section, then $\alpha \approx 2/3$. Thus, for photoproduction $\alpha$ is expected to be in the range $2/3 \le \alpha \le 1.0$. Analysing photoproduction of $\eta$ mesons off nuclei, the dependence of $\alpha$ on the kinetic energy of $\eta$ mesons has been deduced \cite{Roebig-Landau,Mertens} after correction for secondary production processes (Eq.~(\ref{eq:Mertens_cut})) and is shown in Fig.~\ref{fig:sigma_abs} again together with corresponding results for $\omega$ \cite{Kotulla} and $\eta^\prime$ mesons \cite{Nanova_PLB_TA}. The data on $\pi^0$ mesons \cite{Krusche} show that nuclei are transparent for $\pi^0$ mesons with kinetic energies below 30 MeV but they get strongly absorbed when the $\pi^0$ kinetic energy becomes sufficient to excite the $\Delta(1232)$ and higher baryon resonances in collisions with a nucleon. For the $\eta$ meson the $\alpha$ values are close to 2/3, consistent with surface production and the mean free path of $\approx$ 2 fm, deduced above. The $\omega$ meson shows strong absorption as well while the $\eta^\prime$ meson exhibits a much weaker attenuation in normal nuclear matter.
Details for $\eta^\prime$ and $\omega$ mesons will be discussed in Sections \ref{sec:etaprime} and \ref{sec:omega}, respectively.

\subsubsection{\it The real part of the $\eta$ -nucleus potential} 
Although numerous measurements of the excitation function for $\eta$ photoproduction off nuclei have been reported [199-202]
%\cite{Roebig-Landau,Mertens,Yorita,Kinoshita} 
no attempts have been made as yet to extract information on the real part of the $\eta$--nucleus potential in the spirit of Section \ref{sec:real_exci}. No quantitative values  for the $\eta$--nucleus potential can thus be quoted. The interaction is known to be attractive and theoretical expectations will be discussed in the context of possible $\eta$-nucleus bound states in Section \ref{sec:eta_mesic}.

%------------------------------------------------------------------------------------------------------------------------------------------------------------------------------------------------------------------------------------------
%ETAPRIME

\subsection{\it $\eta^{\prime}$--nucleus potential}\label{sec:etaprime}

\subsubsection{\it The $\eta^{\prime}$ -nucleon scattering length} 
The  $\eta^\prime N$ scattering length has been estimated from the study of the $pp \to pp \eta^\prime$ cross section near threshold at COSY  \cite{Moskal_PLB482}. A refined analysis of this reaction, comparing the cross section with that of the $pp \to pp \pi ^0$ reaction, concluded that the scattering length should be of the order of magnitude of that of the $\pi N$ interaction. The analysis of the new COSY-11 data updated the results to values: $Re(a_{p\eta^\prime}) = 0 \pm 0.43 $ fm and $Im(a_{p\eta^\prime}) = 0.37^{+0.4}_{-0.16}$ fm\cite{Czerwinski}. This indicates a rather weak $\eta^\prime N$ interaction. Nevertheless, a lowering of the $\eta^\prime$ mass by 100-150 MeV has been predicted in calculations within the Nambu-Jona-Lasinio model (NJL) \cite{Nagahiro_Hirenzaki, Costa, Nagahiro_PRC74}.

\subsubsection{\it The imaginary part of the $\eta^{\prime}$ -nucleus potential} 
As discussed in Sec.~\ref{sec:imag} the experimental approach to determine the in-medium width and the imaginary part of the meson-nucleus potential (Eq.~(\ref{eq:W})) is the transparency ratio measurement. First experimental data on the  $\eta^\prime$ transparency ratio for several nuclei have been reported by the CBELSA/TAPS Collaboration~\cite{Nanova_PLB_TA}. The obtained in-medium width is $\Gamma(\rho_N=\rho_{0}) \approx  20-30$~MeV corresponding to an inelastic $\eta^\prime N$ cross section of $\sigma_{\rm inel} \approx$ 9-13~mb. A comparison to photoproduction cross sections and transparency ratios measured for other mesons ($\pi,\eta,\omega$) demonstrates the relatively weak interaction of the $\eta^\prime$-meson with nuclear matter. As discussed in Sec.\ref{sec:imag} the two-step meson production could distort the absorption measurement. The effect of these processes, via possible pion-induced reaction $\gamma N \rightarrow \pi N \rightarrow \eta^\prime N$, has been studied by applying a cut on the kinetic energy of the $\eta^\prime$'s ~\cite{Nanova_PLB_TA}. The results show definitely no contribution to the $\eta^\prime$ production via this channel in the observed incident photon energy because $\sigma_{ \pi N \rightarrow \eta^\prime N}$ is only $\approx$ 0.1 mb measured at p$_{\pi} \approx$1.5 GeV/c~\cite{Landolt-B} (see also 
Fig.~\ref{fig:etaprime_A}).

\begin{figure}
\centering
  \includegraphics[width=9cm,clip]{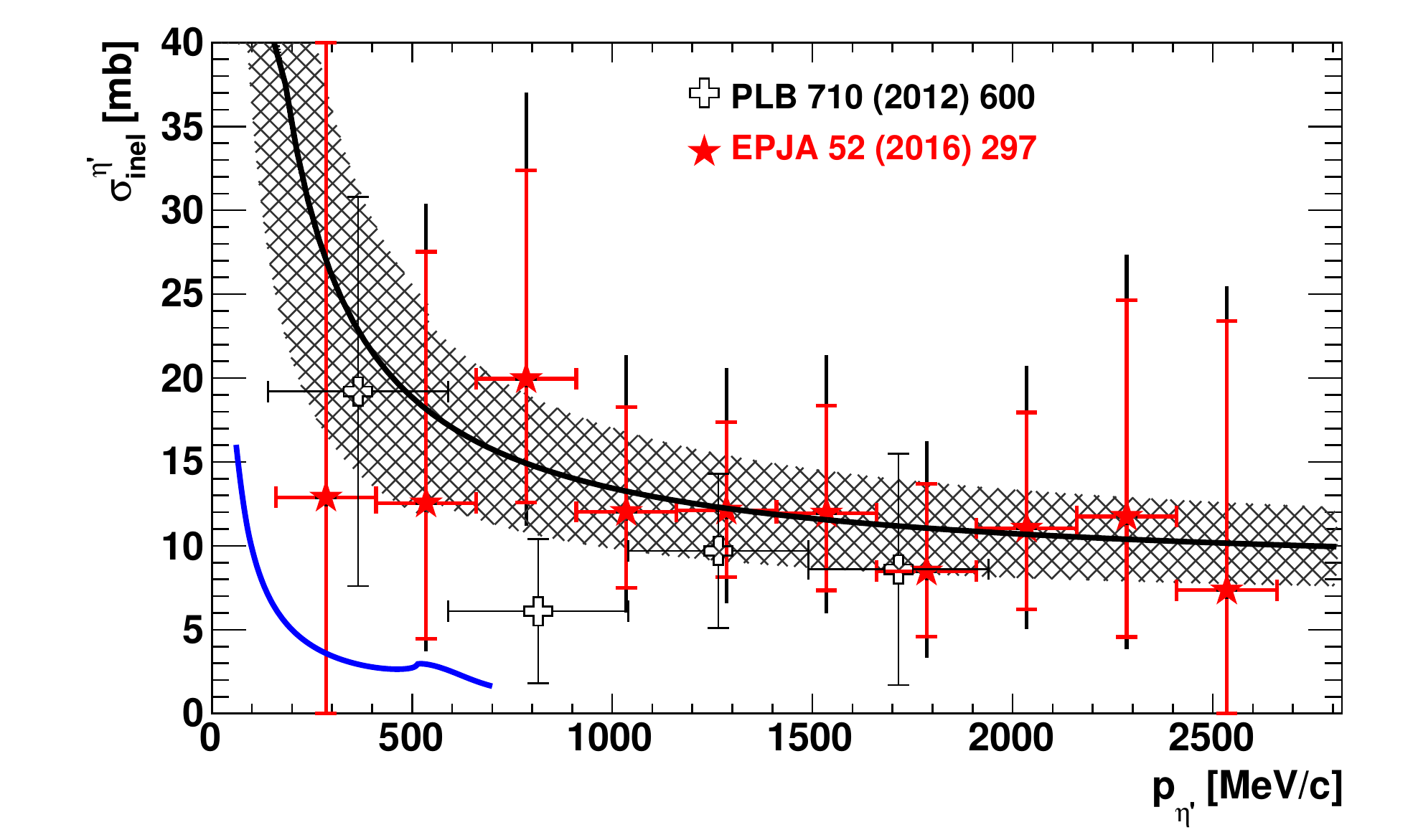} \includegraphics[width=9cm,clip]{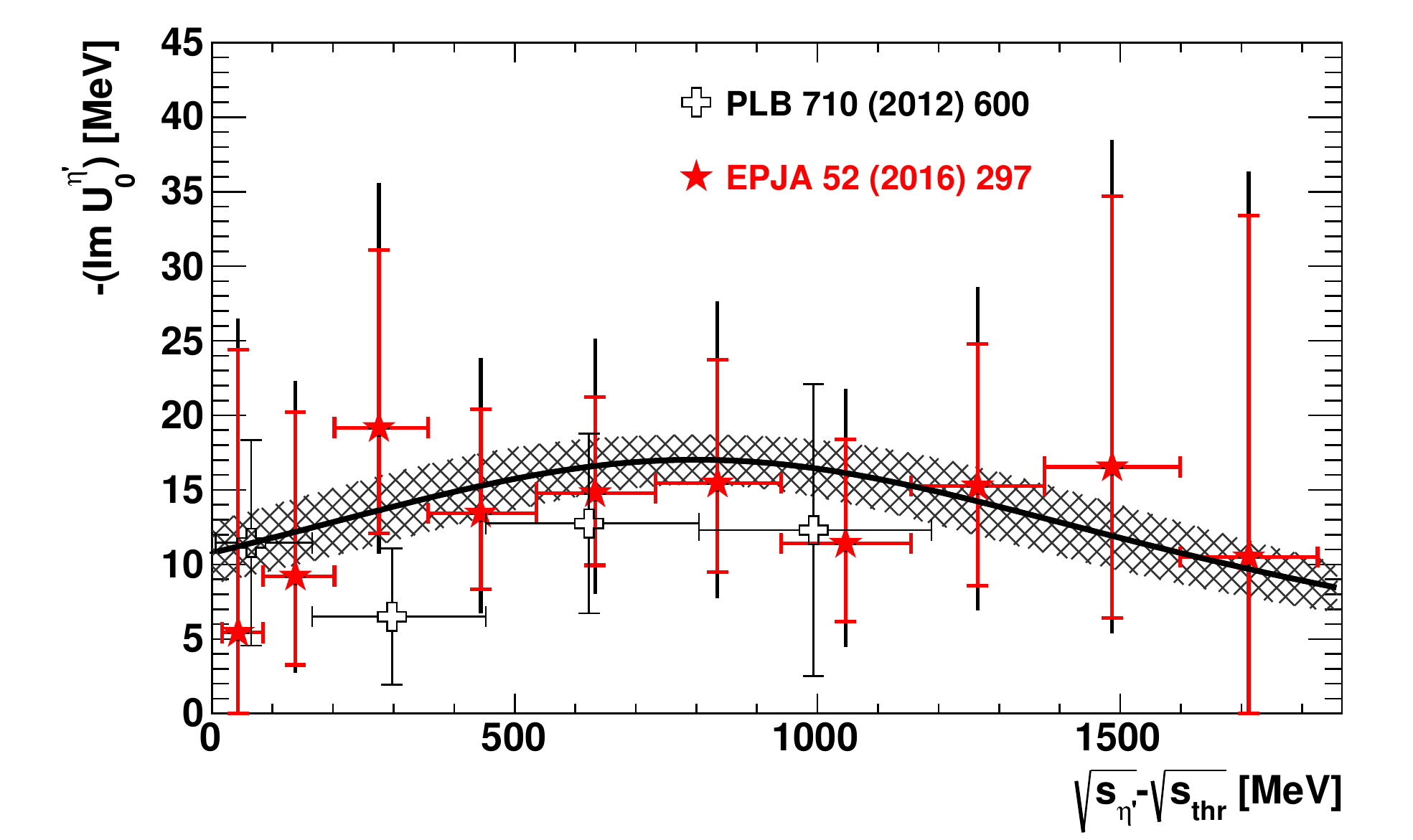}
\caption{(Left) Inelastic $\eta^\prime$-nucleon cross sections deduced from Eq.~(\ref{eq:lindens}) as a function of the meson momentum (red stars)~\cite{Friedrich_EPJA} in comparison to earlier measurements (open crosses) \cite{Nanova_PLB_TA}. The solid black curve is a fit to the data and the shaded area indicates a confidence level of $\pm$1$\sigma$ of the fit curve taking statistical and systematic errors into account. The thick (red) error bars represent the statistical errors. The thin (black) error bars include the systematic errors added in quadrature. The error weighted mean value of the $\eta^\prime$ absorption cross section is (13 $\pm$ 3) mb. The blue curve represents the inelastic $\eta^\prime$-nucleon cross section calculated in \cite{Oset_PLB704}.  (Right) Imaginary part of the $\eta^\prime$-nucleus optical potential as a function of the available energy in the meson-$^{93}$Nb system (red stars)~\cite{Friedrich_EPJA} in comparison to earlier measurements (open crosses) \cite{Nanova_PLB_TA}. The solid curve is a Breit-Wigner fit to the present data. The shaded areas indicate a confidence level of $\pm$1$\sigma$ of the fit curve taking statistical and systematic errors into account. The figures are taken from \cite{Friedrich_EPJA}. With kind permission of The European Physical Journal (EPJ).}
\label{fig:ImU}
\end{figure}

Obviously, it is important to extract information on the in-medium width of the mesons for very low meson momenta, close to the production threshold. Therefore the CBELSA/TAPS Collaboration has studied the momentum dependence of the transparency ratio \cite{Friedrich_EPJA} and extracted the $\eta^\prime$ absorption cross section and the in-medium width of the $\eta^\prime$ meson. The inelastic cross sections $\sigma_{inel}$ has been derived, using Eq.~(\ref{eq:lindens}), and is shown as a function of the $\eta^\prime$ momentum in Fig.~\ref{fig:ImU} (Left). The observed mean value of (13~$\pm$~3) mb is slightly larger but consistent with the earlier result of (10.3 $\pm$ 1.4) mb reported in \cite{Nanova_PLB_TA}. The experimental data are compared to calculations by Oset and Ramos \cite{Oset_PLB704}, within a chiral unitary approach, including $\pi N$ and $\eta N$ coupled channels. The predictions seem to underestimate the experimentally determined inelastic $\eta^\prime$ cross section. This may not be surprising since multi-particle production, probably dominant because of the large $\eta^\prime$ mass, has not been considered in \cite{Oset_PLB704}.

The resulting momentum dependence of the in-medium $\eta^\prime$ width has been converted into the dependence of the imaginary part of the $\eta^\prime$-nucleus potential as a function of the available energy in the meson-$^{93}$Nb system, as presented in \cite{Friedrich_EPJA} . The obtained spectrum is shown in Fig.~\ref{fig:ImU} (Right) compared to the previous measurement with CBELSA/TAPS detector system~\cite{Nanova_PLB_TA}. The data have been fitted and extrapolated towards the production threshold. For the $\eta^\prime$ meson the extrapolation towards the production threshold yields an imaginary potential of -(13 $\pm$ 3(stat)$\pm$3(syst)) MeV, corresponding to an imaginary part of the $\eta^\prime$ scattering length $Im(a_{\eta^\prime N}$) = (0.16 $\pm$ 0.05)~fm. This is actually about a factor two smaller than obtained in the direct determination of the $\eta^\prime N$ scattering length from an analysis of near threshold $\eta^\prime$ production in the $p p \rightarrow p p \eta^\prime$ reaction \cite{Czerwinski}, but almost overlaps within the errors.

\subsubsection{\it The real part of the $\eta^{\prime}$ -nucleus potential} 
Recently many studies have focused on the $\eta^\prime$ meson. Its especially large mass compared to the mass of the other pseudoscalar mesons has to be attributed to chiral and flavor symmetry breaking effects (see Section \ref{sec:motivation}). Due to a reduction of the chiral condensate in the nuclear medium a drop in the U$_{A}(1)$ breaking part of the $\eta^\prime$ mass might be expected \cite{Nagahiro_PRC74,Kwon_PRD86}, causing an $\eta^\prime$ mass shift of $\approx$-120 MeV at nuclear matter density $\rho_{0}$. This prediction is, however, in conflict with earlier calculations within the Nambu-Jona-Lasinio-model which expect almost no change in the $\eta^\prime$ mass as a function of nuclear density \cite{BMetaprime}. Further model calculations, linear $\sigma$ model  \cite{Jido}  and QMC model \cite{Bass}, claim mass shifts of the $\eta^\prime$ of -80, and -40 MeV at $\rho_{0}$, respectively. It is obvious that these contradictory theoretical predictions call for an experimental clarification.

\begin{figure*}
\centering
\includegraphics[width=8.0cm,clip]{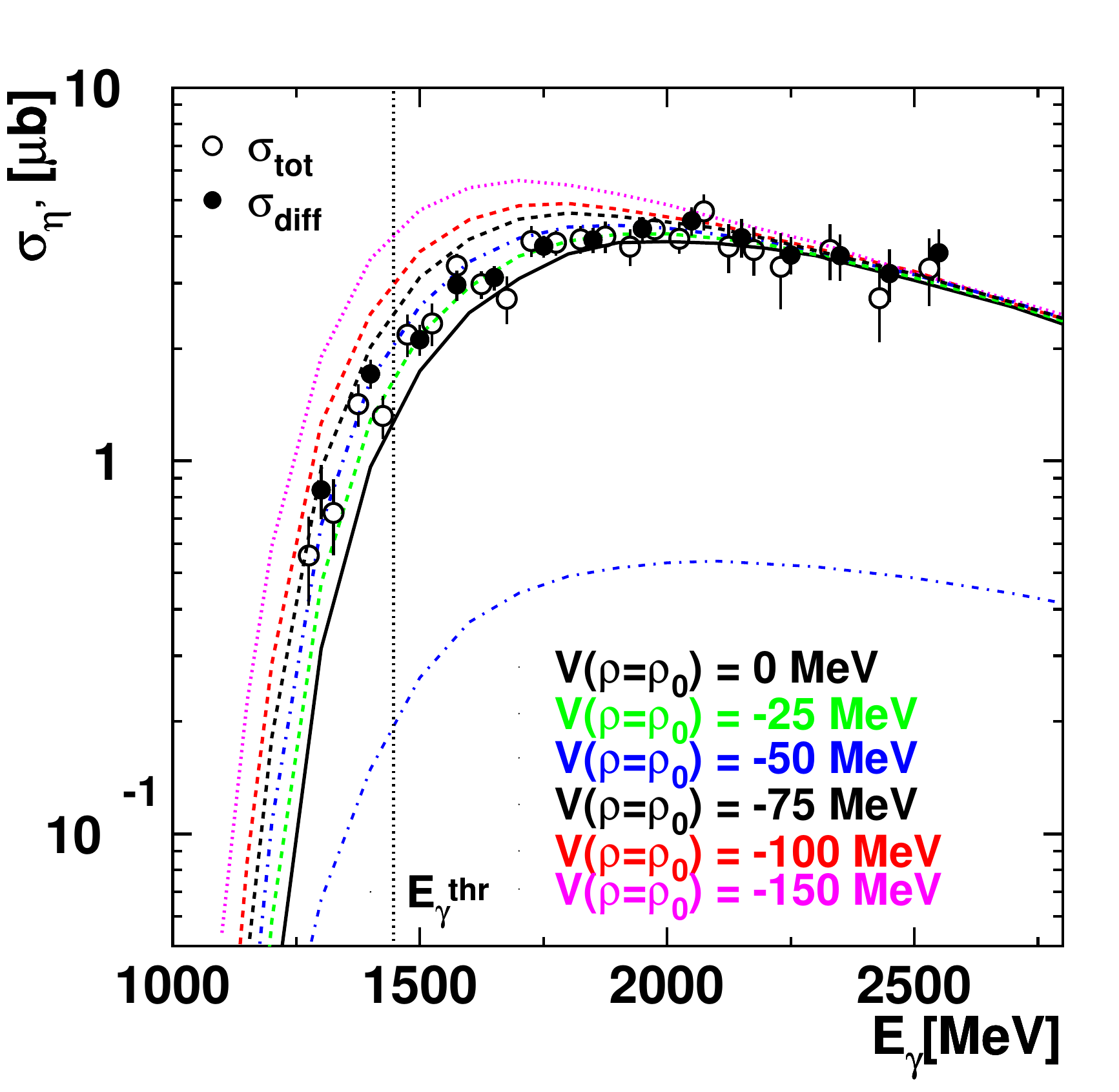} \includegraphics[width=8.0cm,clip]{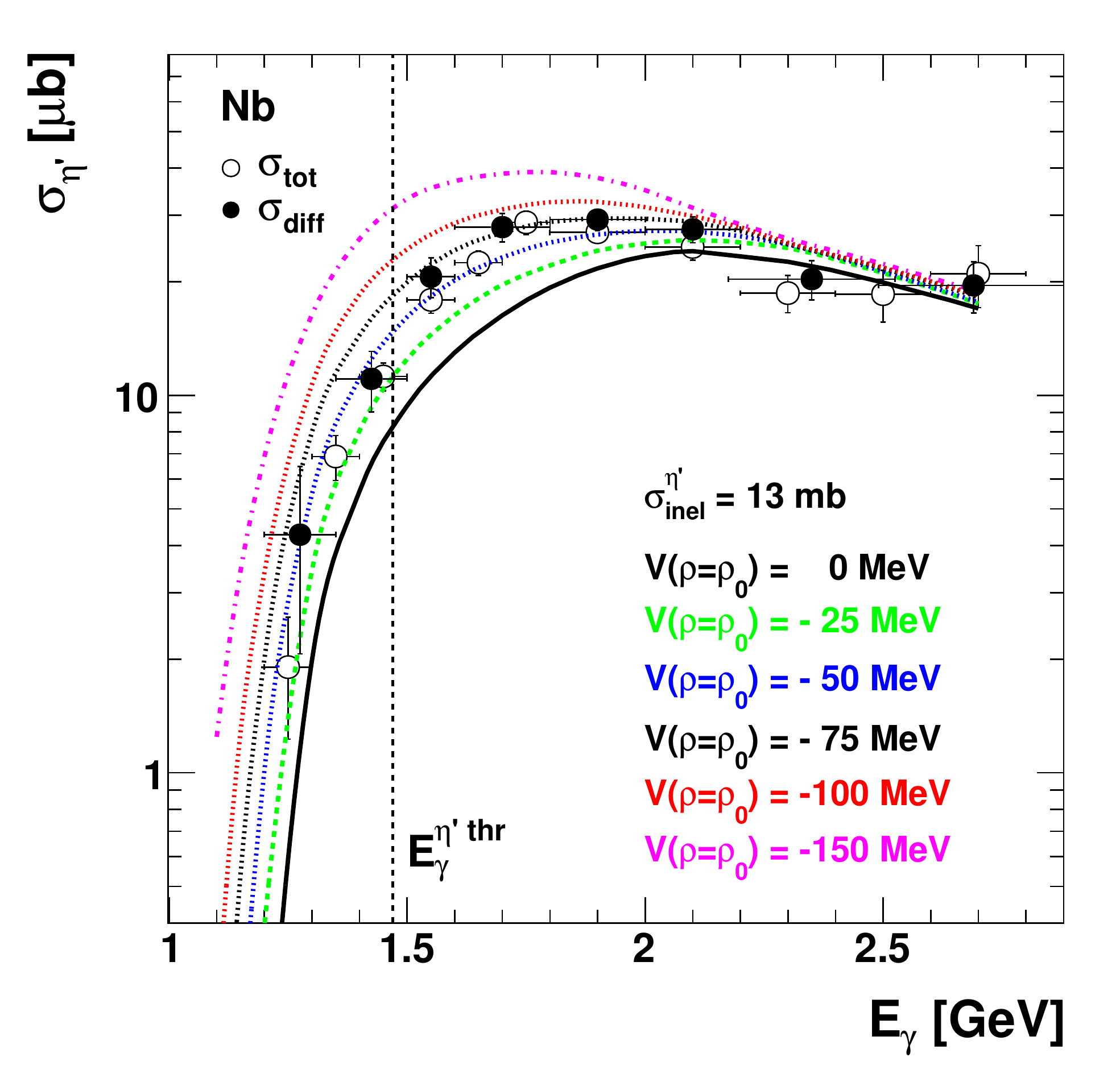}
\caption{Measured excitation function for $\eta^\prime$ meson off ${}^{12}\textrm{C}$ (left) \cite{Nanova_PLB727} and ${}^{93}\textrm{Nb}$ (right) \cite{Nanova_PRC94}, in comparison to theoretical calculations for different scenarios. The experimental data are extracted by integrating the differential cross sections (full circles) and by direct measurement of the $\eta^\prime$ yield in  incident photon energy bins (open circles). The calculations are for $\sigma_{\eta^\prime N}$=11 mb (for C data) and for $\sigma_{\eta^\prime N}$=13 mb (for Nb data), and for potential depths: $V_0$ =0 MeV (black line), -25 MeV (green), -50 MeV (blue), -75 MeV (black dashed), -100 MeV (red) and -150 MeV (magenta) at normal nuclear density, respectively, and using the full nucleon spectral function. The dot-dashed blue curve (in the left spectrum) is calculated for correlated intranuclear nucleons only.}
\label{fig:exc_etapr}
\end{figure*}

The experimental approaches to determine the real part of the potential, which have been applied for $\eta^\prime$ mesons and discussed in Sec.~\ref{sec:real}, are the measurement of the (i) excitation function of the meson and (ii) of the meson momentum distributions. Both these approaches are sensitive to the production point of the meson. The $\eta^\prime$'s as long-lived mesons ($\tau$=1000 fm/c) decay predominantly outside of the nucleus and therefore experimental approaches sensitive to the decay point of the meson, like a line shape analysis, are not applicable. 
The excitation function and momentum distribution of $\eta^\prime$ mesons have been measured in photon induced reactions on $^{12}{}$C~\cite{Nanova_PLB727} and $^{93}{}$Nb~\cite{Nanova_PRC94} in the energy range of 1200-2600 MeV. The experiments have been performed with tagged photon beams from the ELSA electron accelerator using the Crystal Barrel and TAPS detectors. Differential and total $\eta^\prime$ production cross sections have been measured and compared to model calculations using the collision model based on the nucleon spectral function, described in Sec.~\ref{sec:collmod}. Calculations are performed for different scenarios assuming depths of the $\eta^\prime$ real potential at normal nuclear matter density of $V_0$=0, -25, -50, -75, -100 and -150 MeV, respectively, and including $\sigma_{inel}$=11 mb, consistent with the result of transparency ratio measurements \cite{Nanova_PLB_TA}. The measured excitation functions for $\eta^\prime$ mesons on C and Nb solid targets are shown in Fig.~\ref{fig:exc_etapr}. An enhancement of the total cross section below the production threshold (1445 MeV) can be clearly seen in both cases, due to increased phase-space caused by the in-medium mass lowering of the meson (see Sec.~\ref{sec:real}). Within the model, the comparison indicates an attractive  potential of -(40$\pm$6) MeV at normal nuclear matter density from the C data~\cite{Nanova_PLB727} and of -(40$\pm$12) MeV for the $\eta^\prime$-Nb potential. The obtained results do not indicate any dependence of the potential parameters on the nuclear mass number A. 

It has been investigated whether the observed cross section enhancement relative to the $V$=0 MeV case could also be due to $\eta^\prime$ production on dynamically formed compact nucleonic configurations - in particular, on pairs of correlated nucleon clusters - which share energy and momentum. Applying the parametrisation of the spectral function given by~\cite{Efremov_1998} to the theoretical calculations for different scenarios and comparing with the experimental data demonstrated that the correlated high momentum nucleons contribute only about 10-15\% to the $\eta^\prime$ yield in the incident energy regime above 1250 MeV Fig.~\ref{fig:exc_etapr} (left)~\cite{Nanova_PLB727}. The observed cross section enhancement can therefore be attributed mainly to the lowering of the $\eta^\prime$ mass in the nuclear medium.\\

\begin{figure*}
\centering
\includegraphics[width=8cm,clip]{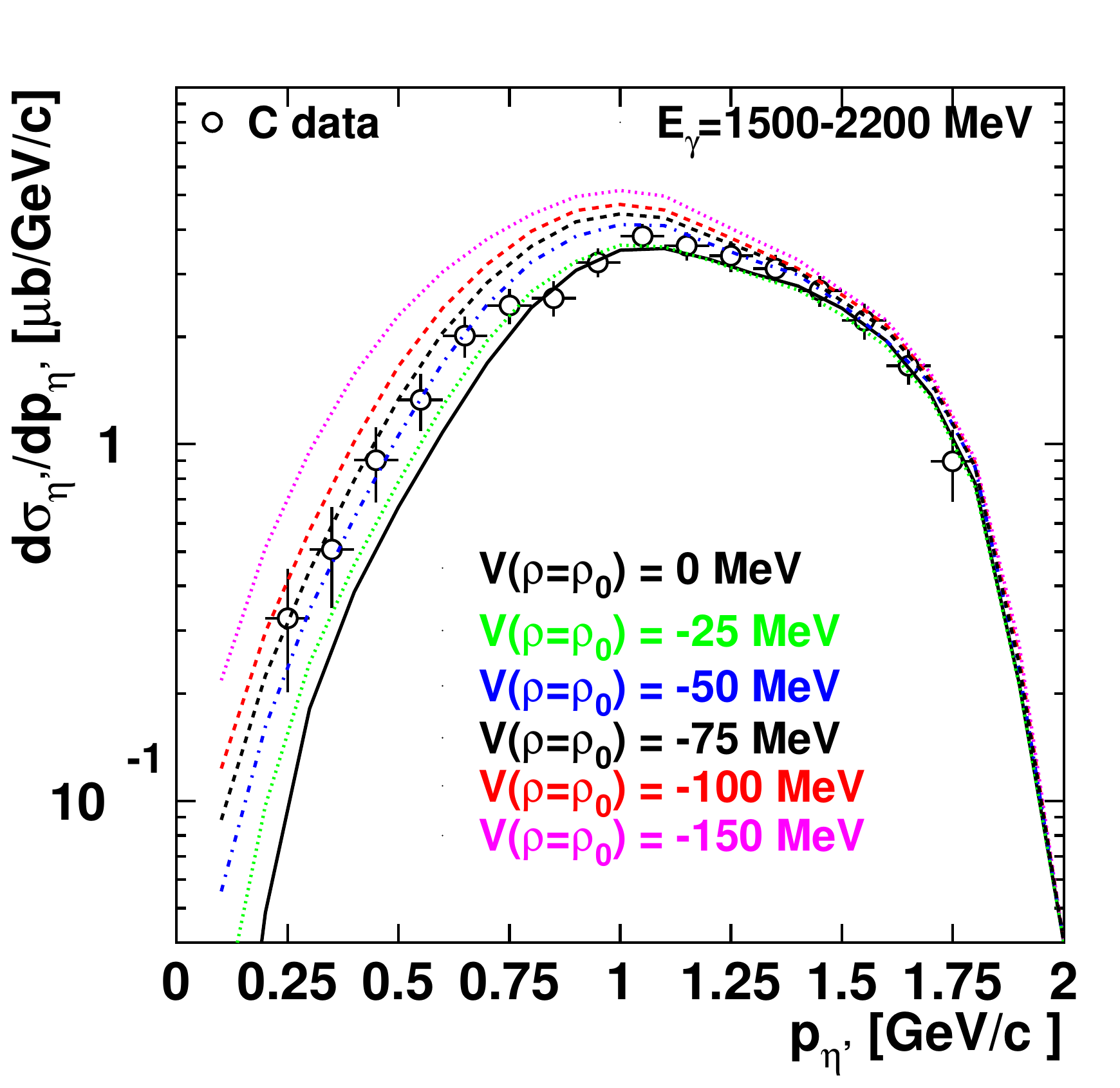} \includegraphics[width=8cm,clip]{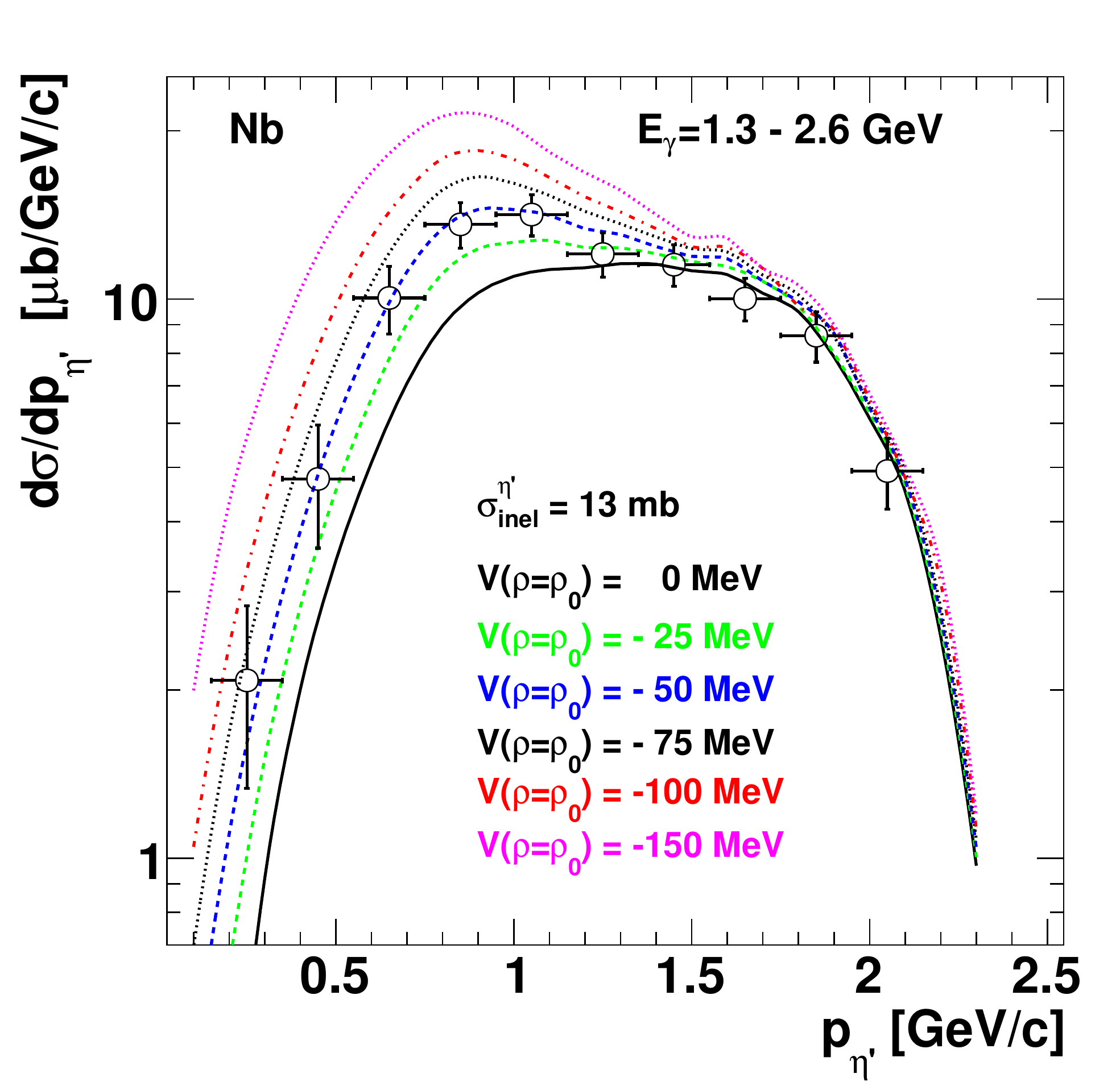}
\caption{Left: Momentum distribution for $\eta^\prime$ photoproduction off C for the incident photon energy range 1500-2200 MeV. The calculations are for
$\sigma_{\eta^\prime N}$=11 mb and have been reduced by a factor 0.75 to match the data at high momenta \cite{Nanova_PLB727}. Right: Momentum distribution for $\eta^\prime$ photoproduction off Nb for the incident photon energy range 1.3-2.6 GeV. The calculations are for $\sigma_{inel}^{\eta^\prime}$=13 mb and have been multiplied by a factor 0.83 \cite{Nanova_PRC94}. In both spectra the theoretical curves are for potential depths $V_0$ = 0, -25, -50, -75, -100 and -150 MeV at normal nuclear density. }
\label{fig:mom_etapr}
\end{figure*}

As a consistency check for the deduced $\eta^\prime$-potential depth, the momentum distribution of $\eta^\prime$ mesons, which is also sensitive to the potential depth, has been investigated by CBELSA/TAPS Collaboration~\cite{Nanova_PLB727, Nanova_PRC94} as well. A comparison of the measured and calculated momentum distributions in the incident photon energy range of 1500-2200 MeV for C and in the incident photon energy range of 1300-2600 MeV for Nb is shown in Fig.~\ref {fig:mom_etapr}~\cite{Nanova_PLB727, Nanova_PRC94}. The momentum resolution of the CB/TAPS detector system is around 25-50 MeV/c, smaller than the chosen bin size of 100 MeV/c, and this does not affect the sensitivity of the potential depth determination. The comparison of data and calculations again seems to exclude strong $\eta^\prime$ mass shifts. A $\chi^{2}$-fit of the data with the calculated momentum distributions for the different scenarios over the full range of incident energies gives a potential depth of -(32$\pm$11) MeV for  $\eta^\prime$-C and  -(45$\pm$20) MeV for $\eta^\prime$-Nb.\\

\begin{figure}
\centering
\includegraphics[width=12cm,clip]{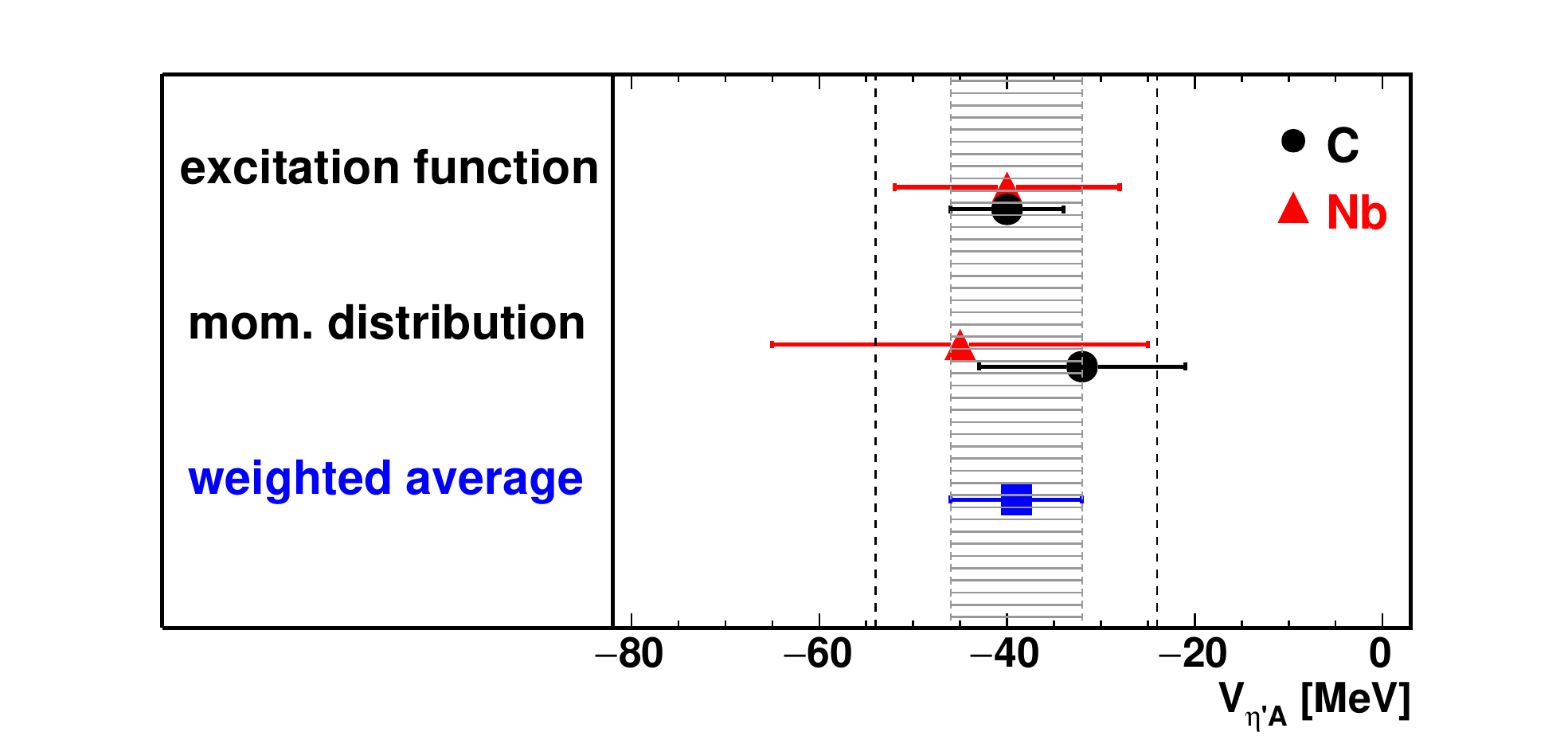}
\caption{Depths of the real part of the $\eta^\prime$ -nucleus potential determined by analyzing the excitation function and the momentum distributions for C \cite{Nanova_PLB727} (full black circles) and for Nb ~\cite{Nanova_PRC94} (red triangles). The weighted overall average is indicated by a blue square and the shaded area. The vertical hatched lines mark the range of systematic uncertainties. The figure is taken from \cite{Nanova_PRC94}.}
\label{fig:concl_etapr}
\end{figure}

The depth of the potential determined for the real part of the $\eta^\prime$-C and $\eta^\prime$-Nb interactions is compared in Fig.~\ref{fig:concl_etapr}~\cite{Nanova_PRC94}. The values deduced by analysis of the excitation functions and the momentum distributions do agree for both nuclei within errors. Obviously there is no evidence for a strong variation of the potential parameters with the nuclear mass number. The combined fit of the C- and Nb-data gives a weighted average of V$(\rho_N=\rho_0) = -(39\pm 7(stat)\pm15(syst))$ MeV reported in~\cite{Nanova_PRC94} and shown in Fig.~\ref{fig:concl_etapr}.
Furthermore, it has been concluded that the mass of the $\eta^\prime $ meson is lowered by 
about 40 MeV in nuclei at saturation density, within the errors quoted for the potential depth. The results for V$(\rho_N=\rho_0)$ are consistent with predictions of the $\eta^\prime$-nucleus potential depth within the QMC model \cite{Bass} and with calculations in \cite{Nagahiro_Oset} but does not support larger mass shifts as discussed in \cite{Costa,Nagahiro_PRC74,Jido}. The results confirm the (indirect) observation of a mass reduction of the $\eta^\prime$ meson in a strongly interacting environment. The attractive $\eta^\prime$-nucleus potential may be sufficient to allow the formation of bound $\eta^\prime$-nucleus states. The search for such states is encouraged by the relatively small imaginary potential of the $\eta^\prime$~of $\approx$ -10 MeV~\cite{Nanova_PLB_TA}. However, because of the relatively shallow $\eta^\prime$-nucleus potential, the search for $\eta^\prime$-mesic states has turned out to be more difficult than initially anticipated on the basis of theoretical predictions \cite{Costa,Nagahiro_PRC74,Jido}.

%\clearpage

%----------------------------------------------------------------------------------------------------------------------------------------------------------------------------------------------------------------------------------------------
%OMEGA
%\clearpage

\subsection{\it The $\omega$--nucleus potential}\label{sec:omega}
\subsubsection{\it The $\omega$--nucleon scattering length}
\begin{figure}[h]
\centering
\includegraphics[width=8cm,clip]{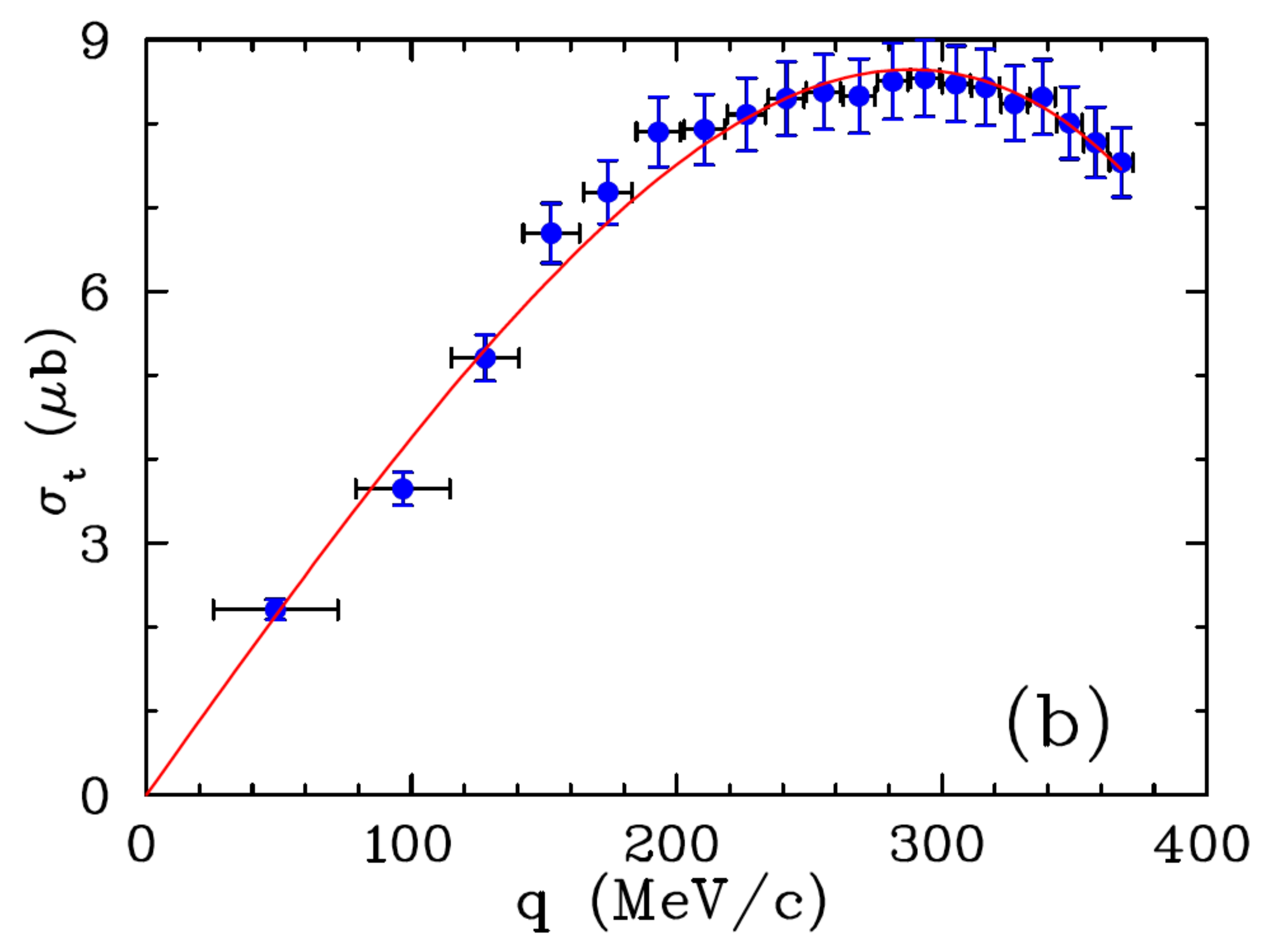}
\caption{Photoproduction cross section for $\omega$ mesons off the proton near threshold as a function of the $\omega$ momentum q in the c.m. system. The red solid curve shows the fit of the data with Eq.~(\ref{eq:sig_thresh}). The figure is taken from \cite{Strakovsky}.}.
\label{fig:Strakovsky}
\end{figure}

The strength of the  $\omega$--nucleon interaction can be extracted from photoproduction experiments off the proton near the meson production threshold which provides information on the
$\omega N$ scattering length. Following Strakovsky et al. \cite{Strakovsky} , the $\omega$ photoproduction cross section near threshold (see Fig.~\ref{fig:Strakovsky}) can be parametrized  as
\begin{equation}
\sigma(q) = a_1 q + a_3 q^3 + a_5 q^5, \label{eq:sig_thresh}
\end{equation}
assuming a contribution of s-, p-, and d-waves. q is the $\omega$ momentum in the c.m. system. Very close to threshold the higher order terms can be neglected and the linear term is determined by the s wave only with the total spin of 1/2 and/or 3/2.
Within the vector meson dominance model (VDM) the cross section near threshold is related to the modulus of the $\omega$--$p$ scattering length $a_{\omega p} $ through \cite{Titov}
\begin{equation}
\sigma (\gamma p \rightarrow \omega p)_{th} = \frac{q}{k} \cdot \frac{4 \alpha \pi^2}{\gamma^2} \cdot \mid a_{\omega p} \mid^2, \label{eq:sig_a}
\end{equation}
where $k$ is the c.m. momentum of the incident photon at the production threshold, $\alpha$ is the fine structure constant and $\gamma = 8.53 \pm 0.14$ is the strength of the
$\gamma - \omega$ coupling, derived from the $\omega \rightarrow e^+ e^-$ decay width. Combining Eqs.~(\ref{eq:sig_thresh}),~(\ref{eq:sig_a}), one obtains
\begin{equation}
\mid a_{\omega p} \mid = \frac{\gamma}{2 \pi} \sqrt{\frac{k a_1}{\alpha}} = (0.82 \pm 0.03)~\mbox{{\rm fm}},
\end{equation}
using the value of $ a_1$ obtained by fitting the experimental excitation function. This result shows an appreciable strength of the $\omega$--$N $ interaction. Since the sign could not be determined it is not clear whether the interaction is attractive or repulsive. Furthermore, the real and imaginary parts
of $a_{\omega p} $ can not be given separately. This information, however, becomes available when studying the $\omega$--nucleus interaction.
\begin{figure}
\centering
  \includegraphics[width=8cm,clip]{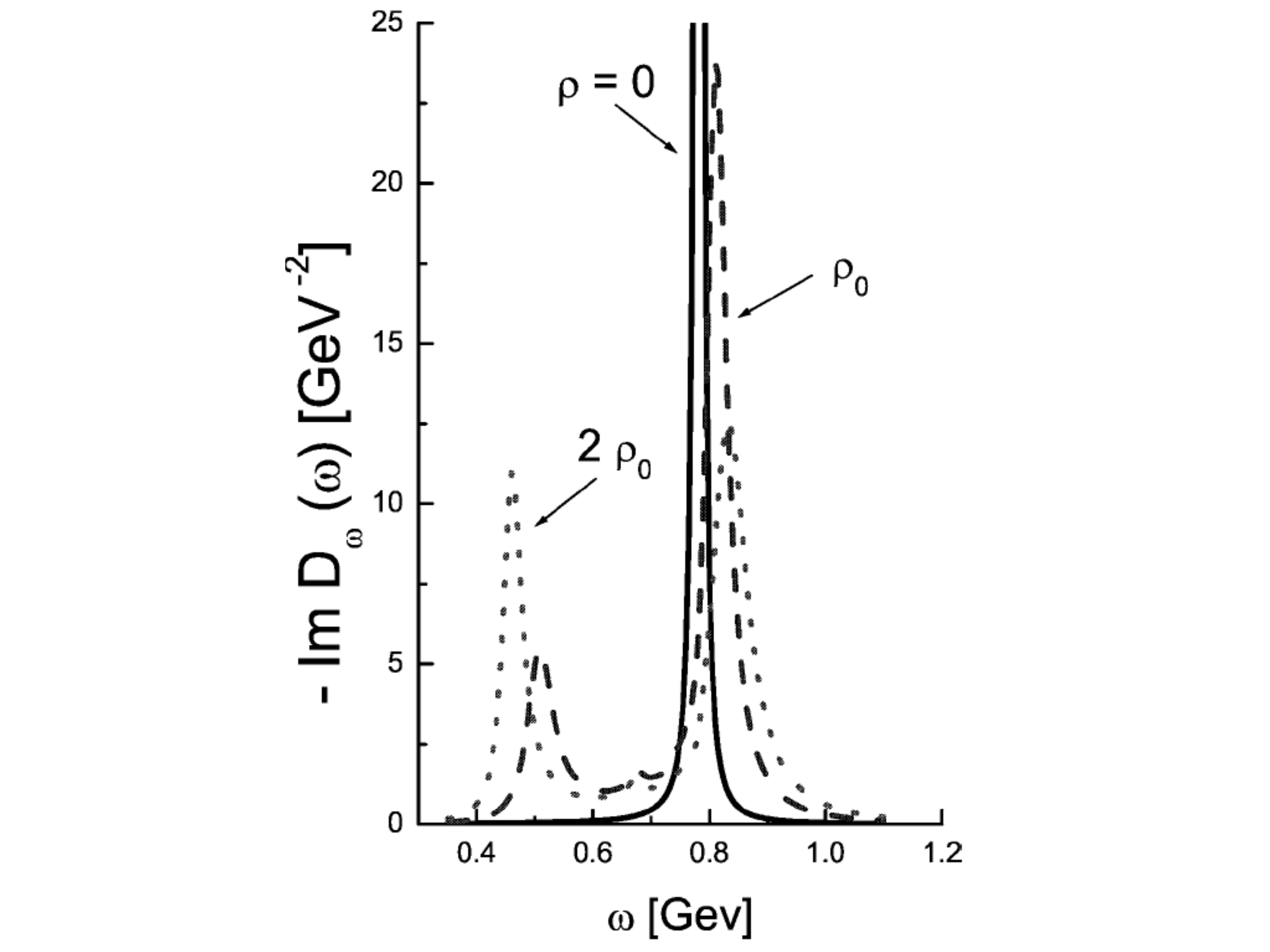} \includegraphics[width=8cm,clip]{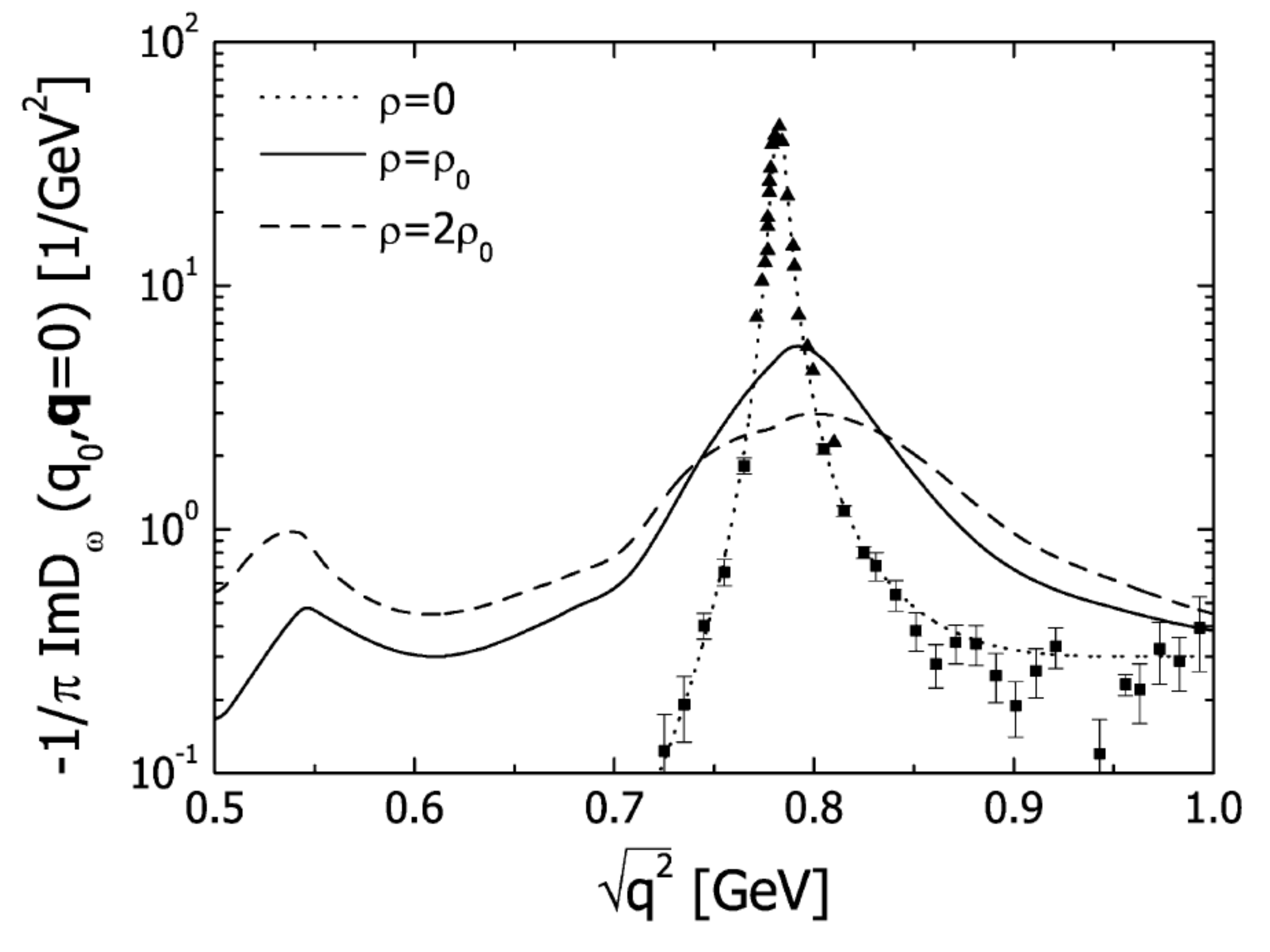}
\caption{(Left) Imaginary part of the $\omega$ meson propagator in nuclear matter compared to the one in the vacuum \cite{Lutz}. (Right) Spectral function of an $\omega$ meson at rest \cite{Muehlich_NPA780}. The results are shown for nucleon densities $\rho =0, \rho = \rho_0 = 0.16~ $fm$^{-3}$, and $\rho = 2 \rho_0 (\rho=\rho_N)$.}
\label{fig:res_coupl}
\end{figure}

\subsubsection{\it The imaginary part of the $\omega$--nucleus potential}\label{sec:omega_imag}
Theoretical predictions for the imaginary part of the $\omega$--nucleus potential range from -20 to -100 MeV (see table \ref{tab:VW_theo}), including calculations starting from an effective Lagrangian \cite{Klingl}, calculations treating explicitly the coupling of the $\omega$ meson to nucleon resonances \cite{Muehlich_NPA780, Lutz}, calculations within the chiral unitary approach \cite{Ramos} and standard hadronic many-body calculations \cite{Cabrera_Rapp}. As an example Fig.~\ref{fig:res_coupl} shows the results obtained in \cite{Muehlich_NPA780, Lutz}.
\begin{figure*}
\centering
  \includegraphics[width=12cm,clip]{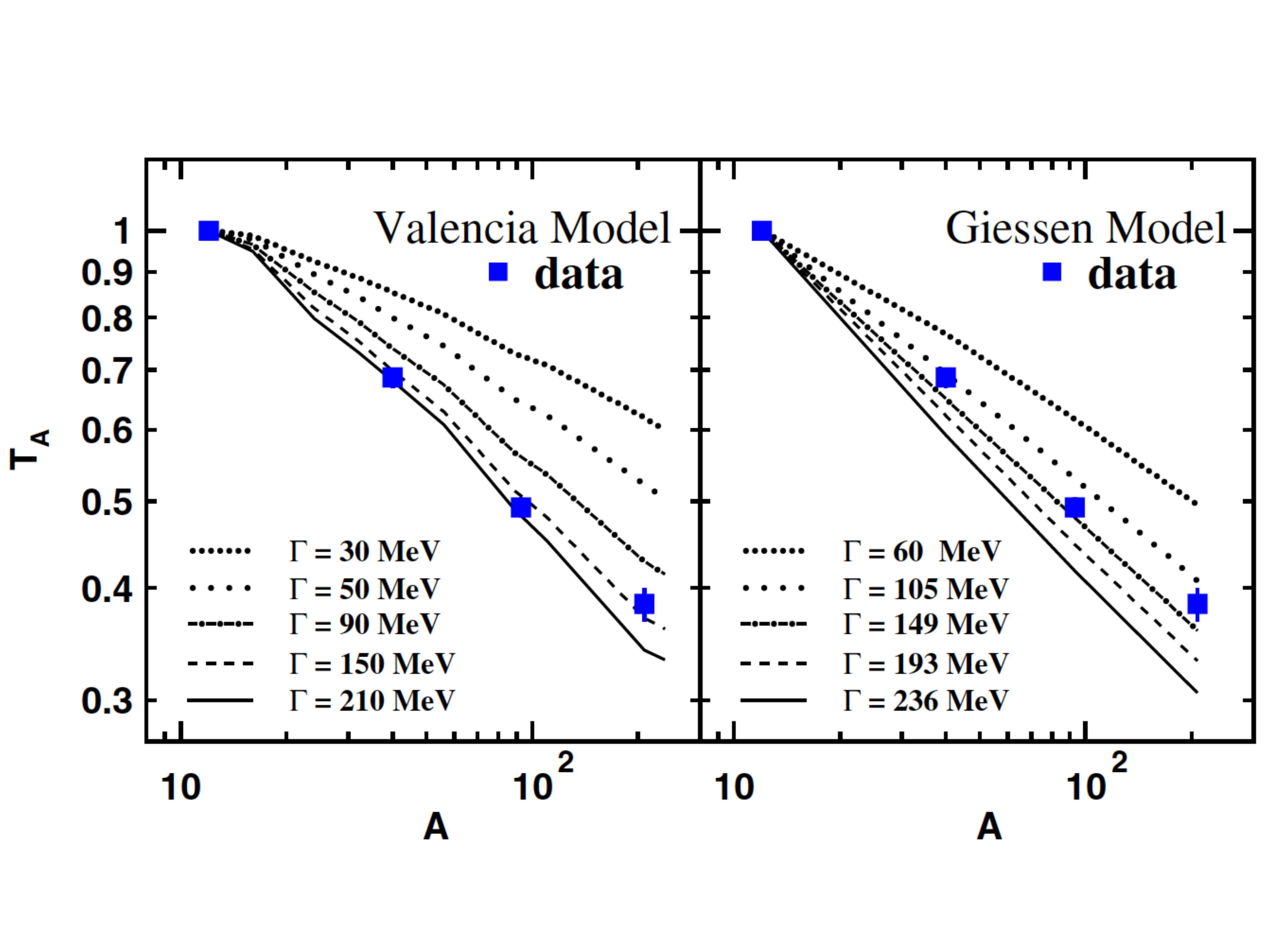} 
\caption{Transparency ratio for $\omega$ mesons normalised to carbon as a function of the nuclear mass number A in comparison with Monte-Carlo calculations \cite{Kaskulov} and GiBUU simulations \cite{Muehlich_Mosel}. The figure is taken from \cite{Kotulla}.}
\label{fig:omega_ta}
\end{figure*}
In both calculations the coupling of the $\omega$ meson to the nucleon resonances leads to a splitting of the $\omega$ strength into an $\omega$ mode and a particles-hole mode at lower mass. In \cite{Muehlich_NPA780} the strength function is, however, dominated by the strong broadening. Cabrera and Rapp \cite{Cabrera_Rapp} and Ramos et al. \cite{Ramos} have studied the $\omega$ in-medium width as a function of the $\omega$ momentum. Both groups attribute the main contribution to the in-medium $\omega$ width to the $\omega \rightarrow \rho \pi$ channel whereby the dressing of the $\rho$ and $\pi$ propagator in the medium is found to be essential. Cabrera and Rapp \cite{Cabrera_Rapp} find only a moderate momentum dependence while Ramos et al. \cite{Ramos} predict an almost linear increase up to $\omega$ momenta of 600 MeV/$c$.

\begin{figure*}
\centering
  \includegraphics[width=9cm,clip]{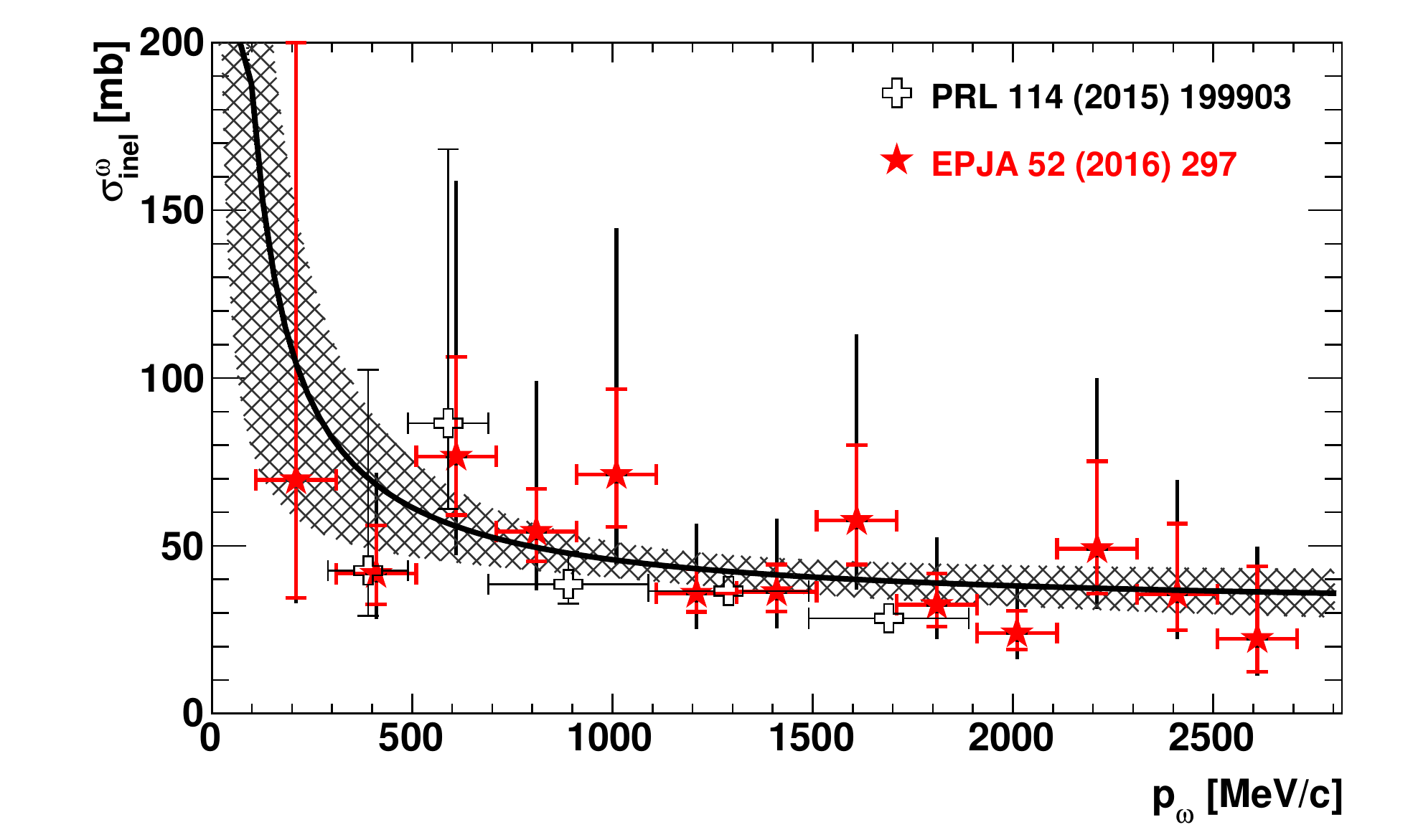} \includegraphics[width=9cm,clip]{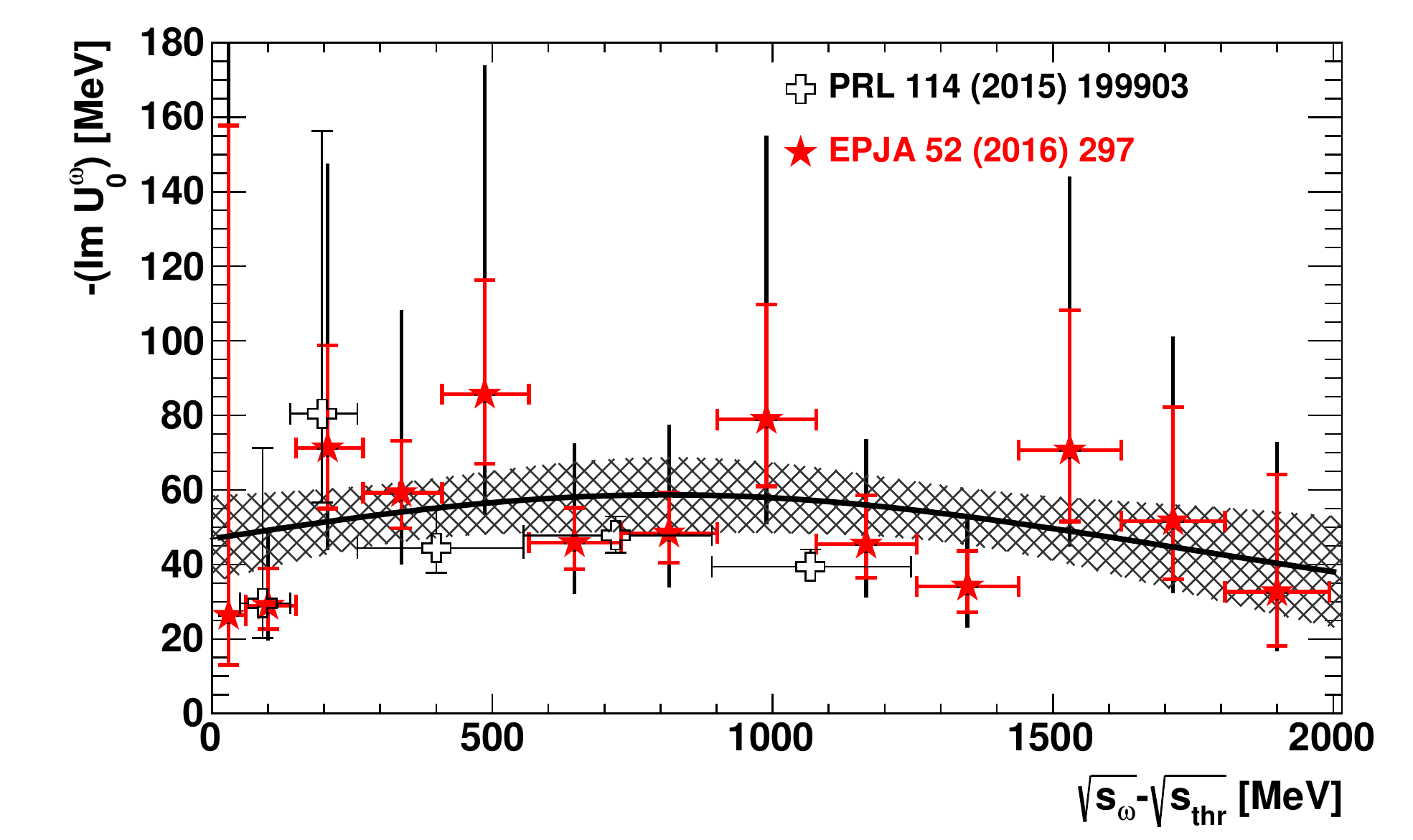}
\caption{(Left) Inelastic $\omega$-nucleon cross sections deduced from Eq. \ref{eq:lindens} as a function of the meson momentum (red stars)  \cite{Friedrich_EPJA} in comparison to earlier measurements (open crosses) \cite{Kotulla}. (Right) Imaginary part of the $\omega$--nucleus optical potential as a function of the available energy in the meson--$^{93}$Nb system (red stars) \cite{Friedrich_EPJA} in comparison to earlier measurements (open crosses) \cite{Kotulla}. The solid curves are Breit-Wigner fits to the present data. The shaded areas indicate a confidence level of $\pm$1$\sigma$ of the fit curve taking statistical and systematic errors into account. The thick (red) error bars represent the statistical errors. The thin (black) error bars include the systematic errors added in quadrature. The figures are taken from \cite{Friedrich_EPJA}. With kind permission of The European Physical Journal (EPJ).}
\label{fig:omega_ImU}
\end{figure*}

Experimentally - as outlined in Section \ref{sec:imag} - the imaginary part of the meson--nucleus potential can be derived from the measurement of the transparency ratio Eqs.~(\ref{eq:transp}),(\ref{eq:transp_C}).
Results of the first measurement for the $\omega$ meson are shown in Fig. \ref{fig:omega_ta}.
A comparison to Monte-Carlo calculations \cite{Kaskulov} and GiBUU simulations \cite{Muehlich_Mosel} gives an in-medium width of 130-150 MeV in the nuclear rest frame for $\omega$ mesons with an average 3-momentum of $\approx 1.1$ GeV/$c$, indicating a strong in-medium broadening compared to the free $\omega$ width of 8.4 MeV \cite{PDG}. More recently the momentum dependence of the transparency ratio has been studied in finer momentum bins \cite{Friedrich_EPJA}. The inelastic $\omega$-nucleon cross sections deduced from Eq.~(\ref{eq:lindens}) is shown in Fig.~\ref{fig:omega_ImU} (Left) as a function of the meson momentum. Extrapolating to vanishing $\omega$ momentum, the imaginary part of the $\omega$--nucleus potential (corresponding to half of the in-medium width) has been determined to -$(48 \pm12$(stat)$ \pm9$(syst)) MeV (cf. Fig. \ref{fig:omega_ImU}, Right). This value is larger than the modulus of the real potential (see below).

A strong absorption of the $\omega$ meson has also been observed in experiments studying the $\omega \rightarrow e^+ e^-$ decay channel. Comparing dilepton invariant mass spectra in the $p + {\rm Nb}$ and $p + p$ reactions at 3.5 GeV, the HADES Collaboration \cite{HADES_omega} observes an appreciable decrease in the strength of the $\omega$ signal.
An even more dramatic absorption of $\omega$ mesons corresponding to an in-medium width of more than 200 MeV has been reported by the CLAS Collaboration \cite{Wood_PRL105}.

\subsubsection{\it The real part of the $\omega$--nucleus potential}
Early predictions of large mass shifts of vector mesons of the order of -100 to -150 MeV \cite{Brown_Rho,Hatsuda_Lee,Klingl,QMC} (see table \ref{tab:VW_theo}) initiated widespread
\begin{figure*}
\centering
  \includegraphics[width=9cm,clip]{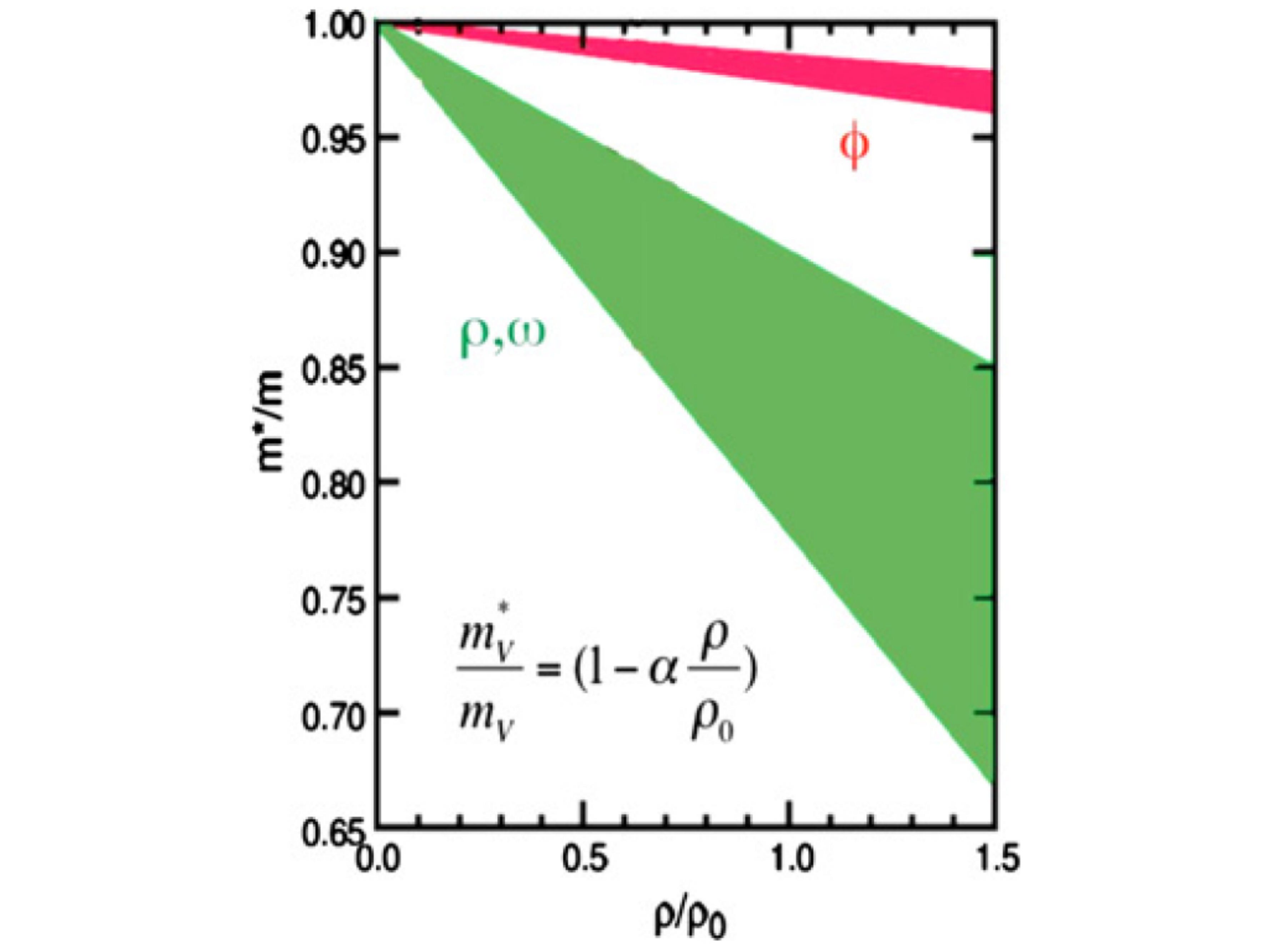} \includegraphics[width=9cm,clip]{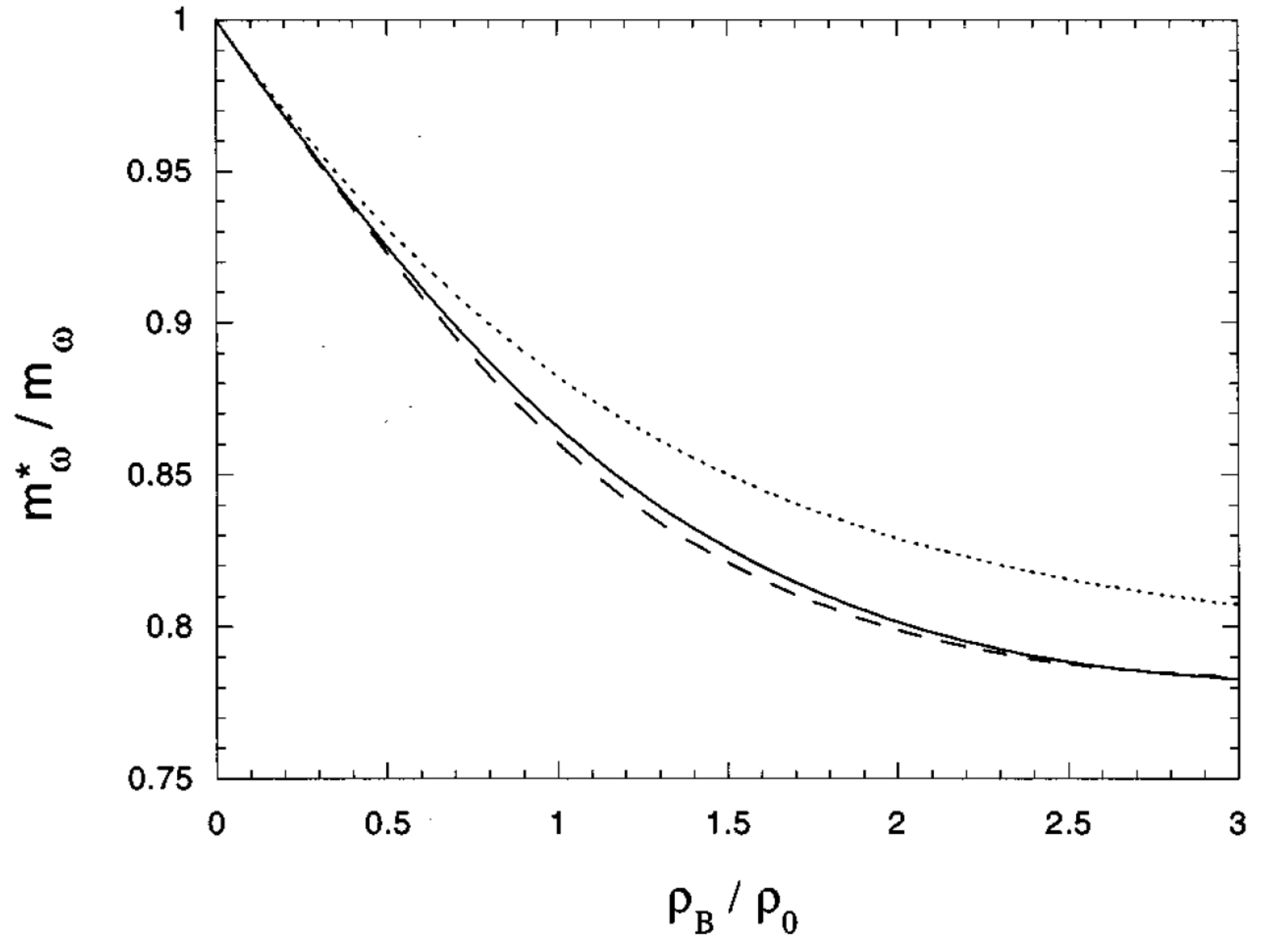}
\caption{(Left) Vector meson masses as a function of nuclear density obtained by QCD sum rules \cite{Hatsuda_Lee} (figure adapted from \cite{Metag}). (Right) Predictions of the $\omega$ mass as a function of the nuclear density for different parameter sets in the QMC model \cite{QMC}.}
\label{fig:omega_drop}
\end{figure*}
theoretical and experimental activities. As an example, results obtained by using QCD sum rules
\cite{Hatsuda_Lee}
and predictions within the QMC model \cite{QMC} are shown in Fig.~\ref{fig:omega_drop}. Other calculations, considering the coupling of the $\omega$ meson to nucleon resonances \cite{Muehlich_NPA780}, however, predict almost no mass shift but a considerable broadening (see Fig.~\ref{fig:res_coupl} (Right)). These conflicting theoretical predictions called for an experimental clarification.

\begin{figure}[h]
\centering
\includegraphics[width=6.7cm,clip]{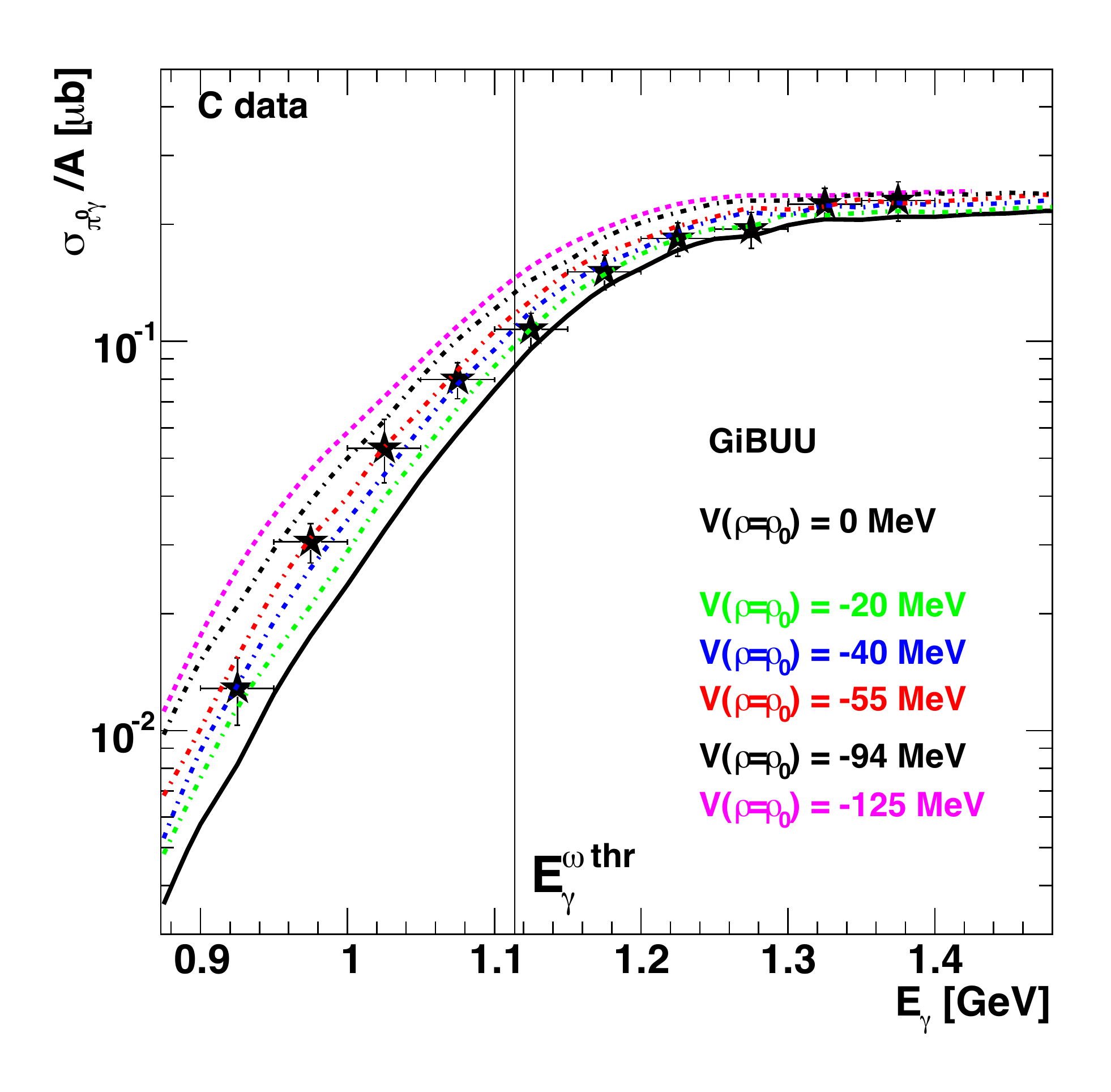}\includegraphics[width=10cm,clip]{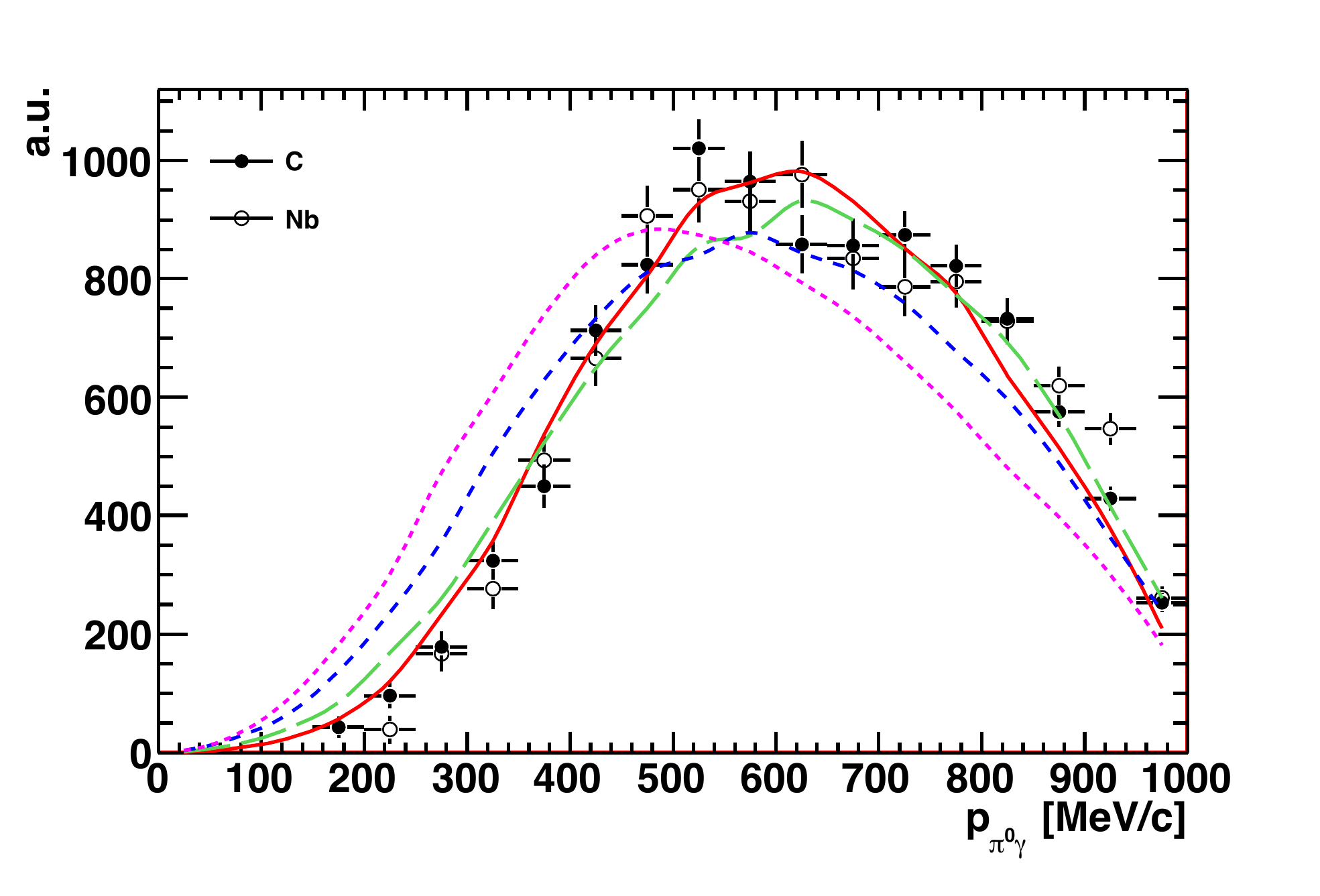}
\caption{(Left) Measured excitation function for $\omega$ meson photoproduction off $^{12}$C
in comparison to GiBUU transport calculations for several in-medium modification scenarios~\cite{Metag,Metag_Hyp}.
(Right) Acceptance corrected $\omega$ momentum distribution for incident photon energies from 900 to 1300 MeV and for $^{12}$C and $^{93}$Nb targets, compared to the theoretical predictions for different in-medium modifications scenarios: no modification (solid red line), collisional broadening (dashed green line), collisional broadening plus mass shift (dashed blue line) and mass shift (magenta line). All distributions are normalised to the same area~\cite{Thiel}. With kind permission of The European Physical Journal (EPJ).}
\label{fig:exc_mom_om}
\end{figure}

The real part of the $\omega$--nucleus potential has been determined by measuring the near threshold excitation function and the momentum distribution for photoproduction off C and Nb. Corresponding results are shown in Fig.~\ref{fig:exc_mom_om}. Fitting the excitation function data with GiBUU simulations for different potential depths gives a potential of $-(42 \pm 17$(stat)$ \pm 20$ (syst)) MeV in contrast to theoretical predictions of large mass shifts of -(100--150) MeV \cite{Klingl}. This result is supported by the  $\omega$ momentum distribution (see Fig.~\ref{fig:exc_mom_om}, Right) which again seems to exclude the scenarios assuming large mass drops in the medium. This data set corresponds to an average $\omega$ momentum of about 600 MeV/$c$.

An attempt has been made in \cite{Friedrich_PLB} to measure the $\omega$ potential depth for even lower momenta which can be accessed by requiring the participant nucleon in coincidence.
\begin{figure*}
\centering
\includegraphics[width=9cm,clip]{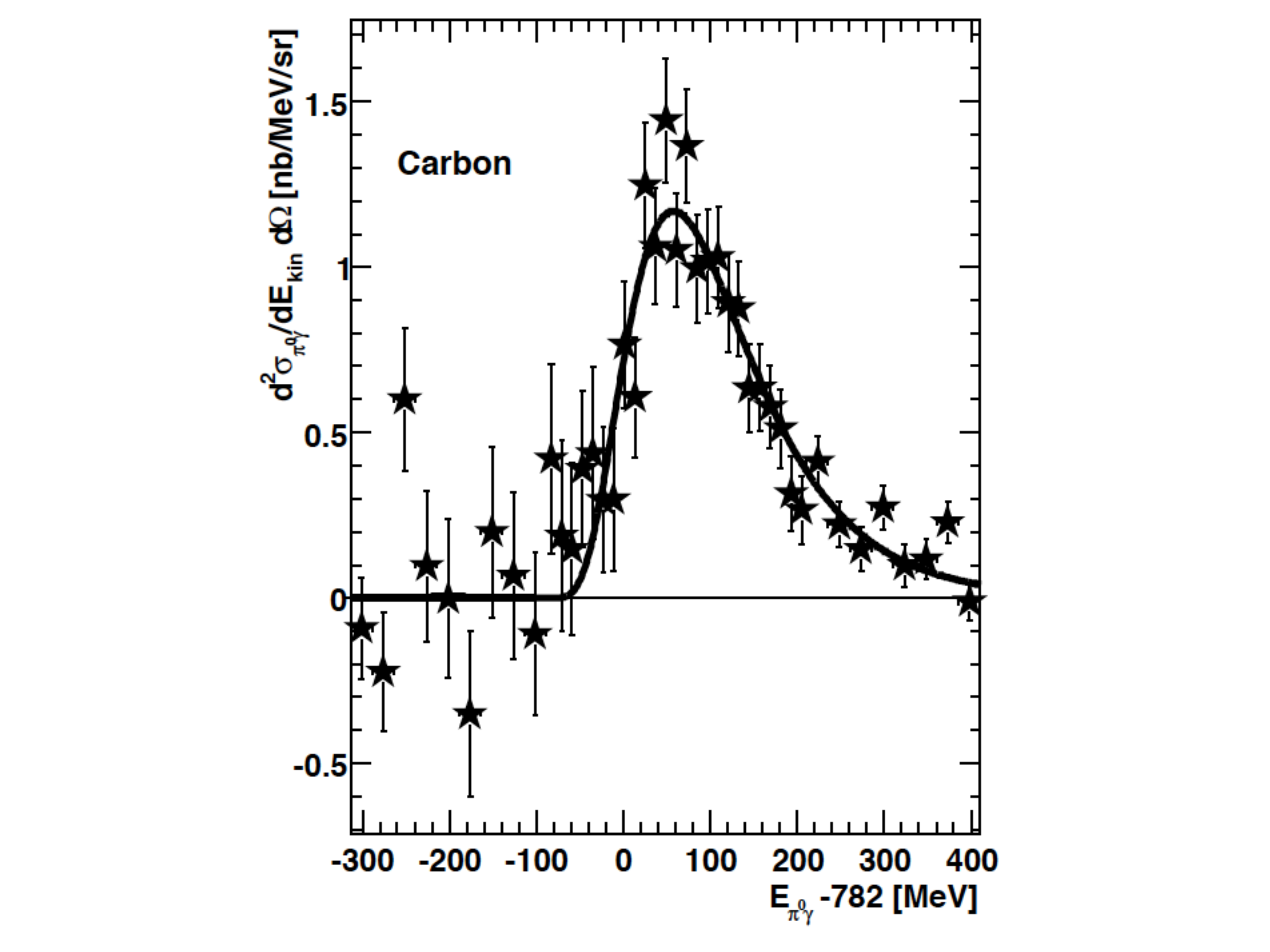}\includegraphics[width=9cm,clip]{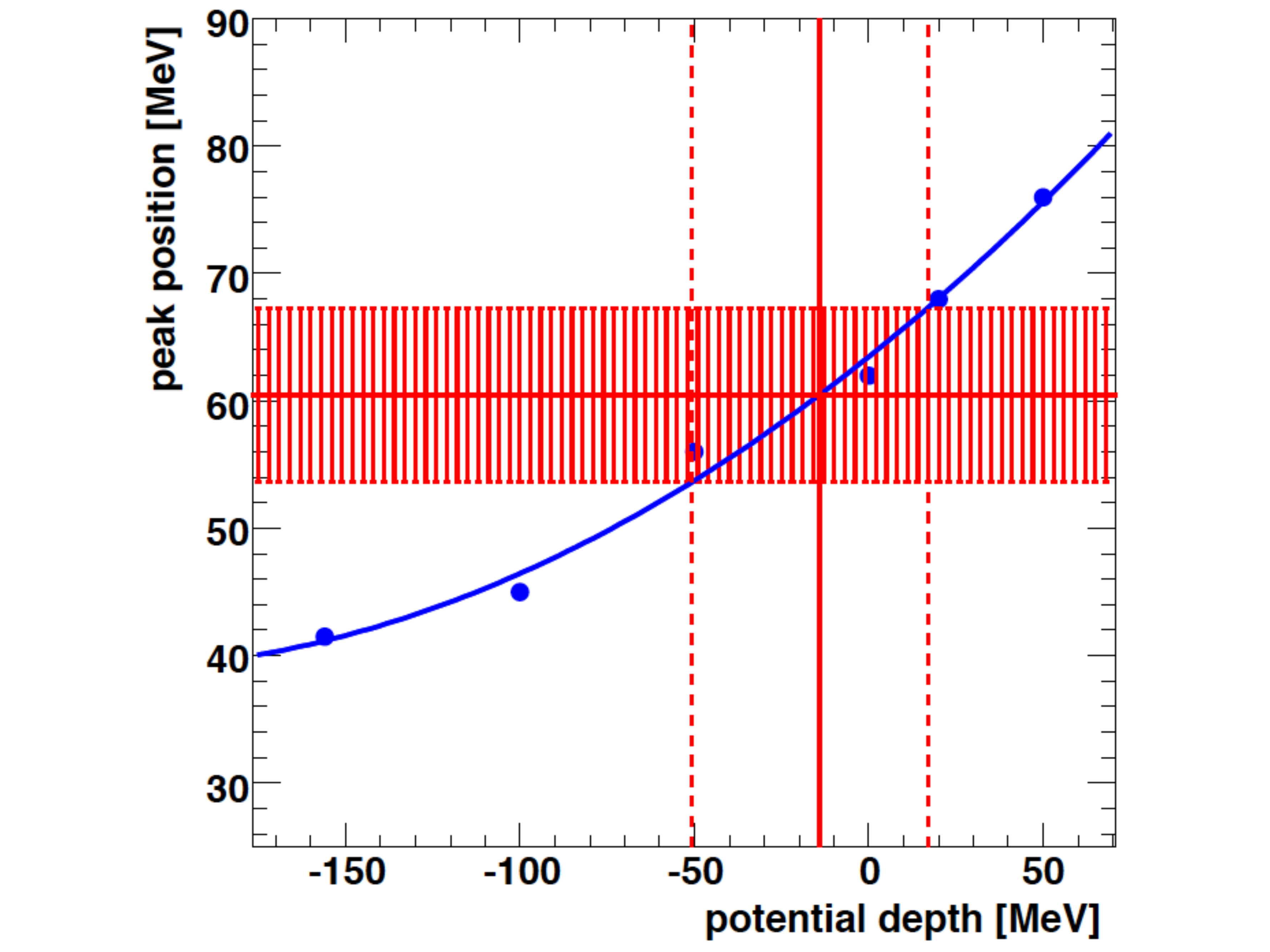}
\caption{(Left) Differential cross section for the photoproduction of $\omega$ mesons off C in coincidence with protons in $\Theta_p = 1^{\circ}-11^{\circ}$ as a function of the total energy of the $\pi^0 \gamma$ pairs minus 782 MeV. The data have been fitted with the Novosibirsk function \cite{Aubert}. (Right) Correlation between the potential depth and the peak position in the kinetic energy distribution. The (blue) points represent the peak position in the kinetic energy distribution for the different scenarios \cite{Friedrich_PLB}. The (blue) solid curve is a fit to the points. The red dashed area corresponds to the peak position of (60.5 $\pm$ 7) MeV. The figures are taken from \cite{Friedrich_PLB}.}
\label{fig:stefan}
\end{figure*}
\begin{figure*}
\centering
\includegraphics[width=11cm,clip]{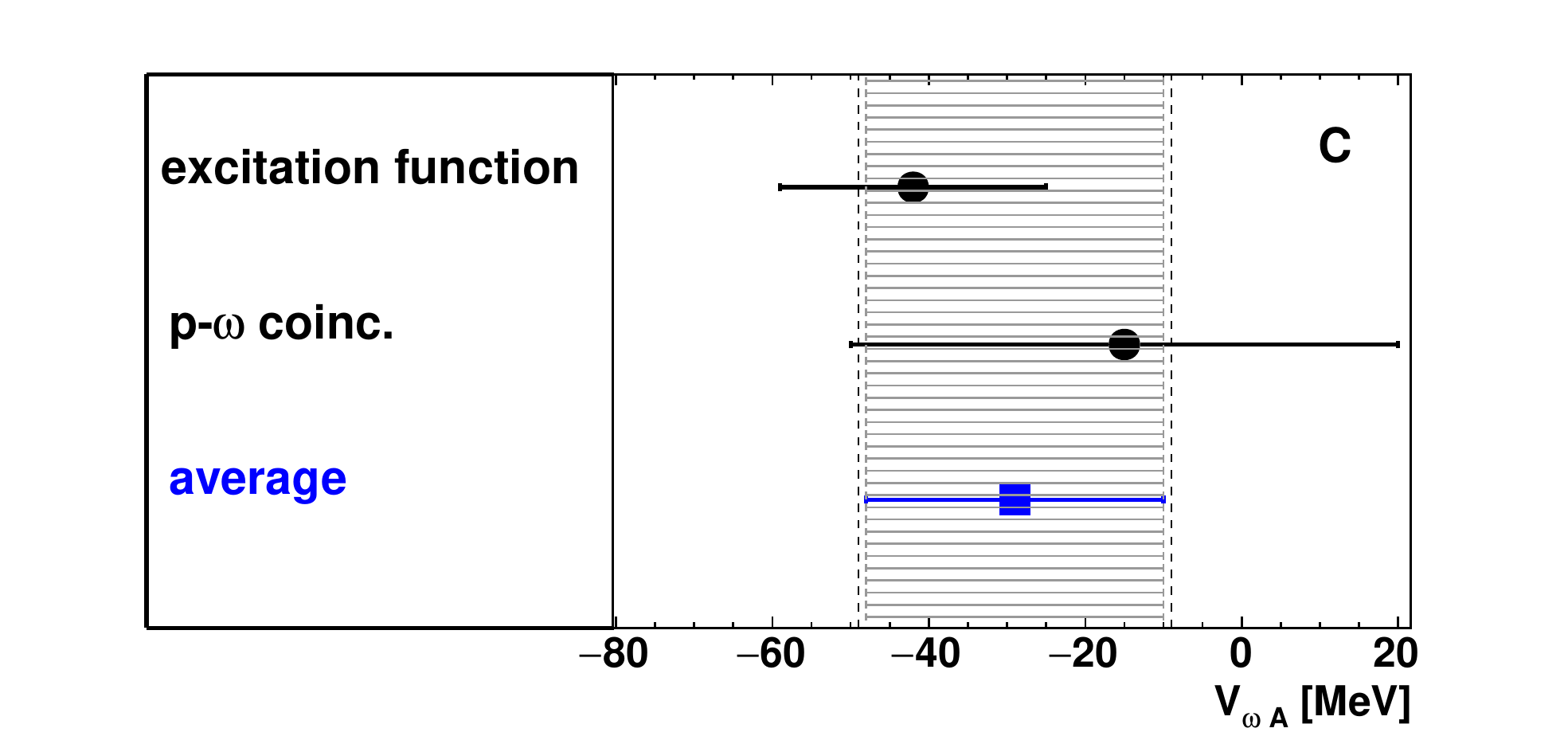}
\caption{Depths of the real part of the $\omega$--nucleus potential determined by analyzing the excitation function and the momentum distributions for C \cite{Nanova_Metag}. The weighted overall average is indicated by a blue square and the shaded area. The vertical hatched lines mark the range of systematic uncertainties. The figure is taken from \cite{Nanova_Metag}.}
\label{fig:omega_concl}
\end{figure*}
Detected at forward angles, it takes over most of the momentum of the incoming photon beam, leaving the $\omega$ meson almost at rest relative to the nucleus. In this situation its momentum (or kinetic energy)
distribution is particularly sensitive to the real part of the meson--nucleus potential: if the interaction is attractive mesons emerging from the nucleus will be further slowed down while for a repulsive interaction the mesons will be accelerated compared to a scenario with vanishing interaction. The peak in the kinetic energy distribution is thus sensitive to the strength and sign of the meson--nucleus potential \cite{Nagahiro_PRC74}, as quantitatively given in Fig.~\ref{fig:stefan} (Right). The left panel of Fig.~\ref{fig:stefan} shows the measured kinetic energy distribution of $\omega$ mesons registered in coincidence with protons detected at $ 1^{\circ}-11^{\circ}$ in photoproduction off $^{12}$C. The kinetic energy distribution peaks at $60.5 \pm 7 $ MeV which - in comparison to Fig.~\ref{fig:stefan} (Right) shows that the interaction is slightly attractive and leads to a potential depth of $V_{\omega A}(\rho_N = \rho_0) = - (15 \pm 35$ (stat) $ \pm 20 $(syst)) MeV. Within the errors this result compares well with the determination of the potential depth from  the excitation function measurement.  The weighted average of both measurements is  $V_{\omega A}(\rho_N = \rho_0) = - (29 \pm 19$ (stat) $ \pm 20 $(syst)) MeV, as shown in Fig.~\ref{fig:omega_concl}. In conclusion, the $\omega$--nucleus attraction is rather weak.

Attempts have been made to extract information on the real and imaginary part of the $\omega$ - nucleus potential from a measurement of the $\omega$ line shape. Such an analysis is subject to the problems discussed in Section \ref{sec:compexp}. Early claims of an $\omega$ in-medium mass drop \cite{Trnka} were not confirmed neither in a refined analysis of the data \cite{Nanova_PRC82} nor in high statistics measurements near the production threshold \cite{Thiel,Nanova_EPJA47}. Ozawa et al.\cite{Ozawa}, analyzing the $\rho,\omega,\phi \rightarrow e^+e^-$ decay, found evidence that spectral shapes of vector mesons are modified at normal nuclear matter density. Naruki et al. \cite{Naruki} reported a drop of the $\rho$ and $\omega$ mass by 9.2$\pm0.2\%$ at normal nuclear matter density without any in-medium broadening of the mesons. This result is surprising since hadrons in the medium must experience inelastic collisions which shorten their time of existence in the medium and thus increase their width, as found in this review for all other mesons. Furthermore, this result is in conflict with results of the CLAS Collaboration who reported no mass shift ( $\le 5 \%$ at 95$\%$ confidence level) for the $\rho$ and $\omega$ meson \cite{Nasseripour,Wood_rho_PRC78}. The results from the line shape analyses thus appear to be inconclusive.

Summarizing the above experimental information on the real and imaginary part of the $\omega$--nucleus interaction one obtains
$V_{\omega A}(\rho_N = \rho_0) = - (29 \pm 19$ (stat) $ \pm 20 $(syst)$) - i (48 \pm12$(stat)$ \pm9$(syst)) MeV. These potential values are not favourable for the search for the $\omega$--nucleus bound states since the potential depth is not very large and the imaginary part of the potential is larger than or comparable to the real one; if there were bound states at all, they would be very broad and thus difficult to separate experimentally from the background (see Section \ref{sec:omega_mesic}).

%---------------------------------------------------------------------------------------------
%PHI

\subsection{\it $\phi$--nucleus potential}\label{sec:phi}
\subsubsection{\it The $\phi$--nucleon scattering length}
As for the $\omega$ meson, the photoproduction of $\phi$ mesons near threshold has been studied to extract the $\phi$--nucleon scattering length \cite{Titov}. At threshold Titov et al. derive a differential cross section of $\frac{d\sigma}{dt} \approx 0.64 \mu b/$GeV$^2$ which is consistent with the estimate for the $\phi$--nucleon scattering length of  $a_{\phi N} = -(0.15 \pm 0.02)$ fm, obtained in QCD sum rule calculations \cite{Koike}. From measurements at  higher incident photon energies ($E_{\gamma}$= 4.6-6.7 GeV) Behrend et al. deduce an inelastic $\phi$--nucleon cross section of $\approx$ 8 mb \cite{Behrend}. Thus the $\phi$--nucleon interaction is found to be rather weak. This is consistent with the rather small mass drop of about -2 to -3$\%$ at nuclear matter density (see Fig.~\ref{fig:omega_drop}) predicted by Hatsuda and Lee \cite{Hatsuda_Lee}, Cabrera et al. \cite{Cabrera_PRC95,Cabrera_Vacas} and Cobos-Martinez et al. \cite{Cobos-Martinez}. More recently an even smaller reduction of the $\phi$ meson mass in the nuclear medium of less than -2 $\%$ has been predicted \cite{Gubler_Weise_PLB,Gubler_Weise_NPA}.

\subsubsection{\it The imaginary part of the $\phi$ - nucleus potential}
\begin{figure*}
\centering
  \includegraphics[width=12cm,clip]{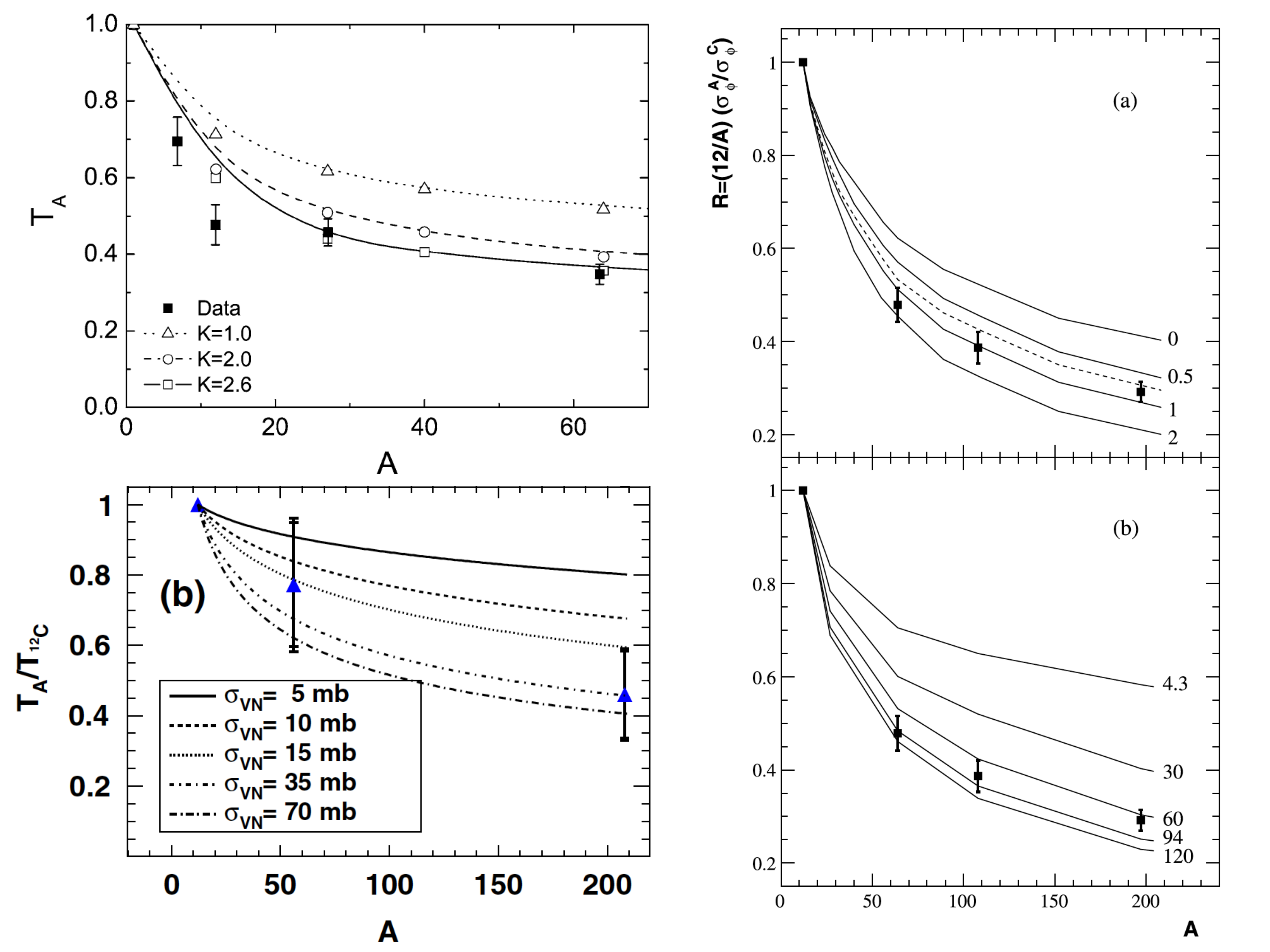}
\caption{(Left top panel) Transparency ratio normalised to carbon reported by Ishikawa et al. \cite{Ishikawa} for photoproduction of $\phi$ mesons in comparison to GiBUU simulations by M\"uhlich and Mosel \cite{Muehlich_Mosel_phi} for different absorption cross sections. The figure is taken from \cite{Muehlich_Mosel_phi}. (Left bottom panel) Transparency ratio for $\phi$ mesons in the $\gamma + A \rightarrow \phi + X \rightarrow e^+e^- + X $ reaction in comparison with Glauber calculations for different in-medium $\phi$ -- nucleon cross sections \cite{Wood_PRL105}.  (Right) Transparency ratio for $\phi$ mesons obtained in p + A collisions \cite{Polyanskiy} compared to calculations by Magas et al. \cite{Magas} (a) and Paryev \cite{Paryev_JPG36} (b) for different in-medium widths. In Fig. (a) the widths are given as multiples of the $\phi$ width at rest of 24 MeV. In Fig.(b) the widths are given in the $\phi$ meson eigen-system. The figure is taken from \cite{Polyanskiy}.}
 \label{fig:phi_TA}
 \end{figure*}

The transparency ratio for the $\phi$ meson has been measured in photoproduction \cite{Wood_PRL105,Ishikawa} and  p + A collisions \cite{Polyanskiy,Hartmann_PRC85} to 
determine the absorption of $\phi$ mesons in nuclei. The data - normalised to the transparency ratio for the carbon target - consistently show a decrease of the transparency to about 40$\%$ for medium mass nuclei and  to 30$\%$ for very heavy nuclei (see Fig.~\ref{fig:phi_TA}). The normalization to carbon reduces the sensitivity of the 
measured cross sections to differences in the initial state interaction for photon- and proton- induced reactions, to secondary production processes and/or uncertainties in the meson production cross section off the neutron. To extract the inelastic in-medium $\phi$ - nucleon cross section and $\phi$ width the data are compared to calculations by M\"uhlich and Mosel \cite{Muehlich_Mosel_phi}, Magas et al.Ê\cite{Magas}, Paryev \cite{Paryev_JPG36}, and Cabrera et al., \cite{Cabrera_NPA733}.
Ishikawa et al. \cite{Ishikawa} deduce an inelastic in-medium $\phi$ - nucleon cross section of $35^{+17}_{-11}$ mb, roughly a factor 4 larger than $\sigma_{\phi N} $ in free space.
The analysis of the same data within the GiBUU transport model yields a similar value of $\approx$ 27 mb \cite{Muehlich_Mosel_phi} for the inelastic $\phi $ - nucleon cross section which corresponds in the linear density approximation (Eq.~(\ref{eq:lindens})) to an in-medium width in the nuclear rest frame of $\approx $ 75 MeV at normal nuclear matter density for an average $\phi$ momentum of 1.8 GeV/c. Wood et al. \cite{Wood_PRL105} deduce an in-medium $\phi$ - nucleon cross section in the range of 16 - 70 mb (see Fig.~\ref{fig:phi_TA} Left bottom panel) within a Glauber model analysis. The transparency ratio data obtained in p + A collisions (see Fig. \ref{fig:phi_TA} Right) by Polyanskiy et al. \cite{Polyanskiy} have been compared to calculations by Magas et al. \cite{Magas} and Paryev \cite{Paryev_JPG36}. Taking into account contributions from two-step production processes which increase the transparency ratio, in-medium widths of $45^{+17}_{-9}$ MeV and $50^{+10}_{-6}$MeV have been deduced, respectively. (The value of  $73^{+14}_{-10}$MeV, quoted in the original literature \cite{Polyanskiy} refers to the 
$\phi$ eigen-system). All these measurements consistently show that - with in-medium widths of the order of 40-60 MeV in the nuclear rest frame -  the $\phi$ meson is strongly broadened in the medium by about an order of magnitude as compared to its free width of 4.3 MeV \cite{PDG}. The imaginary potential of the $\phi$ - nucleus interaction is thus in the range of -(20 - 30) MeV.

These experimental results are close to theoretical predictions. The $\phi$ width increases in the medium because of the opening of inelastic reaction channels. In addition, including the in-medium interaction of the  decay kaons in the nuclear many body system, the $\phi$ spectral function is further broadened. Using a  coupled channel approach based on a chiral effective Lagrangian Klingl et al. predict an in-medium $\phi$ width of 45 MeV \cite{Klingl_PLB1998}. M\"uhlich and Mosel \cite{Muehlich_Mosel_phi} estimate an in medium width of 40 MeV at saturation density. In a more recent calculation Gubler and Weise obtain a $\phi$ width of 45 MeV \cite{Gubler_Weise_PLB,Gubler_Weise_NPA}, emphasising that the  interaction of kaons from $\phi \rightarrow K \bar K $ with the surrounding nuclear medium leads to an asymmetric line shape of the $\phi$ meson peak structure. Within the QMC model a width of 33 - 37 MeV is predicted \cite{Cobos-Martinez}. In addition to the in-medium $\phi \rightarrow \bar K K$ decay, Cabrera et al. \cite{Cabrera_phiN} discuss contributions to the $\phi$ width from the coupling to nucleon resonances yielding an additional broadening of the order of 40-50 MeV.

It should be noted, however, that all these values are much larger than the in-medium width of $\approx$~15~MeV reported in a line shape analysis of the $\phi \rightarrow e^+e^-$ signal by Muto et al. \cite{Muto} who performed a comparative study of $\phi$ meson production off C and Cu nuclei with proton beams of 12 GeV (see below). 

\begin{figure*}
\centering
\includegraphics[width=16.7cm,clip]{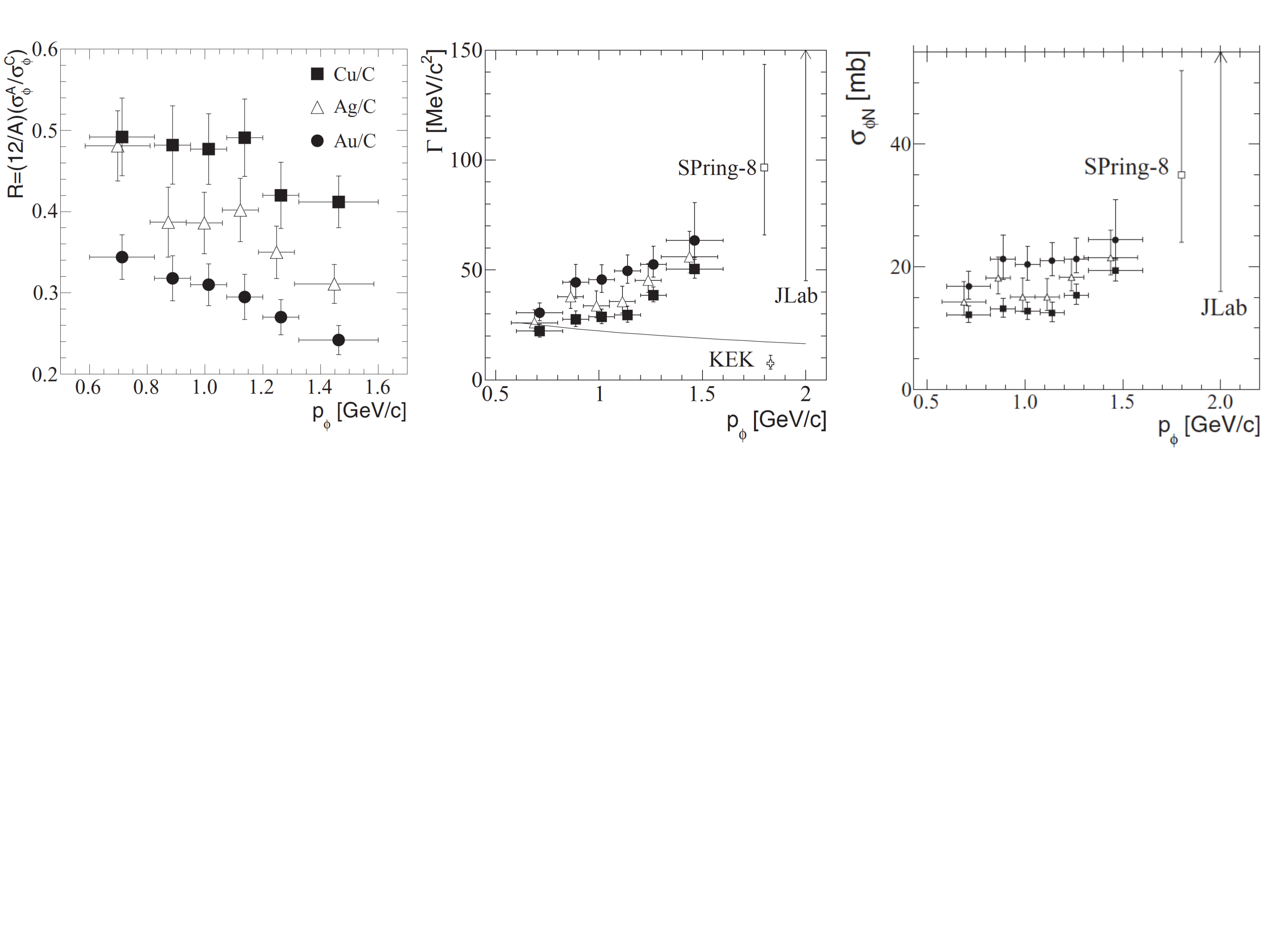}
\vspace{-65mm}\caption{(Left) Momentum dependence of the transparency ratio normalised to carbon for Cu, Ag, and Au targets. (Middle) In-medium width of the $\phi$ meson in the nuclear rest frame at saturation density $\rho_0$ as a function of the $\phi$ momentum. The different symbols refer to different model calculations used in the extraction of the $\phi$ width from the data. (Right) The $\phi N$ absorption cross section as a function of the $\phi$ momentum. The figures are taken from \cite{Hartmann_PRC85}.}
\label{fig:TA_Phi_Hartmann}
\end{figure*}

Hartmann et al. \cite{Hartmann_PRC85} investigated the momentum dependence of the $\phi$ transparency ratio. Fig.~\ref{fig:TA_Phi_Hartmann} shows the transparency ratio measured in proton induced reactions on Cu, Ag, and Au targets, normalised to C. Comparing to model calculations \cite{Paryev_JPG36,Magas} the in-medium $\phi$ width in the nuclear rest frame has been deduced and is found to increase almost linearly with the $\phi$ momentum from 20 to 60 MeV over the momentum range of 0.6-1.6 GeV/$c$, corresponding to an imaginary part of the $\phi$ - nucleus potential in the range of - (10 - 30)  MeV. The effective $\phi N$ absorption cross section, obtained within the linear density approximation (Eq.~(\ref{eq:lindens})), increases from 14 to 25 mb. All data consistently indicate a strong broadening of the $\phi$ meson in the nuclear medium.

\subsubsection{\it The real part of the $\phi$ - nucleus potential}
The only experimental information on the real $\phi$ - nucleus potential and thus on a possible in-medium mass shift of the $\phi$ meson is based on the line shape analysis of the \begin{figure*}
\centering
  \includegraphics[width=14.0cm,clip]{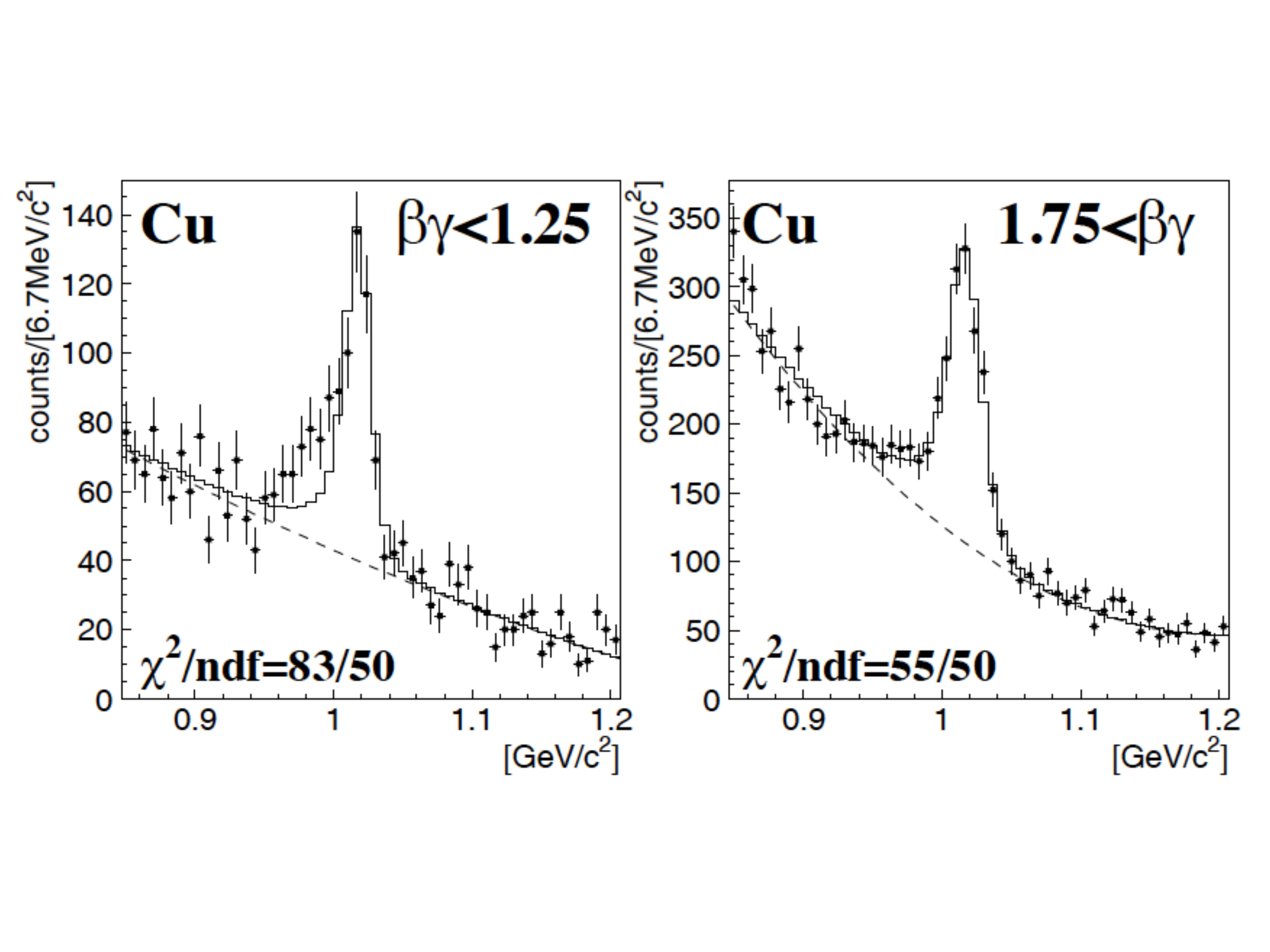}
\vspace{-15mm} \caption{$e^+e^-$ invariant mass distribution near the $\phi$ mass obtained in p+Cu collisions at 12 GeV for slow ($\beta \gamma \le 1.25$) and fast ($\beta \gamma \ge 1.75$) recoiling $\phi$ mesons. The solid histogram represents a fit with the expected $\phi \rightarrow e^+ e^-$ line shape and a quadratic background. The dashed curve represents the background. No difference in line shape is observed for the corresponding measurement on a C target. The figure is taken from \cite{Muto}.}
 \label{fig:KEK325_phi}
\end{figure*}
$\phi \rightarrow e^+e^-$ signal measured in the KEK E325 experiment in p+C, Cu collisions at 12 GeV \cite{Muto}. Mass shifts are expected to be observable only in heavy nuclei and for very slow mesons which have a higher probability to decay within the nucleus than fast ones. For the light C target no difference in the line shape of the invariant mass peak is observed for $\phi$ mesons recoiling with different velocities. For the heavier Cu nucleus Fig.~\ref{fig:KEK325_phi} (Left) shows an excess on the low mass side of the $\phi$ mass peak for slow $\phi$ mesons ($\beta \gamma \le 1.25$).  Analysing this deviation from the expected line shape Muto et al. \cite{Muto} deduce a drop of the $\phi$ mass by 35 MeV and an increase of the $\phi$ width by a factor 3.6 at normal nuclear matter density. The latter value is much smaller than the indirect determinations of the in-medium $\phi$ width, discussed above. A lowering of the $\phi$ mass by 35 MeV is, however, in the range of some theoretical predictions \cite{Hatsuda_Lee,Cabrera_PRC95}, while Klingl et al. \cite{Klingl_PLB1998} and Gubler and Weise \cite{Gubler_Weise_PLB,Gubler_Weise_NPA} expect smaller mass modifications. From the only available measurement, the real part of the $\phi$-nucleus potential is thus $\approx $ -35 MeV.
Since the KEK E325 experiment \cite{Muto} is the only one where an in-medium broadening {\bf and} a mass shift of a meson have been reported, it is of utmost importance to verify this result with better statistics. An experiment (E16) is planned at J-PARC to improve the statistics of the $\phi \rightarrow e^+e^-$ line shape measurement \cite{Yokkaichi} by 2 orders of magnitude, using a new detector system with large angular coverage and improved resolution. $\Phi$ mesons will be produced with 30 GeV protons. Slow $\phi$ mesons going backwards in the center-of-mass system will be selected for this study. The results are highly awaited.
%\clearpage

%_________________________________________________________________________________
%CHARM
%\section{charm}
\section{charm}
\subsection{\it meson--nucleus potentials in the charm sector}\label{sec:charm}
Normal nuclear matter consists of nucleons and thus only of quarks and not of antinucleons (antiquarks). As discussed in Sections \ref{sec:K+},\ref{sec:K0},\ref{sec:K-}, this leads to different properties of kaons $K(q\overline{s})$ and antikaons  $\overline{K}(\overline{q}s)$ (with $q=u,d$) in nuclear matter. Mesons with a light quark ($K^+,K^0$) experience repulsion in nuclear matter while mesons with a light nonstrange antiquark ($K^-,\overline{K^0}$) experience attraction. Analogously, one would expect in the charm sector attraction for $D$ mesons ($D^+(c\overline{d})$ and $D^0(c\overline{u})$) and repulsion for $\overline{D}$ mesons ($D^-(d\overline{c})$ and $\overline{D^0}(u \overline{c})$). These intuitive arguments, however, turn out to be oversimplified. The properties of pseudoscalar $D$ mesons have been investigated in several theoretical studies with partially conflicting results. Some investigations consider hadronic degrees of freedom like self-consistent unitarized coupled channel calculations \cite{Molina_EPJA42}, explicitly incorporating Heavy Quark Spin Symmetry (HQSS) \cite{Garcia-Recio_PLB,Garcia-Recio_PRC85}. These studies have recently been summarized in \cite{Tolos_IJMPE}. Other calculations are based on quark and gluon degrees of freedom like the QMC model \cite{Sibirtsev,Tsushima_PRC59,Tsushima_PRC83} or QCD sum rule analyses \cite{Hayashigaki,Hilger,Suzuki}. For an overview over heavy hadrons in nuclear matter see \cite{Hosaka}.
\begin{figure*}
\centering
  \includegraphics[width=13.0cm,clip]{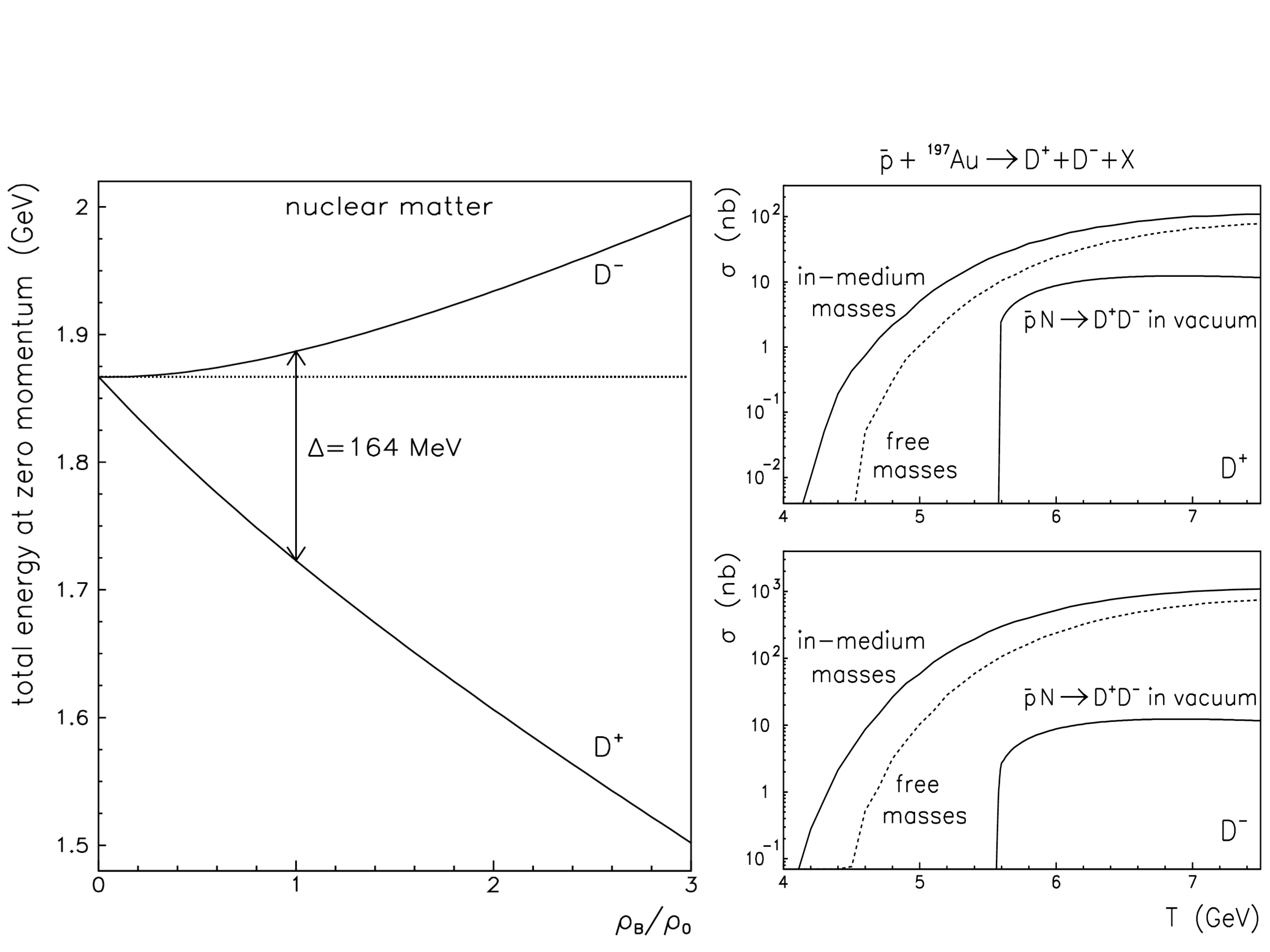}
 \caption{(Left) Total energies of $D^-$ and $D^+$ mesons at zero momentum calculated within the QMC model for nuclear matter, plotted as function of the baryon density in units of the saturation density $\rho_0 = 0.15 $ fm$^{-3}$. (Right) Total cross section for  $D^+$ and $D^-$ meson production in $\overline{p} $ Au annihilation as function of the antiproton energy. The curves represent calculations for free (dashed) and in-medium (solid) masses of the $D$ mesons. For comparison the free $\overline{p} N \rightarrow D^- D^+$ cross section is also shown. The figure is taken from \cite{Sibirtsev}. With kind permission of The European Physical Journal (EPJ).}
 \label{fig:Sibirtsev_D+D-}
\end{figure*}

\begin{figure*}
\centering
  \includegraphics[width=10.0cm,clip]{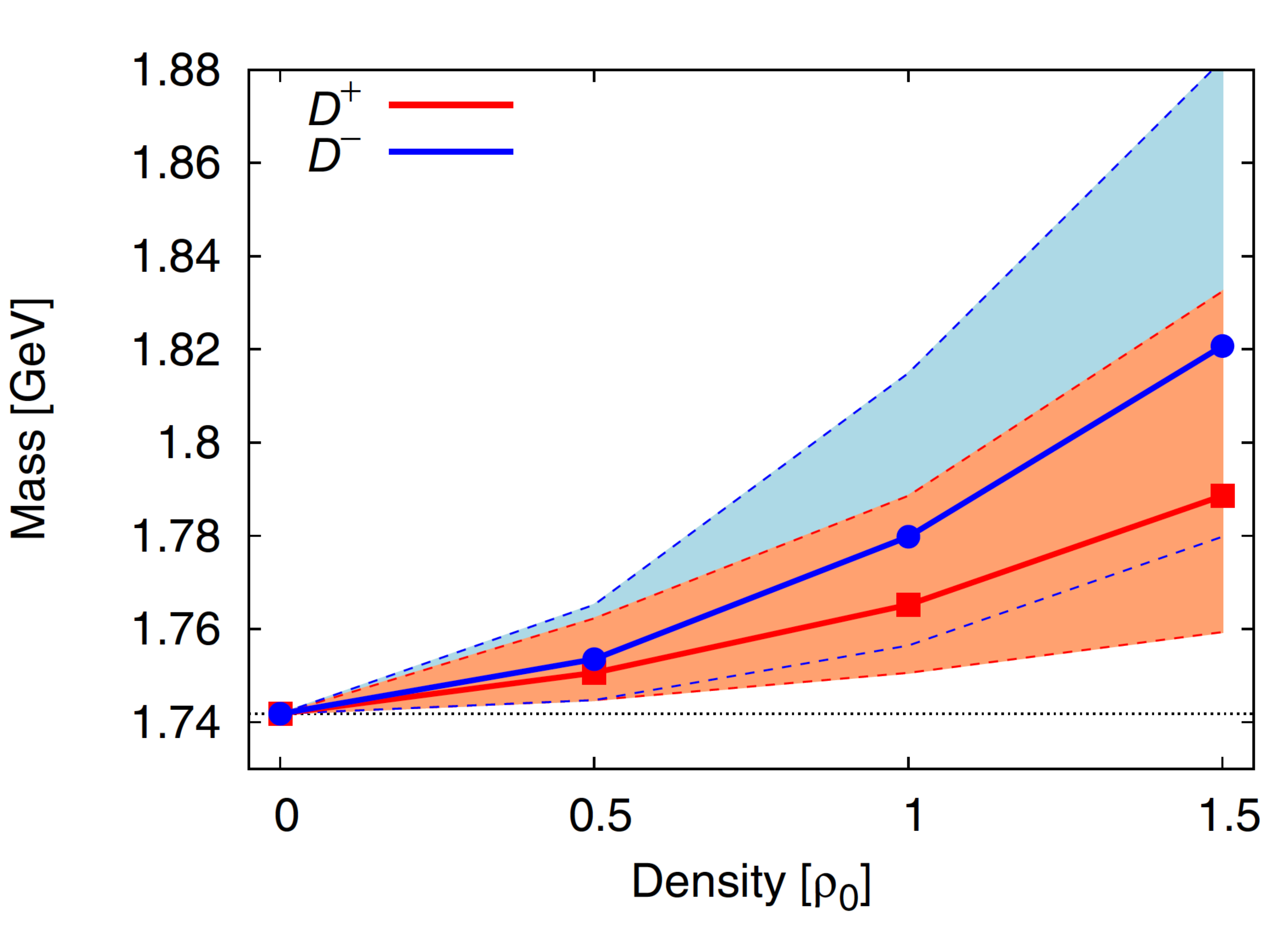}
  \caption{Density dependence of $D^+$ and $D^-$ meson peak positions from a QCD sum rule analysis. The shaded areas indicate the theoretical uncertainties. The figure is taken from \cite{Suzuki}.}
  \label{fig:Suzuki_D}
\end{figure*}

\begin{figure*}
\centering
  \includegraphics[width=8.1cm,clip]{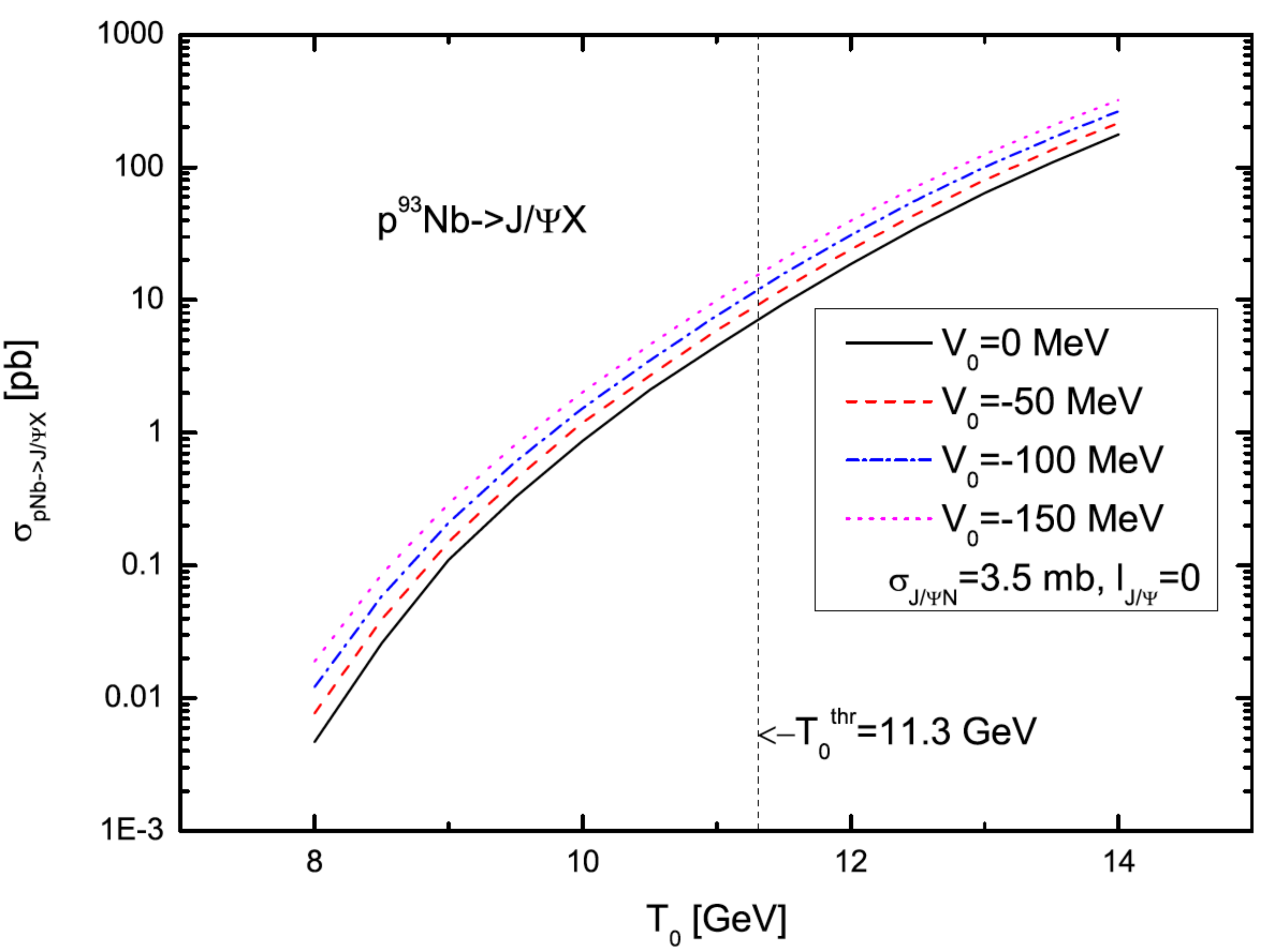}\includegraphics[width=8.0cm,clip]{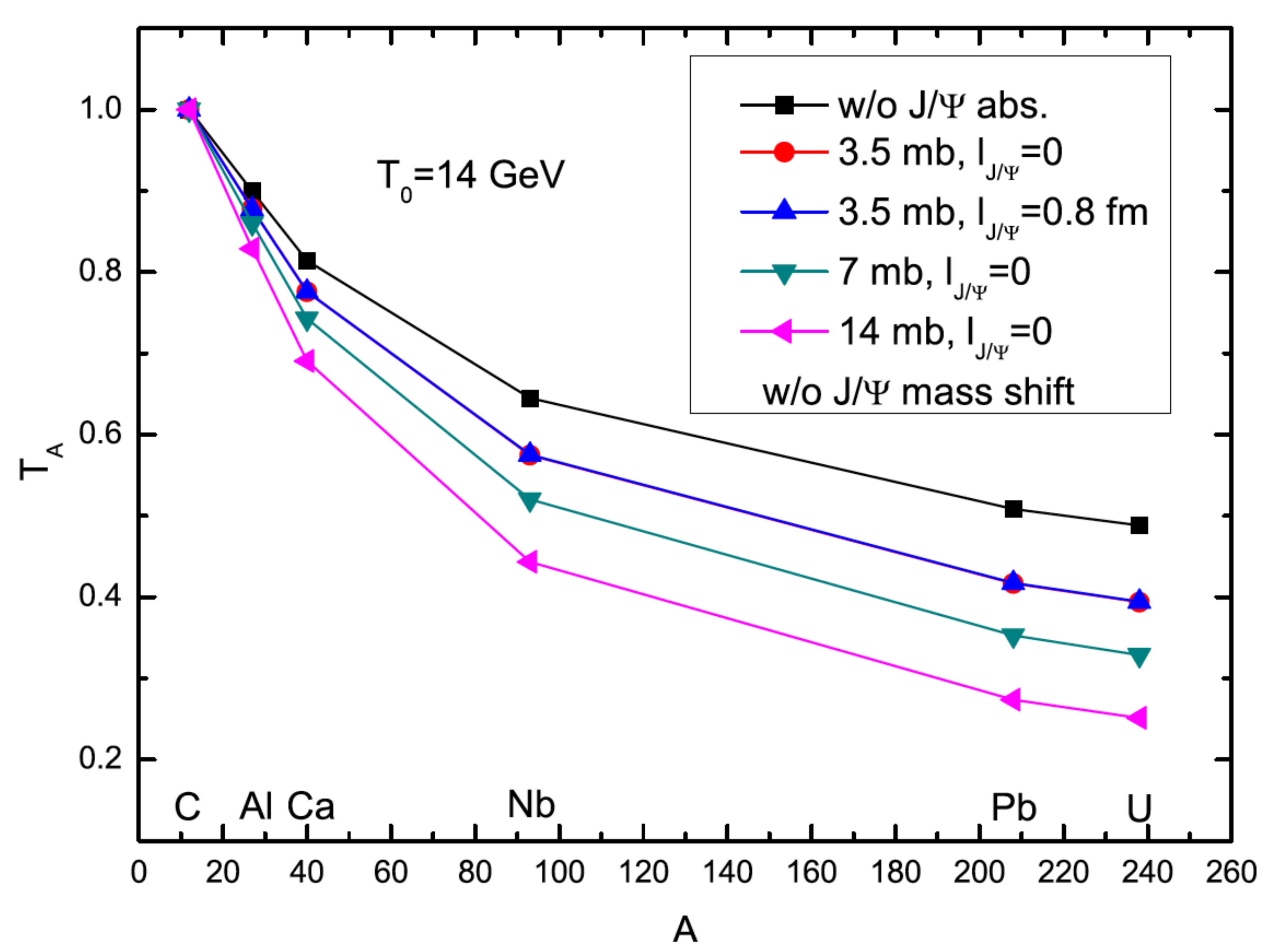}
 \caption{(Left) Excitation function calculated for production of $J/\psi$ mesons in the $p$ + $^{93}$Nb reaction. The curves represent calculations for an absorption cross section 
 $\sigma_{J/\psi N}$ = 3.5 mb and a formation length $l_{J/\psi}$ = 0, assuming $J/\psi$ in-medium mass shifts as depicted in the inset. The vertical dashed line indicates the energy threshold for $J/\psi $ production off the free nucleon. (Right) Transparency ratio $T_A$, normalised to carbon, for $J/\psi$ mesons as function of the nuclear mass number $A$ in the scenario without mass shift for different absorption cross sections $\sigma_{J/\psi N}$ and formation lengths $l_{J/\psi}$ as indicated in the inset. The curves are to guide the eye. The figures are taken from \cite{Kiselev_IJMPE}.}
 \label{fig:Kiselev_Jpsi}
\end{figure*}

In the unitarized coupled channel approach the open-charm in-medium spectral features like $D, \overline D$ masses and widths are calculated using the effective open-charm-nucleon interaction in matter in a self-consistent treatment. The spectral functions exhibit structures due to mixing with resonant-hole states. The potentials for $D$ and $\overline{D}$ mesons are found to be attractive with depths ranging from -15 to -45 MeV but strongly energy dependent close to the open-charm meson mass \cite{Garcia-Recio_PLB}. 

In contrast, using the QMC model, Sibirtsev et al. \cite{Sibirtsev} find a dramatic lowering of the $D^+$ mass by $\approx$ - 150 MeV at normal nuclear matter density, while the $D^-$ meson is shifted up in mass by about 15 MeV, reducing the threshold for $D^+ D^-$ pair production in antiproton-proton annihilation by 135 MeV in nuclei compared to the $D^+ D^-$ production in free $p \overline{p}$ collisions. As for the mesons discussed in the previous sections, such a change in the production threshold, leading to an enhanced $D^+ D^-$ production, could be verified by measuring the excitation function in the $p \overline{p} \rightarrow D^+ D^- $ reaction on nuclei, as suggested in Fig.~\ref{fig:Sibirtsev_D+D-}. The difference in the $D^+$ and $D^-$ meson production rates is due to the $D^+$ absorption in nuclear matter.

QCD sum rule studies have led to somewhat inconsistent results. While Hayashigaki \cite{Hayashigaki} finds a mass shift of the $D$ meson of - 50 MeV at normal nuclear matter density, Hilger et al. \cite{Hilger} obtain an opposite mass shift of +45 MeV. Suzuki et al. \cite{Suzuki} claim a positive mass shift for both $D^+$ and $D^-$ mesons (see Fig.~\ref{fig:Suzuki_D}). For the $D^+ - D^-$ mass difference throughout negative values are predicted at normal nuclear matter density: -65 MeV \cite{Hilger} and -15 MeV \cite{Suzuki}, which are however, much smaller than the mass difference of -164 MeV claimed in the QMC model calculation \cite{Sibirtsev}.

For the $J/\psi (c \overline{c}) $ meson the QMC model gives a mass drop of -16 to -24 MeV at normal nuclear matter density \cite{Krein}. The sensitivity of the excitation function for  $J/\psi $ production in proton- and photon induced reactions off nuclei to in-medium mass shifts of the $J/\psi $ meson has been studied in \cite{Kiselev_IJMPE,Paryev_arXiv} (see Fig.~\ref{fig:Kiselev_Jpsi}). Furthermore, the extraction of the in-medium inelastic $J/\psi N$ cross section from transparency ratio measurements has been discussed which provides information on the imaginary part of the $J/\psi$ -nucleus potential. Molina et al. \cite{Molina_PRC86} calculate the transparency ratio for photon-induced $J/\psi$ - production at 10 GeV as foreseen in the JLAB upgrade and expect that 30-35$\%$ of the $J/\psi$-mesons produced in heavy nuclei are absorbed inside the nucleus.

It is evident that this spread of theoretical predictions calls for an experimental clarification. Such measurements will, however, have to await new experimental possibilities at J-PARC and in the future at FAIR.
%-----------------------------------------------------------------------------------------------------------------------------------

%-----------------------------------------------------------------------------------------------------------------------------------
%TABLES

\section{Compilation of theoretical predictions and experimental results for the real  and imaginary  parts of meson-nucleus potentials}\label{tables}
In this subsection the theoretical predictions and experimental results for the real and imaginary parts of meson-nucleus potentials, discussed in the above sections, have been compiled in two tables. This is meant to provide a quick overview for the reader.

\begin{table}
\begin{center}
\begin{minipage}[t]{16.5 cm}
\caption{Compilation of theoretical predictions for the real ($V_0)$ and imaginary $(W_0)$ meson-nucleus potentials at normal nuclear matter density.}
\label{tab:VW_theo}
\end{minipage}
\begin{tabular}{|c|c|c|c|c|c|c|}
\hline
&&&&&\\[1mm]
meson & $V_0$ [MeV] & $W_0$[MeV] &$\sigma_{inel}$ [mb]  & model & ref.\\[2mm]
\hline
&&&&&\\[1mm]
$K^+$ & $\approx$ +20 & - & - & - & \cite{Schaffner_NPA1997}\\[1.2mm]
$K^+$ & $\approx$ + 40 & - & - & -&\cite{Waas}\\[1.2mm]
$K^+$ & + 36 & - & - & -&\cite{KorpaLutz}\\[1.2mm]
$K^-$ & -(180-200) & -(70-80) & - & NLO30& [7-11]\\[1.2mm]
%\cite{Friedman_Gal_PR,Friedman_Gal_NPA881,Friedman_Gal_NPA899,Gal_NPA914,Friedman_Gal_NPA959}
$K^-$ & $\approx$ -100 & - & - & - & \cite{Schaffner_NPA1997}\\[1.2mm]
$K^-$ & $\approx $ -120 & - & - & - & \cite{Waas}\\[1.2mm]
$K^-$ & -(40-50) & $\approx$ -50 & - & Chiral & \cite{Ramos_Oset}\\[1.2mm]
$K^-$ & -(80-90) & - & - & $\chi$ meson-baryon &\cite{Cieply_PRC2011}\\[1.2mm]
$K^-$ & -(80 -120) & $\approx 30 $ & - & $\chi$ meson-baryon &\cite{Gazda_Mares}\\[1.2mm]
$\eta$ & - & - & $\approx $ 30& - &\cite{Bennhold_Tanabe}\\[1.2mm]
$\eta$ & - & - & $\approx $ 30& - &\cite{Cassing_eta}\\[1.2mm]
$\eta^\prime$ & $\approx$ -150 & - & - & NJL &\cite{Nagahiro_PRC74}\\[1.2mm]
$\eta^\prime$ & $\approx$ 0 & - & - & NJL &\cite{BMetaprime}\\[1.2mm]
$\eta^\prime$ & $\approx$ -80 & - & - & lin $\sigma$ &\cite{Jido}\\[1.2mm]
$\eta^\prime$ & $\approx$ -40 & - & - & QMC &\cite{Bass}\\[1.2mm]
$\omega$ & $\approx$ 0 & - & - & NJL & \cite{Bernard_Meissner}\\[1.2mm]
$\omega$ & $\approx$ -120 & - & - & QCD sum rule &\cite{Hatsuda_Lee}\\[1.2mm]
$\omega$ &$\approx$ -(100-150) & $\approx$ -20  & - & L$_{eff}$ &\cite{Klingl}\\[1.2mm]
$\omega$ & $\approx$ 0 & $\approx$ -30  & - & res. coupl. &\cite{Muehlich_NPA780}\\[1.2mm]
$\omega$ & $\approx$ -30 & $\approx$ -20  & - & res. coupl. &\cite{Lutz}\\[1.2mm]
$\omega$ & - & -(50 - 100)  & - & $\chi$ unitary  &\cite{Ramos}\\[1.2mm]
$\omega$ & - & -(75 - 100)  & - & many body &\cite{Cabrera_Rapp}\\[1.2mm]
$\omega$ &  $\approx$ -100 & & & QMC &\cite{QMC}\\[1.2mm] 
$\phi$ & $\approx$ -30 & - & - & - & \cite{Hatsuda_Lee}\\[1.2mm]
$\phi$ & -35 & -(20-25) & - & - & \cite{Cabrera_PRC95}\\[1.2mm]
$\phi$ & $\approx$ -8 & $\approx$ -15 & - & -& \cite{Cabrera_Vacas}\\[1.2mm]
$\phi$ & $\ge$ -20 & -45 & - & $\chi$ EFT/QCD sum rule& \cite{Gubler_Weise_PLB,Gubler_Weise_NPA}\\[1.2mm]
$D^0$ & -25 & -14 & - & unitarized coupled channel &\cite{Garcia-Recio_PLB}\\[1.2mm]
$D, D^*$ & -62 & - & - & QMC&\cite{Krein}\\[1.2mm]
$D^+$& $\approx$ - 150 & - & - & QMC &\cite{Sibirtsev}\\[1.2mm]
$D^+$ & +23 & - & - & QCD sum rules &\cite{Suzuki}\\[1.2mm]
$D^-$ & +20 & - & - & QMC &\cite{Sibirtsev}\\[1.2mm]
$D^-$ & +38 & - & - & QCD sum rules &\cite{Suzuki}\\[1.2mm]
$J/\psi$ & $\approx$ -(19-24)  & - & - & QMC & \cite{Tsushima_PRC83}\\[1.2mm]
$J/\psi$ & $\approx$ -(16-24)  & - & - & L$_{eff}$ & \cite{Krein}\\[1.2mm]
\hline
\end{tabular}
\end{center}
\end{table}

\begin{table}
\begin{center}
\begin{minipage}[t]{16.5 cm}
\caption{Compilation of experimentally deduced real  ($V_0)$ and imaginary $(W_0)$ meson-nucleus potentials at normal nuclear matter density. The potential values marked with an asterisk have been extrapolated to meson momentum zero, otherwise the potential values have been determined as average over a momentum range, mainly $0 \ll p \le m$. When given separately in the original literature the first error refers to the statistical error and the second one to the systematic error.}
\label{tab:VW_exp}
\end{minipage}

\begin{tabular}{|c|c|c|c|c|c|c|}
\hline
&&&&&&\\[1mm]
meson & $V_0$[MeV] & $W_0$ [MeV] &$\sigma_{inel}$ [mb] & reaction & collaboration & ref.\\[2mm]
\hline
&&&&&&\\[1mm]
$K^+$ & $\approx$ 25  & - & - & p+A &  KaoS & \cite{Scheinast}\\[1.2mm]
$K^+$ & 15-20 & - & - & Ru+Ru, Ni+Ni & FOPI &\cite{Crochet_FOPI}\\[1.2mm]
$K^+$ & 20$\pm3$ & - & - & p+A &ANKE &\cite{Buscher_EPJA22}\\[1.2mm]
$K^+$ & $\approx$ 30  & - & - & Ni+Ni & - &  \cite{Srisawad}\\[1.2mm]
$K^+$ & $20\pm5$ & - & - & Ni+Ni & FOPI &\cite{Zinyuk_PRC90}\\[1.2mm]
$K^0$ & $20\pm5$ & - & - & $\pi^- + A $ & FOPI & \cite{FOPI_K0}\\[1.2mm]
$K^0$ & $\approx 40 $ & - &  -& Ar + KCl & HADES & \cite{HADES_K0}\\[1.2mm]
$K^0$ & $40\pm 5 $ & -  & - &  p + Nb & HADES & \cite{HADES_K0_p}\\[1.2mm]
$K^-$ & $\approx - 80 $ & -  & -& p + A & KaoS & \cite{Scheinast}\\[1.2mm]
$K^-$ & $-(45 - 50) $ & - & -  & Ni + Ni & FOPI& \cite{Zinyuk_PRC90}\\[1.2mm]
$K^-$ & $-60_{-31}^{+50} $ & - & -  & p + A & ANKE & \cite{Kiselev_K-}\\[1.2mm]
$K^-$ & -160 ...-190 & $\approx$ - 60 & -&  $^{12}$C,$^{16}$O$(K^-,N) $& KEK E548 & \cite{Kishimoto}\\[1.2mm]
%$K^-$ & -(180-200) & -(70-80) & - & atomic X-rays & - & \cite{Friedman_Gal_PR,Friedman_Gal_NPA881,Friedman_Gal_NPA899,Gal_NPA914,Friedman_Gal_NPA959}\\[1.2mm]
%$K^-$ & $\approx - 190 $ & -  & -& atomic X-rays & - &  \cite{Friedman_Gal_PR}\\[1.2mm]
$\eta$ &  & -(10 - 30) & 30$\pm6 $ & $\gamma$ + A & A2 & \cite{Roebig-Landau}\\[1.2mm]
$\eta$ & $-(54\pm6)$ & $-(20\pm2)$ & - & p + d & ANKE, COSY11 & \cite{Xie,Oset_priv}\\[1.2mm]
$\eta^\prime$ & - & -(10$\pm$2.5)  & 10.3$\pm$1.4 & $\gamma$ + A & CBELSA/TAPS & \cite{Nanova_PLB_TA}\\[1.2mm]
$\eta^\prime$ &  - &  $ -(13\pm3\pm3)^*$  &13$\pm3$ & $\gamma$ + C, Nb & CBELSA/TAPS & \cite{Friedrich_EPJA}\\[1.2mm]
$\eta^\prime$ & -(37$\pm10\pm10$)&  -& - & $\gamma$ + C & CBELSA/TAPS & \cite{Nanova_PLB727}\\[1.2mm]
$\eta^\prime$ & -(41$\pm10\pm15$)&  -& - & $\gamma$ + Nb & CBELSA/TAPS & \cite{Nanova_PRC94}\\[1.2mm]
$\omega$ & - & -(35 - 50) & $\approx$ 40 & $\gamma$ + A & CBELSA/TAPS & \cite{Kotulla}\\[1.2mm]
$\omega$ &  - & -(48$\pm12\pm9)^*$ & -  & $\gamma$ + C, Nb & CBELSA/TAPS & \cite{Friedrich_EPJA}\\[1.2mm]
$\omega$ & -(29$\pm19\pm20$) & - & - & $\gamma$ + C, Nb & A2 & \cite{Metag_Hyp}\\[1.2mm]
$\omega$ & -(15$\pm35\pm20$) & - & - & $\gamma$ + C & CBELSA/TAPS & \cite{Friedrich_PLB}\\[1.2mm]
$\omega$ & $\approx$ -75 & 0 & - &  p + A & KEK E325 & \cite{Naruki}\\[1.2mm]
$\omega$ &  - &  $\le$ -100 MeV & - & $\gamma$ + A & CLAS & \cite{Wood_rho_PRC78}\\[1.2mm]
$\phi$ & - & -(20 - 30) & - & p + A & ANKE & \cite{Polyanskiy}\\[1.2mm]
$\phi$ & - & -(10 - 30) & 14-25 & p + A & ANKE & \cite{Hartmann_PRC85}\\[1.2mm]
$\phi $ &- & -(23-100) & 16-70 &  $\gamma$ + A & CLAS & \cite{Wood_PRL105}\\[1.2mm]
$\phi$& - &- &  $ 35^{+17}_{-11}$  & $\gamma$ + A & LEPS & \cite{Ishikawa}\\[1.2mm]
$\phi$ & $\approx$-35 & -7.5 &  -&  p + A & KEK E325 & \cite{Muto}\\[1.2mm]
\hline
\end{tabular}
\end{center}
\end{table}

\begin{figure*}
\centering
 \includegraphics[width=17.0cm,clip]{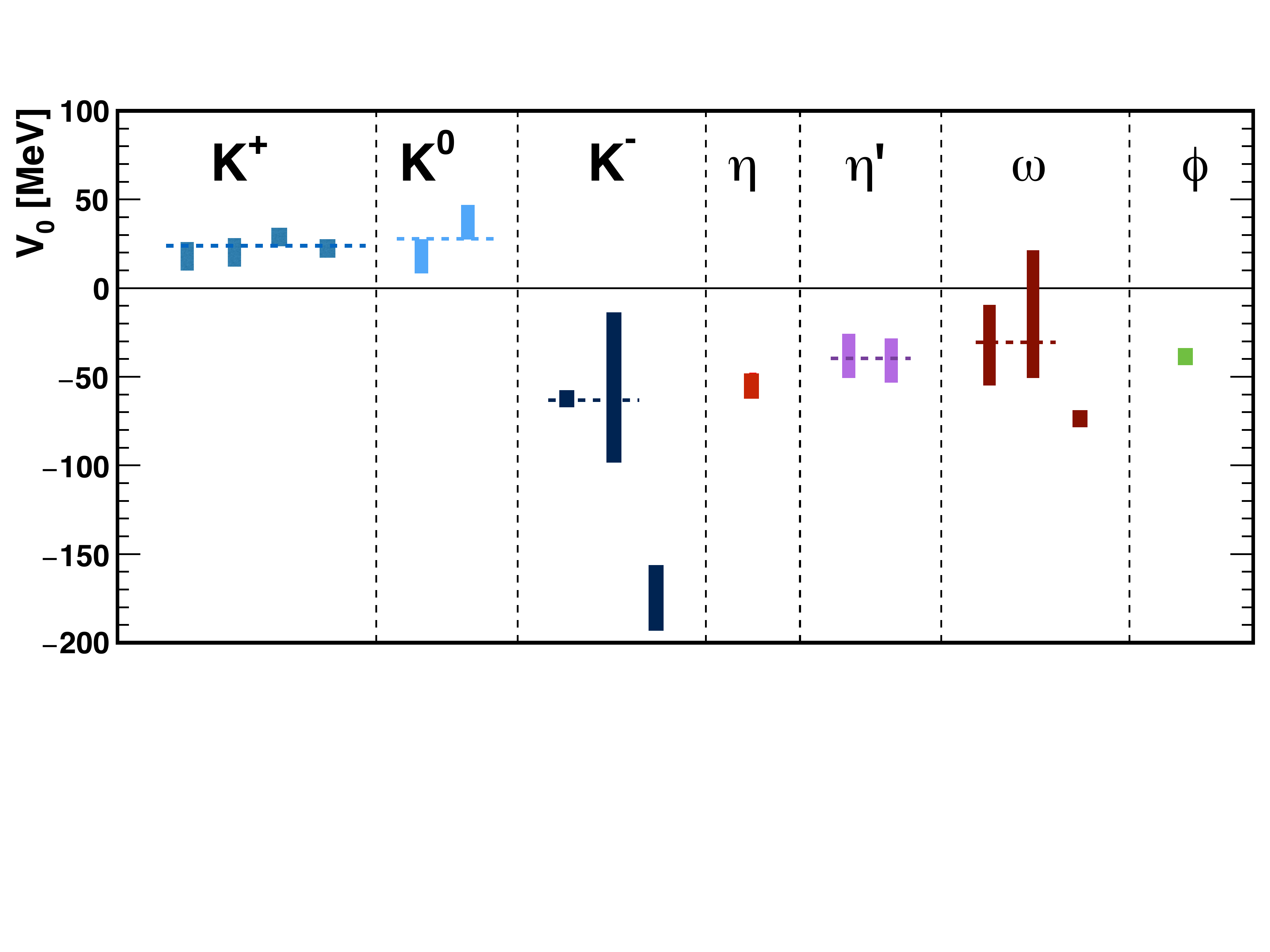}
 \vspace{-30mm} \caption{Compilation of the experimentally determined real part of the meson-nucleus potential for $K^+,K^0,K^-,\eta,\eta^\prime,\omega$ and $\phi$ mesons. The data are taken from table 2. The vertical bars represent the errors or the range of values quoted. The horizontal dashed lines indicate average values. The CLAS results \cite{Wood_PRL105} are not included in the figure because of the large experimental uncertainties quoted by the authors.}
\label{fig:tableplot_V}
\end{figure*}

\begin{figure*}
\centering
 \includegraphics[width=16.0cm,clip]{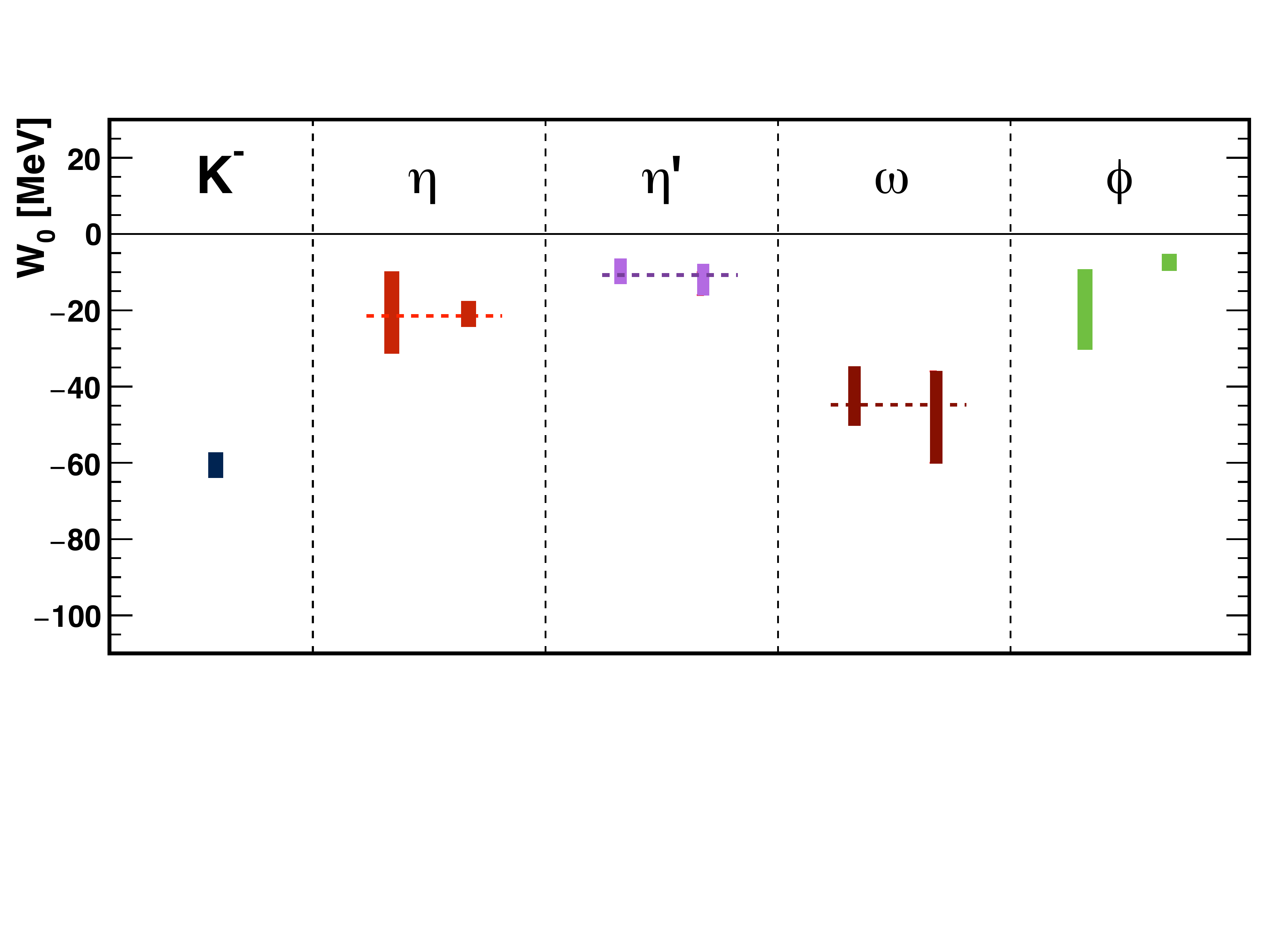}
  \vspace{-30mm}\caption{Compilation of the experimentally determined imaginary part of the meson-nucleus potential for $ K^-, \eta,\eta^\prime,\omega $ and $\phi$ mesons. The data are taken from table 2. The vertical bars represent the errors or the range of values quoted. The horizontal dashed lines indicate average values. The CLAS results \cite{Wood_PRL105} are not included in the figure because of the large experimental uncertainties quoted by the authors.}
\label{fig:tableplot_W}
\end{figure*}

Figs.~\ref{fig:tableplot_V},~\ref{fig:tableplot_W} provide a graphical representation of the experimental data given in table 2.
One sees that the real part of $K^{+}$ and $K^{0}$ - nucleus potentials are weakly repulsive, while the $K^{-}$, $\eta$, $\eta^\prime$, $\omega$ and $\phi$ - nucleus potentials are attractive, however, with widely different strengths. Because of meson absorption in the nuclear medium the imaginary part of meson-nucleus potentials are all negative, again with a large spread. The smallest imaginary part is found for the $\eta^\prime$ mesons.


\begin{figure*}
\centering
 \includegraphics[width=10.0cm,clip]{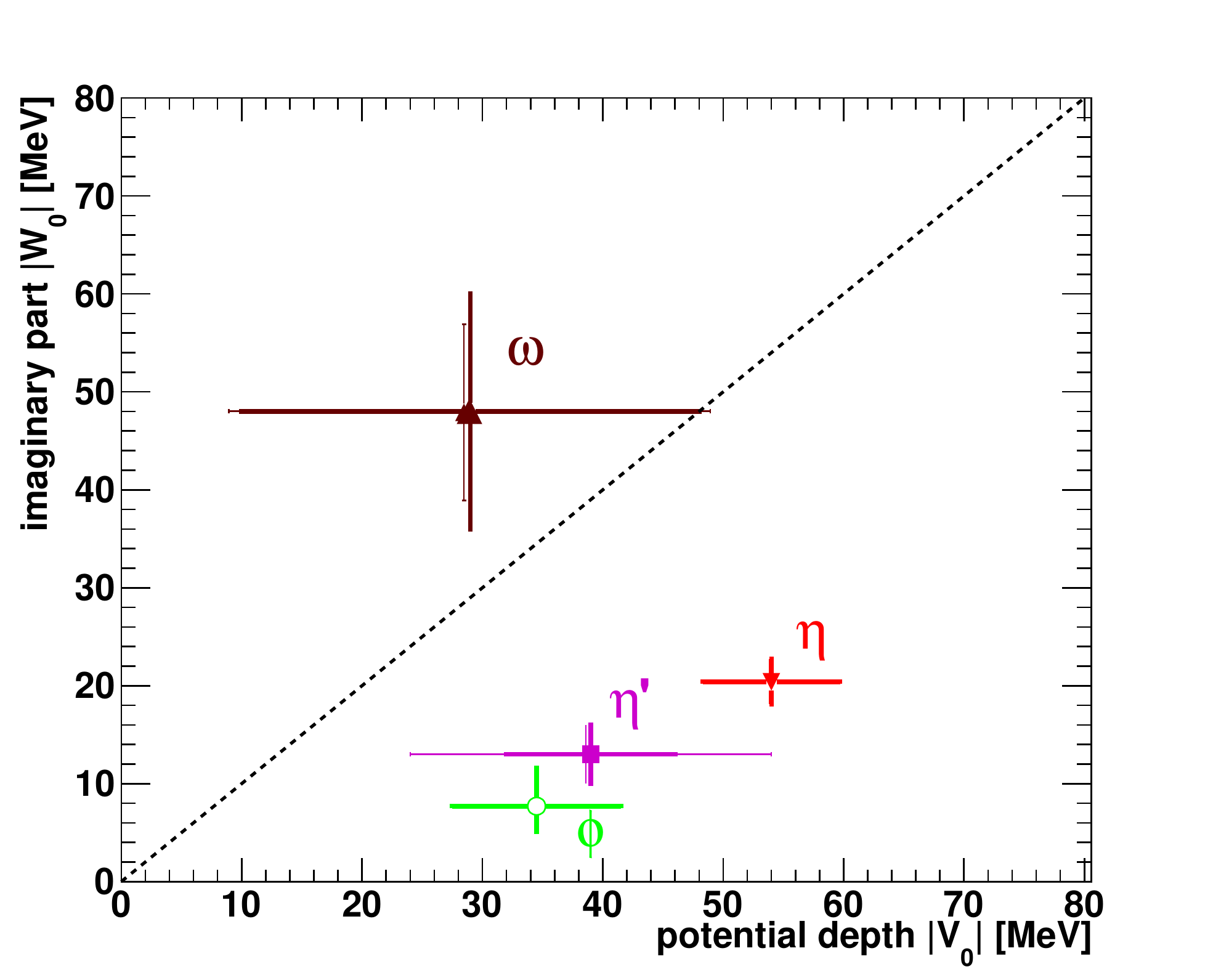}
  \caption{Compilation of the experimentally determined real and imaginary parts of the meson-nucleus potential of $\eta$ \cite{Xie, Oset_priv}, $\eta^\prime$ 
  \cite{Friedrich_EPJA,Nanova_PLB727,Nanova_PRC94}, $\omega$ \cite{Friedrich_EPJA,Friedrich_PLB,Metag}, and $\phi$ \cite{Muto} mesons. Thick (thin) error bars correspond to statistical (systematic) errors. The figure is updated from \cite{Nanova_Metag}.}
\label{fig:eta_etaprime_omega_phi}
\end{figure*}

In many experiments only the real part or only the imaginary part of the meson-nucleus potential have been determined. However, for the $\eta, \eta^\prime, \omega$ and $\phi$ mesons there are experiments which report a real and an imaginary part of the potential. The relative strength of the real and imaginary part is an important information for the observation of meson-nucleus bound states (see Section \ref{sec:mesic}) and can be read off from Fig.~\ref{fig:eta_etaprime_omega_phi}.
For the $\eta,\eta^\prime$ and $\phi$ mesons the modulus of real part $\mid V_0 \mid $ is found to be larger than the modulus of the imaginary part $\mid W_0 \mid$. Note, however, that the small imaginary part reported for the $\phi$ meson in \cite{Muto} is in conflict with the much larger values reported in \cite{Hartmann_PRC85}.

%--------------------------- mesic states --------------------------------------------------------------------------------------
%MESIC STATES

\section{The search for mesic states}\label{sec:mesic}
An attractive meson-nucleus interaction, if strong enough, opens the possibility of having a meson bound in a nucleus to form a short-lived meson-nucleus quasi-bound state. Such states are of great interest in contemporary nuclear and hadronic physics. From the nuclear physics point of view these states are exotic configurations of nuclei as they correspond to states with excitation energies of several hundred MeV up to 1 GeV. For hadron physics the study of these states provides a unique possibility to investigate the properties of mesons and their possible modification at finite nuclear densities. The key to the formation of such a quasi-bound state is to chose a reaction kinematics with low momentum transfer to the meson (recoil-less production). The observation of the mesic state will be facilitated if the potential depth is sizable and large compared to the width of the state, i.e. :
\begin{equation}
\mid V \mid \gg \mid W \mid. \label{eq:real-imag}
\end{equation}
Furthermore the energy spacing of quasi-bound states should be larger than the widths of the states, a condition which becomes more difficult to fulfil in heavier nuclei.

Most of the experiments search for resonance-like structures in the excitation energy range of the residual nucleus where the meson-nucleus bound states are expected either by looking for decays of the mesic states or by missing mass spectroscopy. These measurements, however, suffer from the multi-pion background from competing reactions. An alternative approach is to measure the production of the respective meson near the production threshold, which can normally be identified with a sufficient signal-to-background ratio. The disadvantage is that one has to extrapolate in energy to the mesic pole and will not be able to distinguish between a quasi-bound or anti-bound (virtual) state. In the following sections experimental searches for meson-nucleus bound states for various mesons will be summarised and compared to corresponding theoretical predictions. 
%ÑÑ-----------------------------------------------------------------------------------------------------------------------------

%PIONIC and KAONIC ATOMS
\subsection{\it pionic and kaonic atoms\label{sec:pion_kaon_mesic}}

The existence of pionic and kaonic atoms is clearly established experimentally. Here, one of the atomic electrons is replaced by a negatively charged pion or kaon. These exotic states have been formed by stopping $\pi^-$ or $K^-$ in a target. The mesons are captured in an outer atomic orbit and then cascade down to lower atomic levels by the emission of characteristic X-rays until at some lower principal quantum number the mesons are absorbed due to their interaction with the nucleus.  The lifetimes of $\pi^-$ and $K^-$ mesons are long compared to the stopping times and lifetimes of atomic states so that well-defined exotic atomic states can be formed. The observables of interest are the binding energies and the widths of these states which are affected by the strong interaction in the inner orbits. The level shifts and widths of atomic levels have been analysed to extract information on the meson-nucleus interaction and in particular on its density dependence since the overlap of atomic wave functions with the nucleus covers a wide range of nuclear densities [7-11].
%\cite{Friedman_Gal_PR,Friedman_Gal_NPA881,Friedman_Gal_NPA899,Gal_NPA914}.. 

A breakthrough in these studies has been the observation of deeply bound $\pi^-$ $1s$ and $2p$ states which have been directly populated in recoil-free reactions, avoiding the cascading down from higher orbits [1-4]. The existence of these states had been predicted in [272-274].
%\cite{Friedman_Soff_JPG11,Toki_Yamazaki_PLB213,Toki_NPA501}
%\cite{Gilg_PRC62,Itahashi_PRC62,Geissel_PRL88,Suzuki_PRL92}.
These deeply bound states correspond to atomic orbits very close to the nucleus and are thus more sensitive to the meson-nucleus interaction.
The information on the strong interaction of low energy mesons extracted from these studies has been summarised by Friedman and Gal \cite{Friedman_Gal_PR}. Fitting phenomenological density dependent potentials to $K^-$ atomic data, they deduce deep potentials with a depth of about -200 MeV for $K^-$ mesons in nuclear matter at saturation density, while chiral Lagrangian [142-144,168-171]
%\cite{Ramos_Oset, Lutz_PLB1998, Schaffner_NPA2000,Cieply_NPA2001,Ramos_NPA2001,Tolos_PRC74,Tolos_PRC78}
or meson-exchange calculations \cite{Tolos_NPA690,Tolos_PRC65} predict much shallower $K^-$ - nucleus potentials with a central depth of the order of -50 to -80 MeV. Friedman and Gal \cite{Friedman_Gal_NPA881, Friedman_Gal_NPA899} emphasise that in order to achieve good fits to kaonic atom data the potential constructed from in-medium chirally motivated $K^- N$ amplitudes have to be supplemented by a phenomenological term taking $K^-$ multi-nucleon interactions into account. Note, however, the model dependence of extrapolating potentials to normal nuclear matter density, recently pointed out in \cite{Friedman_Gal_NPA959} and discussed in Section \ref{sec:K-}. It is remarkable that Hirenzaki et al. \cite{Hirenzaki_PRC61} do find an equally good reproduction of the kaonic atom data using a shallow chiral potential.

The existence of deeply bound pionic states has been explained by the superposition of the attractive Coulomb interaction with the repulsive s-wave pion-nucleus interaction such that the pions are bound in a potential pocket at the nuclear surface which gives rise to a halo-like pion distribution around the nucleus \cite{Kienle_Yamazaki}. Kaonic atoms have a different origin. In contrast to the repulsive s-wave $\pi^-$--nucleus interaction, the $K^-$-nucleus interaction is found to be attractive but the strength of the imaginary part of the $K^-$-nucleus interaction suppresses the atomic wave function in the nuclear interior and expels it to the nuclear surface, again leading to a halo-like meson-nucleus configuration \cite{Friedman_Gal_PR}.

This review, however, focuses on the possible existence of meson-nucleus states exclusively bound via the strong interaction. Despite of experimental efforts for over about 30 years no convincing evidence for meson-nucleus states exclusively bound by the strong interaction has as yet been found apart from some promising indications.

\subsection{\it Search for kaonic clusters and $K^-$ nuclear quasi-bound states} \label{sec:K-cluster}

As discussed in Section \ref{sec:K-}, low energy $\overline{K} N$ scattering \cite{Martin} and X-ray spectroscopy of kaonic atoms \cite{Friedman_Gal_PR,Bazzi_PLB704,Bazzi_NPA881,Okada_PLB653,Shi_EPJWC126} have shown that the $\overline {K} N$ interaction is strongly attractive. Similar to the case of the $\eta$ meson, where the strong S$_{11}(1535)1/2^-$ resonance governs the near threshold $\eta N$ interaction, the $\Lambda(1405)1/2^-$ resonance, located 27 MeV below the $\overline{K}$ production threshold, dominates the near threshold anitkaon-nucleon interaction. The nature of the $\Lambda(1405)$ state has been widely discussed in the literature. For a recent summary see \cite{Meissner_Hyodo}. As discussed in Section \ref{sec:K-}, it is considered to be a molecular state emerging from the interference of a $\Sigma \pi$ and a $K^- N $ pole. The interpretation as a $K^- p$ bound state led Akaishi and Yamazaki \cite{Akaishi_Yamazaki,Yamazaki_Akaishi} to predict the existence of kaonic nuclear bound states. The simplest such state would be a bound $K^- N N$ cluster. Depending on the $\overline{K} N$ interaction model, widely scattered binding energies of 10 - 100 MeV and widths of the order of 30 -110 MeV  have been predicted [280-288].
%\cite{Yamazaki_Akaishi,Ikeda_Sato,Shevchenko,Wycech,Dote,Revai,Ikeda,Barnea,Bayar_Oset_PRC88}.

Experimentally, the field is very controversial. Several experiments have reported peak structures about 100 MeV below the $K^- N N$ production threshold. In $\bar p$ $^4$He annihilations at rest the OBELIX collaboration observed a structure in the $\Lambda p$ invariant mass which - if interpreted as the decay of a $K^-pp$ cluster - would correspond to a binding energy of 160$\pm$4.9 MeV with a width of $\le 24\pm8.0$ MeV \cite{OBELIX_NPA789}. Stopping $K^-$ in $^6$Li,$^7$Li, and $^{12}$C targets, the FINUDA Collaboration at DA$\Phi$NE reported a structure in the back-to-back $\Lambda p$ invariant mass spectrum corresponding to a binding energy of 115$^{+6}_{-5}$ MeV and a width of 67$^{+14}_{-11}$ MeV \cite{Agnello_PRL94}. This structure has later been interpreted as being due to $K^-$ two-nucleon absorption followed by final state interactions of the produced particles with the daughter nucleus \cite{Magas_PRC74}. However, later studies by the FINUDA collaboration \cite{Agnello_PLB669} of correlated $\Lambda t$ pairs in the absorption of $K^-$ at rest in light nuclei revealed that even more nucleons may be involved in the absorption process. Studying pp collisions at 2.85 GeV, the DISTO group claimed a structure in the $\Lambda p $ invariant mass with a binding energy of 103 $\pm$ 3 (stat) $\pm$ 5 (syst) MeV and a width of 118 $\pm$ 8 (stat) $\pm$ 10(syst) MeV \cite{Yamazaki}. A "$K^- p p$" like structure in the $\Sigma^0 p$ channel with binding energy of 95$^{+18}_{-17}$(stat) $^{+30}_{-21}$(syst) MeV and width of 162$^{+87}_{-45}$(stat)$^{+66}_{-78}$(syst) MeV was reported by the E27 experiment at J-PARC, investigating the d($\pi^+, K^+)$ reaction at 1.69 GeV/$c$ \cite{Ichikawa_2015}. In contrast, no significant structures were found by the HADES Collaboration in $p + p$ reactions [295-297]
%\cite{HADES_PLB742,Fabbietti,Epple_Fabbietti}
 nor in a photon induced inclusive reaction studied at LEPS \cite{Tokiyasu}. 

\begin{figure*}
\centering
 \includegraphics[width=9.0cm,clip]{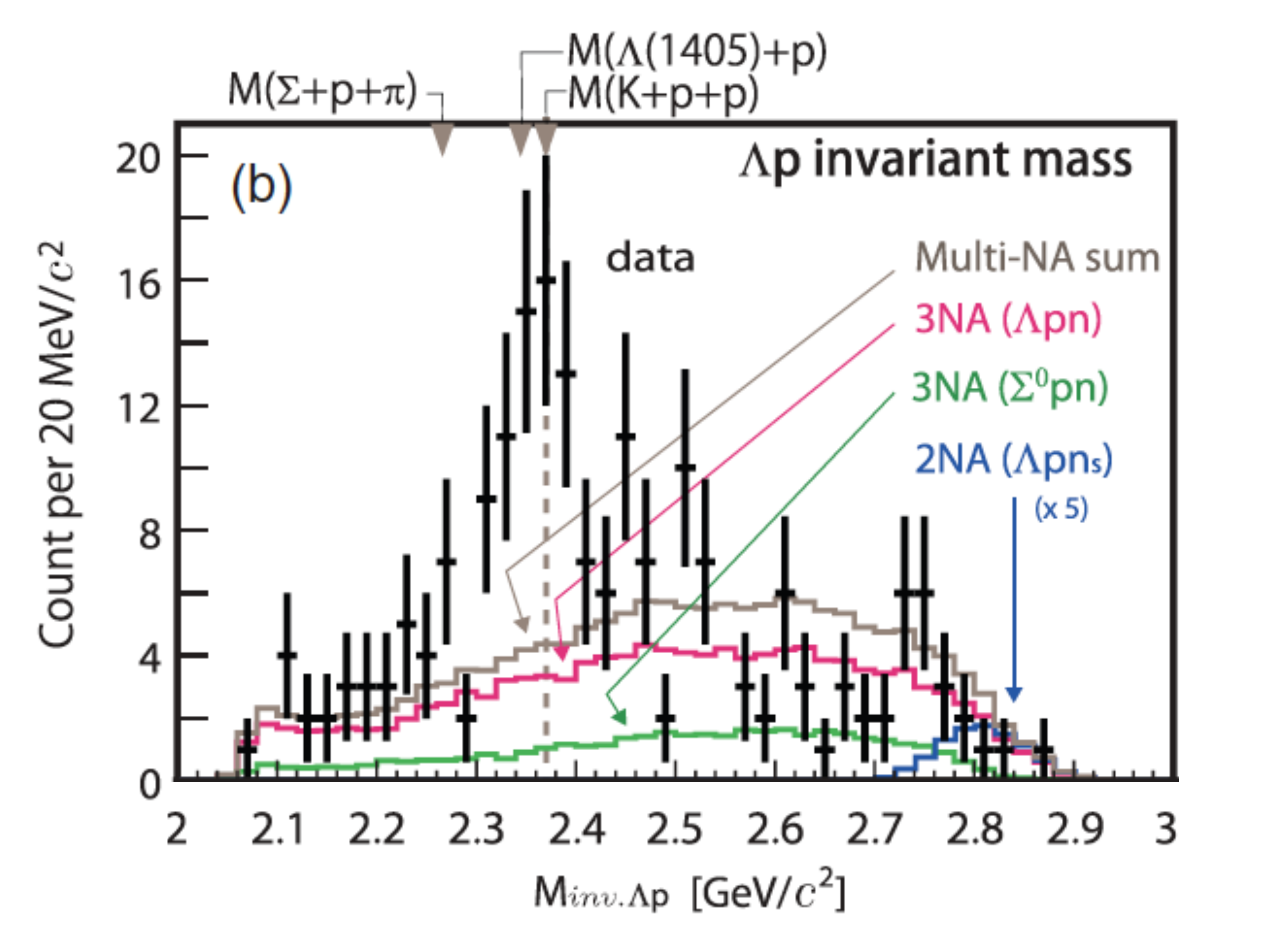}\includegraphics[width=8.0cm,clip]{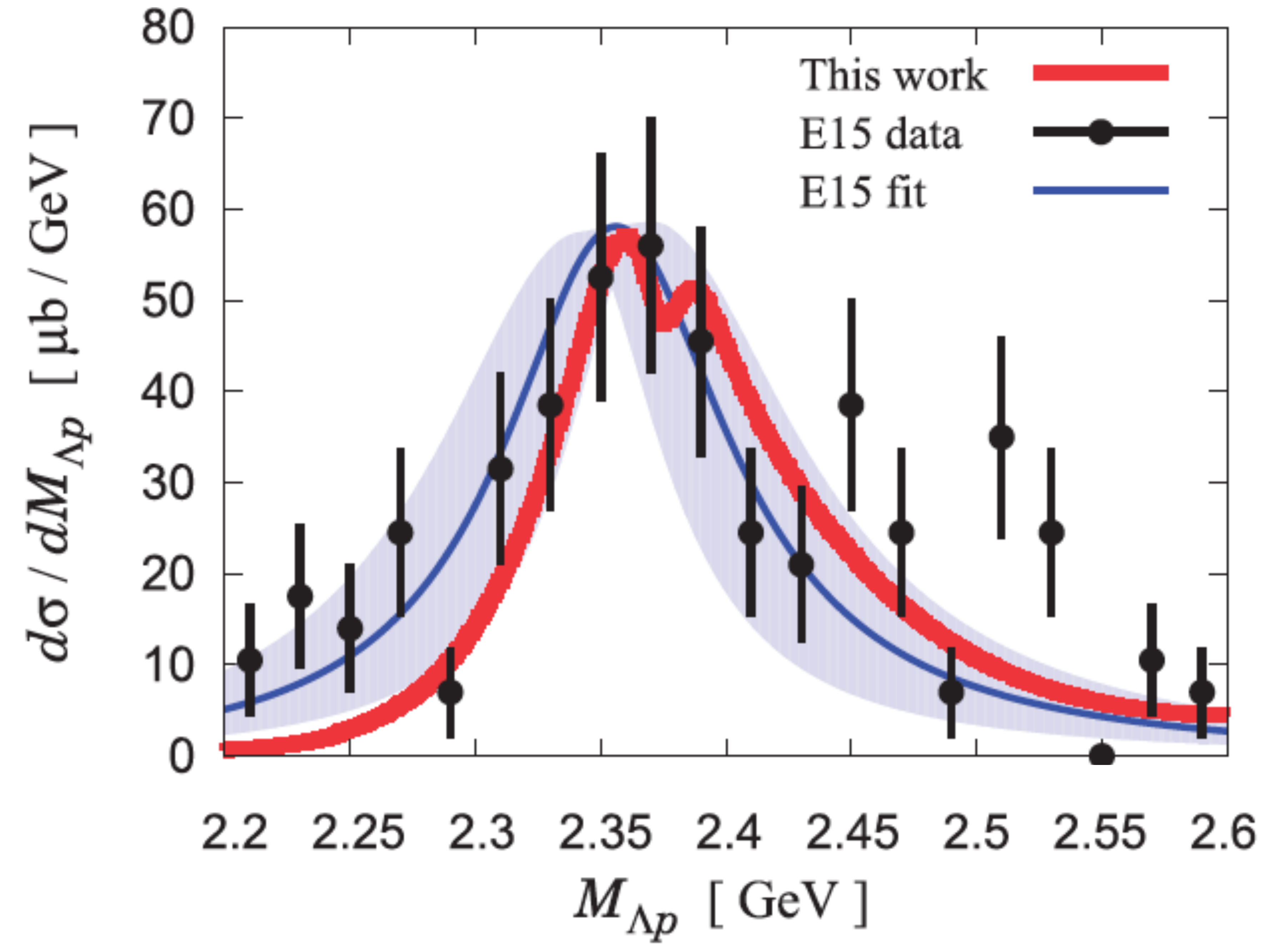}
  \caption{(Left) $\Lambda p $ invariant mass distribution for the in-flight $^3$He$(K^-, \Lambda p) n $ reaction at a $K^-$ momentum of 1.0 GeV/$c$ with simulated spectra for $K^-$ absorption on three or two nucleons. The figure is taken from \cite{Sada}. (Right) The same data in comparison with a calculated $\Lambda p$ invariant mass distribution, assuming  contributions from the decay of a bound $\overline{K} N N$ state (lower peak) and from the formation of an unbound $\Lambda(1405) p $ system (higher peak). The figure is taken from \cite{Sekihara_PTEP2016}. }
\label{fig:K-pp}
\end{figure*}

On the other hand, more recently the E15 experiment at J-PARC has reported a structure in the $\Lambda p $ invariant mass near the $K^- p p$ threshold in the in-flight $^3$He$(K^-, \Lambda p) n $ reaction at a $K^-$ momentum of 1.0 GeV/$c$ \cite{Sada}, as shown in Fig.~\ref{fig:K-pp}. The $\Lambda p n$ final-state is reconstructed from the momenta of the charged particles ($\pi^- p p) $ measured in a cylindrical drift chamber while the neutron was kinematically identified as missing mass particle. The structure in the $\Lambda p$ invariant mass resides
on a background of events attributed to kaon absorption on three- and/or two- nucleons. A Dalitz-plot analysis shows that this structure is associated with forward going neutrons and with a small momentum transfer to the $\Lambda p$ system. The structure has a mass of 2355$^{+6}_{-8}$(stat)$ \pm$ 12 (syst) MeV and a width of $ 110^{+19}_{-17}$(stat)$ \pm$ 27 (syst)  MeV and could thus be a signal of a $\overline{K} N N$ bound state, located $\approx$ 15 MeV below the $K^- p p$ threshold. This interpretation is supported by a recent theoretical study \cite{Sekihara_PTEP2016}, assuming two possible reaction scenarios: (i) the $\Lambda(1405)$ resonance is generated after the emission of an energetic neutron from the absorption of the initial $K^-$ without forming a bound state with the remaining proton; (ii) after the emission of the energetic neutron a $\overline{K} N N $ bound state is formed, decaying subsequently into a $\Lambda p $ pair. The contributions from both scenarios give rise to a two-peak structure in the $\Lambda p$ invariant mass distribution as shown in 
Fig.~\ref{fig:K-pp} (Right) where the calculations are compared to the experimental data of Fig.~\ref{fig:K-pp} (Left). With the statistics of the present data this two-peak structure cannot be resolved but overall the shape of the experimental invariant mass spectrum is well described by the calculation. A high statistics follow-up run will hopefully bring further insight.
\begin{figure*}
\centering
 \includegraphics[width=15.0cm,clip]{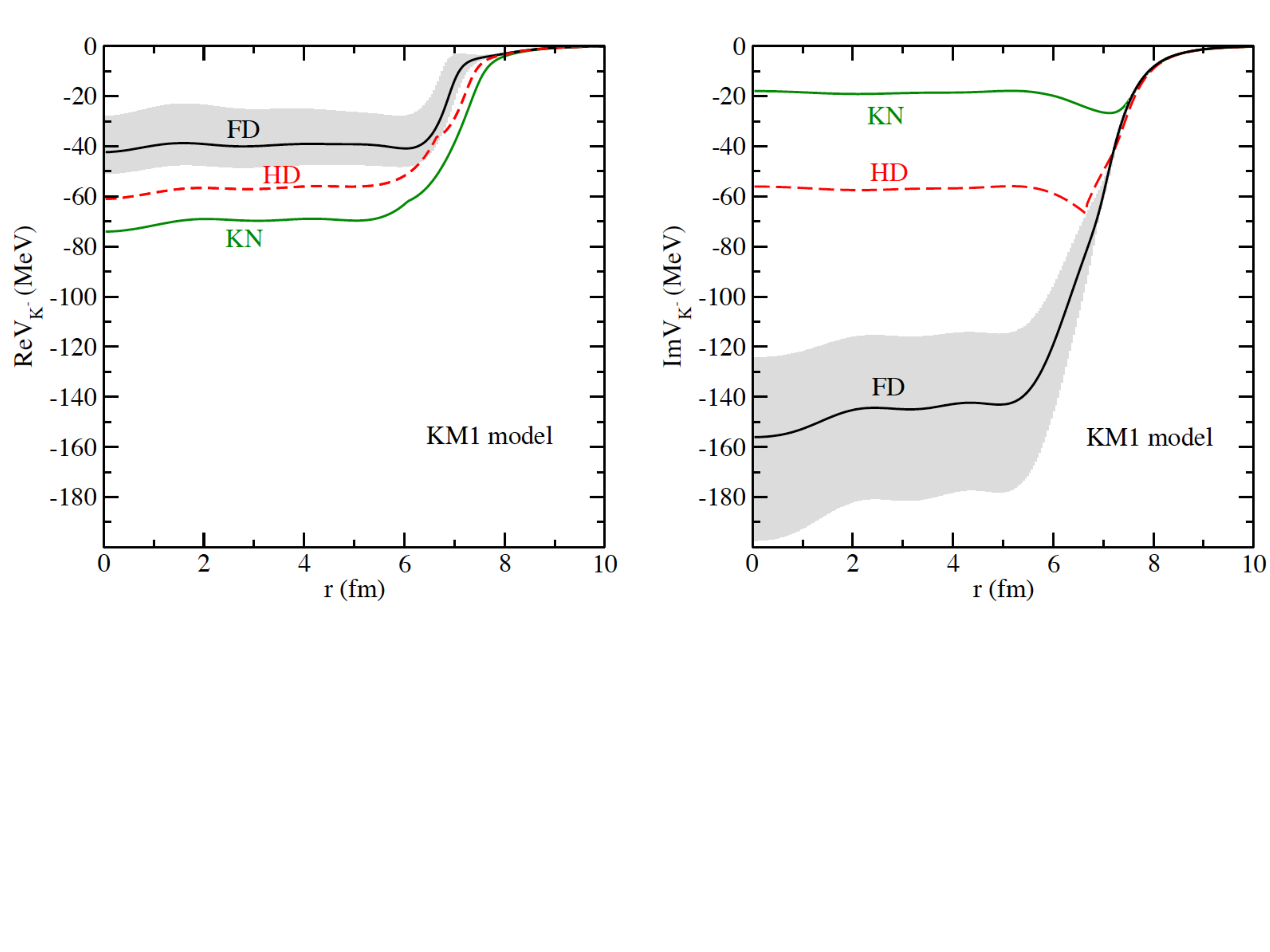}
 \vspace{-35mm}   \caption{ The real (Left) and imaginary (Right) parts of the $K^- -^{208}$Pb potential, calculated self-consistently for the single-nucleon $K^-$ potential (KN) and two different versions (FD, HD) of the density dependence of the $K^-$ multi-nucleon potential. The shaded areas indicate the uncertainties. The KM1 model applies the Ikeda-Hyodo-Weise chiral $K^- N$ potential \cite{Ikeda_Hyodo_Weise_PLB706,Ikeda_Hyodo_Weise_NPA881} plus a phenomenological term proportional to ($\rho/\rho_0)^{\alpha}$ with $\alpha$ = 1, as shown in Fig.~\ref{fig:dens_dep} of this review. The figure is taken from \cite{Hrtankova_Mares}.}
\label{fig:K-pot}
\end{figure*}

The multi-nucleon kaon absorption which accounts for the background in Fig.~\ref{fig:K-pp} plays a decisive role in the search for $K^-$ - nuclear quasi bound-states, as recently pointed out in \cite{Hrtankova_Mares}. Hrt\' ankov\' a and Mare\v s find that $K^-$ - nuclear states in many-body nuclear systems, if they exist at all, will have huge widths, considerably exceeding the $K^-$ binding energies (see Fig.~\ref{fig:K-pot}) because of the large multi-nucleon kaon absorption contributions. This will make the observation of $K^-$ -mesic states very difficult. In fact, the search for $K^--$ nucleus bound states pursued by the AMADEUS Collaboration \cite{AMADEUS} has as yet not provided evidence for such exotic states but for the importance of $K^-$ multi-nucleon absorption [303-305].
%\cite{Scordo_EPJWC130,Scordo_AIPCP1735,Marton_1704}.
%-----------------------------------------------------------------------

%-----------------------------------------------------------------------------------------------------------------------------------------------------------------------------------------------------------------------------------------
%------------------------------------------------------------------------------------------------------------------------------------------------------------------------------------------------------------------------------------------
\subsection{\it Search for $\eta$ -mesic states\label{sec:eta_mesic}}
%ETA-MESIC
As discussed in Section \ref{sec:eta}, the large scattering length and the strong s-wave attraction due to the near threshold $S_{11}(1535)1/2^-$ resonance led Haider and Liu to predict the existence of exotic states with an $\eta$ meson bound to a nucleus in 1986. Motivated by this work many experimental searches for these exotic states and numerous theoretical investigations have been performed since then which have recently been comprehensively summarised in dedicated reviews \cite{Machner,Kelkar}. Here, only the main results and most recent developments will be discussed. 

\begin{figure*}
\centering
  \includegraphics[width=15.0cm,clip]{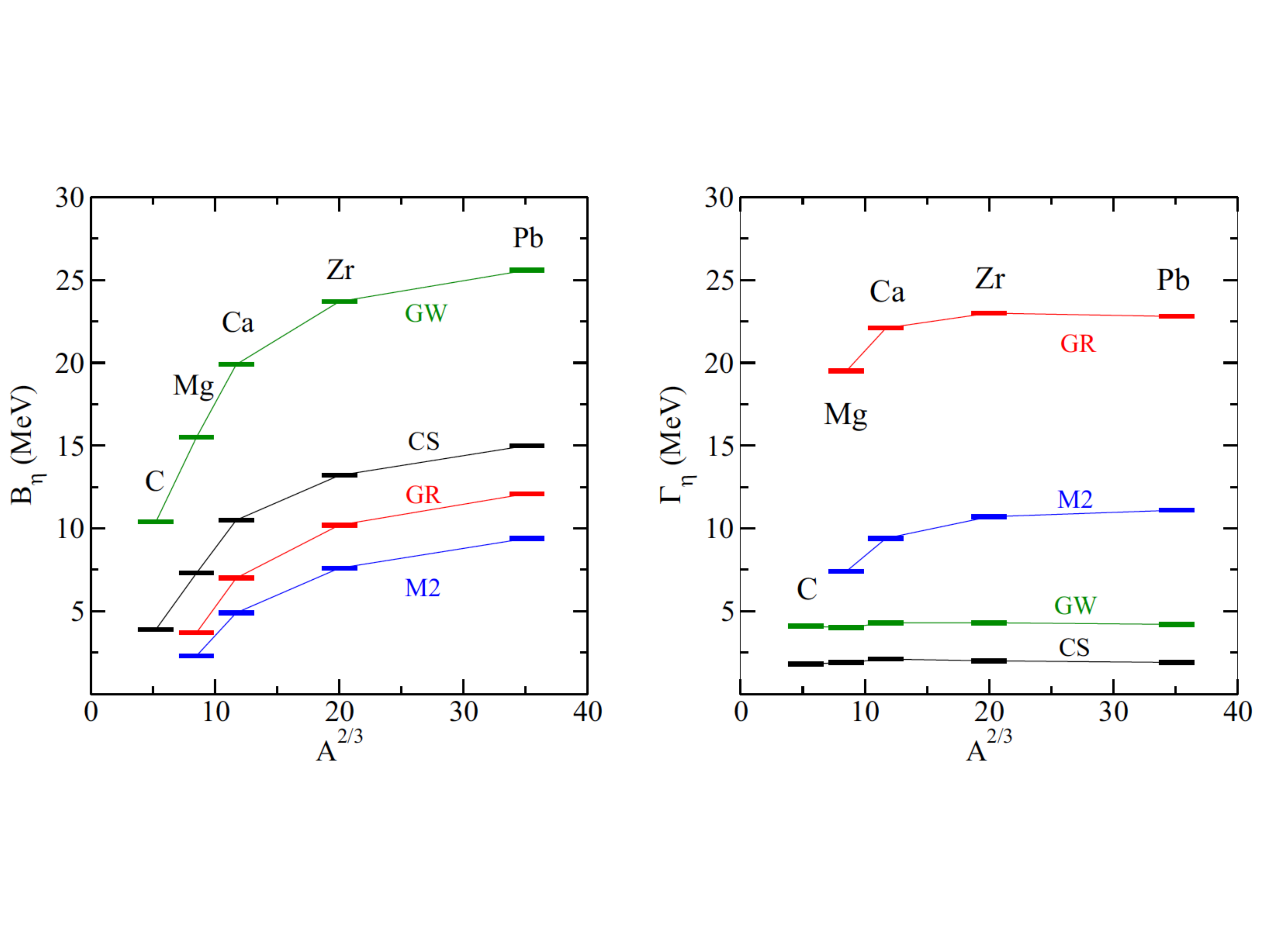}
 \vspace{-15mm}  \caption{(Left) Binding energies and (Right) widths of 1$s$ $\eta$-nuclear states in selected nuclei calculated self-consistently using different $\eta N$ scattering amplitudes shown in Fig.~\ref{fig:eta_scat} of this review. The figure is taken from \cite{Mares_eta}.}
\label{fig:eta_mesic}
\end{figure*}

The model dependence of the $\eta N$ scattering length, discussed in Section \ref{sec:eta}, manifests itself in a broad spread of recent predictions for binding energies and widths of $\eta$--nuclear states shown in Fig.~\ref{fig:eta_mesic}.
The binding energies increase with the nuclear mass number A and saturate at larger values of A, depending on the model for the $\eta N$ amplitude. For some models the widths are only 2-4 MeV, making an observation feasible, provided a suitable production reaction is found. Other models give too large widths or are too weak to generate $\eta$--nucleus bound states. It should be noted that all predicted widths are considerably smaller than the widths extracted from $\eta$ production reactions at above threshold energies (see \ref{sec:eta_imag}) since they are calculated from the subthreshold imaginary amplitudes shown in Fig.~\ref{fig:eta_scat} of the present review.
 \begin{figure*}
\centering
  \includegraphics[width=11.0cm,clip]{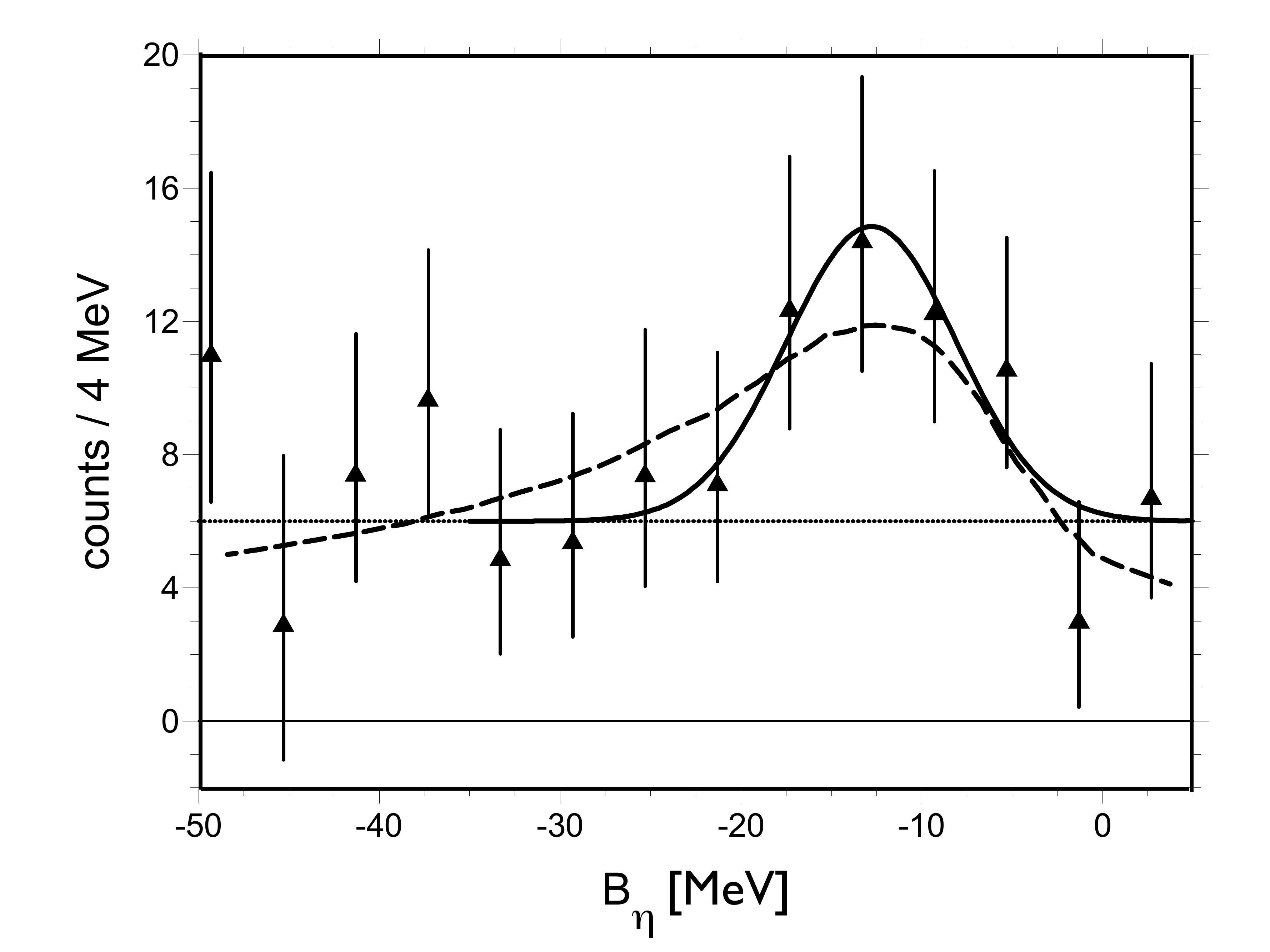}
\caption{Counts for the $p + ^{27}$Al $ \rightarrow^3$He~p~$\pi^-$ X reaction as a function of the excess energy in the $\eta  \otimes ^{25}$Mg system \cite{Budzanovski,Moskal}. The solid curve represents a fit to the data with a Gaussian and a constant background.  The dashed curve is a fit with a Breit-Wigner function taking the interference with a non-resonant reaction amplitude into account \cite{Haider_Liu_COSY}. The figure is adapted from \cite{Machner}.}
\label{fig:eta_3He_Mg}
\end{figure*}

Experimental searches of structures in the expected bound state region were undertaken with pion- \cite{Chrien}, photon- [307-309],
%\cite{Sokol,Pfeiffer,Pheron}, 
proton- \cite{Budzanovski,Moskal} and deuteron beams [312-315],
%\cite{Afanasiev,Adlarson_PRC87,Moskal_APPB41,Adlarson_NPA959}, 
however, with inconclusive results. The strongest claim for the discovery of an $\eta$-mesic state has been made by the COSY-GEM Collaboration \cite{Budzanovski}. They have studied the $p + ^{27}$Al $ \rightarrow ^3$He$ + \eta \otimes^{25}$Mg$ \rightarrow^3$He~p~$\pi^-$ X reaction at a proton energy of 1745 MeV.  In this two-nucleon transfer reaction the forward going $^3$He was detected with high resolution in the Big-Karl magnetic spectrometer and $\eta$ mesic states with binding energies 0 MeV $\le B_{\eta} \le 30$ MeV could be produced under almost recoil-free conditions with a momentum transfer $q \le 30$ MeV/$c$. The decay of $\eta$ mesic states was expected to occur via the $\eta n \rightarrow N^*(1535) \rightarrow \pi^- p$ channel, leading to a $\pi^-$ and a proton emitted almost back-to-back in the laboratory system. Under these conditions the corresponding excitation energy spectrum shows an enhancement about 13 MeV below the $\eta$ production threshold with a width of $\approx$ 10 MeV(see Fig.~\ref{fig:eta_3He_Mg}), while no structure is observed without the back-to-back requirement. This observation is remarkably close to the prediction of a 1s $\eta$-mesic state in Mg with a binding energy of 12.6 MeV \cite{Garcia_Recio_PLB550}. In a refined analysis of the COSY-GEM data, Haider and Liu \cite{Haider_Liu_COSY} take into account that the $\eta$ produced in the intermediate state can also be inelastically scattered by the residual nucleus and emerge as a pion, without being captured by the nucleus. Allowing for the interference with this non-resonant reaction amplitude  a binding energy of $B_{\eta} \approx$ 6.5 - 8.0 MeV is obtained (see  Fig.~\ref{fig:eta_3He_Mg}). 

\begin{figure*}
\centering
  \includegraphics[width=9.0cm,clip]{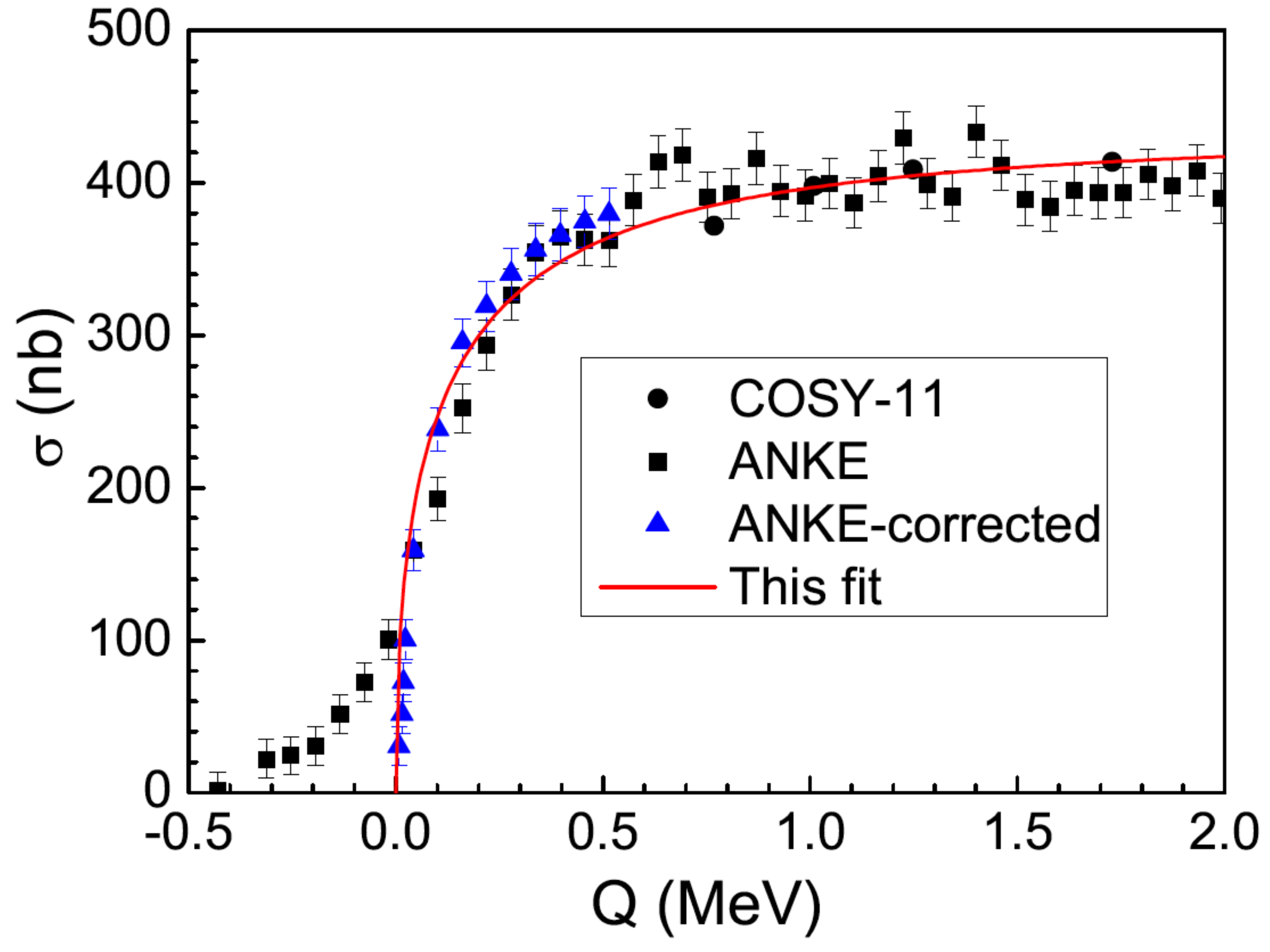} \includegraphics[width=9.0cm,clip]{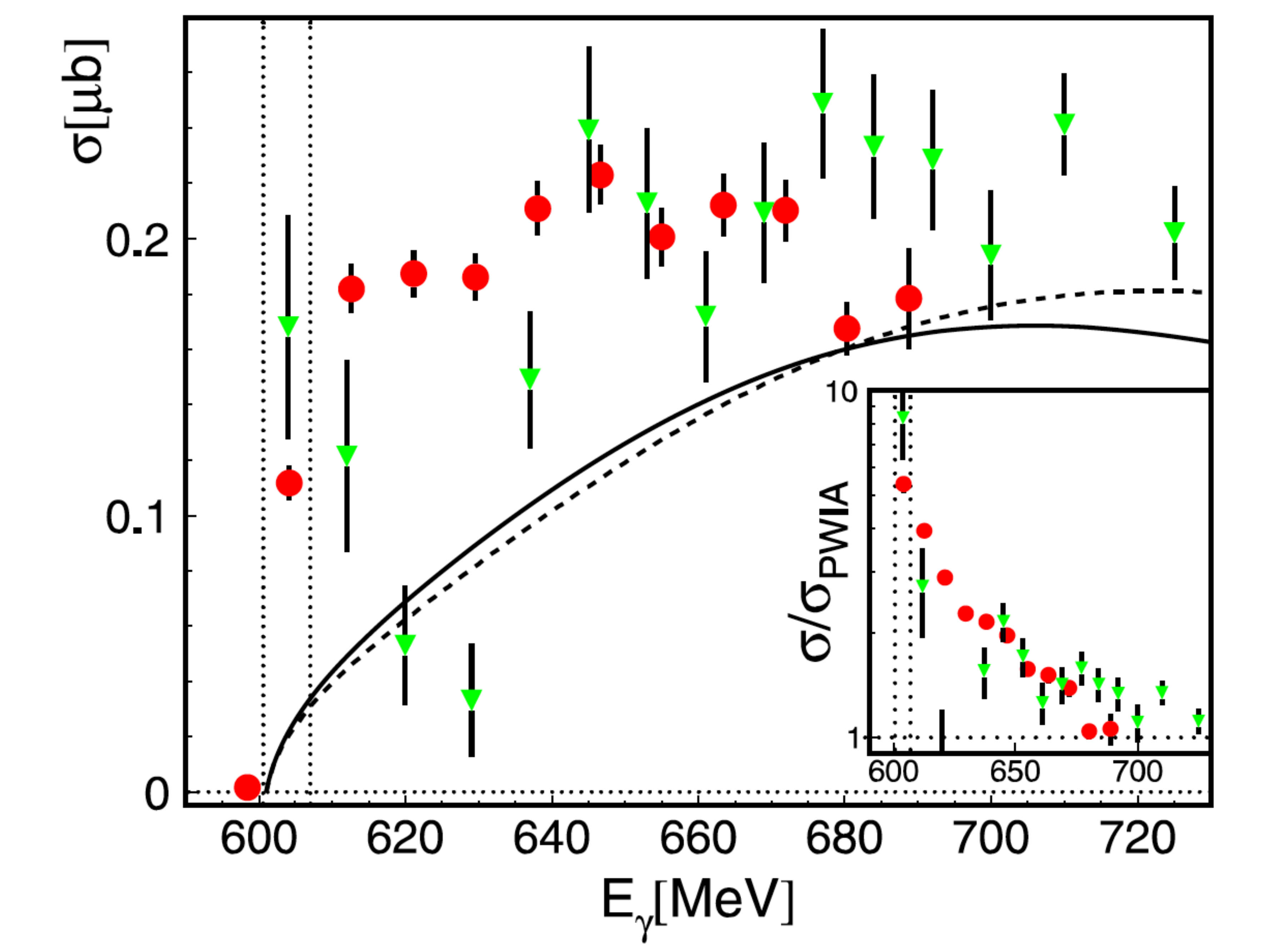}
   \caption{(Left) Total cross section for the $d p \rightarrow \eta ^3$He reaction below an excess energy of 2 MeV. The data are from \cite{Smyrski,Mersmann}. The curve represents a fit to the data performed by \cite{Xie}. The figure is taken from \cite{Xie}. (Right) Total cross section for the $\gamma ^3$He$ \rightarrow \eta ^3$He reaction (red points) from \cite{Pheron} compared to data (green triangles) from \cite{Pfeiffer}. Solid (dashed) curves represent plane wave impulse approximation (PWIA) calculations with a realistic (isotropic) angular distribution for the $\gamma n \rightarrow n \eta$ reaction. Insert: ratio of measured and PWIA cross sections. The figure is taken from \cite{Pheron}.}
\label{fig:eta_3He_COSY}
\end{figure*}

The $\eta ^3$He system has also been studied in $\eta$ production experiments close to the threshold in the reactions $d + p \rightarrow \eta ^3He$ \cite{Smyrski,Mersmann,Adam_PRC75} and $\gamma + ^3$He$ \rightarrow  \eta + ^3$He \cite{Pfeiffer,Pheron}. Corresponding excitation functions are shown in Fig.~\ref{fig:eta_3He_COSY}. In both reactions a very sharp rise of the cross section above the threshold was observed which was taken as an indication for the possible existence of an $\eta ^3$He quasi-bound state extremely close to threshold \cite{Mersmann}. The $d + p$  data which are much superior in statistics have been analysed by Wilkin et al. \cite{Wilkin_PLB654} and more recently by Xie et al. \cite{Xie}. While Wilkin et al. report a pole in the production amplitude at $Q_0$ = -0.3 MeV very close to the threshold with a width of 0.4 MeV, Xie et al. find a Breit-Wigner type structure at a mass of -0.3 MeV with a width of 3 MeV, but a pole in the continuum. This is at the verge of binding. From their analysis of the ANKE and COSY 11 data Xie et al. deduce an attractive $\eta ^3$He potential of U = [-(54$\pm$6)-i(20$\pm$2)] MeV \cite{Oset_priv}. For the onset of real binding in the $\eta ^3$He and $\eta^4$He systems Barnea et al. \cite{Barnea_PLB771} find in pionless effective field theory calculations that deeper potentials would be required corresponding to a real part of the $\eta N$ scattering length close to 1 fm and exceeding 0.7 fm, respectively. 

Selecting correlated $\pi^0 p$ pairs, a peak in the $\pi^0 p$ invariant mass spectrum was observed in the $\gamma ^3$He$ \rightarrow \pi^0 p X$ reaction and interpreted as evidence for the decay of a mesic $\eta \otimes ^3$He state \cite{Pfeiffer}. In a later experiment \cite{Pheron} this signal was confirmed with much improved statistical significance, but could be attributed to a structure arising from quasi-free pion production. In an attempt to search for $\eta ^4$He mesic states, back-to-back pion-nucleon pairs were also measured in the $d + d \rightarrow \eta \otimes ^4$He $ \rightarrow \pi^0 n ^3$He, $\pi^- p ^3$He reaction \cite{Adlarson_PRC87,Adlarson_NPA959}. No unambiguous signal was found and upper limits on the cross sections of $\approx $ 3 nb for the $\pi^0 n ^3$He and $\approx $ 6 nb for the $\pi^- p ^3$He channel were deduced, respectively.

30 years of experiments without an absolutely convincing and unambiguous signal have shown how cumbersome the search for $\eta$-mesic states really is. New approaches and higher statistics measurements are needed to make progress. Recently high statistics data on the p + d reaction have been taken by the WASA-at-COSY Collaboration dedicated to the search for $\eta ^3$He bound states in three decay channels: $ppp\pi^-, ^3$He $2 \gamma$, and $ppn$. The data analysis is still in progress \cite{Moskal}. In addition, the $\pi^0 \eta$ photoproduction off $^4$He has recently been measured at MAMI in an attempt to identify $\eta - ^4$He bound states \cite{Krusche_priv}; also here the data analysis is ongoing. Utilising the high intensity $\pi$ beam at J-PARC a high statistics search for  $\eta$-mesic nuclei is going to be performed \cite{Fujioka}.

% \begin{figure*}
%\centering
 % \includegraphics[width=9.0cm,clip]{figures/eta_3He_COSY_zoom.pdf}
%\caption{}
%\label{fig:eta_3He_COSY_zoom}
%\end{figure*}

%\begin{figure*}
%\centering
 % \includegraphics[width=9.0cm,clip,angle=90]{figures/COSY-GEM_Haider_Liu.pdf}
%\caption{}
%\label{fig:eta_3He_COSY_zoom}
%\end{

%--------------------------------------------------------------------------------------------------------------------------------------------
%ETAPRIME-MESIC

\subsection{\it Search for $\eta^{\prime}$ -mesic states\label{sec:etaprime_mesic}}

\begin{figure}[h]
\centering
  \includegraphics[width=10cm,clip]{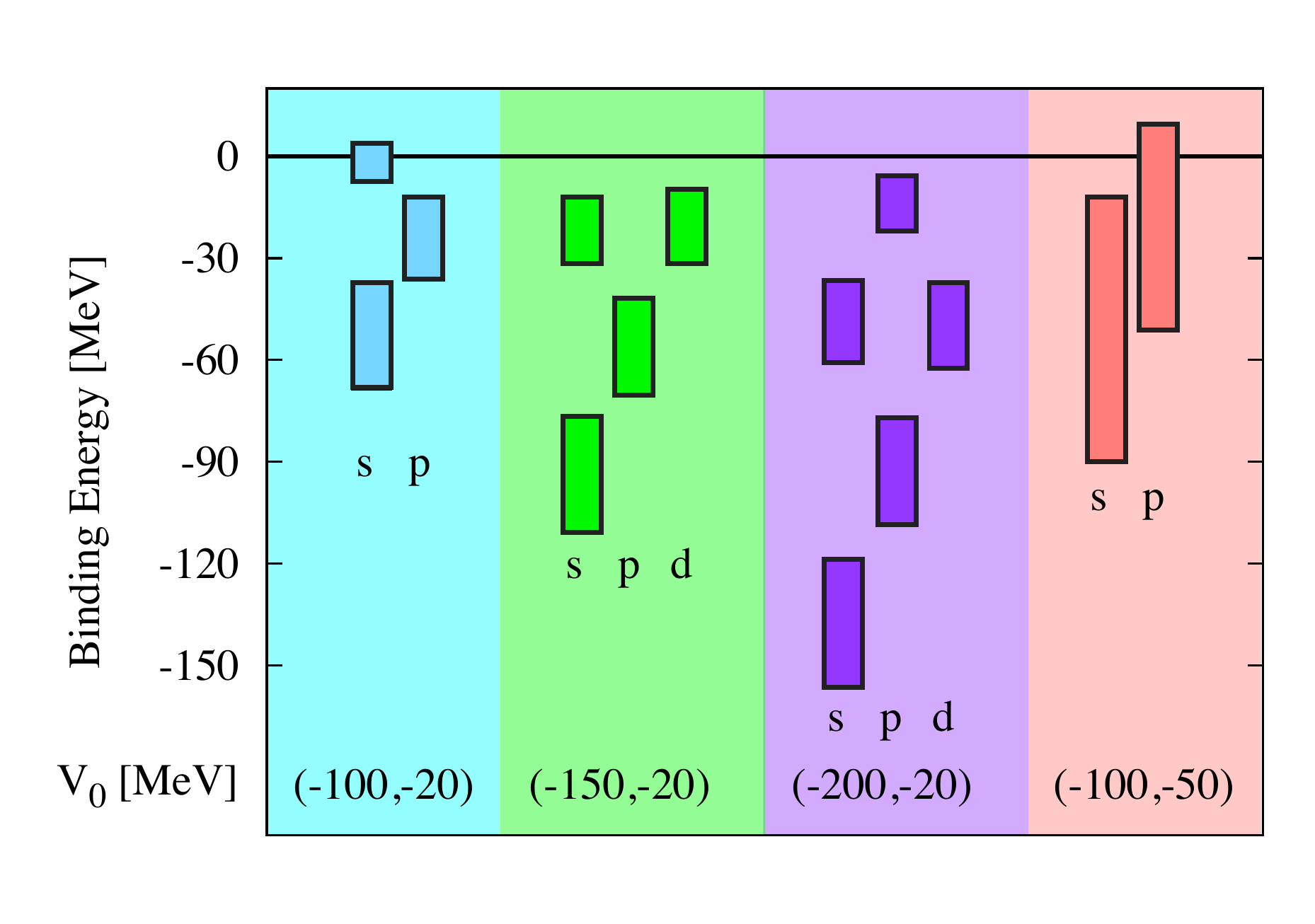}\includegraphics[width=9cm,clip]{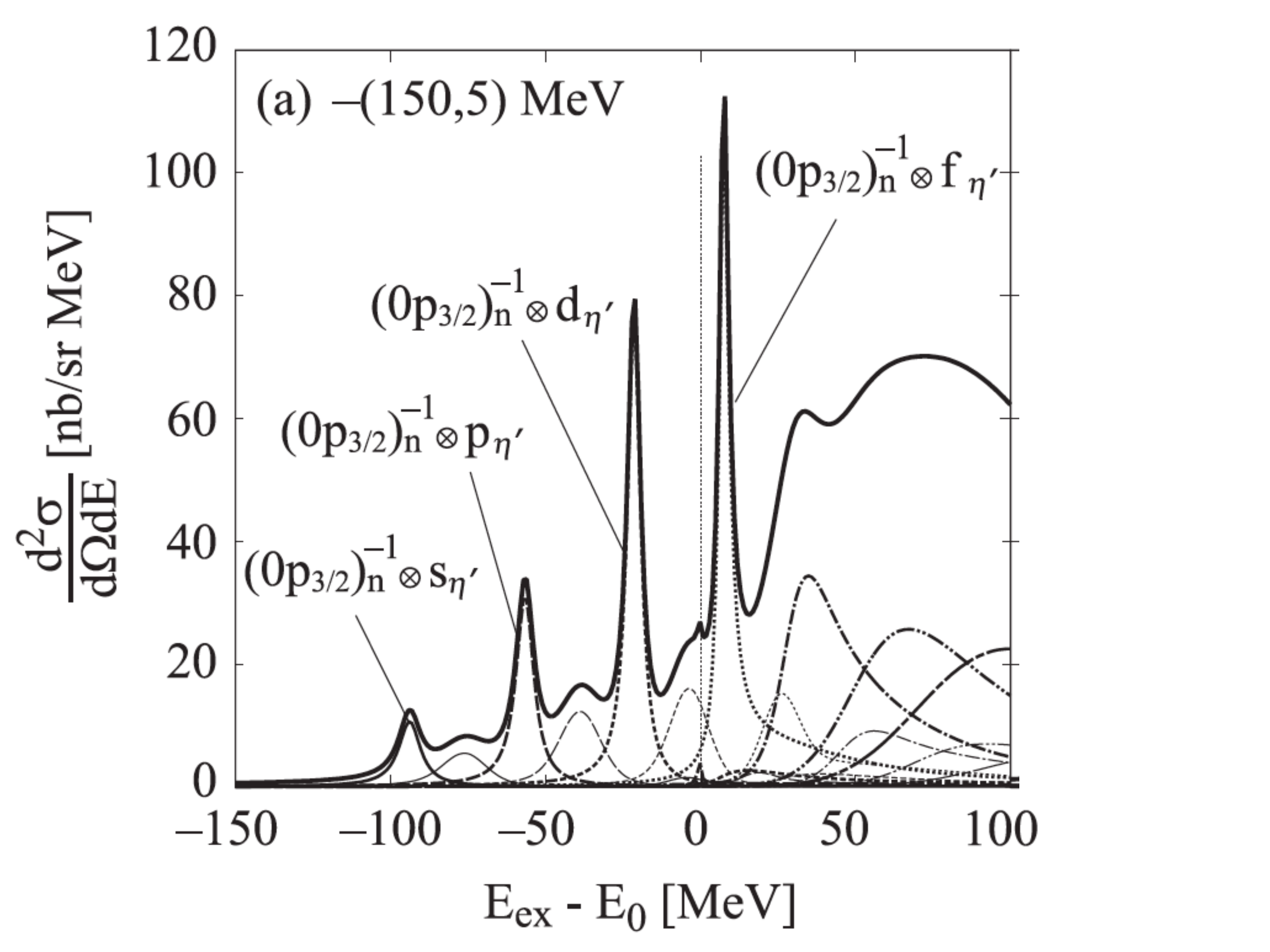}
\caption{(Left) Bound state spectra of the  $\eta^\prime$ meson in $^{11}C$ in units of MeV. Boxes denote the binding energies and indicate the widths of the bound states. The letters s, p and d label the angular momentum states. The potential of the $\eta^\prime$ meson in the nucleus is assumed in the form of Eq.~\ref{eq:potential_effective} with the potential depths at normal nuclear matter density V$_{0}$=-100, -150, and -200 MeV with a fixed imaginary potential W$_{0}$ = -20 MeV. Also shown is the result with W$_{0}$ = -50 MeV and V$_{0}$=-100 MeV \cite{Jido_PRC85}. (Right) Calculated spectra of the $^{12}$C(p,d)$^{11}$C$\otimes\eta^\prime$ reaction for the formation of $\eta^\prime$-nucleus systems with proton kinetic energy $T_{p}$=2.5 GeV and deuteron angle $\theta_{d}$=0$^{0}$ as a function of the excitation energy E$_{ex}$. The $\eta^\prime$ production threshold is E$_{0}$. The $\eta^\prime$-nucleus optical potential is taken to be (V$_{0}$, W$_{0}$)=-(150, 5) MeV. The thick solid line shows the total spectrum and dashed lines indicate subcomponents. The neutron-hole states are indicated as $(nl_{j})^{-1}_{n}$ and the $\eta^\prime$ states as $l_{\eta^\prime}$  \cite{Nagahiro_PRC87}.}
\label{fig:etapr_bound}
\end{figure}

As discussed in Sec.~\ref{sec:etaprime} the mass reduction of the $\eta^\prime$ meson corresponds to an attractive interaction between an $\eta^\prime$ meson and a nucleus. Furthermore, due to the larger real part of the $\eta^\prime$nucleus potential compared to its imaginary part the $\eta^\prime$ meson has become a promising candidate for the search  of $\eta^\prime$-bound states. Binding energies of $\eta^{\prime}$ -mesic states have been calculated based on the NJL model \cite{Nagahiro_Hirenzaki,Nagahiro_PRC74,Jido_PRC85,Nagahiro_PRC87} and on a chiral unitary model \cite{Nagahiro_Oset}. As an example, bound-state spectra of the $\eta^\prime$-mesic states assuming an attractive potential with a real part of V$_{0}$ = -100, -150, -200 MeV and an imaginary part W$_{0}$ = -20 MeV are shown in Fig.~\ref{fig:etapr_bound} (Left) \cite{Jido_PRC85}. There are several bound states in a small nucleus ($^{11}$C) due to the strong attraction, well separated  because of the assumed relatively small imaginary part of the $\eta^\prime$-nucleus potential. In contrast, as it can be seen in this figure, potentials having an imaginary part comparable with the real part,  V$_{0}$ = -100 MeV and W$_{0}$ = -50 MeV, provide bound states with larger widths than the binding energy and are overlapping. In this case it would be hard to observe bound states as clear peaks in the formation spectra.
Reactions like  $(\gamma, p)$, $(\pi^{+}, p)$, $(p, d)$ have been proposed for the formation of  $\eta^{\prime}$ -mesic states for nuclei such as $^{12}C$ and $^{40}Ca$. The calculated spectrum of the $^{12}$C(p, d)$^{11}$C$\otimes\eta^\prime$ reaction for the formation of the $\eta^\prime$-nucleus system at a proton kinetic energy energy of 2.5 GeV and at deuteron angle $\theta_{d} = 0^{0}$ is shown in Fig.~\ref{fig:etapr_bound} (Right) \cite{Nagahiro_PRC87} as a function of the excitation energy and for the $\eta^\prime$- potential parameters (V$_{0}$,W$_{0}$)=-(150, 5) MeV. One can see clearly separated peaks corresponding to $\eta^\prime$ bound states in the $s$, $p$ and $d$ states. Spectra for other combinations of (V$_{0}$,W$_{0}$) and different $\eta^\prime$ scattering lengths have been calculated showing that the width of each peak becomes wider as W$_{0}$ and/or $\mid a_{p\eta^\prime}\mid$ increase.
It has been pointed out that the spectra for nuclei heavier than $^{39}$Ca will exhibit an overlap of different nucleon-hole-$\eta^\prime$ configurations that smear out the individual peaks and the structures get less prominent \cite{Nagahiro_Oset,Jido_PRC85}.\\

These theoretical predictions and the experimental data \cite{Nanova_PLB727} encouraged the search for $\eta^{\prime}$ bound states. 
The first pioneering experiment searching for $\eta^\prime$ bound states was performed in 2014 at the Fragment Separator (FRS) at GSI using the $^{12}$C(p,d) reaction at an incident proton energy of 2.5 GeV \cite{Tanaka}. Similar to the experiment for pionic atoms \cite{Suzuki_PRL92}, ideas for the formation of $\eta^\prime$-mesic states in almost recoil-less kinematics have been developed and studied in simulations \cite{Kenta_PTP}. The proton beam hits the $^{12}$C target and ejects a deuteron with high momentum, while the $\eta^\prime$ meson produced with low momentum could be bound to the $^{11}$C. In the experiment only the deuteron momentum distribution 
\begin{figure}[h]
\centering
  \includegraphics[width=8cm,clip]{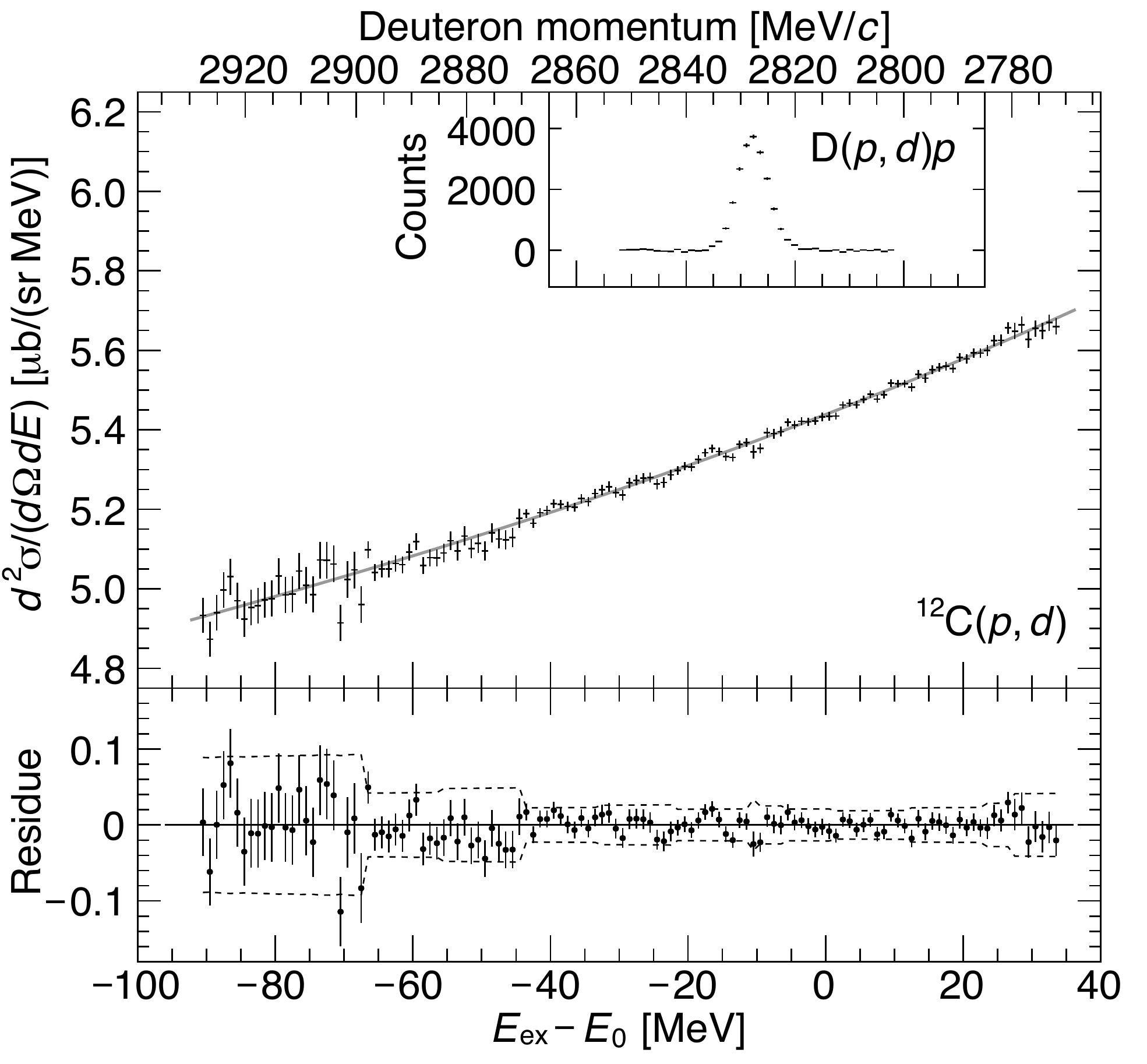}\includegraphics[width=12cm,clip]{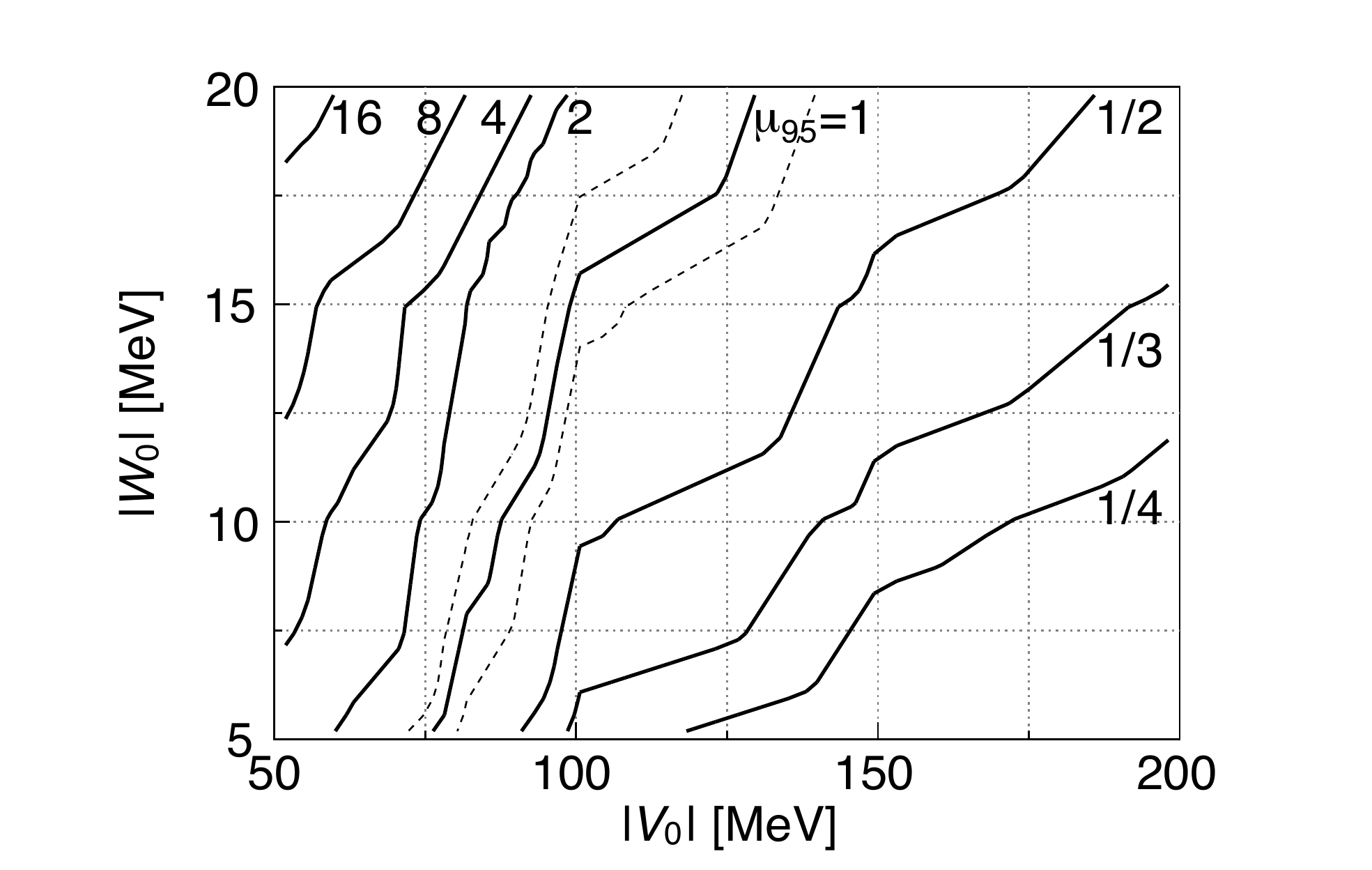}
\caption{Left: (top panel) Excitation spectrum of $^{11}C$ measured in the $^{12}C(p,d)$ reaction at a proton energy of 2.5 GeV. The abscissa is the excitation energy relative to the $\eta^{\prime} $ production threshold $E_0 = 957.78$ MeV.  The gray curve represents a fit of the data with a third-order polynomial. The upper horizontal axis is the deuteron momentum scale. (Inset) Deuteron momentum spectrum measured in the elastic $D(p,d) p $ reaction using a 1.6 GeV proton beam. (Bottom panel) Fit residues with envelopes of 2 standard deviations. Right: Contour plot of $\mu_{95}$ (solid curves), upper limit of the scale parameter $\mu$ at the 95$\%$ confidence level, on a plane of real and imaginary potential parameters $ (V_0 ,  W_0 )$. The limits have been evaluated for the potential parameter combinations ($V_0,W_0$) in (-50,-100,-150,-200) x (-5,-10,-15,-20) and (-60,-80) x (-5,-10,-15) MeV and linearly interpolated in between. Dashed curves show a band of $\mu_{95}$ =1 contour indicating systematic errors. Regions for $\mu_{95} \le 1$ are excluded by the analysis. The figure is taken from \cite{Tanaka}.}
\label{fig:Tanaka}
\end{figure}
\begin{figure}[h]
\centering
  \includegraphics[width=10cm,clip]{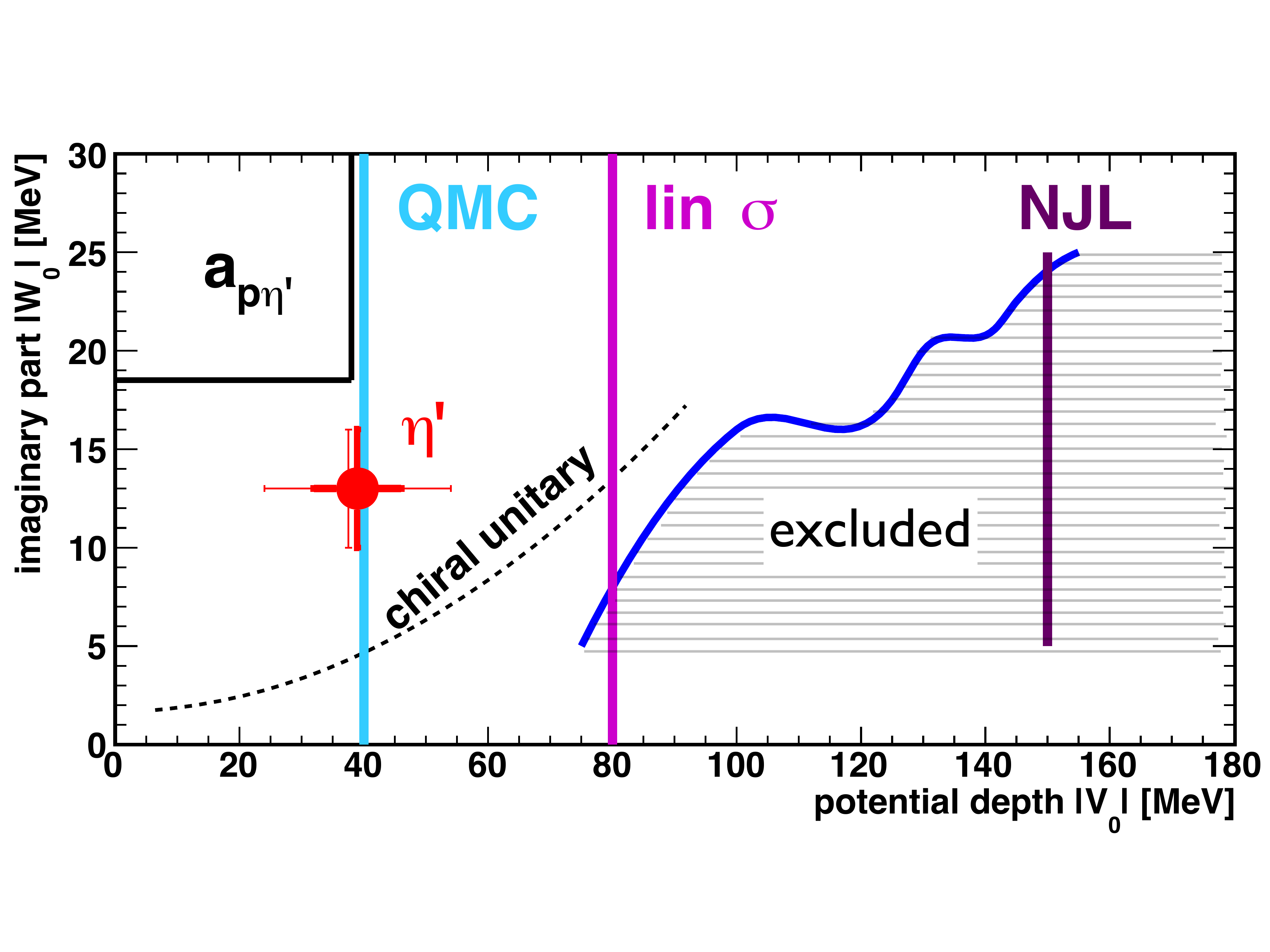}
\vspace{-10mm}\caption{The imaginary part versus the real part of the $\eta^\prime$-nucleus potential. The vertical lines denote the model calculations for the real part of the $\eta^\prime$ optical potential: (from left to right) QMC model \cite{Bass}, linear $\sigma$ model \cite{Jido} and NJL model \cite{Nagahiro_PRC74}. The dashed curve is the result of a study in the chiral unitary approach, accounting for $\eta^\prime$ absorption by pairs of nucleons \cite{Nagahiro_Oset}. The box labeled $a_{p \eta^\prime}$ indicates the potential parameters corresponding to the $p \eta^\prime$ scattering length reported by  Czerwinski et al. \cite{ Czerwinski}. The red point with thick (statistic) and thin (systematic) error bars represents the experimental result for V$_{0}$ and W$_{0}$ for the $\eta^\prime$-nucleus potential measured by CBELSA/TAPS \cite{Friedrich_EPJA, Nanova_PRC94, Nanova_Metag}. The blue line is the $\mu_{95}$=1 line in Fig.~\ref{fig:Tanaka} (Right) and the dashed area represents the region of V$_{0}$,W$_{0}$ parameter combinations excluded by the experiment of Tanaka et al. \cite{Tanaka}.}
\label{fig:comparison}
\end{figure}
has been measured, applying missing mass spectrometry. The FRS has been used as spectrometer with a specially developed ion optical setting. In Fig.~\ref{fig:Tanaka} (Left) the measured excitation spectrum of the $^{12}$C(p,d) reaction near the $\eta^\prime$ emission threshold is shown \cite{Tanaka}. Unfortunately, no narrow structure has been observed in spite of the extremely good statistical sensitivity. For positive excitation energies the increasing contributions from quasi-free $\eta^\prime$ production is observed. An upper limit for the formation cross section of $\eta^\prime$-mesic nuclei of $\approx$20 nb/(sr MeV) near the threshold has been deduced. The spectrum has been compared with the theoretical spectra from \cite{Nagahiro_PRC87} for different potential parameters ($V_{0}, W_{0}$) and is shown in Fig.~\ref{fig:Tanaka} (Right). 
For each potential parameter combination the spectrum has been fitted by a polynomial describing the background and a theoretical spectrum scaled by a factor $\mu$, folded with the spectral resolution $\sigma_{exp}$ ($\sigma_{exp}=2.5\pm$0.1 MeV). Upper limits of the scale parameter $\mu_{95}$ have been evaluated at the 95$\%$ confidential level (CL). As can be seen in Fig.~\ref{fig:Tanaka} (Right), a strongly attractive potential of $V_{0}\approx$-150 MeV predicted by the NJL model \cite{Nagahiro_PRC74} is rejected for a relatively shallow imaginary potential \cite{Friedrich_EPJA}.

The experimental results of Fig.~\ref{fig:Tanaka} (Right) are shown again in Fig.~\ref{fig:comparison} in comparison to the theoretical predictions for $\eta^\prime$-nucleus potentials based on NJL \cite{Nagahiro_PRC74}, linear $\sigma$ \cite{Jido} and QMC \cite{Bass} models, and the experimental potential values determined by CBELSA/TAPS \cite{Friedrich_EPJA, Nanova_PRC94, Nanova_Metag}. In conclusion the FRS at GSI experiment had only a limited sensitivity for a relatively weak attraction as predicted by the QMC model \cite{Bass} and deduced by the $\eta^\prime$ photoproduction experiments \cite{Friedrich_EPJA, Nanova_PRC94, Nanova_Metag}. The shallow potential depth, nevertheless larger than the absorption width, requires more sensitive experiments for the observation of bound structures. One possibility to improve the sensitivity is to suppress the background by detecting in coincidence the decay particles of the $\eta^\prime$-mesic nuclei. A semi-exclusive measurement  has been considered for the Super-FRS at FAIR \cite{Nagahiro_PRC87, Tanaka}. 
An alternative approach is the photoproduction of $\eta^\prime$ mesons in the $^{12}$C($\gamma$,p) reaction, again in almost recoil-free kinematics. The predictions based on NJL model by \cite{Nagahiro_Hirenzaki,Nagahiro_PRC74} and the estimated formation cross section motivated experiments at LEPS2 facility (Spring8) \cite{Muramatsu_BGOegg, Tomida_BGOegg} and BGO-OD at ELSA \cite{Nanova_Metag, Metag_BGO}, where the missing mass spectroscopy is combined with detecting the decay particles of the $\eta^\prime$-mesic states. The BGO \cite{Bantes_BGO} and BGO-egg \cite{Tomida_BGOegg} detectors are well suited for detecting photons from decay particles of the $\eta^\prime$-mesic state, e.g. $\eta^\prime N \rightarrow \eta N \rightarrow \gamma  \gamma N$ \cite{Oset_PLB704,Nagahiro_PRC87}. 
As an alternative, the $(\pi^{\pm},p)$ reaction has been proposed for the formation of $\eta^\prime$-mesic states for a measurement at J-PARC \cite{Nagahiro_Oset}. Sofar no experimental results are available from LEPS2, BGO-OD and J-PARC.

%\clearpage
%----------------------------------------------------------------------------------------------------------------------------------------------------------------------------------------------------------------------------------------------
%OMEGA-MESIC

\subsection{\it Search for $\omega$ -mesic states\label{sec:omega_mesic}}
\begin{figure*}
\centering
 \includegraphics[width=10cm,clip]{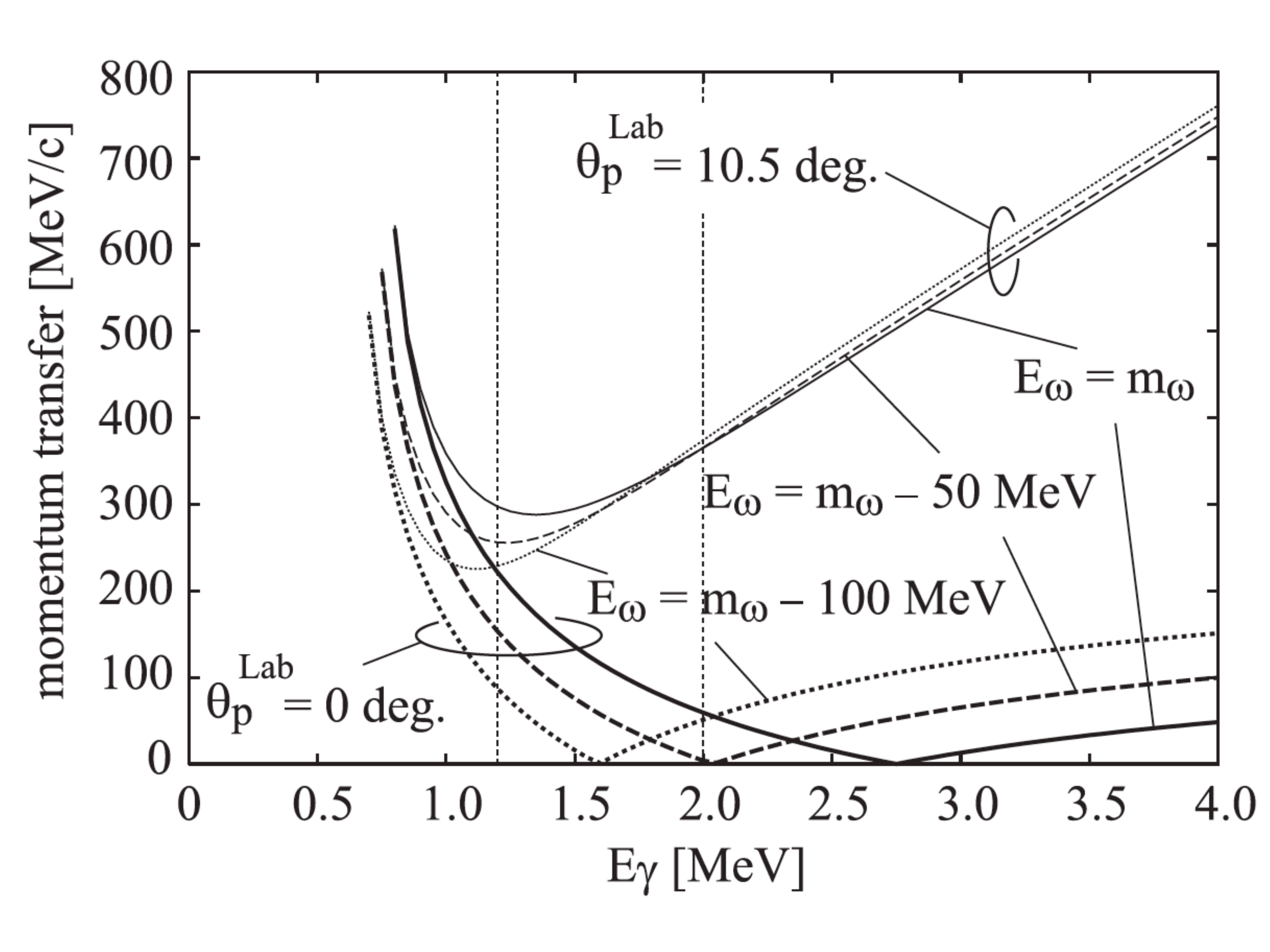} 
\caption{Momentum transfers as a function of the incident photon energy E$_{\gamma}$ in the $(\gamma,p)$ reaction. The solid, dashed, and dotted lines show the momentum transfers at $\omega$ energies $E_{\omega} = m_{\omega}, E_{\omega} = m_{\omega} -50$ MeV, and $E_{\omega} = m_{\omega} -100$ MeV, respectively. The thick and thin lines refer to $\theta_{lab} = 0^\circ$ and 10.5$^\circ$, respectively. The figure is taken from \cite{Kaskulov_PRC75}.}
\label{fig:omega_rest}
\end{figure*}

The experimental results discussed in Section \ref{sec:omega} have shown that the $\omega$--nucleus attraction is quite weak. Nevertheless, motivated by theoretical predictions for the existence of $\omega$--nucleus bound states in [332-334],
%\cite{Marco_Weise,Nagahiro_NPA761,Kaskulov_PRC75}, 
an experimental search for $\omega$-mesic states has been performed \cite{Friedrich_PLB}. As proposed in [332-334],
%\cite{Marco_Weise,Nagahiro_NPA761,Kaskulov_PRC75}, 
the $^{12}$C ($\gamma, p)$ reaction has been studied in recoil-less kinematics. The optimum incident photon energy needed to produce an $\omega$ at rest in the nucleus is around 2.75 GeV. At this incident photon energy the proton detected at forward angles takes over the momentum of the incident photon, leaving the $\omega$ at rest in the laboratory system (see Fig.~\ref{fig:omega_rest}) so that it can be captured by the nucleus in case of an attractive $\omega$-nucleus interaction. If the $\omega$ mass in the nucleus were reduced by 100 MeV, the optimum incident energy would be lowered to around  E$_{\gamma} = 1.6$ GeV. For an incident energy of around 1.2 GeV the momentum transfer to the $\omega$ meson is lower than 300 MeV/c (comparable to the average nucleon Fermi momentum) for all envisaged mass drops as long as the outgoing proton is confined to laboratory angles of 1$^\circ$ - 11$^\circ$. This is the kinematics chosen for the CBELSA/TAPS experiment \cite{Friedrich_PLB}. Protons were identified in the TAPS forward wall and $\pi^0 \gamma$ pairs from $\omega$ mesons or the decay of a quasi-bound $\omega$ mesic state were detected with the Crystal Barrel detector. This approach allowed the simultaneous study of quasi-free $\omega$ production above the threshold as well as the search for decays of $\omega$-nucleus states in the bound state regime below the production threshold.

\begin{figure*}
\centering
 \includegraphics[width=8.5cm,clip]{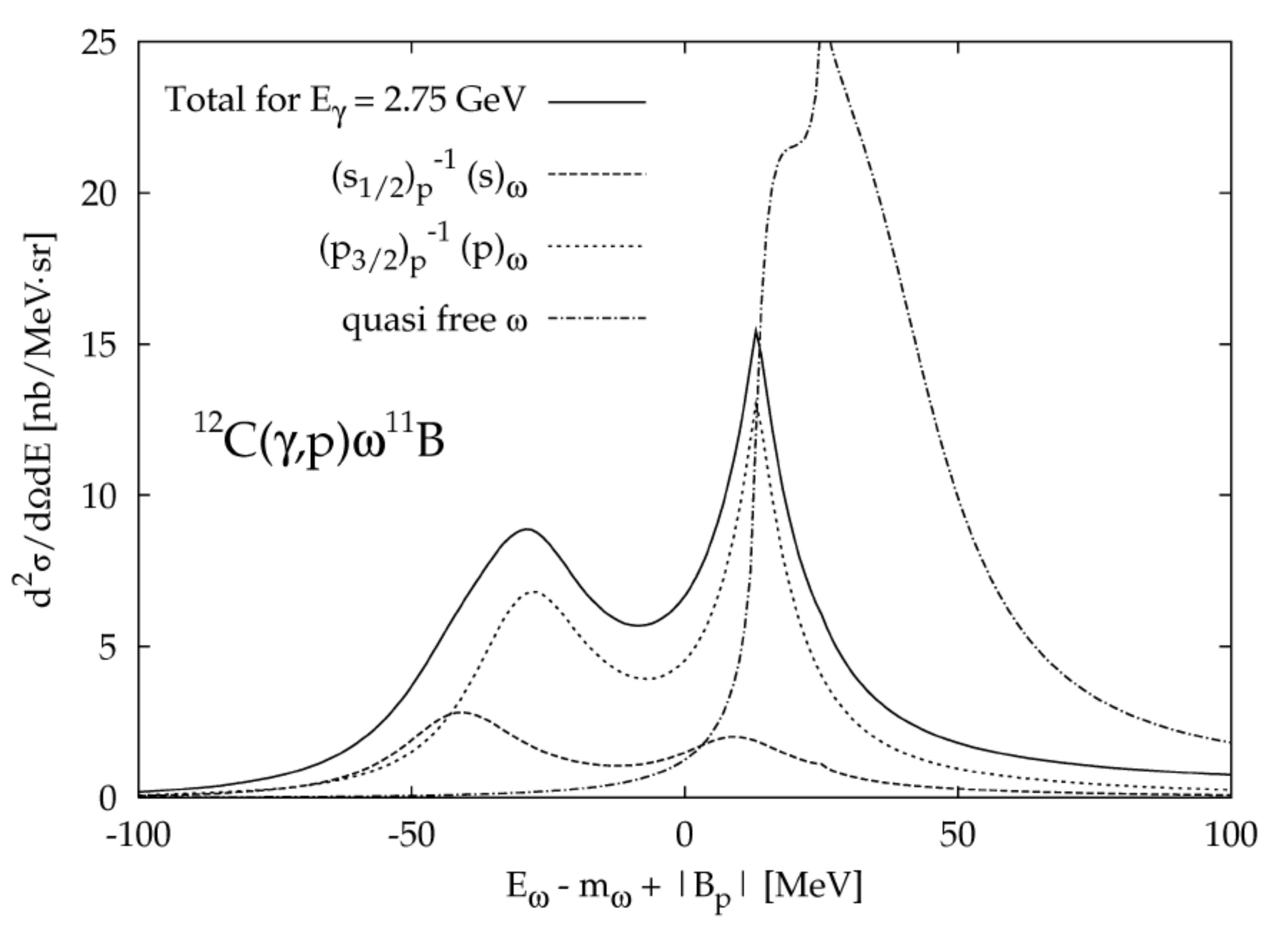} \includegraphics[width=8.5cm,clip]{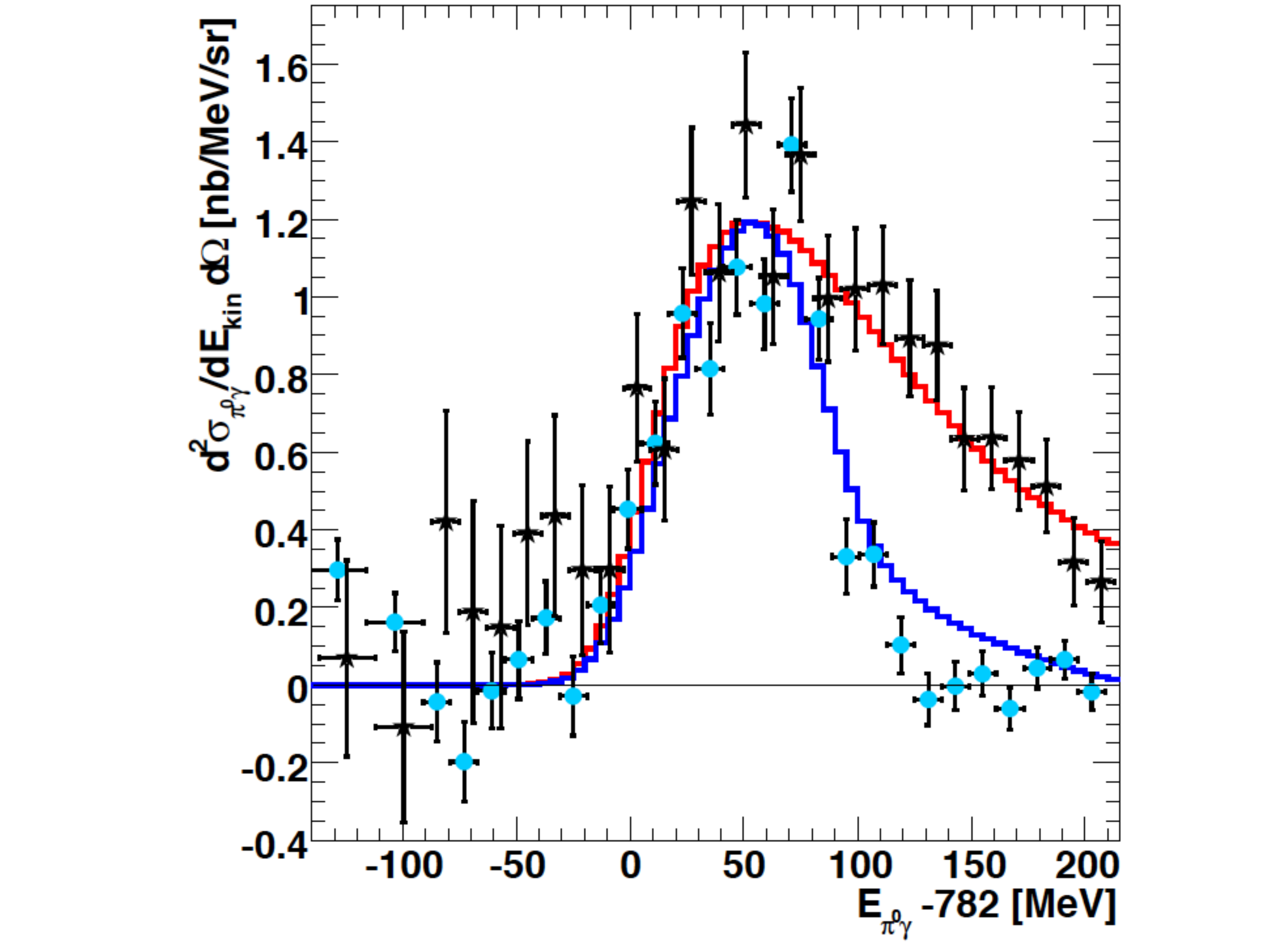} 
\caption{(Left) Missing energy spectra for the $^{12}$C$(\gamma, p) \omega \otimes ^{11}$B reaction at $E_{\gamma}$ = 2.75 GeV. Dotted lines represent the contributions from two particular combinations of bound $\omega$ and proton-hole states. The quasi-free $\omega$ production cross section is also shown. The figure is taken from \cite{Marco_Weise}. (Right) Kinetic energy distribution of the $\omega$ meson off Carbon (black stars) compared with the kinetic energy distribution of the $\omega$ meson off the free proton (full blue circles). The LH$_2$ data are normalised to the C data in the peak of the total energy distribution. The experimental distributions are compared to Monte Carlo simulations (LH$_2$ : blue histogram; C: red histogram), taking the Fermi motion of nucleons into account for the C target. All distributions request the detection of a proton in the polar angular range $1^{\circ}-11^{\circ}$ and are normalised to the fitted peak height for C. The Monte Carlo simulations are folded with the experimental resolution of $\sigma_E\approx$ 16 MeV.  The figure is taken from \cite{Friedrich_PLB}.}
\label{fig:bound_om}
\end{figure*}

 Fig.~\ref{fig:bound_om} (Left) shows the theoretically predicted excitation energy spectrum of the residual nucleus $^{11}$B near the $\omega$ production threshold, decomposed into contributions from the $\omega$ in s or p states bound to different proton hole states. The corresponding experimental spectrum is shown in Fig.~\ref{fig:bound_om} (Right). There is some yield of correlated $\pi^0 \gamma$ pairs in the bound state region but no statistically significant structure can be observed. This may not be surprising in view of the strong in-medium broadening of the $\omega$ meson discussed in Section \ref{sec:omega_imag}. Correcting for the effective branching ratio for in-medium $\omega \rightarrow \pi^0 \gamma$ decay, modified by the increased in-medium width to about 1.5$\%$, the $\pi^0 \gamma$ yield in the range -90 $\le$ E$_{\pi^0 \gamma} - 782$ MeV $\le $ - 20 MeV corresponds to a population cross section of (22 $\pm$ 7) nb MeV$^{-1}$sr$^{-1}$ which is comparable to the theoretically expected formation cross section of $\omega$-mesic states. The tailing in the 
total energy distribution into the bound state region may, however, simply be a consequence of the large in-medium broadening of the $\omega$ meson.
Conclusive results can only be obtained in an experiment with much higher statistics. Such an experiment, using the $A(\pi^-, n) \omega$ reaction and looking for $\pi^0 \gamma$ pairs from $\omega$ decays and decays of $\omega$-nucleus bound states is in preparation, taking advantage of the high intensity $\pi^-$ beams at J-PARC \cite{Ozawa_E26}.

%---------------------------------------------------------------------
%PHI-MESIC

\subsection{\it Search for $\phi$ -mesic states\label{sec:phi_mesic}}
\begin{figure*}
\centering
 \includegraphics[width=18cm,clip]{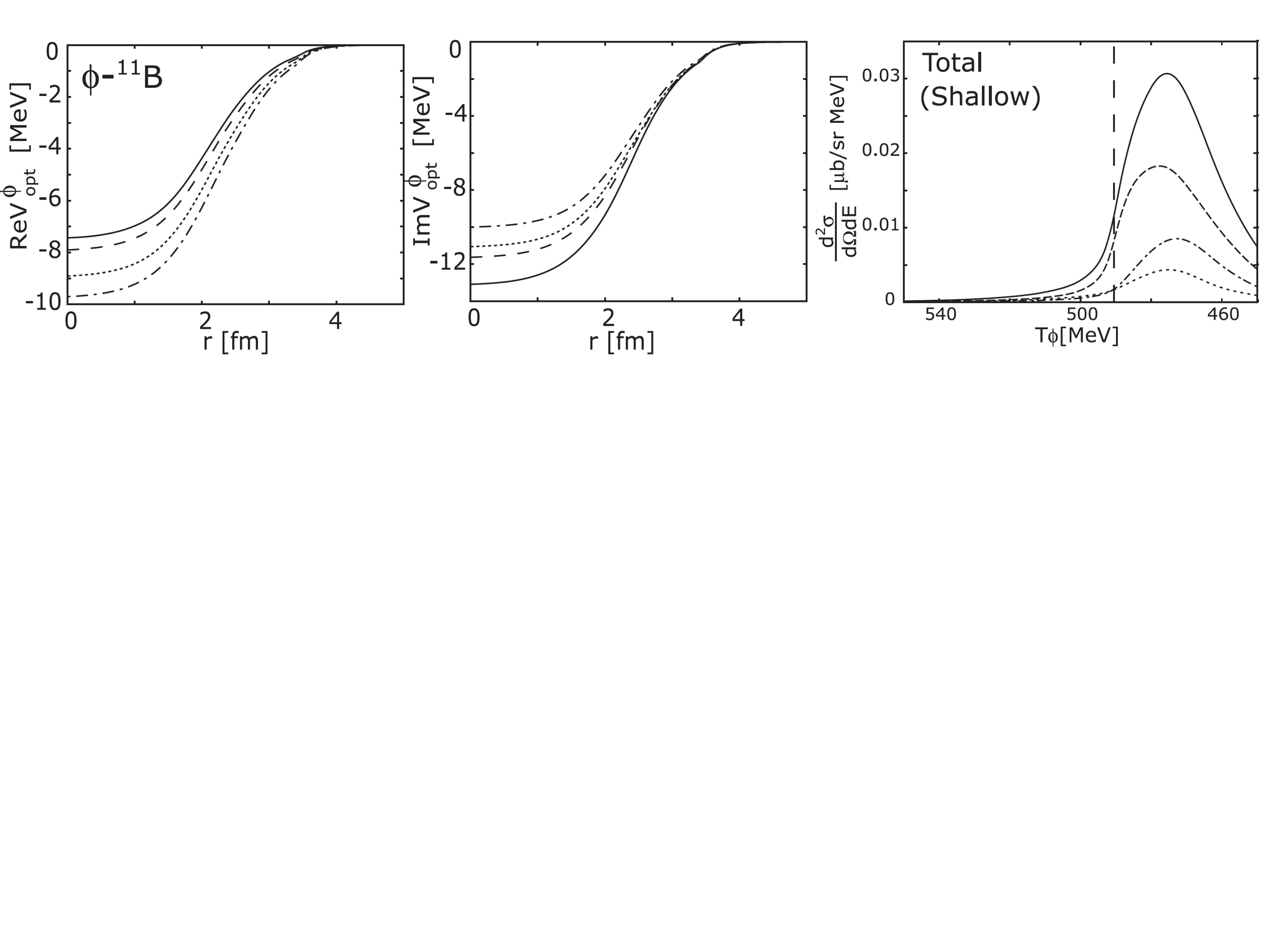} 
\vspace{-80mm}\caption{Theoretical predictions for the real (Left) and the imaginary part (Middle) of the $\phi$ meson optical potential as a function of the radial distance r for the $\phi - ^{11}$B system, obtained from the $\phi$ self-energy reported in \cite{Cabrera_Vacas}. The solid, dashed, dotted, and dotted-dashed curves indicate the potential strength for the $\phi$ meson energies Re (E -m$_{\phi})$ = 0 MeV, -10 MeV, -20 MeV, and - 30 MeV, respectively. (Right) Differential cross section for the formation of $\phi$--nucleus bound states plotted as a function of the $\phi$ meson energy in the $^{12}C(\overline{p}, \phi) $ reaction at p$_{\overline{p}} = 1.3 $ GeV/$c$ for the $\phi$- nucleus potential given in the left and middle part of the figure. The dashed and dotted curves represent the contributions from different proton hole states. The vertical line indicates the $\phi$ meson production threshold. The figures are taken from \cite{Sekihara}.} 
\label{fig:phi_pot}
\end{figure*}

The discussion in Section \ref{sec:phi} has shown that the $\phi$--nucleus attraction is expected to be rather weak. Some theoretical calculations give even smaller potential depths than the experimentally claimed mass lowering of about 30 MeV at normal nuclear matter density \cite{Muto}. Fig.~\ref{fig:phi_pot} shows the calculated real and imaginary part of the $\phi ^{11}B$ - potential which is too shallow to support a $\phi$--nucleus bound state. Bound states are only expected for heavier nuclei. Only when the strength of the real potential is scaled up by a factor 4 to match the reported experimental mass drop \cite{Muto}, bound $\phi$--nucleus states may also be expected in lighter nuclei. 
Theoretical predictions for the imaginary part of the potential are comparable to the real part (see Fig.~\ref{fig:phi_pot}, Middle), implying rather broad states with a width in the range of $\approx 20 $ MeV \cite{Sekihara}. Similar results have recently been obtained within the QMC model \cite{Cobos-Martinez_1705.06653}. This leads to structure-less excitation energy spectra exhibiting some tailing into the bound state region as shown in Fig.~\ref{fig:phi_pot} (Right). 

In spite of these not really encouraging predictions, an interesting experiment (E29) \cite{Ohnishi} with an antiproton beam has been proposed at J-PARC to study $\phi$-mesic states. Using double $\phi$ production off nuclei, the forward going $\phi$ meson in the primary reaction $\overline{p}p \rightarrow  \phi \phi$ is used to tag the production of the backward going $\phi$ meson which may be captured by the nucleus. The $\phi$ - nucleus bound state may decay via the process $\phi + p \rightarrow K^+ + \Lambda$ where the reaction products are expected to emerge almost back-to-back. By registering the $\phi \rightarrow K^+ K^-$ decay of the forward going $\phi$ meson in coincidence with the back-to-back emitted $\Lambda$ and K$^+$, the formation as well as the decay of the $\phi$-nucleus state can be measured. $^{12}$C and Cu are to be used as targets. It will be interesting to see the results of this experiment.

%--------------------------------------------------------------------------------------------------------------------------------------------------------------------------------

%-------------------------------------------------------------------------------------------
%CHARM-MESIC
\subsection{\it Search for mesic states in the charm sector \label{sec:charm_mesic}}
Already in 1990 Brodsky et al. \cite{Brodsky} predicted the existence of ($c \overline{c}$ ) - nucleus bound states due to a QCD van der Waals type attractive interaction arising from multiple-gluon exchange. More recently the possible existence of $J/\psi$ - nucleus bound states has been addressed in quark-meson coupling model calculations \cite{Krein}, predicting binding energies in the range of 5-15 MeV. Quenched lattice QCD calculations find potentials for the $\eta_c - N$ and $J/\psi - N $ systems which are weakly attractive at short distances and exponentially screened at large distances \cite{Kawanai_Sasaki}.  More recent lattice QCD calculations report binding energies of $\le 40 $ MeV \cite{Beane}.

\begin{figure*}
\centering
 \includegraphics[width=11.8cm,clip]{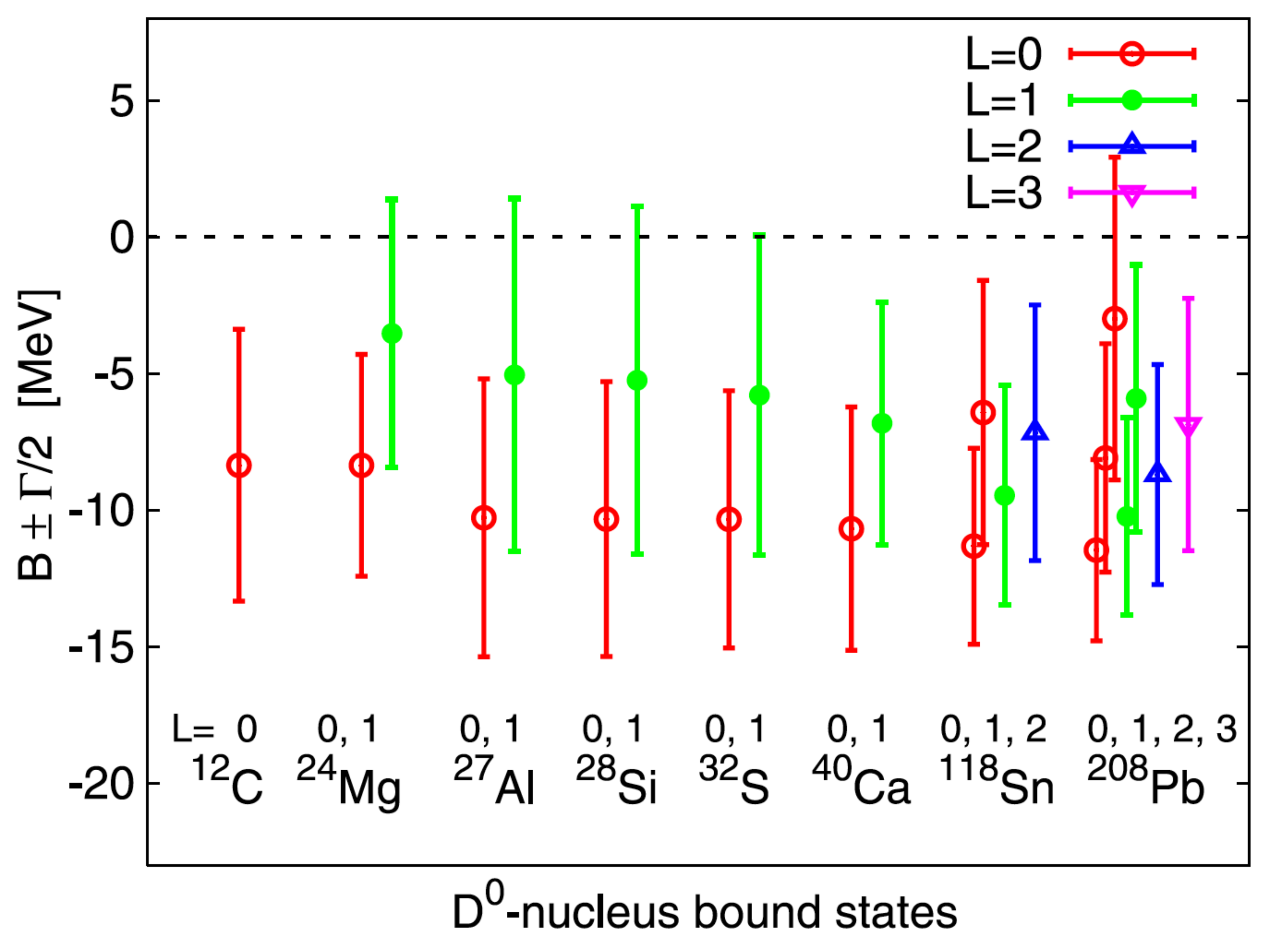} 
\caption{Binding energies and widths for different $D^0$ - nucleus states, predicted in unitarized coupled-channel calculations. The figure is taken from \cite{Garcia-Recio_PLB}.}
\label{fig:D-mesic}
\end{figure*}

Using unitarized coupled channel calculations, Garcia-Recio et al. \cite{Garcia-Recio_PLB} investigated also the existence of mesic nuclei with open charm mesons. As shown in Fig.~\ref{fig:D-mesic} they predict binding energies of $D^0$ - nuclear states of the order of 10 MeV with comparable widths throughout the nuclear chart. The Coulomb interaction prevents bound states for $D^+$ mesons. For $\overline{D}$ mesons not only $D^-$ but also $\overline{D^0}$ mesons are predicted to bind to nuclei \cite{Garcia-Recio_PRC85}.

Experimentally, the main problem is the large momentum transfer associated with charm production. It is highly unlikely that momenta higher than 1 GeV/$c$ can be taken up by the whole nucleus through Fermi-motion or re-scattering effects. The probability for a nucleus to change momentum and stay intact is given by the square of its form factor $F_A^2(q_A^2)$. The question of minimising the momentum transfer has been studied by Faessler \cite{Faessler}. Antiproton induced reactions provide maximum energy release at low momentum transfer. As most favourable reaction he investigates reactions of the type:
\begin{equation}
 \overline{p} p \rightarrow X Y \label{eq:double D production}
\end{equation}
If one of the particles $X,Y$ goes forward, the other one will go backwards in the center-of-mass system and will thus have a small momentum in the laboratory so that it can be captured by the nucleus. In the high energy limit , i.e. for energies $\sqrt{s} \gg m_X$ the minimum momentum of particle $X$ is
\begin{equation}
p_{min}(X) \approx \frac{m_X^2 - m_p^2}{2 m_p}\label{eq:pmin}
\end{equation}
In case of $D$ meson pair production the laboratory momentum of the backward (in c.m.) produced $D$ meson is still as large as 1.4 GeV/c according to Eq.~(\ref{eq:pmin}). This kinematics has also been considered by Yamagata-Sekihara et al. \cite{Sekihara_PLB} who calculate formation spectra for $D^-\otimes^{11}$B and $D^0\otimes^{11}$B with differential cross sections in the pb/(sr MeV) range for an antiproton beam impinging on a $^{12}$C target. Larger cross sections might be achieved in reactions producing $\overline{D^*}D$ meson pairs such as 
\begin{equation}
\overline{p} + p \rightarrow D^{*-} + D^+,
$$
$$
D^{*-} + (Z,A) \rightarrow \pi^0 + D^- \otimes (Z,A)
\end{equation}
After pion emission the charmed meson may be slow and can get trapped by the nucleus. The PANDA detector will be highly suited for such investigations making use of the high quality antiproton beams at FAIR \cite{PANDA}.

%--------------------------------------------------------------------------------------------------------------------------------------------------------------------------------

\section{Summary and Conclusions}
In this review we have compiled experimental results on the interaction of $K^+, K^0, K^-, \eta, \eta^\prime, \omega $ and $\phi$ mesons with nuclei, reported in photon-, proton-, and pion-induced reactions and heavy-ion collisions, and compared them to theoretical predictions. We have focused the discussion on the energy regime near the production threshold where under optimised kinematic conditions the mesons may be so slow that they could be captured by the nucleus to form a meson-nucleus bound state if there were sufficient attraction. We have confronted the experimental results with theoretical predictions and have found that in many cases the experimentally observed in-medium modifications are smaller than theoretically predicted. 

All mesons exhibit a broadening in the nuclear medium and thus a non-zero imaginary potential, ranging from about -13 MeV for the $\eta^\prime$ meson to about -50 MeV for the $\omega$ meson at normal nuclear matter density and low meson momenta. In most of the cases the broadening has been indirectly inferred from transparency ratio measurements and the observation of meson absorption via inelastic channels which reduce the meson in-medium "lifetime". A direct observation of an in-medium broadening has only been reported for the $\phi$ meson by the KEK E325 experiment \cite{Muto}, described in Section \ref{sec:phi}, and for the $\rho \rightarrow e^+e^- \cite{Nasseripour}, \mu^+ \mu^-$ \cite{Arnaldi} decays of the short-lived $\rho$ meson, not discussed in this review. 

Regarding the real part of the meson-nucleus potential, there seems to be consensus that the K$^+$ and K$^0$ meson experience a repulsive potential of about 20-40 MeV at normal nuclear matter density, while the $K^-$ and $\eta^\prime$ mesons feel an attractive potential, leading to a mass drop by about 60 and 40 MeV, respectively. For the $\eta$ and $\omega$ meson no evidence for strong mass modifications has been established. The $\phi$ meson is the only case where a mass shift and a broadening has been claimed, however, only in one experiment. It is of utmost importance to verify this experimental finding.

The determination of meson-nucleus potentials has paved the way for the search for meson-nucleus bound states, exclusively bound by the strong interaction. These states are of interest from the nuclear physics point of view as they correspond to highly excited nuclear states with excitation energies of the order of several hundred MeV up to 1 GeV. For hadron physics these states provide a unique laboratory for studying the properties of mesons and their possible modification in a strongly interacting environment at finite nuclear density in a quasi-static configuration.The formation and observation of such states is favoured by a strong real part of the meson-nucleus potential corresponding to a deep potential well. Furthermore, the modulus of the imaginary part of the potential should be much smaller than the modulus of the real part, leading to relatively narrow bound states which do not overlap and can more easily be  distinguished from background experimentally. Mesic states have been searched for the $K^-, \eta, \omega$ and $\eta^\prime$ meson. Experimental searches have so far not provided definitive and unambiguous experimental evidence for such states, although several indications have been reported. Most of the experiments suffer from low statistics or limited sensitivity because of insufficient background suppression. A new generation of experiments is required where the formation of the mesic state can be tagged and identified and its decay simultaneously measured via a characteristic decay mode. Current efforts focus on the $K^-, \eta$, $\eta^\prime$ and $\phi$ meson. Results from such  experiments planned at J-PARC and GSI/FAIR are eagerly awaited.

\section{Acknowledgement}
We would like to thank all collaborators and colleagues in the field of meson-nucleus interactions who have given helpful advice during the preparation of this review, in particular Wolfgang Cassing, Laura Fabbietti, Bengt Friman, Avraham Gal, Satoru Hirenzaki, Kenta Itahashi, Bernd Krusche, Pawel Moskal, Eulogio Oset, Laura Tolos and Wolfram Weise. Special thanks go to Ulrich Mosel for his close and long-time collaboration and many stimulating discussions on in-medium physics. Some of the data on $\eta,\eta^\prime$ and $\omega$ mesons, summarised in this review, were taken together with the colleagues from the CBELSA/TAPS collaboration in Bonn and the A2 collaboration in Mainz. These experiments would not have been possible without the support from the Deutsche Forschungsgemeinschaft within SFB/TR16. 

%\clearpage

\end{document}